\documentclass[12pt,a4paper,amsmath,amssymb,floatfix,twoside,openright]{article}

\usepackage{graphicx}
\usepackage{color}
\usepackage{amsmath}
\usepackage{amssymb}
\usepackage{pifont}
\usepackage{bm}
\usepackage{mathrsfs} 
\usepackage{xspace}
\usepackage{pstricks}
\usepackage{german}

\usepackage[english]{babel}
\usepackage{currvita}

\begin{document}

\newcommand{\captionfont}{\it}
\newcommand{\dzero}{D0\xspace} 
\newcommand{\runi}{Run~I\xspace}
\newcommand{\runii}{Run~II\xspace}
\newcommand{\SM}{Standard Model\xspace}
\newcommand{\alljets}{all-jets\xspace}

\newcommand{\W}{\ensuremath{W}\xspace}
\newcommand{\Z}{\ensuremath{Z}\xspace}
\newcommand{\Higgs}{\ensuremath{H}\xspace}
\newcommand{\e}{\ensuremath{e}\xspace}
\newcommand{\uquark}{\ensuremath{u}\xspace}
\newcommand{\antiuquark}{\ensuremath{\overline{u}}\xspace}
\newcommand{\dquark}{\ensuremath{d}\xspace}
\newcommand{\antidquark}{\ensuremath{\overline{d}}\xspace}
\newcommand{\squark}{\ensuremath{s}\xspace}
\newcommand{\antisquark}{\ensuremath{\overline{s}}\xspace}
\newcommand{\cquark}{\ensuremath{c}\xspace}
\newcommand{\anticquark}{\ensuremath{\overline{c}}\xspace}
\newcommand{\bquark}{\ensuremath{b}\xspace}
\newcommand{\antibquark}{\ensuremath{\overline{b}}\xspace}
\newcommand{\tquark}{\ensuremath{t}\xspace}
\newcommand{\antitquark}{\ensuremath{\overline{t}}\xspace}
\newcommand{\pizero}{\ensuremath{\pi^0}\xspace}
\newcommand{\Kshort}{\ensuremath{K^0_{\rm s}}\xspace}
\newcommand{\PDz}{\ensuremath{D^0}\xspace}
\newcommand{\PKm}{\ensuremath{K^-}\xspace}
\newcommand{\Ppip}{\ensuremath{\pi^+}\xspace}
\newcommand{\jpsi}{\ensuremath{J/\psi}\xspace}
\newcommand{\epem}{\ensuremath{\e^+ \e^-}\xspace}
\newcommand{\mupmum}{\ensuremath{\mu^+ \mu^-}\xspace}
\newcommand{\lplm}{\ensuremath{\ell^+ \ell^-}\xspace}
\newcommand{\tauptaum}{\ensuremath{\tau^+ \tau^-}\xspace}
\newcommand{\emu}{\ensuremath{e\mu}\xspace}

\newcommand{\glueglue}{\ensuremath{gg}\xspace}
\newcommand{\qqbar}{\ensuremath{q\bar{q}}\xspace}
\newcommand{\qqbarprime}{\ensuremath{q\bar{q}'}\xspace}
\newcommand{\ppbar}{\ensuremath{p\bar{p}}\xspace}
\newcommand{\ccbar}{\ensuremath{c\bar{c}}\xspace}
\newcommand{\bbbar}{\ensuremath{b\bar{b}}\xspace}
\newcommand{\ttbar}{\ensuremath{t\bar{t}}\xspace}
\newcommand{\ttj}{\ensuremath{t\bar{t}j}\xspace}
\newcommand{\wjets}{\ensuremath{W}\ensuremath{+}{\text jets}\xspace}
\newcommand{\wjetsshort}{\ensuremath{W}\!\ensuremath{+}\!{\text jets}\xspace}
\newcommand{\wljets}{\ensuremath{\W(\rightarrow \ell\nu)\ensuremath{+}{\text jets}}\xspace}
\newcommand{\wjjjj}{\ensuremath{\W jjjj}\xspace}
\newcommand{\wcjjj}{\ensuremath{\W cjjj}\xspace}
\newcommand{\wccjj}{\ensuremath{\W\ccbar jj}\xspace}
\newcommand{\wbbjj}{\ensuremath{\W\bbbar jj}\xspace}
\newcommand{\wlj}{\wjjjj}  
\newcommand{\whj}{\ensuremath{\W hf}\xspace}
\newcommand{\ww}{\ensuremath{\W\W}\xspace}
\newcommand{\wz}{\ensuremath{\W\Z}\xspace}
\newcommand{\zz}{\ensuremath{\Z\Z}\xspace}
\newcommand{\wfourpshort}{\ensuremath{\W\!+\!{\rm\text 4p}}\xspace}  
\newcommand{\Zgammastartwopshort}{\ensuremath{\Z/\gamma^*}\ensuremath{+}{\text 2p}\xspace}  
\newcommand{\WWtwopshort}{\ensuremath{\ww}\ensuremath{+}{\text 2p}\xspace}  
\newcommand{\Wthreepshort}{\ensuremath{\W}\ensuremath{+}{\text 3p}\xspace}  
\newcommand{\bbbarfourpshort}{\ensuremath{\bbbar\!+\!{\rm\text 4p}}\xspace}  
\newcommand{\sixp}{\ensuremath{\rm 6p}\xspace}
\newcommand{\ejets}{\ensuremath{e}\ensuremath{+}{\text jets}\xspace}
\newcommand{\mujets}{\ensuremath{\mu}\ensuremath{+}{\text jets}\xspace}
\newcommand{\ljets}{\ensuremath{\ell}\ensuremath{+}{\text jets}\xspace}
\newcommand{\etmiss}{\ensuremath{E \kern-0.6em\slash_{T}}\xspace}
\newcommand{\etmissx}{\ensuremath{E \kern-0.6em\slash_{x}}\xspace}
\newcommand{\etmissy}{\ensuremath{E \kern-0.6em\slash_{y}}\xspace}
\newcommand{\ptmiss}{\ensuremath{p \kern-0.45em\slash_{T}}\xspace}
\newcommand{\ptmissx}{\ensuremath{p \kern-0.45em\slash_{x}}\xspace}
\newcommand{\ptmissy}{\ensuremath{p \kern-0.45em\slash_{y}}\xspace}
\newcommand{\ptmissvec}{\ensuremath{\vec{p} \kern-0.45em\slash_{T}}\xspace}
\newcommand{\etuncl}{\ensuremath{E_{T}^{\rm uncl}}\xspace}
\newcommand{\ptuncl}{\ensuremath{p_{T}^{\rm uncl}}\xspace}
\newcommand{\cent}{\ensuremath{\mathcal C}\xspace}
\newcommand{\apla}{\ensuremath{\mathcal A}\xspace}
\newcommand{\spher}{\ensuremath{\mathcal S}\xspace}
\newcommand{\Htwoprime}{\ensuremath{H_{T2}'}\xspace}
\newcommand{\ktminp}{\ensuremath{K_{T,{\rm min}'}}\xspace}
\newcommand{\pt}{\ensuremath{p_{T}}\xspace}
\newcommand{\pz}{\ensuremath{p_{z}}\xspace}
\newcommand{\Et}{\ensuremath{E_{T}}\xspace}
\newcommand{\et}{\ensuremath{E_{T}}\xspace}
\newcommand{\Ht}{\ensuremath{H_{T}}\xspace}
\newcommand{\etrecoil}{\ensuremath{E_{T,{\rm recoil}}}\xspace}

\newcommand{\dsigma}{\ensuremath{{\rm d}\sigma}\xspace}
\newcommand{\sigmaP}{\ensuremath{\sigma_{\!P}}\xspace}
\newcommand{\dsigmaP}{\ensuremath{{\rm d}\sigmaP}\xspace}
\newcommand{\sigmaPobs}{\ensuremath{\sigmaP^{\rm obs}}\xspace}
\newcommand{\sigmaobs}{\ensuremath{\sigma^{\rm obs}}\xspace}
\newcommand{\pevt}{\ensuremath{L_{\rm evt}}\xspace}
\newcommand{\psgn}{\ensuremath{L_{\ttbar}}\xspace}
\newcommand{\pbkg}{\ensuremath{L_{\rm bkg}}\xspace}
\newcommand{\pbkgi}{\ensuremath{L_{{\rm bkg}\ i}}\xspace}
\newcommand{\Ntag}{\ensuremath{n_{\rm tag}}\xspace}
\newcommand{\psgnb}{\ensuremath{L_{\ttbar}^{\Ntag}}\xspace}
\newcommand{\pevtb}{\ensuremath{L_{\rm evt}^{\Ntag}}\xspace}
\newcommand{\mtop}{\ensuremath{m_{t}}\xspace}
\newcommand{\mtopME}{\ensuremath{m_{t}^{ME}}\xspace}
\newcommand{\mtopID}{\ensuremath{m_{t}^{ID}}\xspace}
\newcommand{\jes}{\ensuremath{JES}\xspace}
\newcommand{\jesME}{\ensuremath{\jes^{ME}}\xspace}
\newcommand{\jesID}{\ensuremath{\jes^{ID}}\xspace}
\newcommand{\mt}{\ensuremath{m_{t}}\xspace}
\newcommand{\mH}{\ensuremath{m_{H}}\xspace}
\newcommand{\mZ}{\ensuremath{m_{Z}}\xspace}
\newcommand{\yt}{\ensuremath{y_{t}}\xspace}
\newcommand{\ftop}{\ensuremath{f_{\ttbar}}\xspace}
\newcommand{\ftopb}{\ensuremath{f_{\ttbar}^{\Ntag}}\xspace}
\newcommand{\ftopbest}{\ensuremath{f_{\ttbar}^{\rm best}}\xspace}
\newcommand{\fttj}{\ensuremath{f_{\ttbar j}}\xspace}
\newcommand{\Gtop}{\ensuremath{\Gamma_{t}}\xspace}
\newcommand{\Gt}{\ensuremath{\Gamma_{t}}\xspace}
\newcommand{\mW}{\ensuremath{m_{W}}\xspace}
\newcommand{\MW}{\ensuremath{M_{W}}\xspace}
\newcommand{\GW}{\ensuremath{\Gamma_{W}}\xspace}
\newcommand{\mb}{\ensuremath{m_{b}}\xspace}
\newcommand{\esig}{\ensuremath{\epsilon_{\rm sig}}\xspace}
\newcommand{\eqcd}{\ensuremath{\epsilon_{\rm QCD}}\xspace}
\newcommand{\fPDF}{\ensuremath{f_{\rm PDF}}\xspace}
\newcommand{\fbarPDF}{\ensuremath{\bar{f}_{\rm PDF}}\xspace}
\newcommand{\msbar}{\ensuremath{\overline{\rm MS}}\xspace}

\newcommand{\JES}{\ensuremath{JES}\xspace}
\newcommand{\DeltaJES}{\ensuremath{\Delta_{\rm JES}}\xspace}
\newcommand{\mtopreco}{\ensuremath{\mtop^{\rm reco}}\xspace}
\newcommand{\mjjreco}{\ensuremath{m_{\rm jj}^{\rm reco}}\xspace}
\newcommand{\Lxy}{\ensuremath{L_{xy}}\xspace}

\newcommand{\subbellnu}{\ensuremath{{b\ell\nu}}\xspace}
\newcommand{\subellnu}{\ensuremath{{\ell\nu}}\xspace}
\newcommand{\subbdubar}{\ensuremath{{\overline{b}d\overline{u}}}\xspace}
\newcommand{\subdubar}{\ensuremath{{d\overline{u}}}\xspace}
\newcommand{\subbbar}{\ensuremath{\overline{b}}\xspace}
\newcommand{\subb}{\ensuremath{{b}}\xspace}

\newcommand{\thad}{\ensuremath{t_{had}}\xspace}
\newcommand{\Whad}{\ensuremath{W_{had}}\xspace}
\newcommand{\bhad}{\ensuremath{b_{had}}\xspace}
\newcommand{\tlep}{\ensuremath{t_{lep}}\xspace}
\newcommand{\Wlep}{\ensuremath{W_{lep}}\xspace}
\newcommand{\blep}{\ensuremath{b_{lep}}\xspace}

\newcommand{\pqone}{\ensuremath{p_{d}}\xspace}
\newcommand{\pqtwo}{\ensuremath{p_{\overline{u}}}\xspace}
\newcommand{\pvecqone}{\ensuremath{\vec{p}_{d}}\xspace}
\newcommand{\pvecqtwo}{\ensuremath{\vec{p}_{\overline{u}}}\xspace}
\newcommand{\pmagqone}{\ensuremath{|\pvecqone|}\xspace}
\newcommand{\pmagqtwo}{\ensuremath{|\pvecqtwo|}\xspace}
\newcommand{\mbellnu}{\ensuremath{m_{\subbellnu}}\xspace}
\newcommand{\mellnu}{\ensuremath{m_{\subellnu}}\xspace}
\newcommand{\chatbell}{\ensuremath{{\hat c}_{b\ell}}\xspace}
\newcommand{\mbdubar}{\ensuremath{m_{\subbdubar}}\xspace}
\newcommand{\mdubar}{\ensuremath{m_{\subdubar}}\xspace}
\newcommand{\chatbbard}{\ensuremath{{\hat c}_{\overline{b}d}}\xspace}
\newcommand{\mthad}{\mbdubar} 
\newcommand{\mtlep}{\mbellnu} 
\newcommand{\mtophad}{\mthad}
\newcommand{\mtoplep}{\mtlep}
\newcommand{\EWhad}{\ensuremath{E_{\subdubar}}\xspace}
\newcommand{\pvecWhad}{\ensuremath{\vec{p}_{\subdubar}}\xspace}
\newcommand{\pmagWhad}{\ensuremath{|\pvecWhad|}\xspace}
\newcommand{\mWhad}{\mdubar} 
\newcommand{\mWlep}{\mellnu} 
\newcommand{\ptoplep}{\ensuremath{p_{\subbellnu}}\xspace}
\newcommand{\pvectoplep}{\ensuremath{\vec{p}_{\subbellnu}}\xspace}
\newcommand{\pxtoplep}{\ensuremath{p^x_{\subbellnu}}\xspace}
\newcommand{\pytoplep}{\ensuremath{p^y_{\subbellnu}}\xspace}
\newcommand{\pztoplep}{\ensuremath{p^z_{\subbellnu}}\xspace}
\newcommand{\Etoplep}{\ensuremath{E_{\subbellnu}}\xspace}
\newcommand{\pxtophad}{\ensuremath{p^x_{\subbdubar}}\xspace}
\newcommand{\pytophad}{\ensuremath{p^y_{\subbdubar}}\xspace}
\newcommand{\ptmu}{\ensuremath{p_{T,\mu}}\xspace}
\newcommand{\qoverptmu}{\ensuremath{\left(q/\pt\right)_\mu}\xspace}
\newcommand{\qoverptmurec}{\ensuremath{\qoverptmu^{\rm rec}}\xspace}
\newcommand{\qoverptmugen}{\ensuremath{\qoverptmu^{\rm ass}}\xspace}
\newcommand{\qmu}{\ensuremath{q_\mu}\xspace}
\newcommand{\ptj}{\ensuremath{p_{T,j}}\xspace}
\newcommand{\etj}{\ensuremath{E_{T,j}}\xspace}
\newcommand{\pzbnu}{\ensuremath{p_{z,b\nu}}\xspace}

\newcommand{\DeltaR}{\ensuremath{\Delta{\cal R}}\xspace}

\newcommand{\pb}{\ensuremath{p_{b}}\xspace}
\newcommand{\pvecb}{\ensuremath{\vec{p}_{b}}\xspace}
\newcommand{\Elep}{\ensuremath{E_{\ell}}\xspace}
\newcommand{\plep}{\ensuremath{p_{\ell}}\xspace}
\newcommand{\pveclep}{\ensuremath{\vec{p}_{\ell}}\xspace}
\newcommand{\pxlep}{\ensuremath{p^x_{\ell}}\xspace}
\newcommand{\pylep}{\ensuremath{p^y_{\ell}}\xspace}
\newcommand{\pzlep}{\ensuremath{p^z_{\ell}}\xspace}
\newcommand{\Enu}{\ensuremath{E_{\nu}}\xspace}
\newcommand{\pnu}{\ensuremath{p_{\nu}}\xspace}
\newcommand{\pnux}{\ensuremath{p_{\nu,x}}\xspace}
\newcommand{\pnuy}{\ensuremath{p_{\nu,y}}\xspace}
\newcommand{\pnubar}{\ensuremath{p_{\overline{\nu}}}\xspace}
\newcommand{\pnubarx}{\ensuremath{p_{\overline{\nu},x}}\xspace}
\newcommand{\pnubary}{\ensuremath{p_{\overline{\nu},y}}\xspace}
\newcommand{\pblepnu}{\ensuremath{p_{\subb\nu}}\xspace}
\newcommand{\Eblepnu}{\ensuremath{E_{\subb\nu}}\xspace}
\newcommand{\pmagb}{\ensuremath{|\vec{p}_{b}|}\xspace}
\newcommand{\pvecblepnu}{\ensuremath{\vec{p}_{\subb\nu}}\xspace}
\newcommand{\pmagblepnu}{\ensuremath{|\pvecblepnu|}\xspace}
\newcommand{\alphablepblepnu}{\ensuremath{\alpha_{\subb,\subb\nu}}\xspace}
\newcommand{\Eblep}{\ensuremath{E_{\subb}}\xspace}
\newcommand{\pblep}{\ensuremath{p_{\subb}}\xspace}
\newcommand{\pvecblep}{\ensuremath{\vec{p}_{\subb}}\xspace}
\newcommand{\pmagblep}{\ensuremath{|\pvecblep|}\xspace}
\newcommand{\pbhad}{\ensuremath{p_{\subbbar}}\xspace}
\newcommand{\pvecbhad}{\ensuremath{\vec{p}_{\subbbar}}\xspace}
\newcommand{\pmagbhad}{\ensuremath{|\pvecbhad|}\xspace}
\newcommand{\pvecnu}{\ensuremath{\vec{p}_\nu}\xspace}
\newcommand{\pvecmu}{\ensuremath{\vec{p}_\mu}\xspace}
\newcommand{\pmagmu}{\ensuremath{|\pvecmu|}\xspace}
\newcommand{\thetamu}{\ensuremath{\theta_\mu}\xspace}
\newcommand{\pznu}{\ensuremath{p_{\nu,z}}\xspace}
\newcommand{\alphabhadWhad}{\ensuremath{\alpha_{\subbbar, \subdubar}}\xspace}
\newcommand{\alphaqoneqtwo}{\ensuremath{\alpha_{d, \overline{u}}}\xspace}
\newcommand{\alphaqonebhad}{\ensuremath{\alpha_{d, \subbbar}}\xspace}
\newcommand{\alphaqtwobhad}{\ensuremath{\alpha_{\overline{u}, \subbbar}}\xspace}
\newcommand{\Ebhad}{\ensuremath{E_{\subbbar}}\xspace}
\newcommand{\Eqone}{\ensuremath{E_{d}}\xspace}
\newcommand{\Eqtwo}{\ensuremath{E_{\overline{u}}}\xspace}
\newcommand{\pxnu}{\ensuremath{p^x_{\nu}}\xspace}
\newcommand{\pynu}{\ensuremath{p^y_{\nu}}\xspace}
\newcommand{\uxnu}{\ensuremath{u^x_{\nu}}\xspace}
\newcommand{\uynu}{\ensuremath{u^y_{\nu}}\xspace}
\newcommand{\uvecblep}{\ensuremath{\vec{u}_{\subb}}\xspace}
\newcommand{\uxblep}{\ensuremath{u^x_{\subb}}\xspace}
\newcommand{\uyblep}{\ensuremath{u^y_{\subb}}\xspace}
\newcommand{\uzblep}{\ensuremath{u^z_{\subb}}\xspace}
\newcommand{\pxblep}{\ensuremath{p^x_{\subb}}\xspace}
\newcommand{\pyblep}{\ensuremath{p^y_{\subb}}\xspace}
\newcommand{\pzblep}{\ensuremath{p^z_{\subb}}\xspace}
\newcommand{\Mhad}{\ensuremath{M_{had}}\xspace}
\newcommand{\subbellnubdubar}{\ensuremath{{b\ell\nu}\overline{b}d\overline{u}}\xspace}
\newcommand{\Ettbar}{\ensuremath{E_{\subbellnubdubar}}\xspace}
\newcommand{\pzttbar}{\ensuremath{p^z_{\subbellnubdubar}}\xspace}

\newcommand{\qone}{\ensuremath{\xi_1}\xspace}
\newcommand{\qtwo}{\ensuremath{\xi_2}\xspace}
\newcommand{\qgeneral}{\ensuremath{\xi}\xspace}
\newcommand{\xpdf}{\ensuremath{x}\xspace}

\newcommand{\ca}{\ensuremath{\rm ca}\xspace}
\newcommand{\wa}{\ensuremath{\rm wa}\xspace}
\newcommand{\fca}{\ensuremath{f_{\ca}}\xspace}
\newcommand{\Sca}{\ensuremath{S_{\ca}}\xspace}
\newcommand{\Swa}{\ensuremath{S_{\wa}}\xspace}

\newcommand{\lambdaqcd}{\ensuremath{\Lambda_{\rm QCD}}\xspace}

\newcommand{\madgraph}{{\sc madgraph}\xspace}
\newcommand{\alpgen}{{\sc alpgen}\xspace}
\newcommand{\pythia}{{\sc pythia}\xspace}
\newcommand{\herwig}{{\sc herwig}\xspace}
\newcommand{\vecbos}{{\sc vecbos}\xspace}
\newcommand{\evtgen}{{\sc evtgen}\xspace}
\newcommand{\QQ}{{\sc qq}\xspace}
\newcommand{\tauola}{{\sc tauola}\xspace}
\newcommand{\minuit}{{\sc minuit}\xspace}
\newcommand{\vegas}{{\sc vegas}\xspace}
\newcommand{\gnu}{{\sc gnu}\xspace}
\newcommand{\gsl}{{\sc gsl}\xspace}
\newcommand{\geant}{{\sc geant}\xspace}
\newcommand{\cteqfivel}{{\sc cteq5l}\xspace}
\newcommand{\factorizationscalesquared}{\ensuremath{\mu_F^2}\xspace}
\newcommand{\renormalizationscalesquared}{\ensuremath{\mu_R^2}\xspace}
\newcommand{\processscalesquared}{\ensuremath{Q^2}\xspace}

\newcommand{\m}  {\ensuremath{\mathrm{m}}\xspace}
\newcommand{\cm} {\ensuremath{\mathrm{cm}}\xspace}
\newcommand{\mm} {\ensuremath{\mathrm{mm}}\xspace}
\newcommand{\mum}{\ensuremath{\mathrm{\mu m}}\xspace}
\newcommand{\ipb}{\,\ensuremath{\rm pb^{-1}}\xspace}
\newcommand{\ifb}{\,\ensuremath{\rm fb^{-1}}\xspace}
\newcommand{\keV}{\,\ensuremath{\mathrm{ke\kern-0.1em V}}\xspace}
\newcommand{\MeV}{\,\ensuremath{\mathrm{Me\kern-0.1em V}}\xspace}
\newcommand{\GeV}{\,\ensuremath{\mathrm{Ge\kern-0.1em V}}\xspace}
\newcommand{\GeVc}{\,\ensuremath{\mathrm{Ge\kern-0.1em V}}\xspace}
\newcommand{\GeVcc}{\,\ensuremath{\mathrm{Ge\kern-0.1em V}}\xspace}
\newcommand{\TeV}{\,\ensuremath{\mathrm{Te\kern-0.1em V}}\xspace}
\newcommand{\GHz}{\ensuremath{\mathrm{GHz}}\xspace}
\newcommand{\kHz}{\ensuremath{\mathrm{kHz}}\xspace}
\newcommand{\Hz}{\ensuremath{\mathrm{Hz}}\xspace}

\newcommand{\Eref}[1]{(\ref{#1})}
\newcommand{\Fref}[1]{Figure~\ref{#1}}
\newcommand{\Rref}[1]{Reference~\cite{#1}}
\newcommand{\Sref}[1]{Section~\ref{#1}}
\newcommand{\Tref}[1]{Table~\ref{#1}}
\newcommand{\Aref}[1]{Appendix~\ref{#1}}

\newcommand{\pandm}[2]{^{+#1}_{-#2}}
\newcommand{\pandmenspacel}[2]{^{+#1}_{\phantom{0}-#2}}
\newcommand{\pandmenspaceu}[2]{^{\phantom{0}+#1}_{-#2}}
\newcommand{\pandmenspaceul}[2]{^{\phantom{0}+#1}_{\phantom{0}-#2}}

\newcommand{\WAerrstat}          {\ensuremath{\pm 1.2}\xspace} 
\newcommand{\WAerrjes}           {\ensuremath{\pm 1.4}\xspace} 
\newcommand{\WAerrsignal}        {\ensuremath{\pm 0.9}\xspace} 
\newcommand{\WAerrbackground}    {\ensuremath{\pm 0.3}\xspace} 
\newcommand{\WAerrfit}           {\ensuremath{\pm 0.3}\xspace} 
\newcommand{\WAerrMC}            {\ensuremath{\pm 0.3}\xspace} 
\newcommand{\WAerrUNMI}          {\ensuremath{\pm 0.1}\xspace} 
\newcommand{\WAerrphysics}       {\ensuremath{\pm 1.0}\xspace} 
\newcommand{\WAerrdetector}      {\ensuremath{\pm 1.4}\xspace} 
\newcommand{\WAerrtotal}         {\ensuremath{\pm 2.1}\xspace} 

\newcommand{\MEberrstatnojeslong}{\ensuremath{\pm2.5\enspace}\xspace} 
\newcommand{\MEberrjes}          {\ensuremath{^{+3.2}_{-3.7}}\xspace} 
\newcommand{\MEberrjeslong}      {\ensuremath{+3.2~-\hspace{-0.65ex}3.7}\xspace} 
\newcommand{\MEbjeserrstat}    {\ensuremath{^{+0.035}_{-0.032}}\xspace} 
\newcommand{\MEbjeserrstatlong}{\ensuremath{+0.035~-\hspace{-0.65ex}0.032}\xspace} 
\newcommand{\MEberrsgnmod} {\ensuremath{\pm0.46}\xspace} 
\newcommand{\MEberrbkgmod} {\ensuremath{\pm0.40}\xspace} 
\newcommand{\MEberrpdf}    {\ensuremath{+0.16~-\hspace{-0.65ex}0.39}\xspace} 
\newcommand{\MEberrbjes}   {\ensuremath{\pm 0.56}\xspace} 
\newcommand{\MEberrbclepbr}{\ensuremath{\pm0.05}\xspace} 
\newcommand{\MEberrjespt}  {\ensuremath{\pm 0.19}\xspace} 
\newcommand{\MEberrbresp}  {\ensuremath{+0.63~-\hspace{-0.65ex}1.43}\xspace} 
\newcommand{\MEberrtrg}    {\ensuremath{+0.08~-\hspace{-0.65ex}0.13}\xspace} 
\newcommand{\MEberrbtagging}{\ensuremath{\pm 0.24}\xspace}
\newcommand{\MEberrftop}   {\ensuremath{\pm0.15}\xspace} 
\newcommand{\MEberrqcd}    {\ensuremath{\pm0.29}\xspace} 
\newcommand{\MEberrmccalib}{\ensuremath{\pm 0.48}\xspace} 
\newcommand{\MEberrsyst}   {\ensuremath{^{+1.2}_{-1.8}}\xspace} 
\newcommand{\MEberrtotal}  {\ensuremath{^{+4.3}_{-4.9}}\xspace} 
\newcommand{\MEberrsystlong}   {\ensuremath{+1.2~-\hspace{-0.65ex}1.8}\xspace} 
\newcommand{\MEberrtotallong}  {\ensuremath{+4.3~-\hspace{-0.65ex}4.9}\xspace} 

\pagenumbering{Roman}
\clearpage
\begin{titlepage}
\begin{center}{\Huge\bf Precision Measurements\vspace{1ex}\\ of the Top Quark Mass}
\end{center}

\vspace{3ex}

\begin{center}Frank Fiedler\\
Ludwig-Maximilians-Universit\"at M\"unchen
\end{center}

\begin{center}Habilitation thesis\\
28 February 2007
\end{center}

\vspace{12ex}

\begin{abstract}
The experimental status of measurements of the top quark mass is 
reviewed.
After an introduction to the definition of the top quark mass and 
the production and decay of top quarks, an in-depth 
comparison of the analysis techniques used in top quark mass
measurements is presented, and 
the systematic uncertainties on the top quark mass are discussed
in detail.
This allows the reader to understand the experimental issues in the 
measurements, their limitations, and potential future improvements,
and to comprehend the inputs to and formation of the current world average 
value of the top quark mass.
Its interpretation within the 
frameworks of the Standard Model and of models beyond it are presented.
Finally, future prospects for measurements of the top quark mass
and their impact on our understanding of particle physics are
outlined.
\end{abstract}
\end{titlepage}

\clearpage
\begin{titlepage}
\begin{flushright}
F\"ur Grit, Lukas und Julia
\end{flushright}
\end{titlepage}

\clearpage
\pagenumbering{Roman}
\tableofcontents 
\clearpage
\pagestyle{headings}
\pagenumbering{arabic}
\section{Introduction}
\label{introduction.sec}
\begin{center}
\begin{tabular}{p{15cm}}
{\it The top quark is the heaviest known elementary particle.  
  While it has not yet been possible to answer the question {\em why}
  its mass is so large, the precise measurements of the top quark mass
  that have become available since its discovery have already greatly
  improved constraints on our picture of nature; for example they have
  made predictions of the mass of the as yet undiscovered Higgs boson
  possible.
  This report first defines the top quark mass and then describes in 
  detail the techniques used to measure it.
  This is followed by a description of the systematic uncertainties. 
  The current world average value of the top quark mass is presented,
  and the constraints it provides on elementary particle physics
  models are shown.
  Finally, potential future improvements of the precision on the top quark mass
  are outlined.}
\end{tabular}
\end{center}

Of all known elementary fermions, the top quark has by far the largest
mass.
This renders the top quark unique from a theoretical standpoint:
The top quark Yukawa coupling is close to unity, which may be a hint
that the top quark mass is related with electroweak symmetry breaking.
Via loop contributions,
the masses of the \W boson, the top quark, and the yet undiscovered Higgs boson are
interrelated so that the Higgs mass (which is not predicted in the
Standard Model of elementary particle physics) may be
constrained from precise measurements of the \W boson and top quark
masses~\cite{bib-Zbible,bib-LEPEWWG}.
Experimentally, on the other hand, the top quark is unique as it is the only
quark that does not hadronize because its lifetime is too short~\cite{bib-pdg}; it is
therefore possible to directly measure the properties of the quark
instead of a hadron containing the quark of interest.

Long before the discovery of the top quark, its 
existence as the up-type partner of the bottom quark
had been postulated within the Standard Model,
and its mass could be predicted from precision 
measurements of electroweak observables.
Currently, indirect constraints within the \SM yield a top quark mass
value of 
$\mtop=178\,\pandmenspacel{12}{9}\,\GeV$~\cite{bib-LEPEWWG}\footnote{Throughout
this report, the convention $\hbar=1$, $c=1$ is followed.  Charge conjugate
processes are included implicitly.  Top quark masses quoted are pole
masses unless noted otherwise -- see Section~\ref{theory.mtopdef.sec}
for a definition of the pole mass.}.
The top quark was finally discovered~\cite{bib-topdiscovery} by the CDF and
\dzero experiments in proton-antiproton collisions at
the Fermilab Tevatron Collider.
Since then, measurements of the top quark mass have been performed
both during \runi of the Tevatron in the 1990s at a proton-antiproton
center-of-mass energy of $\sqrt{s}=1.8\,\TeV$~\cite{bib-topmassruni} 
and during the ongoing
\runii at an increased center-of-mass energy of $1.96\,\TeV$ and with
larger data sets~\cite{bib-topmassrunii,bib-CDFupdates,bib-Dzeroupdates}.
Their average value of 
$\mtop = 171.4 \pm 2.1 \,\GeV$~\cite{bib-TEVEWWG}
is in striking agreement with the indirect prediction, thus
supporting the \SM as the theory of nature.
Innovative measurement techniques have made this precision possible, 
which already surpasses the original expectations for Tevatron
\runii~\cite{bib-CDFruniiTDR}.
Within the \SM, a value of the Higgs boson mass close to the current
lower exclusion limit is favored~\cite{bib-LEPEWWG}.

The Tevatron experiments have performed many more measurements of top
quarks.
The total cross section for top-antitop pair
production~\cite{bib-ttbarxsruni,bib-ttbarxsrunii} is consistent
with the predictions from QCD~\cite{bib-xsprediction}, using the above top
quark mass as input.
No evidence for effects beyond those predicted in the \SM has been
found in production and decay of top 
quarks~\cite{bib-searchesfornonsmeffects-runi,bib-Whelicity,bib-othersearchesfornonsmeffects}.
A recent review of top quark measurements can be found in~\cite{bib-arnulf}.

To date, the Tevatron Collider still provides
the only possibility to produce top quarks.
In the near future, the LHC proton-proton collider will start
operation, which is expected to provide much larger samples of top
quark events.
While the measurement of the top quark mass will be subject to very
similar systematic uncertainties, it can be assumed that the large
data samples will allow for a further reduction of the error.
However, only a linear \epem collider scanning the \ttbar production
threshold will allow for an order of magnitude improvement of the
precision.

This paper provides an overview of current measurements of the top
quark mass at hadron colliders, focusing on the Tevatron \runii
results.
The purpose of this document is twofold:
\begin{list}{$\bullet$}{\setlength{\itemsep}{0.5ex}
                        \setlength{\parsep}{0ex}
                        \setlength{\topsep}{0ex}}
\item 
to review the current status of top quark
mass measurements, compare the assumptions made in the
various analyses, discuss the limiting systematic
uncertainties together with potential future improvements, and 
to give an overview of the interpretation of the measurements; and 
\item
to provide a detailed description of the measurement techniques
developed and used so far for the measurement of the top quark mass,
not only to complement the information on the physics results, but
also as a reference for the development of future measurements
(of the top quark or other particles).
\end{list}

The general structure of the paper is as follows:
Section~\ref{theory.sec} gives a
brief summary of definitions of the top quark mass and discusses the
relevance of measurements of the top quark mass for elementary
particle physics.
Section~\ref{toprodec.sec} then outlines the production mechanisms for
top quarks at hadron colliders and the event characteristics.
The steps needed to obtain a set of data events with which to measure
the top quark mass are described in Sections~\ref{reco.sec}
(reconstruction of top quark events) and~\ref{detcalib.sec} (detector
calibration).

An overview of the different techniques (template, Matrix Element,
and Ideogram methods) to determine the top quark mass
from such a set of calibrated data events is given in
Section~\ref{methods.sec}.
The principle of template based measurements and examples 
using different event
topologies are discussed in 
Section~\ref{templatemeasurements.sec}.
Section~\ref{memeasurements.sec} gives an in-depth description of 
the Matrix Element method, and
the Ideogram method is described in Section~\ref{idmeasurements.sec}.
The fitting procedure to determine the top quark mass is discussed in
Section~\ref{massfit.sec}.

The current world average of the top quark mass is already
dominated by systematic uncertainties.
The different sources of systematic uncertainties and the estimation 
of the size of the corresponding effects are discussed in
Section~\ref{systuncs.sec}.
Section~\ref{resinterp.sec} then summarizes the current knowledge
of the top quark mass and the interpretation of these results 
and outlines possible future
developments. 
Section~\ref{conclusion.sec} summarizes and concludes the paper.

\clearpage
\section{Definition and Relevance of the Top Quark Mass}
\label{theory.sec}
\begin{center}
\begin{tabular}{p{15cm}}
{\it The definition of the mass of a particle may seem trivial.  
  However,
  when used in conjunction with a quark it is in fact by no means
  obvious how ``mass'' should best be defined.
  This section introduces different possible definitions
  and states in general terms which kind of
  measurement determines which mass.
  The section then outlines how the precise knowledge of the top quark
  mass improves our understanding of elementary particles and the
  description of their interactions within the Standard Model of
  particle physics.}
\end{tabular}
\end{center}

\subsection{Definitions of the Top Quark Mass and Measurement Concepts}
\label{theory.mtopdef.sec}
In general, ``the'' mass $m$ of a particle is only defined within a
theory or model in which it occurs as a parameter.
The mass of a particle can then be determined through a
comparison of measurements with the predictions of the theory
(the validity of the mass value obtained is then restricted to this
particular theory).
While it is straightforward to find a suitable definition of the mass
of a color-neutral particle, there are several possibilities for
defining the mass of a (color-charged) quark.
This section illustrates the underlying concepts and 
defines how the word mass is used
in conjunction with the top quark in the remainder of this report.
See Reference~\cite{bib-pdg} for more detailed
reviews of Quantum Chromodynamics (QCD), quark masses, and 
top quark physics.

For each quark, a mass parameter is introduced in the 
QCD Lagrangian.
(In the Standard Model, the value of this parameter is proportional
to the Yukawa coupling of the quark to the Higgs boson.)
The value depends on the renormalization scheme and the
renormalization scale $\mu$.
At high energies, the QCD coupling constant $\alpha_s$ is small, and
observables are typically calculated in perturbation theory, commonly 
applying the \msbar renormalization scheme.
(The \msbar scheme is used by the Particle Data Group to report all
quark masses except the top quark mass.)

For an observable (i.e., non-colored) particle, the position of the 
pole in the propagator defines the mass.
In perturbative QCD, this pole mass can also be used as a definition
of quark masses.
However, the pole mass cannot be used to arbitrarily high accuracy:
Because of confinement (i.e., because of non-perturbative effects in
QCD), the full quark propagator does not have a pole.  
This is true even for the top quark which does not hadronize before
decaying.
The general argument is presented in a very intuitive way in
Reference~\cite{bib-scottwillenbrock}.
The relation between the pole mass and \msbar mass is known to three 
loops, see~\cite{bib-pdg} and references therein, but there
necessarily remains an uncertainty of order \lambdaqcd
in the pole mass~\cite{bib-scottwillenbrock}.

Different definitions of the pole mass are used.
An unstable particle can generally be described by a Breit-Wigner
resonance~\cite{bib-pythia}
\begin{equation}
  \label{breitwigner_constantwidth.eqn}
    f(s) \sim \frac{s}{\pi} \,
                     \frac{ \tilde{m}\tilde{\Gamma} }
                          {   (s-\tilde{m}^2)^2
                            + (\tilde{m}\tilde{\Gamma})^2 }
  \ ,
\end{equation}
where $s=p^2$ is the squared four-momentum of one particle, 
and the properties of the resonance are described by a constant
width $\tilde{\Gamma}$ and the corresponding (pole) mass $\tilde{m}$.
It is possible to absorb higher-order corrections into the pole mass
definition.
For example, for the experimental determination of the $Z$ boson mass 
an $s$-dependent width is used to describe the resonance, with
the term $(\tilde{m}\tilde{\Gamma})$ replaced by
$(s\Gamma/m)$.
To accomodate the same experimental data, different numerical values
of the mass parameter are needed in the two approaches; for the $Z$
boson the relation between the two parameter values is given 
by~\cite{bib-LEPEWWG}
\begin{equation}
  \label{zmassdefinitiondifference.eqn}
    m_Z = \tilde{m}_Z \sqrt{ 1 + \frac{\tilde{\Gamma}_Z^2}{\tilde{m}_Z^2} }
        \approx \tilde{m}_Z + 34.20\,\MeV
  \ .
\end{equation}

Similarly, different definitions are possible for the top quark mass.
Measurements of the top quark mass at a hadron collider rely 
on comparisons of the data with simulated events, and thus it is important
to state the definition adopted in the simulation which is used in the 
measurement.
The two simulation programs used most commonly in current measurements are
\alpgen~\cite{bib-alpgen}, which uses
fixed widths in propagators, and \pythia~\cite{bib-pythia}, where
a factor $(1-2.5\,\alpha_s(s)/\pi)$ is included
for top quarks to approximate loop corrections.
The energy dependence of $\alpha_s$ in principle introduces a
difference between the two definitions; this is however negligible
compared to the intrinsic uncertainty of 
order \lambdaqcd.

To determine the top quark mass defined in any given scheme, 
one has to find observables
measurements of which can be compared to theory predictions which in
turn depend on this top quark mass.
In practice, there are three fundamentally different approaches:
\begin{list}{$\bullet$}{\setlength{\itemsep}{0.5ex}
                        \setlength{\parsep}{0ex}
                        \setlength{\topsep}{0ex}}
\item
{\bf Indirect constraints from electroweak measurements:}
Even before the first direct observation of top quarks, indirect constraints
were obtained from fits of the Standard Model prediction as a function
of the top quark mass to precision measurements of electroweak
observables~\cite{bib-Zbible,bib-LEPEWWG}.
This method of course has the drawback that it is not an actual
discovery of the top quark, and that the mass value is only valid
within the Standard Model (or in other theories whose predictions do
not significantly differ from those of the \SM).

\item
{\bf Reconstruction of top quark decay products:}
Today and in the near future, top quarks are and will be produced at
the hadron colliders Tevatron and LHC, allowing for a
direct measurement of the top quark mass from
the reconstructed decay products.
The momenta of the decay products are related according to
\begin{equation}
  \label{massfromfourmomenta.eqn}
    \mt(i)^2
  =
    p_{t}(i)^2
  =
    \left( \sum_{j} p_j(i) \right)^2
  \ ,
\end{equation}
where $p$ denotes the 4-momentum of a particle and the sum is over all 
decay products $j$ of the top quark $t$ in a specific event $i$.
A measurement based on the momenta of the decay products 
thus ideally corresponds to a measurement of
the pole mass since the squared sum of four-momenta as given in 
Equation~(\ref{massfromfourmomenta.eqn}) 
enters in the denominator 
\begin{equation}
  \label{propagatorterm.eqn}
    {p_t^2 - \mt^2 + i\mt\Gt}
\end{equation}
of the propagator term.
Individual measurements differ in how an observable that is related
with the top quark mass is constructed from the measured decay
products, and the situation is more complicated for measurements 
relying on complex techniques like
the Matrix Element or Ideogram methods discussed in 
Sections~\ref{memeasurements.sec} and~\ref{idmeasurements.sec}.
In the most precise measurements in the \ljets channel, 
the experimental information comes to a very large extent 
from the invariant mass of the reconstructed top quark decay products;
thus the measured value can be expected to correspond (most closely) to
the pole mass, but this issue has not yet been studied in detail.

In contrast to the other quarks (up, down, charm, strange, and
bottom), the top quark decays before forming hadrons~\cite{bib-pdg}.
This makes a direct measurement of the top {\em quark} mass (instead of
a hadron mass) possible; hadronization only affects the decay 
products of the top quark and leads to jet formation, 
cf.\ Section~\ref{toprodec.topeventtopologies.sec}.

Top quark mass measurements based on the decay products 
are valid not only within the Standard Model but in 
any model which does not introduce significant changes to those
features of top quark production and decay that are used 
in the measurement.
However, the results are subject to an intrinsic uncertainty of 
order \lambdaqcd as mentioned above.

\item
{\bf\boldmath\ttbar threshold scan:}
In the long-term future, it will be desirable to 
determine the top quark mass based on a definition that is not 
subject to the uncertainty on the pole mass, even though 
the current combined experimental 
uncertainty is almost a magnitude larger.
The best-known example is the measurement of the cross section 
for top-antitop pair production near threshold at a future \epem
collider.
This experimentally very clean measurement could be related to 
theory predictions that are calculated as a function of a top quark mass
parameter that can be translated into the \msbar mass with much
smaller uncertainty~\cite{bib-topmassdefinitionsforilc}.
The principle of the measurement is analogous to the determination 
of the \W boson mass from the measurement of the \ww production cross
section at threshold at LEP2.
\end{list}

This report focuses on the techniques, current results, and prospects of
top quark mass measurements at the Tevatron, where the mass is
reconstructed from the properties of the decay products.
Consequently, the pole mass
definition is implicitly assumed throughout the remainder of this
report unless noted otherwise.
This is consistent with the conventions of the Tevatron Electroweak Working
Group~\cite{bib-TEVEWWG} and the Particle Data Group~\cite{bib-pdg}.

\subsection{Relevance of the Top Quark Mass within the Standard Model}
\label{theory.relevance.sec}
\begin{figure}
\begin{center}
\includegraphics[width=0.3\textwidth]{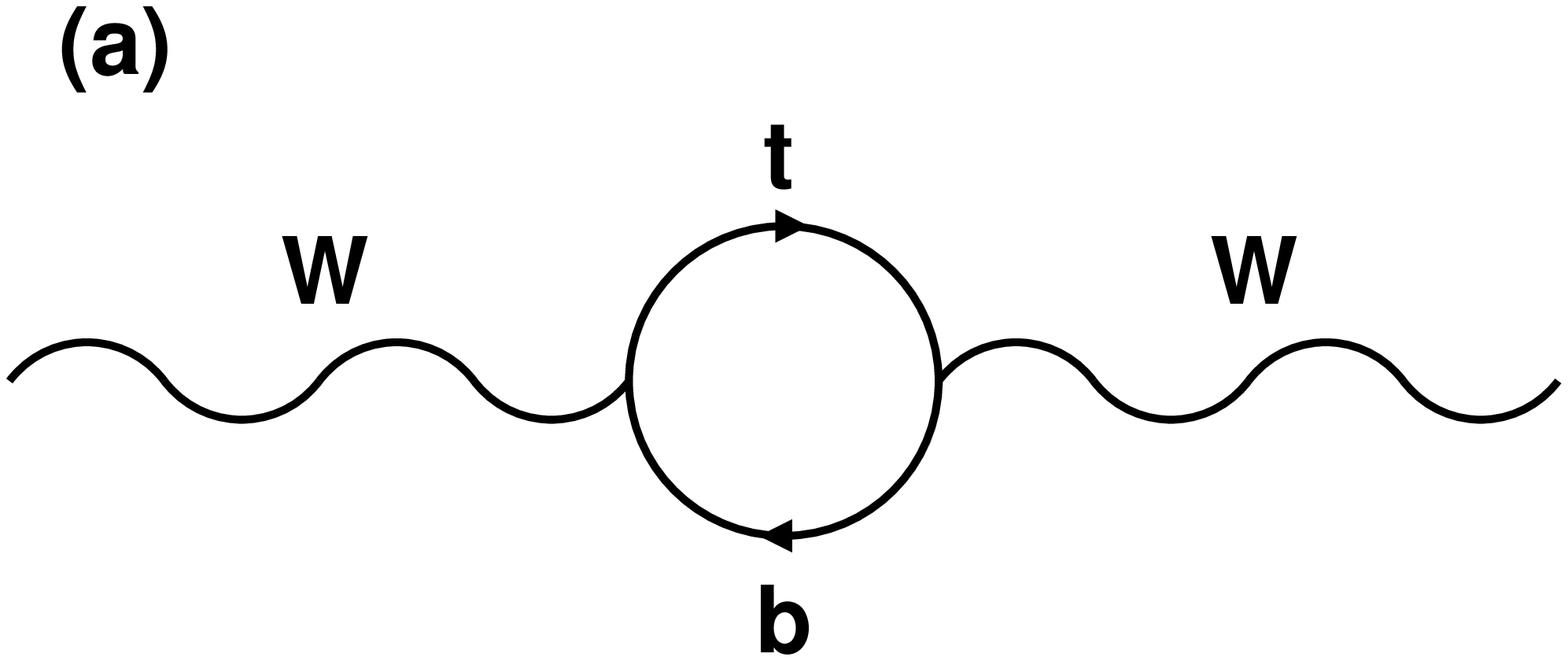}
\includegraphics[width=0.3\textwidth]{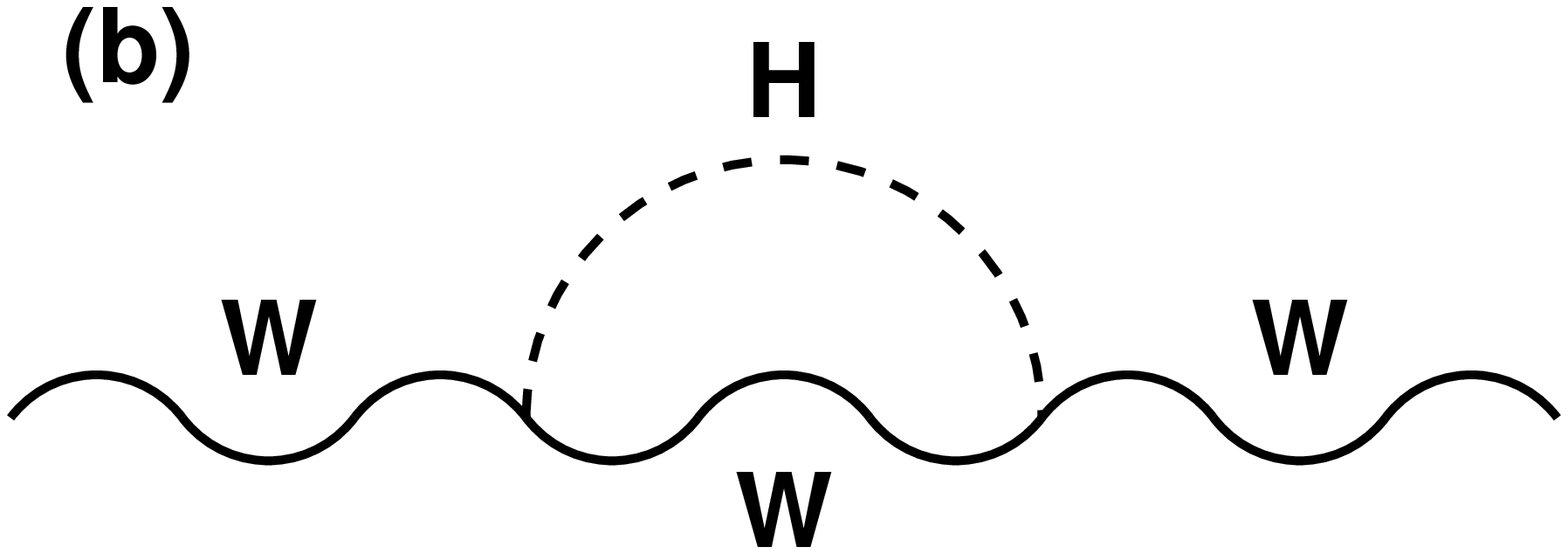}
\includegraphics[width=0.3\textwidth]{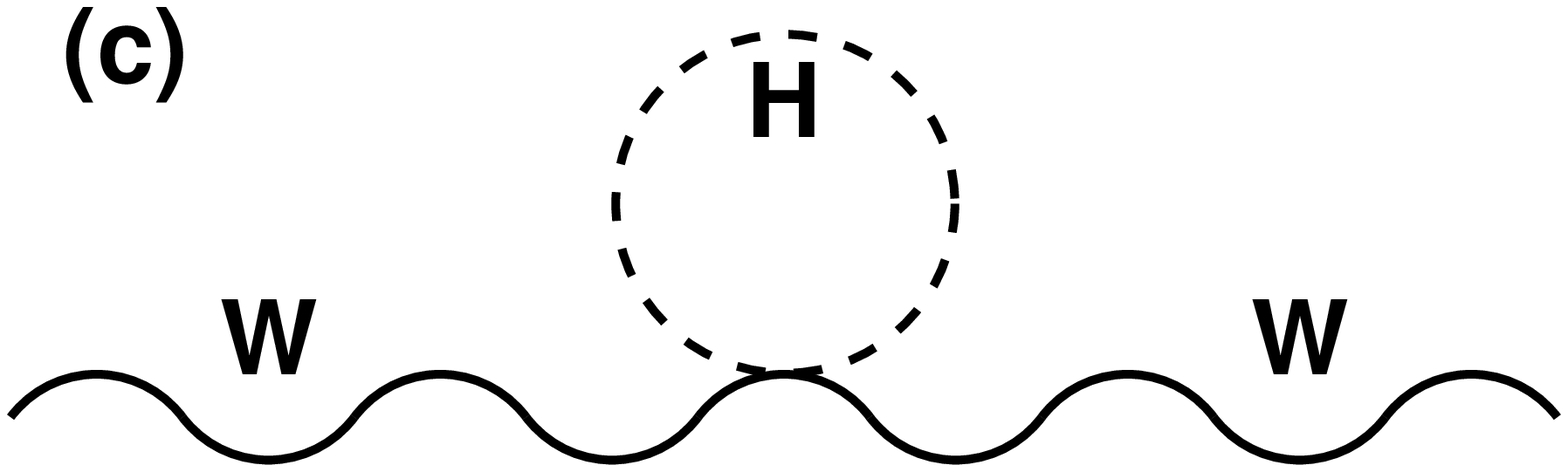}
\caption{\captionfont\label{wpropagator.fig}Feynman diagrams of loop
  processes that lead to a dependence of the \W boson propagator on
  (a) the top quark mass and (b, c) the Higgs boson mass.}
\end{center}
\end{figure}

In perturbation theory, predictions for observables receive contributions
from loop diagrams, where particles contribute even
if they are too massive to be produced on shell.
The size of these corrections to leading-order predictions depends on 
the values of the masses of the particles in the loops.
Of particular importance for Standard Model fits is the dependence of 
the \W boson mass on the top quark and Higgs boson masses.
The lowest-order diagram leading to the dependence on the top quark
mass is shown in Figure~\ref{wpropagator.fig}(a), those resulting in the Higgs
mass dependence in Figures~\ref{wpropagator.fig}(b) and (c).
The corrections that arise from these diagrams are quadratic in the top
quark mass,
but only logarithmic in
the Higgs boson mass
(yielding a much weaker dependence).

Since the dependence on the Higgs boson mass is weak, measurements
of the \W mass (and of other electroweak observables) lead to indirect
constraints on the top quark mass.
This led to predictions of the mass of the top quark before its actual
discovery, as already outlined in Section~\ref{theory.mtopdef.sec}.
Also, precise measurements of both the
\W boson and top quark masses result in constraints on the Standard
Model Higgs boson mass.
In the following sections, the experimental 
measurements of the top quark mass are
discussed in detail.
The interpretation of the current results within the Standard Model
(and models beyond the \SM) is then further
discussed in Section~\ref{resinterp.interp.sec}.

\clearpage
\section[Top Quark Production and Decay at Hadron Colliders]{Top\,Quark\,Production\,and\,Decay\,at\,Hadron\,Colliders}
\label{toprodec.sec}
\begin{center}
\begin{tabular}{p{15cm}}
{\it Top quarks can be studied best when produced on shell in
  a collider experiment.
  This is currently only possible at the Fermilab Tevatron
  proton-antiproton collider near Chicago.
  In the near future, the LHC proton-proton collider at CERN near Geneva
  will produce large numbers of top quarks.
  This section describes the properties of events produced in
  reactions involving top quark decays.}
\end{tabular}
\end{center}

In this section, the mechanisms for top quark production in
hadron collisions (\ppbar or $pp$) are described.  
Events containing a \ttbar pair are used to measure the top
quark mass, and thus the different topologies of these events, which
depend on the top quark decays, are discussed.
The relevant background processes are also described.

\subsection{Top Quark Production}
\label{toprodec.topproduction.sec}
\begin{figure}
\begin{center}
\includegraphics[width=0.5\textwidth]{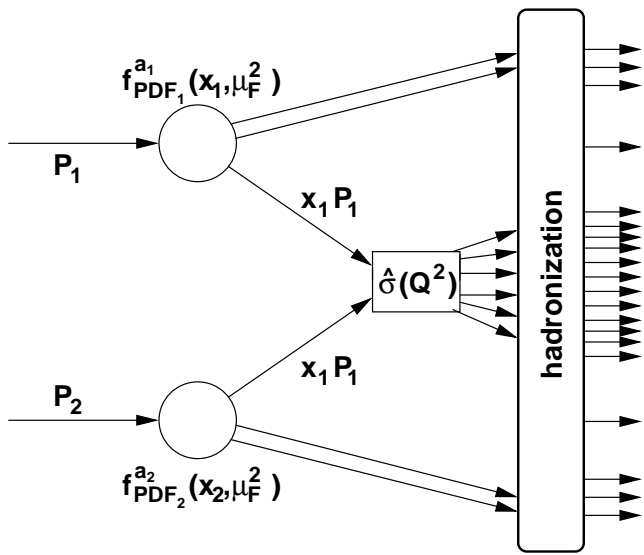}
\caption{\captionfont\label{factorization.fig}Schematic drawing illustrating the
  concept of factorization.  
  Shown is a collision of two hadrons
  leading to a hard-scattering process at a scale \processscalesquared.
  This hard interaction is initiated by two partons of momenta
  $\xpdf_1 P_1$ and $\xpdf_2 P_2$, where $P_1$ and $P_2$ are the
  momenta of the colliding hadrons.
  The partonic cross section $\hat\sigma$ of the hard interaction can
  be calculated 
  perturbatively, based on the renormalization and factorization scales
  \renormalizationscalesquared and 
  \factorizationscalesquared.
  The factorization scale is also used to evaluate the 
  parton distribution functions \fPDF, which parametrize 
  the probabilities to find the partons $a_1$ and $a_2$ inside the 
  colliding hadrons.
  If the hard interaction involves the production of top quarks, their
  decays are included in its description, since the top quark lifetime
  is so short that no top hadrons are formed.
  The observable final-state particles are then formed in a
  hadronization process which again cannot be calculated
  perturbatively, but is independent of the hard interaction.}
\end{center}
\end{figure}

Because of the large top quark mass, high energies are required to
produce top quarks, and the production processes (including those
proceeding via the strong interaction) can be described
in perturbation theory.
The internal
structure of the colliding hadrons is resolved, and top quarks are
thus produced in a hard-scattering process of two constituent partons
(quarks/antiquarks or gluons) inside the hadrons.
The description of the reaction factorizes into the modeling of the constituents
of the incoming hadrons, of the hard-scattering process yielding the 
top quarks (and also describing their subsequent decay), and of the 
formation of the observable final-state particles.
A schematic illustration of this factorization scheme is given in 
Figure~\ref{factorization.fig}.

\begin{figure}
\begin{center}
\includegraphics[width=0.65\textwidth]{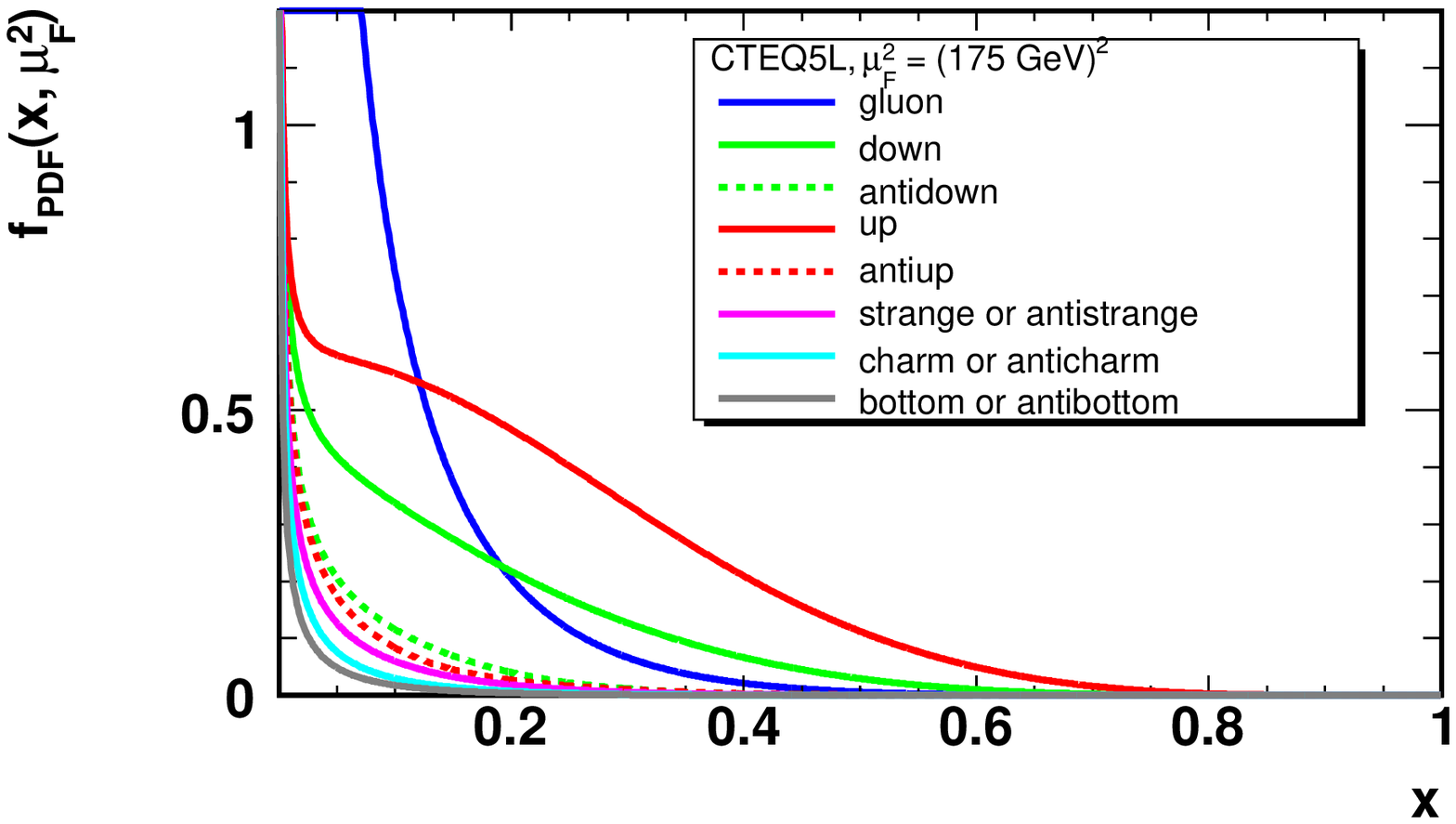}
\caption{\captionfont\label{cteq5l.fig}The \cteqfivel
  parametrization~\cite{bib-CTEQ5L} of the distribution functions for
  different parton species in the proton as a function of the momentum
  fraction $\xpdf$ of the proton carried by the parton, for a
  factorization scale $\factorizationscalesquared=(175\,\GeV)^2$.}
\end{center}
\end{figure}

To calculate the (differential) cross section for top quark
production, a factorization scale $\factorizationscalesquared$ is introduced to separate
the hard-scattering partonic cross section from the modeling of the 
constituents of the proton/antiproton.
The latter is independent of the hard-scattering process, and parton
distribution functions (PDFs) 
$f_{\rm PDF}^a (\xpdf, \factorizationscalesquared)$ are
introduced that describe the probability density to find a parton $a$
(quark or antiquark of given flavor or gluon) with longitudinal
momentum fraction \xpdf inside a colliding proton.
The PDFs cannot be calculated, and are determined in fits
to experimental data.
As an example, the \cteqfivel parametrization~\cite{bib-CTEQ5L}
is shown in Figure~\ref{cteq5l.fig} for a scale of 
$\factorizationscalesquared=(175\,\GeV)^2$ (a common choice used in
current measurements for the 
description of top quark production).
Even though experimental observables cannot depend on the factorization
scale, the PDFs (and the hard-scattering cross section) depend on the 
value of \factorizationscalesquared chosen, and an overall dependence
remains if calculations are not done to infinite order in perturbation
theory.
In the following sections, the dependence on the factorization scale is not
mentioned explicitly, and the symbol $\fPDF^a(\xpdf)$ is used.
To assess the systematic uncertainty related to the choice of 
factorization scale, experiments compare the results of simulations 
based on different values for the scale.

There are two main mechanisms for top quark production at hadron
colliders: top-antitop pair production via the strong interaction, 
and single top production via the electroweak interaction.
Single top production has only recently been
observed~\cite{bib-singletop}, and this process is not (yet) used to measure
the top quark mass.
Consequently, the emphasis of this section is on \ttbar pair
production.

\begin{figure}
\begin{center}
\includegraphics[width=0.95\textwidth]{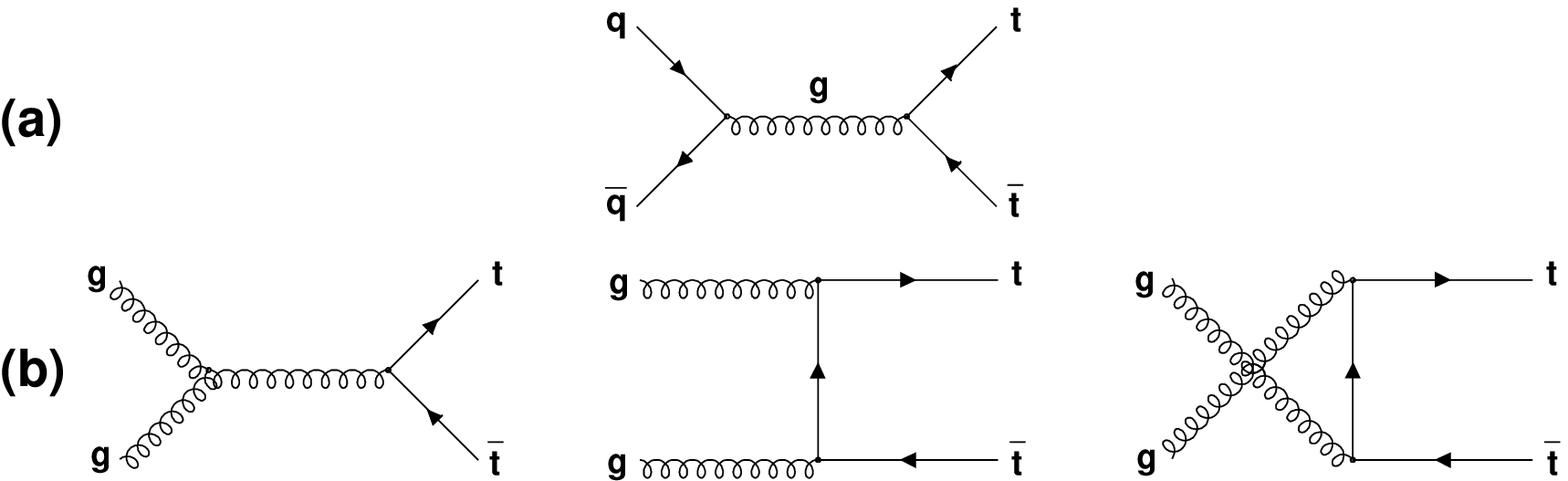}
\caption{\captionfont\label{pairproductiondiagrams.fig}Leading-order Feynman diagrams of the 
  hard-scattering processes that lead to \ttbar production at a hadron
  collider: (a) $\qqbar\to\ttbar$, (b) $\glueglue\to\ttbar$.}
\end{center}
\end{figure}

The leading-order Feynman diagrams for the hard-scattering process
of \ttbar production are shown in \Fref{pairproductiondiagrams.fig}.  
They apply to both proton-antiproton (Tevatron) and proton-proton
(LHC) collisions.
When contributions from higher-order diagrams are
included, renormalization of divergent quantities becomes necessary.
This leads to the introduction of 
another scale, the renormalization scale
$\renormalizationscalesquared$.
In practice, the factorization and renormalization scales are 
often chosen to be equal.

To obtain the \ttbar production cross section in hadron collisions, 
the partonic cross section $\hat\sigma$ must be folded with the
appropriate parton distribution functions $\fPDF^a(\xpdf)$,
integrated over all possible initial-state parton momenta, and then 
summed over all contributing initial-state parton species:
\begin{equation}
  \label{factorization.eqn}
    \sigma(P_1, P_2)
  =
    \!
    \sum_{a_1,a_2}
    \int
    \!
    {\rm d}\xpdf_1 {\rm d}\xpdf_2
    f_{PDF_1}^{a_1}\left(\xpdf_1, \factorizationscalesquared \right)
    f_{PDF_2}^{a_2}\left(\xpdf_2, \factorizationscalesquared \right)
    \hat\sigma\left(\xpdf_1 P_1, \xpdf_2 P_2,
                    \alpha_s\left(\factorizationscalesquared\right),
                    \frac{\processscalesquared}{
                          \renormalizationscalesquared}
              \right)
  \, ,
\end{equation}
where $P_1$ and $P_2$ are the momenta of the incoming hadrons, 
the sum is over all possible combinations of parton species
$a_1$ and $a_2$ that can initiate the hard interaction,
and the hard-scattering
cross section $\hat\sigma$ depends on their momenta, the 
factorization scale, and the ratio of the scale
\processscalesquared of the 
hard interaction and the renormalization scale.
Resulting Standard Model predictions for the \ttbar production cross 
section at the Tevatron and LHC are listed in
Table~\ref{productioncrosssections.table}.
At the Tevatron, in proton-antiproton
collisions at $\sqrt{s}=1.96\,\TeV$, the quark-antiquark induced
process dominates.  
At the LHC, in proton-proton collisions at $\sqrt{s}=14\,\TeV$, 
the fraction \xpdf of the
proton momentum carried by the colliding partons may be much smaller.
Because the gluon PDF is much larger at small \xpdf than the quark PDFs,
the gluon induced process dominates at the LHC.  
The overall \ttbar 
cross section at the LHC is two orders of magnitude larger than that at the 
Tevatron.

\begin{table}[htbp]
\begin{center}
\begin{tabular}{c c l @{} l}
\hline
\hline
  Channel
&
  \begin{tabular}{@{}c@{}}\phantom{a}\vspace{-2ex}\\
                          Tevatron \runii:\\
                          \ppbar collisions,\\
                          $\sqrt{s}=1.96\,\TeV$ 
  \end{tabular}
&
  \multicolumn{2}{@{}c@{}}{\begin{tabular}{@{}c@{}}LHC:\\
                                                   $pp$ collisions,\\
                                                   $\sqrt{s}=14\,\TeV$ 
                           \end{tabular}}
\vspace{-2ex}\\
\\
\hline
\vspace{-2ex}\\
  \ttbar pair production
&
  $5.8\enspace\hspace{1ex}{\text -}\hspace{1ex}7.4\enspace\,{\rm pb}$~\cite{bib-xsprediction}
&
  $830\phantom{.0}\hspace{0.8ex}\pandm{50\phantom{.0}}{40\phantom{.0}}$
&
  ${\rm pb}$~\cite{bib-ttbarxsatLHC}
\vspace{2ex}\\
  \begin{tabular}{@{}r@{}} single top, s-channel\\
                           single antitop, s-channel
  \end{tabular}
&
  $0.98\pm0.04\,{\rm pb}$~\cite{bib-singletoptevatronkidonakis}
&
  \begin{tabular}{@{}l@{}r@{}}
      $\enspace\enspace7.2\hspace{0.8ex}\pandm{\phantom{0}0.6}{\phantom{0}0.5}$
    \\
      $\enspace\enspace4.0\hspace{0.8ex}\pandm{\phantom{0}0.1}{\phantom{0}0.2}$
  \end{tabular}
&
  \begin{tabular}{@{}l@{}r@{}}
      ${\rm pb}$~\cite{bib-singletoplhckidonakis}
    \\
      ${\rm pb}$~\cite{bib-singletoplhckidonakis}
  \end{tabular}
\vspace{2ex}\\
  \begin{tabular}{@{}r@{}} single top, t-channel\\
                           single antitop, t-channel
  \end{tabular}
&
  $2.2\enspace\pm0.1\enspace\,{\rm pb}$~\cite{bib-singletoptevatronkidonakis}
&
  \begin{tabular}{@{}l@{}r@{}}
      $146\phantom{.0}\pm 5\phantom{.0}$
    \\
      $\enspace89\phantom{.0}\pm 4\phantom{.0}$
  \end{tabular}
&
  \begin{tabular}{@{}l@{}r@{}}
      ${\rm pb}$~\cite{bib-singletoplhckidonakis}
    \\
      ${\rm pb}$~\cite{bib-singletoplhckidonakis}
  \end{tabular}
\vspace{2ex}\\
  \begin{tabular}{@{}l@{}}
      single top+antitop, $W$+\tquark production
    \\
      \phantom{.}\vspace{-2.5ex}
  \end{tabular}
&
  \begin{tabular}{@{}l@{}}
      $0.26\pm0.06\,{\rm pb}$~\cite{bib-singletoptevatronkidonakis}
    \\
      \phantom{.}\vspace{-2.5ex}
  \end{tabular}
&
  $\enspace82\phantom{.0}\pm 8\phantom{.0}$
&
  \begin{tabular}{@{}l@{}r@{}}
      ${\rm pb}$~\cite{bib-singletoplhckidonakis}
    \\
      \phantom{a}\vspace{-2.5ex}
  \end{tabular}
\\
\hline
\hline
\end{tabular}
\caption{\captionfont\label{productioncrosssections.table}Predicted top
  quark production cross sections for various processes at the
  Tevatron and LHC.
  The predictions are at next-to-leading
  order, including threshold corrections from soft gluons.
  All values are quoted for an assumed top quark mass of $175\,\GeV$.
  For the dependence of the cross sections on the top quark mass
  hypothesis see Figure~\ref{xsmassconsistency.fig} (\ttbar production) and 
  References~\cite{bib-singletoptevatronkidonakis,bib-singletoplhckidonakis}
  (single top/antitop production).
  The range of \ttbar cross sections quoted for the 
  Tevatron includes PDF uncertainties (which have been found to 
  be dominant by studying the variations of the CTEQ6~\cite{bib-CTEQ6Mvar}
  and MRST~\cite{bib-MRST} parametrizations) while the
  LHC uncertainty is only based on a variation of the renormalization
  scale.
  At the Tevatron, the relative contributions of \qqbar and \glueglue
  induced process are roughly $85\%$ and $15\%$; at 
  the LHC these numbers are about $10\%$ and $90\%$, respectively.
  Wherever the cross sections for single top and antitop production
  are equal, the sum of the cross sections for both processes is
  listed.
  This is the case for single top/antitop production 
  at the Tevatron because it is a proton-antiproton collider.
  The cross sections for production of a \W boson in association
  with a top or antitop quark are equal also at the LHC because the 
  \bquark and \antibquark PDFs are equal.
  The single top cross sections quoted are similar to the 
  next-to-leading order values published in~\cite{bib-singletopatnlo}.}
\end{center}
\end{table}

\begin{figure}
\begin{center}
\includegraphics[width=0.65\textwidth]{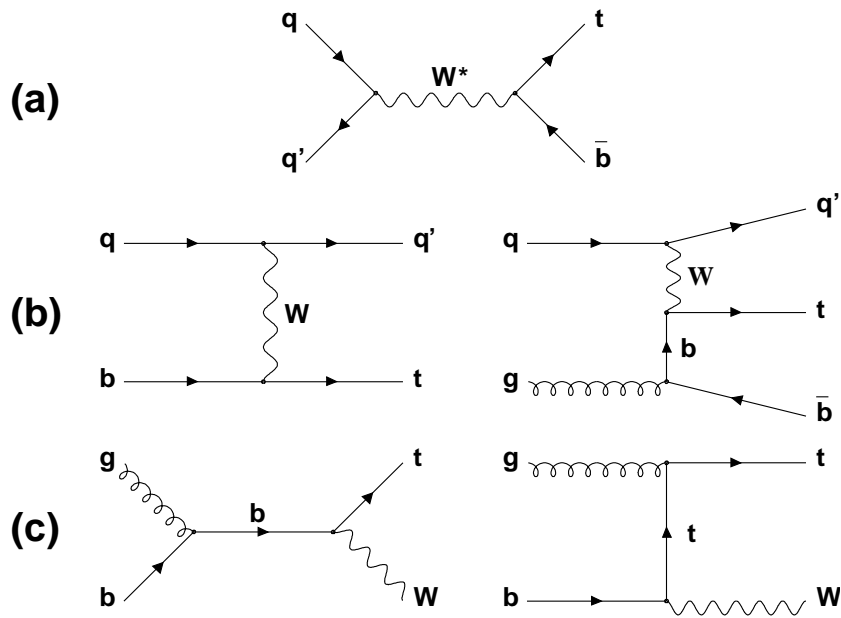}
\caption{\captionfont\label{singleproductiondiagrams.fig}Leading-order Feynman diagrams of the 
  hard-scattering processes that lead to single top production at a hadron
  collider: (a) s-channel, (b) t-channel, (c) W+t associated production.}
\end{center}
\end{figure}

Production of single top quarks via the electroweak interaction is
expected to proceed via three different channels.  
Predictions for the Standard Model cross sections are given in
\Tref{productioncrosssections.table}, and 
\Fref{singleproductiondiagrams.fig} shows the
leading-order diagrams for the three processes.
The remainder of this report focuses on \ttbar pair production.

\subsection{Top Quark Decay and Event Topologies}
\label{toprodec.topeventtopologies.sec}

In the Standard Model, top quarks decay almost exclusively to a
\bquark quark and a \W boson~\cite{bib-pdg,bib-LHCtopreport}, and
the top quark decay width being much larger than \lambdaqcd, no
top quark hadronization takes place.
Therefore, the event topology of a \ttbar event is determined by the 
decays of the two \W bosons.  
The \bquark quarks and quarks from hadronic \W decays hadronize and
are reconstructed as jets in the detector.
The presence of final-state neutrinos is signalled by missing transverse 
energy \etmiss, defined as
the magnitude of the transverse momentum vector \ptmissvec needed to balance the
event in the plane perpendicular to the beam direction.

Commonly, the event topologies are 
classified as dilepton, lepton+jets (\ljets), and \alljets topologies.
These three categories exclude events with one or more tauonic \W
decays, which are more difficult to reconstruct and provide less
mass information than corresponding events with electronic or muonic
\W decays because of the additional neutrinos from $\tau$ decays.
In this report, the
word ``lepton'' always refers to an electron or muon
unless otherwise mentioned.

\begin{figure}
\begin{center}
\includegraphics[width=0.6\textwidth]{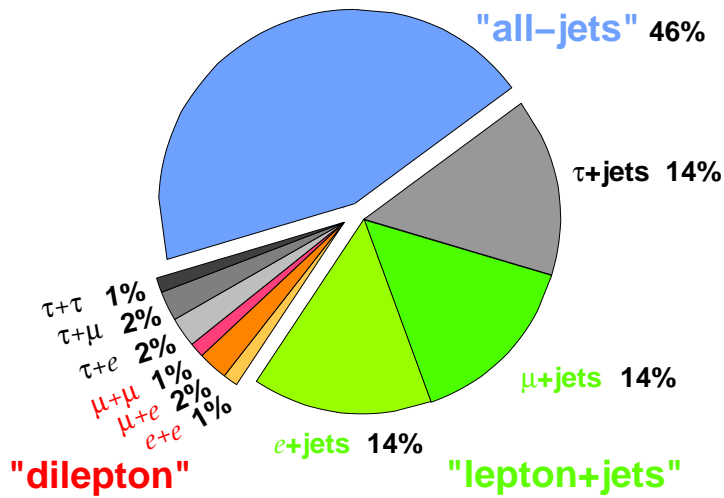}
\caption{\captionfont\label{piechart.fig}Relative abundance of the \ttbar
  event topologies, calculated from the \W branching fractions listed in 
  Reference~\cite{bib-pdg}.
  The figure has been taken from~\cite{bib-annheinson}, and the values
  have been updated.
  Note the rounding errors; the total ``dilepton'' and ``lepton+jets''
  branching fractions are about $5\%$ and $29\%$, respectively.}
\end{center}
\end{figure}

In the following, the characteristics of the topologies used for top
quark mass measurements are discussed,
the main backgrounds are listed, and the consequences for measurements
of the top quark mass are mentioned.  
The relative abundance of events in the various topologies is shown 
schematically in Figure~\ref{piechart.fig}.

\begin{itemize}
\item{\bf Dilepton Events:}
\label{toprodec.topeventtopologies.dileptoneventopology.sec}
In about $5\%$ of \ttbar events, both \W bosons decay into an
electron or a muon plus the corresponding neutrino.
These so-called dilepton events are characterized by two
oppositely charged isolated energetic leptons, two energetic \bquark
jets, and missing transverse energy due to the two neutrinos from the 
\W decay.

Because of the two charged leptons, these events are 
relatively easy to select.
The largest physics background is from production of a \Z boson (decaying to 
\epem or \mupmum) in association with two jets.  
This background affects
only the dielectron and dimuon channels and can be reduced by requiring
that the invariant dilepton mass be inconsistent with the \Z mass.
Correspondingly, the \emu channel is very clean; here, the main physics
background is from $\Z\to\tauptaum$ decays where the \Z boson is produced in 
association with two jets.  
Instrumental background where a hadronic
jet with a leading $\pizero\to\gamma\gamma$ decay is misidentified as an 
isolated electron is also important at the Tevatron experiments.

In spite of the small backgrounds the statistical information on the 
top quark mass that can be extracted per dilepton event is limited
because the event
kinematics is underconstrained when the top quark mass is treated as
an unknown.
The 4-momenta of the 6 final-state particles are fully specified by 
24 quantities; the 6 masses are known, and the 3-momenta of four
particles (the two jets and the two charged leptons) are measured in
the detector.  
Additional constraints can be obtained by assuming 
transverse momentum balance of the event (2), the known masses of 
the \W bosons (2), and by imposing equal top and antitop quark masses 
(1 constraint).  
This leads to 23 quantites that are known, measured,
or can be assumed.  
The event kinematics could therefore only be
solved if the value of the top quark mass itself were also assumed.
Consequently, to measure the top quark mass, additional information is
used, e.g.\ the relative probabilities for different
configurations of final-state particle momenta.

\item{\bf Lepton+Jets Events:}
\label{toprodec.topeventtopologies.lepjetseventtopology.sec}
Those $29\%$ \ttbar events with one $\W\to\e\nu$ or $\W\to\mu\nu$
and one hadronic \W boson decay are called lepton+jets events.
They contain one energetic isolated lepton,
four energetic jets (two of which are \bquark jets), 
and missing transverse energy.

The main background is from events where a leptonically decaying \W is 
produced in association with four jets.  
Multijet background where one
jet mimicks an isolated electron also plays a role.

In lepton+jets events, the transverse
momentum components of the one neutrino can be obtained from the 
missing transverse momentum, and the event kinematics is
overconstrained when assuming equal masses of the top and antitop
quarks and invariant $\ell\nu$ and $\qqbarprime$ masses equal to the
\W boson mass.  
The measurement of the top quark mass is however 
complicated by the fact that the association of measured jets with
final-state quarks is not known. 
The number of possible 
combinations and also the background can be reduced when \bquark
jets are identified (\bquark-tagging).  

Today, the lepton+jets topology yields the most precise top quark mass 
measurements.

\item{\bf All-Jets Events:}
\label{toprodec.topeventtopologies.alljetseventtopology.sec}
In $46\%$ of \ttbar events both \W bosons decay hadronically, yielding
6 energetic jets, no charged leptons, 
and no significant missing transverse energy.

The background from multijet production is large (and cannot easily
be modeled with Monte Carlo generators).  
It can be reduced with 
\bquark-tagging information, which is also important to reduce
combinatorics in the jet-quark assignment.
\end{itemize}

The aim is to measure the top quark mass in all three categories
in order to cross-check the measurements and to search for signs
of effects beyond the Standard Model.
The above picture could be changed if non-Standard Model particles
with masses below the top quark mass exist.  
An example are
top quark decays to a \bquark quark and a charged Higgs boson in 
supersymmetric models:
Depending on the parameters of the model, charged Higgs decays could
alter the relative numbers of events in the different \ttbar
event topologies or lead to events with 
extra jets in the final state~\cite{bib-LHCtopreport}.

\clearpage
\section{Event Reconstruction and Simulation}
\label{reco.sec}
\begin{center}
\begin{tabular}{p{15cm}}
{\it The previous section gave an overview of the production of top
  quarks at hadron colliders and of the topologies of top quark
  events.
  This section describes how top quark events are reconstructed in the
  detector.
  It also briefly introduces the simulation of events.}
\end{tabular}
\end{center}

To measure the top quark mass, \ttbar events must first be identified
online as potentially interesting and saved for further analysis.
The \ttbar decay products (charged lepton(s), jets, and missing transverse
energy from the neutrino(s)) are then reconstructed.
The top quark mass is obtained from the energies/momenta and
directions of the decay products measured in the detector.

A brief overview of the CDF and \dzero detectors at the Tevatron is given
in Section~\ref{reco.detectors.sec}.
In Section~\ref{reco.trigger.sec}, 
the trigger requirements used at CDF and \dzero for
the different \ttbar event topologies are presented, and 
Section~\ref{reco.reco.sec} briefly discusses the
reconstruction and selection of electrons, muons, and jets
and the identification of $b$ quark jets.
Section~\ref{reco.simulation.sec} describes the 
simulation of events used to verify and calibrate the 
techniques for the top quark mass measurements.
The detector calibration and the determination of the 
detector resolution are described in 
Section~\ref{detcalib.sec}.

\subsection{The CDF and \dzero Detectors}
\label{reco.detectors.sec}
\begin{figure}
\begin{center}
\includegraphics[width=1.14\textwidth]{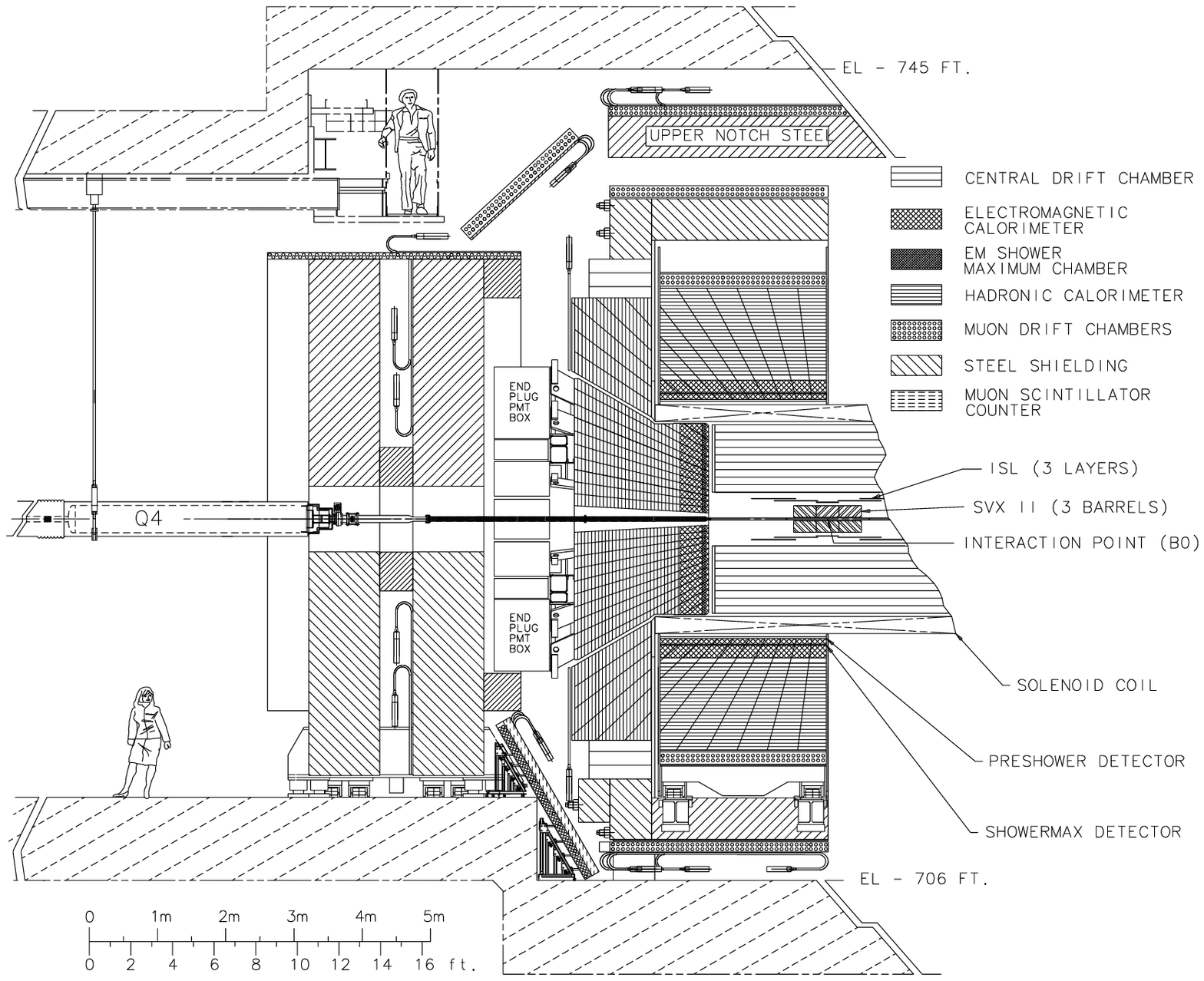}
\includegraphics[width=0.95\textwidth]{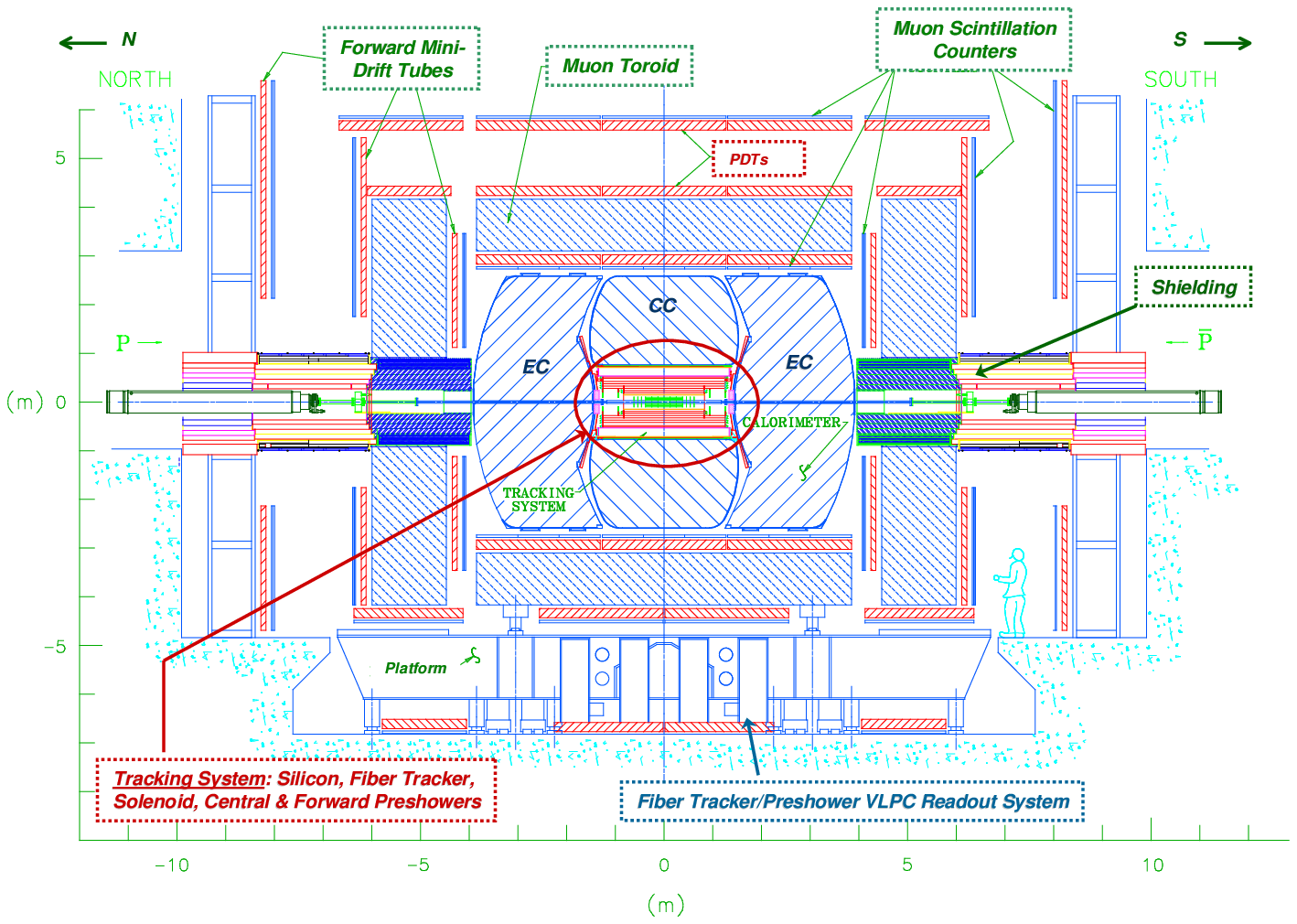}
\caption{\captionfont\label{detectors.fig}Schematic drawings of the 
  CDF~\cite{bib-CDFdetectordrawing} (top) and
  \dzero~\cite{bib-Dzerodetectordrawing} (bottom) detectors during
  Tevatron \runii.}
\end{center}
\end{figure}

The CDF and \dzero \runii detectors are described in detail
elsewhere~\cite{bib-CDFdet,bib-Dzerodet}.
Both detectors have the standard cylindrical setup
of a general-purpose collider detector.
From the interaction region in the center of the detector, particles
first traverse the tracking detector surrounding the beam pipe.
Here, the trajectories of 
charged particles and their transverse momenta are measured.
The tracking detector can be subdivided into a silicon microvertex
detector 
needed for precise primary and secondary vertex
reconstruction and a larger-volume tracking chamber providing the
lever arm to reconstruct the transverse momentum from the curvature of
the track in a solenoidal magnetic field.
The calorimeters are used to measure the energy and direction of
electrons, photons, and hadronic jets.
They are adapted to the different properties of both electromagnetic and
hadronic showers.
Finally, the calorimeters are surrounded by tracking detectors which
serve to identify muons, which are the only charged particles that
traverse the calorimeter without being absorbed.
Schematic drawings of both CDF and \dzero are shown in
Figure~\ref{detectors.fig}.
Both experiments employ a three-layer trigger system that allows
for an online selection of events for further analysis.
All subdetectors, their readout electronics, and the trigger system
are adapted to the Tevatron bunch crossing frequency of 
$1/(396\,{\rm ns})$.

As far as details of some of the subdetectors are concerned,
CDF and \dzero differ significantly.
However, the general functionality is very similar, and both
experiments
reconstruct charged leptons, hadronic jets, secondary decay vertices,
and missing transverse energy which are then used to select \ttbar
candidate events and measure the top quark mass.
The experiments use a coordinate system centered at the 
interaction point with the $z$ axis along the
beam pipe.
Directions are expressed in terms of the
azimuthal angle $\phi$ around the beam pipe and the pseudorapidity
$\eta=-\ln\left(\tan(\theta/2)\right)$, where $\theta$ is the polar angle relative
to the $z$ axis.

Of the integrated luminosity of more than $2\,\ifb$ delivered to each
of CDF and \dzero, up to $1\,\ifb$ has been used so far in top quark
mass measurements.
In comparison, \runi measurements were based on integrated
luminosities of the order of $100\,\ipb$.
A total integrated luminosity per experiment of $4-9\,\ifb$ is
expected until the end of \runii of the Tevatron.

\subsection{Trigger Strategies}
\label{reco.trigger.sec}

Triggering \ttbar event candidates that involve at least one leptonic
\W decay is relatively straightforward because of the presence of 
an isolated electron or muon with large transverse energy or momentum.
The presence of energetic jets can
be used as an additional trigger criterion.

To identify dilepton candidate events, both CDF and \dzero require
the events to be triggered by the presence of a high-\Et electron or 
high-\pt muon~\cite{Abulencia:2005uq,Abazov:2005yt}.
While CDF requires one electron or muon, in the \dzero analysis
two charged leptons in the first-level trigger and one or two 
(depending on the channel) charged leptons in the high-level triggers
are required.

In the \ljets event topology, CDF also relies exclusively on the 
charged lepton trigger~\cite{Abulencia:2005aj}.
The \dzero experiment requires a charged lepton and a jet, both with 
large transverse momentum or energy, to be
found in the trigger~\cite{bib-me}.

Triggering \ttbar events in the \alljets channel is more
difficult because of the large QCD multijet background.
The CDF analysis~\cite{bib-CDFallhadxs} uses a trigger that 
requires at least four jets and a minimum scalar sum of transverse energies,
$\Ht$, of at least 125~GeV.
In the \alljets channel, the \dzero experiment has performed
a measurement of the \ttbar cross section~\cite{bib-D0allhadxs}, but
not yet of the top quark mass.

The characteristics of dilepton and \ljets \ttbar events are
distinctive, so typical trigger efficiencies are around 90\% or above 
(see for example~\cite{Acosta:2005zd}).
In the \alljets channel, the CDF experiment quotes a trigger
efficiency of 85\%~\cite{bib-CDFalljetsxspaper}.
In general, the trigger requirements and therefore also the efficiencies
vary as conditions are adjusted to changing instantaneous luminosity.
The efficiencies are measured in the 
data as outlined in Section~\ref{detcalib.sec} as a function of the 
momenta of reconstructed particles (charged leptons, jets) in the event.
The overall probability for a simulated event to pass the trigger
conditions is obtained as the weighted average of the trigger
efficiencies, taking into account the relative integrated luminosity
for which each trigger condition was in
use~\cite{bib-D0ljetsbtagxspaper}.
The trigger efficiency depends on the top quark mass, mainly
because of the $\pt$ or $\Et$ cuts imposed in the trigger, 
and this effect must be taken into account in the mass measurement.

\subsection{Reconstruction and Selection of Top Quark Decay Products}
\label{reco.reco.sec}

The offline reconstruction of the events selected by the trigger
criteria aims at (1) further reducing the backgrounds and (2)
reconstructing the momenta of the \ttbar decay products as precisely
as possible to obtain the maximum information on the top quark mass.
In this section, the reconstruction and selection of isolated
energetic charged leptons, of energetic jets, and of the missing
transverse energy in \ttbar event candidates are discussed.
Also, the different possibilities for the identification of bottom-quark
jets are described.

\subsubsection{Charged Lepton Selection}
\label{reco.reco.leptonsel.sec}
Electrons are identified by a charged particle track pointing at
an electromagnetic shower in the calorimeter.
Additional criteria are then applied~\cite{bib-me,bib-CDFljetstopoxspaper}:
Background from mis-identified hadrons is reduced based on the
ratio of the energy measured in the electromagnetic and hadronic
calorimeter, the shower
shape, and on the quality of the match between
the calorimeter shower and the charged particle track.
CDF in addition vetos electrons from photon conversion processes.
Non-isolated electrons, e.g.\ from semielectronic heavy-hadron 
decays in jets, are rejected by isolation criteria that impose
a maximum calorimeter energy in a cone around the electron.

Muons traverse the calorimeter and leave a track both in the 
central tracking chamber and in the muon chambers.
The following criteria are applied to select muons from \W decay
in \ttbar events~\cite{bib-me,bib-CDFljetstopoxspaper}:
Background from mis-identified hadrons is reduced based on the 
distance between the central track extrapolated to the muon 
chambers and the muon chamber track.
In addition, CDF requires the energy deposit in the calorimeter to 
be consistent with that of a minimum ionizing particle, and rejects
muons with too large a distance of closest
approach in the transverse plane, $d_0$, to the beam spot.
Cosmic ray muons are rejected based on timing information.
As for electrons, non-isolated muons, e.g.\ from semimuonic heavy-hadron 
decays in jets, are rejected by isolation criteria requiring 
a maximum calorimeter energy in a cone around the muon not to be exceeded.
The \dzero experiment in addition imposes a similar isolation
criterion based on the transverse momenta of tracks in a cone around 
the muon direction.

Finally, a fiducial and kinematic selection is applied.
To ensure reliable electron reconstruction in the calorimeter, 
electron candidates must be well within the central or forward
calorimeters, excluding the overlap regions around $|\eta|\approx1$.
Some analyses exclude electrons in the forward calorimeter.
The pseudorapidity range within which muons can be identified is 
limited by the acceptance of the tracking chamber.
Typically, electrons (muons) are required to have a transverse
energy (momentum) larger than a cut value between 15 and 25~GeV,
depending on the analysis.
Here, the calibrated energy and momentum values are
used; the detector calibration is described in
Section~\ref{detcalib.sec}.

\subsubsection{Primary Vertex Reconstruction}
\label{reco.reco.pvtx.sec}
The position of the primary vertex is needed in order to compute
the jet directions and to identify bottom quark jets using 
secondary vertex information.
While the position of the hard interaction in the transverse plane
(``beam spot'') is well determined, the interaction region extends
over tens of centimeters along the beam line.
Tracking information is used to measure the $z$ position of the 
primary vertex for each event.
Since there may be multiple interactions per event, the vertex
associated with the \ttbar decay has to be identified.
This is done based on reconstructed charged lepton 
information (CDF analyses involving charged leptons), 
or the vertex most consistent
with the \ttbar decay is selected among the
candidates~\cite{bib-CDFljetsbtagxspaper,bib-D0ljetsbtagxspaper}.

\subsubsection{Jet Reconstruction and Selection}
\label{reco.reco.jetsel.sec}
The final-state quarks in \ttbar events are reconstructed as jets,
using a 
cone algorithm~\cite{Abe:1991ui,bib-D0conejetalgorithm}
with radius
$\DeltaR = \sqrt{(\Delta\eta)^2 + (\Delta\phi)^2} = 0.4$ (CDF)
or 0.5 (\dzero).  
The jet transverse energy is defined using the primary vertex
position described in the previous section.
The \dzero experiment applies cuts to select well-measured
jets~\cite{bib-me}, and both CDF and \dzero ensure that calorimeter 
energy deposited by electron candidates is not used in the jet
reconstruction.
A minimum number of jets within a fiducial calorimeter volume of 
typically $|\eta|<2.0$ (CDF,~\cite{bib-CDFljetstopoxspaper}) or $|\eta|<2.5$ 
(\dzero,~\cite{bib-me}) and with a (calibrated) transverse energy above a cut
value of typically 15 or 20~GeV is required.
The calibration of the calorimeter energy scale is discussed in 
Section~\ref{detcalib.sec}.

\subsubsection{Missing Transverse Energy}
\label{reco.reco.met.sec}
Neutrinos can
only be identified indirectly by the imbalance of the event in the
transverse plane.
A feature of lepton+jets and dilepton \ttbar events is thus
significant missing transverse energy $\etmiss$.
The missing transverse momentum is reconstructed from the 
vector sum of all calorimeter objects, i.e.\ using finer granularity
than the reconstructed jets and thus taking into account also small
additional energy deposits~\cite{bib-me,bib-CDFljetstopoxspaper}.
The missing transverse momentum vector is corrected for the energy scale of
jets and for muons in the event.
For the selection of lepton+jets events typically a
missing transverse energy of $\etmiss>20\,\GeV$ is required; the cut value for
dilepton analyses is usually higher.

The unclustered transverse energy $\etuncl$ is defined as the 
magnitude of the vector sum of transverse energies of all calorimeter objects that are
not assigned to a jet or charged lepton.

\subsubsection{Identification of Bottom Quark Jets}
\label{reco.reco.btagging.sec}

A $\ttbar$ event contains two bottom quark jets, while jets in
background events predominantly originate from light quarks or
gluons. 
This is why the signal to background ratio is significantly
enhanced after the requirement that at least one of the jets is $b$-tagged.
In addition, the number of relevant assignments of reconstructed
jets to final-state quarks ({\em jet-parton assignments}) can be 
considerably reduced with $b$-tagging information.

Three different signatures can in principle be used to identify 
bottom-quark jets:
\begin{itemize}
\item
The presence of an explicitly reconstructed secondary vertex 
corresponding to the decay of the bottom-flavored hadron, 
\item
a low probability for all charged particle tracks in the jet to 
come from the primary event vertex (which again implies the existence
of a displaced secondary decay vertex), or
\item
the presence of a charged lepton within the jet from a semileptonic
bottom or charm hadron decay.
\end{itemize}
To date, for measurements of the top quark mass using $b$ tagging, explicit
secondary vertex reconstruction is used, which proceeds as
follows~\cite{bib-D0ljetsbtagxspaper,bib-CDFljetsbtagxspaper}.
Tracks in the jet passing a $\pt$ cut are selected if they have 
significant impact parameter relative to the primary event vertex.
CDF rejects poorly reconstructed tracks based on the hits and the 
track fit $\chi^2$; \dzero rejects tracks from $\Kshort$ and $\Lambda$
decays and requires that the impact parameter of any track used in 
secondary vertex finding have a positive projection onto the jet axis
(negative when determining the mistag efficiency, see below).
Jets are called {\em taggable} if they contain at least two 
tracks that pass these criteria.
These tracks are used to form secondary vertices; if a vertex is found
with a large positive decay length significance
$L_{xy}/\sigma(L_{xy})$ ($>3$ for CDF and $>7$ for \dzero) the jet
is called $b$-tagged.
The distance $L_{xy}$ in the $xy$ plane between primary and secondary vertex is multiplied
by the sign of the cosine of the angle $\phi$ between the vector
pointing from the primary to the secondary vertex and the jet momentum 
vector.
While a large positive value of $L_{xy}$ is a sign for a decay of a 
long-lived particle, the distribution of negative values contains 
information about the $L_{xy}$ resolution.
Jets tagged with negative $L_{xy}$ are used in the determination of 
the mistag efficiency, i.e.\ the efficiency with which non-\bquark
quark jets are erroneously tagged, see Section~\ref{detcalib.btagging.sec}.

\subsection[Backgrounds and \ttbar Event Selection]{\boldmath Backgrounds and \ttbar Event Selection}
\label{reco.backgrounds.sec}
Two types of background have to be distinguished:
(1) physics background where all final-state particles are produced 
but in a different reaction; generally these processes will not involve
top quarks, but misassignment of top quark events to the wrong event
topology also
has to be taken into account; and (2) instrumental background, where
part of the event is mis-reconstructed.  
At a hadron collider, instrumental background mainly involves jets
that lead to wrongly identified isolated leptons.
Together with the backgrounds, a general outline of the event
selection for the different \ttbar topologies is given below; concrete
examples of event selection criteria are described more fully later
together with the top quark mass measurements.

\subsubsection{Dilepton Events}
\label{reco.backgrounds.dilepton.sec}
Physics background in the dilepton channel arises from all processes
leading to a final state with two charged leptons of opposite charge
and two jets.
For the $ee$ and $\mu\mu$ channels, the largest background is
from Drell-Yan events containing two additional jets.
These events can be efficiently removed by requiring a minimum charged
lepton \pt (to remove low-mass resonances), inconsistency of
the dilepton invariant mass with the \Z mass, and significant
missing transverse energy.
For all dilepton channels, $\Z/\gamma^* \to\tau\tau$ events with two
leptonic $\tau$ decays as well as diboson events (the \ww cross
section is largest, but \wz events also have to be taken into
account) with leptonic \W decay remain.
For the dilepton channels as well as the other channels,
misidentification of \ttbar events containing tauonic \W decays with
subsequent leptonic $\tau$ decay has to be accounted for.

Instrumental background in the dilepton channel arises mainly from 
events with one leptonic \W decay and three jets, one of which is 
mis-identified as another lepton.
Jets can appear as isolated electrons if they contain a leading
$\pi^0\to\gamma\gamma$ decay, resulting in large electromagnetic energy
deposition in the calorimeter, possibly with a track pointing at it
from conversion ($\gamma\to\epem$) of one of the photons, and only
little surrounding jet activity.
Additional contributions come from semileptonic bottom or charm hadron
decays within jets.

Leptons from $\tau$ decays and jets not from top quark decay have
mostly small transverse energies.
To select \ttbar dilepton event candidates, the experiments thus
typically require two charged leptons of opposite charge with large
\et and spatially isolated from jet activity, two large-\et jets, and
significant missing transverse energy.
Most of the remaining background can be removed by requiring jets to be
\bquark-tagged; however, this is often not desirable for small data
samples.

\subsubsection{Lepton+Jets Events}
\label{reco.backgrounds.ljets.sec}
Leptonic \W decays produced in association with jets, which lead to
instrumental background for dilepton events, are the main physics
background for \ttbar events in the \ljets channel.
Another physics background is from electroweak single top
production with additional jets.
Diboson events contribute when in contrast to above, one leptonic \W
decay occurs together with another hadronic weak boson decay.
Background from events with a leptonic \Z decay can be removed by
rejecting events with more than one isolated energetic charged lepton.
Similarly, background from $\Z/\gamma^* \to\tau\tau$ events arises if 
one $\tau$ decays leptonically and the other hadronically.

Instrumental background in the \ljets channel is due to QCD multijet
events with at least five jets, one of which is mis-identified as a
lepton as described above.

Lepton+jets \ttbar events are selected by requiring one 
isolated charged lepton with large \et, 
normally four large-\et jets at least one of
which is \bquark-tagged (both requirements can be relaxed), and
significant missing transverse energy.

\subsubsection{All-Jets Events}
\label{reco.backgrounds.alljets.sec}
The overwhelming background in the \alljets channel is from QCD
multijet events that contain six or more reconstructed jets.
Most of this background does not contain \bquark jets, and the
kinematic properties of the jets differ slightly from those of jets in
signal events.
The selection relies on a combination of \bquark tagging and kinematic
criteria. 
Since the QCD multijet process cannot be reliably simulated and the total
background has to be estimated from the data, there is no need to 
explicitly account for individual subdominant background processes.

\subsection{Jet-Parton Assignment}
\label{reco.jetpartonassignment.sec}
In most analyses, in particular those based on explicit top quark mass
reconstruction, the reconstructed jets need to be assigned to the
final-state quarks from the \ttbar decay to measure the top quark mass.
Depending on the \ttbar topology, different numbers of possible
jet-parton assignments have to be considered; for \alljets events,
90 different assignments have to be distinguished.
In \ljets and \alljets events, the number of relevant assignments can
be reduced when \bquark-tagged jets are present, which are likely to
be direct top quark decay products.

A further complication arises when additional jets are present in the
event.
Since jets from initial-state radiation, from the underlying event
(interactions involving the proton or antiproton remnant), or
from additional hard interactions in the same beam crossing typically
have small transverse energy \et, many analyses consider the $n$
highest-\et jets as \ttbar decay products, where $n=2,4,6$ in the
dilepton, \ljets, and \alljets topologies, respectively.

The issue of jet-parton assignment is further discussed in
Sections~\ref{templatemeasurements.sec}, \ref{memeasurements.sec},
and~\ref{idmeasurements.sec} together with the individual analyses.

\subsection{Simulation}
\label{reco.simulation.sec}
Monte Carlo simulated events are used for several purposes in the
analyses: 
\begin{list}{$\bullet$}{\setlength{\itemsep}{0.5ex}
                        \setlength{\parsep}{0ex}
                        \setlength{\topsep}{0ex}}
\item
to
compare measured and simulated distributions in order to check the detector;
\item
to determine the detector resolution;
\item
to optimize the selection and determine the fraction of signal events
in the selected data sample;
\item
to calibrate the methods for measuring the top quark mass; and 
\item
to compare the top quark mass uncertainty obtained in 
the data with the value expected for the measured fraction of signal events.
\end{list}
Simulation programs are based on the factorization scheme
(cf.\ Section~\ref{toprodec.topproduction.sec}), and in general, 
separate program libraries can be used to model the hard interaction,
additional gluon and photon radiation in the initial and final state, 
the parton distribution functions, hadronization, decays of unstable
particles, and the detector response.
Interference between different processes populating the same
experimental final state is usually neglected\footnote{An exception are
Drell-Yan events, where interference between photon and \Z exchange
is included.}.
This is a good approximation since the final-state
color, flavor, and spin configurations are in general
different: For example, \ljets \ttbar production 
can only interfere with those \wjets 
events that contain a \bbbar pair and two
additional quarks (but no hard gluons) in the final state.

The simulation used so far in the Tevatron analyses 
is based on leading-order matrix elements to
describe the hard process.
The Monte Carlo generators \pythia~\cite{bib-pythia},
\herwig~\cite{bib-herwig},
or \alpgen~\cite{bib-alpgen} are used to generate the hard 
parton-scattering process in \ttbar events and background events involving
weak vector bosons (\wjets events; \ww, \wz, and \zz 
events; single top production; and Drell-Yan events in association
with jets).
These generators are interfaced to leading-order parton 
distribution functions, in general \cteqfivel~\cite{bib-CTEQ5L}.
Leading-order calculations of total cross sections have
large uncertainties, and where possible, absolute production rates are
scaled to accommodate the data, so that only the prediction of
relative cross sections is taken from the simulation.

The simulation of the hard-scattering process 
is interfaced with \pythia or \herwig to simulate
initial- and final-state gluon radiation.
Matching procedures have been developed to ensure that
the phase space regions covered by 
hard gluon radiation and by gluon emission included in the matrix element
calculations do not overlap.
\pythia or \herwig are also used to model
fragmentation and hadronization, and are interfaced
with \evtgen~\cite{bib-evtgen} or
\QQ~\cite{bib-QQ} and \tauola~\cite{bib-tauola} to simulate heavy 
hadron and tau lepton decays.
The simulated events are passed through a detailed simulation of the
detector response based on \geant~\cite{bib-geant} and are then subjected to 
the same reconstruction and selection criteria as the data.
A detailed general discussion of the event simulation process can be found
in~\cite{bib-generalMC}, and a list of programs used for top quark 
measurements is given in~\cite{bib-topquarkMC}.

Depending on the instantaneous luminosity, it
is possible that more than one \ppbar or $pp$ collision takes place
in one bunch crossing.
To simulate this effect, minimum bias events (events with only very 
loose trigger requirements) are recorded and superimposed on the
simulated events.
Similarly, pileup of signals from collisions in subsequent bunch
crossings is simulated by overlaying events recorded with a random
trigger.

Background not involving any leptons from vector boson decay (QCD
multijet background) is not modeled using Monte Carlo simulation,
but estimated from the data using events with non-isolated
leptons~\cite{bib-D0ljetsbtagxspaper,bib-matrixmethod} and/or little
$\etmiss$~\cite{bib-CDFljetsDLMmass}. 
An exception is one CDF analysis in the \alljets final state, where 
\alpgen is used to model the multijet
background~\cite{bib-CDFallhadID}.

The reconstructed energies and momenta in the simulation 
are smeared such that the detector resolution agrees with that 
of the actual data.
The modeling of kinematic distributions in the simulation is
then checked.
Signal events are generated for various assumed top quark masses in
order to calibrate the measurement methods.

\clearpage
\section{Detector Calibration}
\label{detcalib.sec}
\begin{center}
\begin{tabular}{p{15cm}}
{\it To measure the top quark mass it is not sufficient to merely
  select \ttbar event candidates.
  An accurate understanding of how the detector responds to 
  the decay products in \ttbar events is also indispensable.
  It is only this second step that allows to relate the properties of
  the events to the value of the top quark mass.
  The procedures with which the experiments calibrate the detector
  response are outlined in this section.}
\end{tabular}
\end{center}

An accurate calibration of the energy/momentum scale and resolution
for the reconstructed particles used to measure the top quark mass
is crucial.
Also, even though the measured top quark mass does not directly
depend on the absolute detector efficiency, the dependence
of the efficiency on particle energies/momenta and pseudorapidities
must be known, too.
In this section, the calibration procedures used by the Tevatron 
experiments are introduced.
It is worth noting that usually a large fraction of the analysis work needed
in a top quark mass measurement is related to detector
calibration.

\subsection{Charged Leptons}
\label{detcalib.leptons.sec}

The reconstruction of electrons and muons can be calibrated using
$\Z\to\epem$ and $\Z\to\mupmum$ decays.
In addition, information from $\W\to e\nu$ events, cosmic ray muons, 
and $\ccbar$ and $\bbbar$ resonance decays can be used.
These events have the advantage that they can be identified with 
low backgrounds, and that the measurement of one particle or by one
detector system can be cross-checked with another.
The electromagnetic calorimeter yields the most precise measurement of
the energy of energetic electrons, while the central tracking chamber
is used to measure the muon (transverse) momentum.

The transverse momentum {\em scale} for energetic 
muons is adjusted such
that the reconstructed \Z mass reproduces the known value.
Additional information on the momentum scale is obtained from the 
lower-mass resonance
decays $J/\psi\to\mupmum$ and $\Upsilon(1S)\to\mupmum$.
The reconstructed invariant mass distribution of $\Z\to\mupmum$ decays
obtained with the CDF experiment is shown in
Figure~\ref{cdfz.fig}(a)~\cite{bib-cdfwmass}.
An example of further studies of the momentum scale
is given in \Fref{d0zmumu.fig}~\cite{bib-d0zmumu}, which shows the
$\Z\to\mupmum$ mass distribution for \dzero data in events
where (1) both muons are isolated and (2) one muon fails the isolation 
criteria, indicating the presence of Bremsstrahlung.
\begin{figure}
\begin{center}
\includegraphics[width=0.5\textwidth]{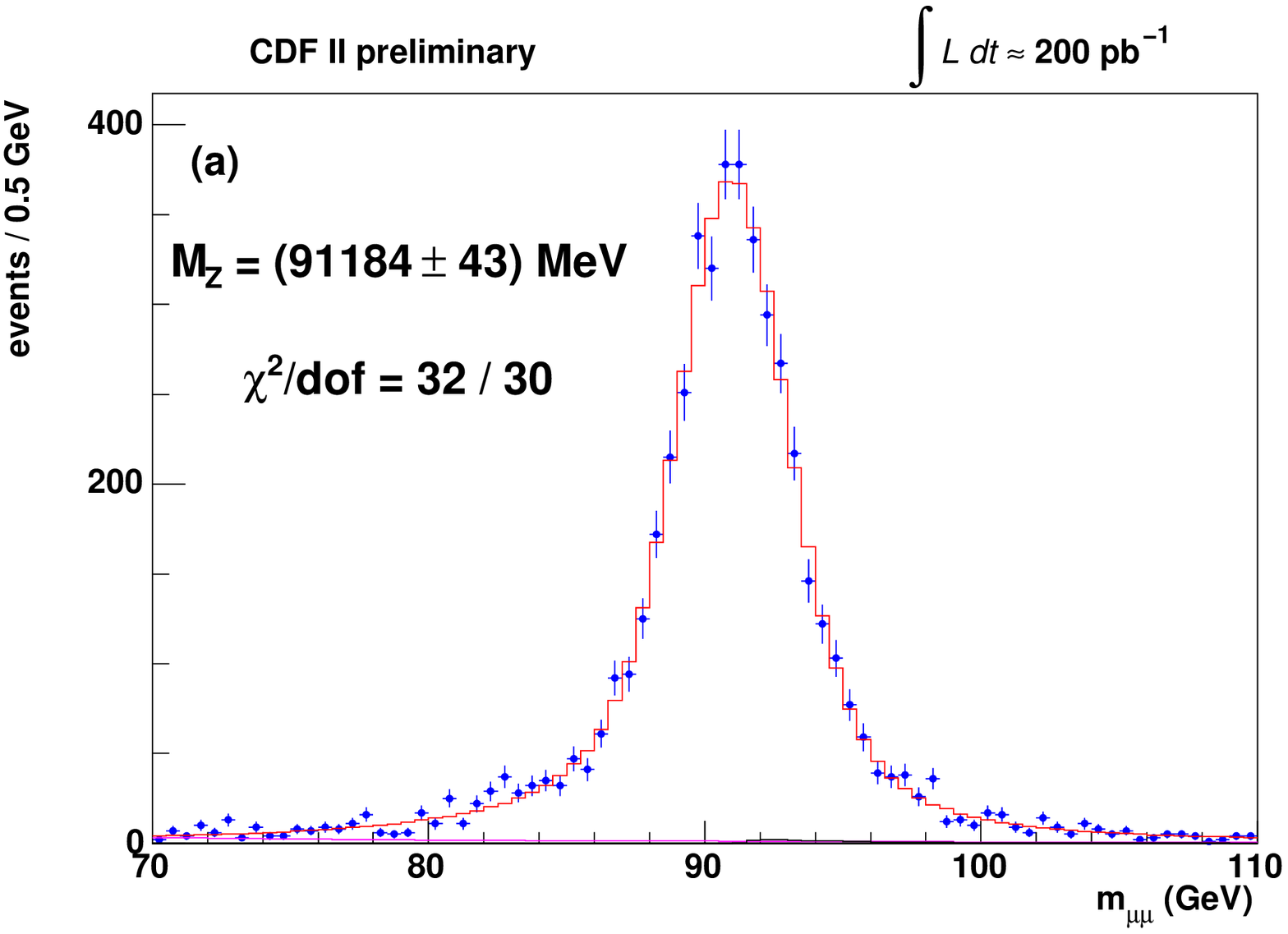}
\hspace{-5ex}
\includegraphics[width=0.5\textwidth]{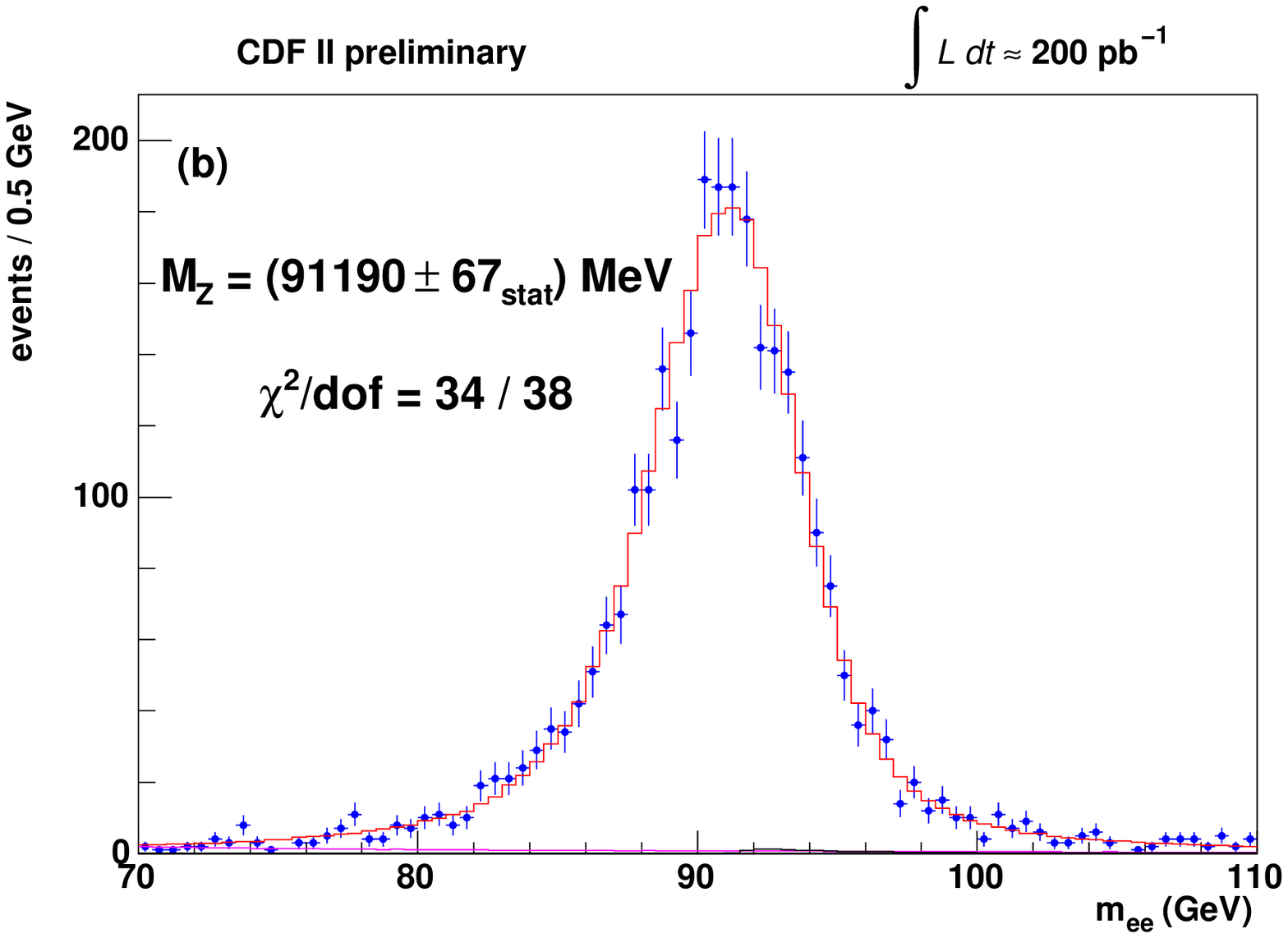}
\caption{\captionfont\label{cdfz.fig}Distributions of invariant dilepton masses
  for (a) $\Z\to\mupmum$ and (b) $\Z\to\epem$ decays reconstructed with
  the CDF detector~\cite{bib-cdfwmass}.}
\end{center}
\end{figure}
\begin{figure}
\begin{center}
\includegraphics[width=0.5\textwidth]{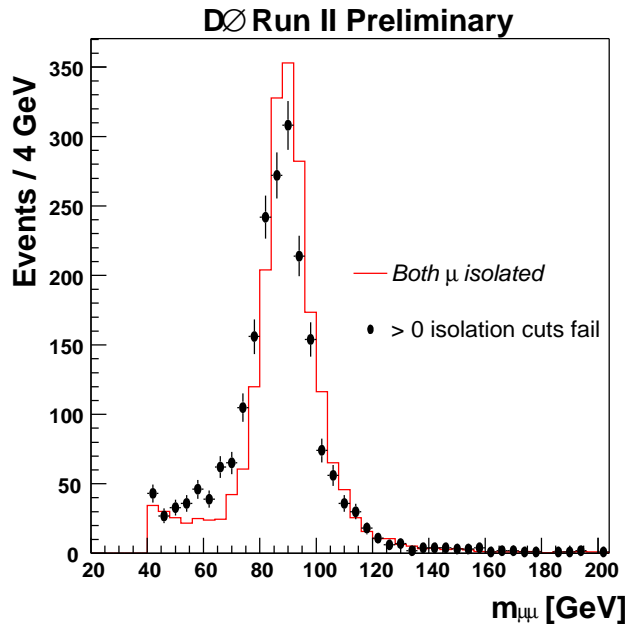}
\caption{\captionfont\label{d0zmumu.fig}Distribution of the invariant $\mupmum$ mass for selected
  $\Z\to\mupmum$ events in the \dzero data for isolated muons
  (histogram) and events where at least one muon does not pass all
  isolation cuts (points with error bars, scaled to the same number of
  entries)~\cite{bib-d0zmumu}.}
\end{center}
\end{figure}
The energy {\em scale} for energetic electrons is set with the 
reconstructed $\Z\to\epem$ invariant mass distribution.
The distribution
obtained by the CDF experiment is shown in Figure~\ref{cdfz.fig}(b).
Additional input is obtained from a comparison of 
reconstructed electron energy and track momentum in $\W\to e\nu$
decays as discussed below.
The resulting uncertainties in the calibration of the absolute  muon
momentum and electron energy scales are negligible for top quark mass
measurements (compared with the jet energy scale
uncertainties, see below).

The energy/momentum {\em resolution} can be studied using $\Z\to\epem$
and $\Z\to\mupmum$ events, too.
Also, a
cosmic ray muon traversing the center of the detector is
reconstructed as two muons, and the distribution of the difference 
between the two reconstructed momenta yields additional information
on the momentum resolution.
Similarly, since the calorimeter energy measurement is more precise
at high energies than the track momentum, a comparison between the 
two quantities in clean samples of isolated electrons can be made
to cross-check the track momentum resolution.
\Fref{cdftrackresolution.fig} shows the results of these
studies with CDF data~\cite{Abulencia:2005ix}.
\begin{figure}
\begin{center}
\includegraphics[width=0.47\textwidth]{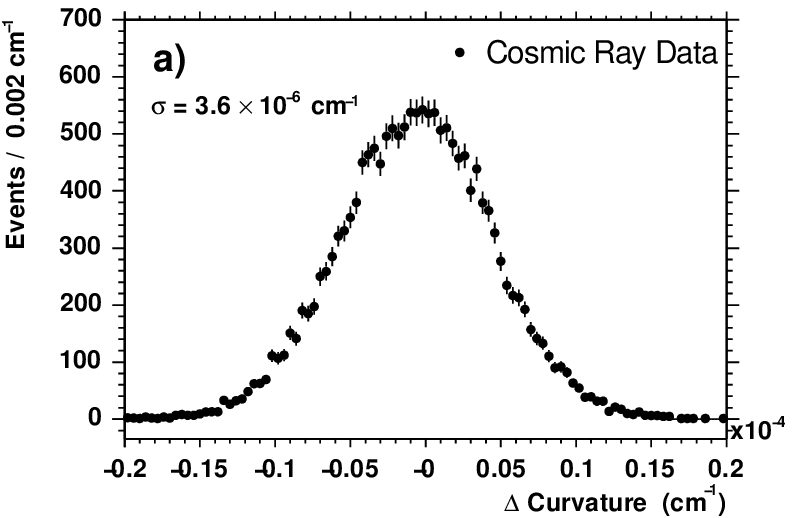}
\includegraphics[width=0.45\textwidth]{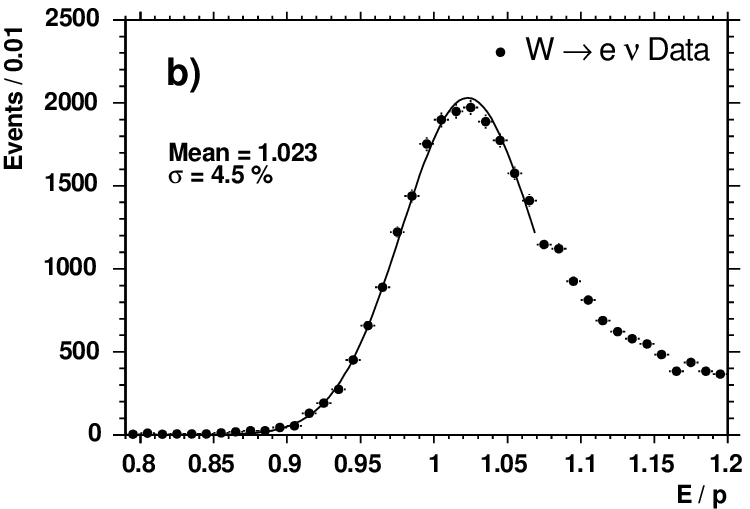}
\caption{\captionfont\label{cdftrackresolution.fig}Distribution of the difference in curvature for the two
  tracks in a CDF cosmic ray event (a), yielding a measurement of the
  momentum resolution.  Distribution of the energy divided by momentum
  in $\W\to e\nu$ events at CDF (b) together with a Gaussian fit in the range
  $0.8<E/p<1.08$.  Electrons with with significant
  Bremsstrahlung at large $E/p$ are excluded from the 
  fit; their abundance is a measure of the
  amount of detector material.}
\end{center}
\end{figure}

The {\em efficiency} to reconstruct an electron or muon 
can be factorized into several contributions:
trigger efficiency, 
tracking efficiency, 
the efficiency to identify the track as electron or muon, 
and the efficiency of further criteria like isolation cuts.
All individual efficiencies are measured in the data using 
$\Z\to\lplm$ events 
(CDF determines the tracking efficiency with $\W\to e\nu$ candidate
events using calorimeter-only selection criteria), see for 
example~\cite{bib-d0zmumu,Abulencia:2005ix,bib-Dzerotagandprobe}.
The concept of the tag-and-probe method in $\Z\to\lplm$ events is 
visualized in \Fref{tagandprobe.fig}:
A clean sample of $\Z\to\lplm$ events is obtained using a selection
where the criterion under investigation is not applied to one of the 
leptons.
The fraction of selected events where this lepton also passes the 
additional criterion is then a measure of the efficiency.
\begin{figure}
\begin{center}
\includegraphics[width=0.45\textwidth]{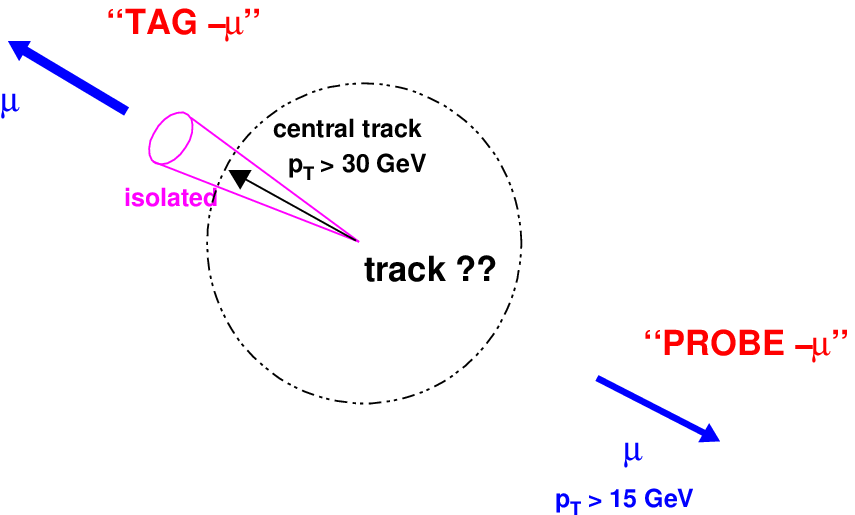}
\includegraphics[width=0.45\textwidth]{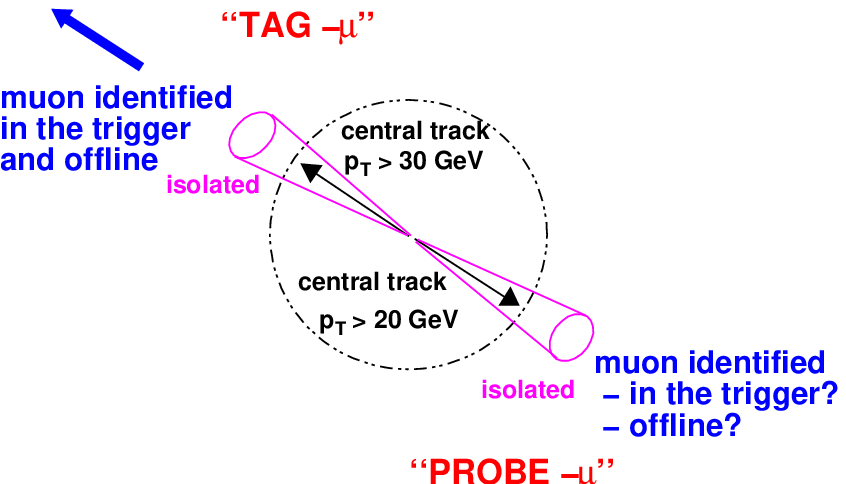}
\caption{\captionfont\label{tagandprobe.fig}Schematic illustration of the tag-and-probe method
  to measure the tracking efficiency (left) and the efficiency of the 
  charged lepton identification in the trigger and 
  off\-line~\cite{bib-d0zmumu}.}
\end{center}
\end{figure}

The calibration of top quark mass measurements relies heavily on the
quality of the detector {\em simulation}.
The simulation is tuned (and an additional scaling and smearing is
applied where necessary) to reproduce the position and width of the
$\Z\to\lplm$ invariant mass peak.
This can for example become necessary when the description of the 
detector material or alignment in the simulation does not fully
reproduce reality.
Also, the efficiency in the simulation may have to be scaled.

\subsection{Hadronic Jets}
\label{detcalib.jets.sec}

For the same reasons as outlined above in the section about charged
leptons, it is crucial to have a precise knowledge of the jet energy
scale and resolution, and to accurately reproduce them in the
simulation.
In most analyses, the measurement of the top quark mass 
relies to a large extent
on the reconstructed jet energies.
However, 
the energy scale and resolution of jets is more difficult to determine
experimentally than that of charged leptons.
Therefore, current
top quark mass measurements are systematically dominated by the 
knowledge of the absolute jet energy scale~\cite{bib-TEVEWWG},
and for a given sample and analysis technique the 
statistical error on the top quark mass is dominated by the jet energy
resolution (see below for a more detailed discussion).

In the following, the determination of the jet energy scale, the 
relevance of the jet energy resolution, and the agreement between 
data and Monte Carlo simulation are discussed.

\subsubsection{Overall Jet Energy Scale}
\label{detcalib.jets.jes.sec}
For the determination of the top quark mass, the momentum vectors of 
the quarks in the final state are needed.
However, the detectors measure particle jets, and their directions and 
energies are taken as a measure of the quark momentum.
While the direction of the initial quark is quite well reproduced by 
the jet direction, the correspondence between jet and quark energies
is more involved.
This correspondence is established in two steps: 
\begin{enumerate}
\item
First, the 
energy of the {\em measured jet} is related to the true energy of the
{\em particle jet}.
This step depends on detector effects and on the jet algorithm used.
\item
Second, the {\em quark} energy is inferred from the {\em particle jet}
energy.
This second step only involves the effects of fragmentation and
hadronization and is thus independent of the experimental setup.
Depending on the analysis, this relation can be established via
Monte Carlo models or via a parametrization with transfer functions.
\end{enumerate}
In this section, the correction procedures applied by the two
Tevatron experiments to obtain particle jet energies are outlined;
for details, see~\cite{bib-CDFruniijes,bib-D0runijes}.
The transition to quark energies is regarded as part of each specific
top quark mass measurement and is described later in
Sections~\ref{templatemeasurements.sec}-\ref{idmeasurements.sec}
together with the individual analyses.

The transition from measured to true particle jet energies requires
several corrections:
\begin{itemize}
\item
{\bf Energy Offset \boldmath$E_O$:}
Before corrections are made, the energy scale for the electromagnetic
calorimeter is set such that the $\Z\to\epem$ peak is correctly
reproduced, as described in Section~\ref{detcalib.leptons.sec}.
Contributions from detector
noise, energy pile-up from previous bunch crossings, additional 
interactions in the same bunch crossing (``multiple interactions''),
and the underlying event, i.e.\ reactions of partons in the proton
and antiproton other than those that initiated the $\ppbar\to\ttbar$
process, are then subtracted from the measured jet energy.
The correction for this energy offset $E_O$ depends on the jet
algorithm and parameters (e.g.\ the cone size), the pseudorapidity,
and the instantaneous luminosity.
The \dzero experiment determines it from energy densities in minimum
bias events.

\item
{\bf Calorimeter Response \boldmath$R$:}
The second correction concerns the calorimeter response.
There is no straightforward way to determine the response 
with a resonance similar to the procedure applied for electrons and
muons based on leptonic \Z decays as described in
Section~\ref{detcalib.leptons.sec}, because hadronic decays of single \W 
or \Z bosons cannot be distinguished experimentally from QCD dijet
events.
(An exception are hadronic \W decays in \ttbar events, which are 
discussed below.)

The response to hadronic jets can therefore only be measured with events
where a jet is balanced by another object for which the detector
response is known.
The \dzero experiment uses $\gamma$+jet events, taking the photon
energy scale from $\Z\to\epem$ events.
In these events, the so-called 
{\em missing $\Et$ projection fraction method}
allows to measure the calorimeter response from the $\pt$ 
imbalance~\cite{bib-D0runijes}:
For an ideal detector, the photon 
transverse momentum
$\pt^\gamma$ and the transverse momentum of the hadronic recoil
$\pt^{\rm had}$ are expected to be balanced.
However, before calibration of the calorimeter response an 
overall transverse momentum imbalance $\ptmissvec\neq\vec0$ may be observed:
\begin{equation}
\label{eq:metprojfracbalance}
    R^\gamma    \vec{p}_T^{\,\gamma}
  + R^{\rm had} \vec{p}_T^{\,\rm had}
  = - \ptmissvec
  \ .
\end{equation}
The missing transverse momentum vector is corrected
for the electromagnetic calorimeter response $R^\gamma$ determined from 
$\Z\to\epem$ events.
After that, the hadronic response is obtained as
\begin{equation}
\label{eq:metprojfrac}
    R^{\rm had}
  = 1 + \frac{\ptmissvec^{\,\,\rm corr} \cdot \vec{p}_T^{\,\gamma}}
             {\left( \vec{p}_T^{\,\gamma} \right)^2 }
  \ .
\end{equation}

In events with one photon and exactly one jet, the jet response can be
identified with the hadronic response $R^{\rm had}$.
The calorimeter response depends on the jet energy
and pseudorapidity;
in particular, the response for jets in the overlap regions between
the central and endcap calorimeters at $|\eta^{\rm jet}|\approx 1$ is different 
from that for jets fully contained in one of the calorimeters.
These effects are taken into account by measuring the response as 
a function of both pseudorapidity and estimated jet energy.
Since the energy resolution for jets is broad, the true jet energy 
in $\gamma$+jet events is estimated from the photon transverse
energy $\Et^\gamma$ and jet pseudorapidity $\eta^{\rm jet}$ as
\begin{equation}
\label{eq:eprime}
    E'
  = \Et^\gamma \cosh(\eta^{\rm jet})
  \ .
\end{equation}

The CDF experiment first measures 
the dependency of the response on the position in the detector (as
for \dzero, the
response is not expected to be uniform because of gaps between the 
individual parts of the calorimeter and because of their different
responses); after applying these {\em $\eta$ dependent corrections}
the absolute jet energy scale is determined 
from a Monte Carlo
simulation of the detector and cross-checked with 
results of the missing $\Et$ projection fraction method
described above~\cite{bib-CDFruniijes}.
The simulation is tuned to model the response to single particles by
comparing the calorimeter energy and track momentum measurements
for single tracks,
using both test beam data and CDF data taken during Tevatron \runii.
Because of the limited tracking in the forward regions, this procedure
is used for the central calorimeter only, and the forward calorimeter
response is determined relative to the one for the central
calorimeter.
The $\eta$ dependent corrections are obtained by balancing dijet events
and are shown in Figure~\ref{CDFetadependentcorrections.fig}.
Because the simulation only describes the data well for values of 
$|\eta|$ up to
about 1.4, separate corrections are derived for data and \pythia Monte
Carlo simulation; \herwig events are not used because of the large discrepancies
for $|\eta|>1.4$ and $\pt<55\,\GeV$.
An indirect determination of the response for jets, inferred from 
the momenta of the tracks within the jet, 
is shown in Figure~\ref{CDFresponse.fig}.

\begin{figure}
\begin{center}
\includegraphics[width=0.8\textwidth]{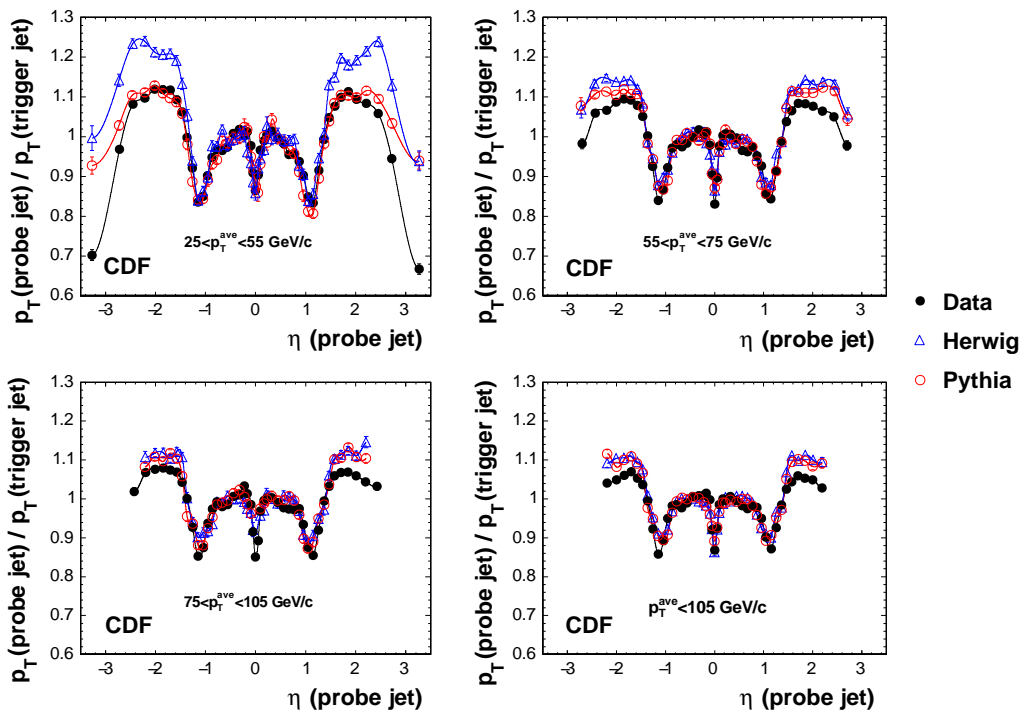}
\caption{\captionfont\label{CDFetadependentcorrections.fig}The CDF experiment uses dijet events with a trigger jet within
  $0.2<|\eta|<0.6$ to obtain $\eta$ dependent corrections to the jet
  energies.
  Shown is the ratio of the second (probe) jet \pt and the trigger
  jet \pt as a function of probe jet pseudorapidity for various
  average jet \pt regions and for data and Herwig and Pythia simulated
  events as explained in the figure~\cite{bib-CDFruniijes}.}
\end{center}
\end{figure}

\begin{figure}
\begin{center}
\includegraphics[width=0.6\textwidth]{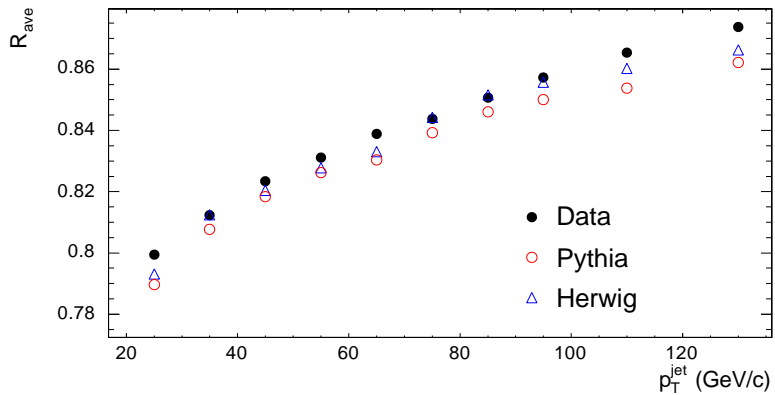}
\caption{\captionfont\label{CDFresponse.fig}The response for jets in the CDF experiment as a function of
  the jet transverse momentum, for data as well as simulated events.
  The response is determined indirectly from the track
  momenta, and the deviation of the jet response values from $1.0$ is
  due to the calorimeter response to hadrons being smaller than 
  unity~\cite{bib-CDFruniijes}.}
\end{center}
\end{figure}

\item
{\bf Showering Correction \boldmath$S$:}
The first two corrections are specific to each experiment and yield jet
energies that are independent of the experimental setup, but still 
depend on the jet finding algorithm.
In general, not all energy deposits belonging to the jet are 
assigned to it by the
jet algorithm, and thus a fraction of the energy is not accounted
for in the measured jet energy.
The energy fraction assigned to the jet is a function of the jet
algorithm and its parameters, the jet energy itself, and the 
pseudorapidity.
\end{itemize}

The particle jet energy $E^{\rm corr}$ is thus obtained from the
raw measured energy $E^{\rm meas}$ as
\begin{equation}
\label{eq:jescorrection}
    E^{\rm corr}
  = \frac{ E^{\rm meas}(a) - E_O\left(a, {\cal L}, \eta \right) }
         {   R\left(a, E^{\rm meas}, \eta \right) 
           \ S\left(a, E^{\rm meas}, \eta \right) }
  \ ,
\end{equation}
where $E_O$, $R$, and $S$ are the three corrections described
above, depending on the jet algorithm and its parameters, denoted by
$a$, the instantaneous luminosity ${\cal L}$, the pseudorapidity
$\eta$, and the jet energy itself.
An example of the different contributions to the uncertainty on the
jet energy scale is given in \Fref{CDFjesunc.fig} for the CDF
experiment.
\begin{figure}
\begin{center}
\includegraphics[width=0.6\textwidth]{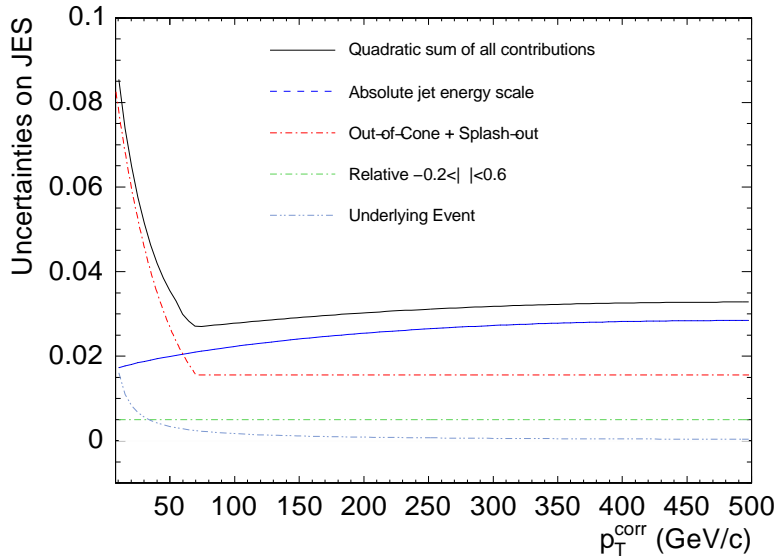}
\caption{\captionfont\label{CDFjesunc.fig}The different contributions
  to the jet energy scale
  uncertainty at the CDF experiment as a function of the corrected jet
  $\pt$~\cite{bib-CDFruniijes}.}
\end{center}
\end{figure}
This uncertainty on the overall energy scale for jets leads to the dominant 
systematic error on the top quark mass unless the scale is determined
simultaneously with the top quark mass from the same events.

Hadronic \W decays in \ttbar events provide a means of calibrating the 
energy scale for light-quark jets with the same event sample for which the 
calibration is needed to measure the top quark mass.
Such an {\em in situ} calibration is very attractive experimentally
since one becomes independent of uncertainties due to e.g.\ the photon
selection, the jet flavor composition of $\gamma$+jet events, or Monte Carlo
simulation.
However, at the Tevatron the size of the \ttbar event samples is
not sufficient to calibrate the jet energy scale
as a function of 
pseudorapidity and energy.
Therefore, all jet energy corrections described above are still
applied, and in situ calibration is then used only to determine the
overall energy scale for all jets.
With this approach, the largest part of the jet energy scale
systematic error on the top quark mass can still be absorbed in an
increased statistical uncertainty.
If desired, the information on the overall scale parameter 
from $\gamma$+jet events or Monte Carlo
calibration can be used as an external
prior to further reduce the uncertainty.
Analyses using in situ calibration are presented in detail in 
Sections~\ref{templatemeasurements.kinrec.sec}, \ref{memeasurements.sec},
and~\ref{idmeasurements.sec}.

\subsubsection{Bottom-Quark Jet Energy Scale}
\label{detcalib.jets.bjes.sec}
Even for a given momentum of the parton initiating a jet,
both the frequency with which the various hadron species are produced
and their momentum spectra are different for quark jets of different
flavor or gluon jets.
The experiments in general distinguish between bottom-quark and 
light-flavor jets, where the latter includes any jet that is not initiated
by a bottom quark.

Because the particle momentum spectrum differs, the ratio of
electromagnetic to hadronic energy is different, thus leading to a 
different response for bottom-quark and light-flavor jets.
A further correction is necessary for jets containing neutrinos 
which are not measured at all, and for muons which only deposit a 
small fraction of their energy in the calorimeter.
This correction is relevant for jets
containing semileptonic heavy hadron decays.
An explicit correction can be applied for jets in which the charged
lepton is identified inside the jet (only muons are used at the 
moment).
The response for bottom-quark jets without an identified muon will 
still be shifted due to unidentified semimuonic and semielectronic 
heavy hadron decays.
The showering correction will in principle be different for bottom-
and light-flavor jets as well, due to the mass of the decaying bottom
hadron.

In practice, the full jet energy corrections are derived as described
above for light-flavor jets (for example, most $\gamma$+jet events
will not contain bottom-quark jets).
For bottom-quark jets, additional corrections are applied to this jet
energy scale, and systematic uncertainties are quoted both for the 
overall (light-flavor) jet energy scale and for the relative scale
for bottom-quark and light-flavor jets; for details, see for 
example~\cite{Abulencia:2005aj,bib-me}.

\subsubsection[Jet Energy Scale Corrections Specific to \ttbar Events]{\boldmath Jet Energy Scale Corrections Specific to \ttbar Events}
\label{detcalib.jets.ttbarcorrections.sec}
In addition to the general corrections described so far, the CDF
experiment applies specific corrections to the energy
scale of jets in \ttbar events.
These corrections account for the \pt spectra and jet
flavors encountered in \ttbar events, which are different from those
of the events for which the general corrections have been derived
(the jets in \ttbar events are initiated by quarks, two of which are
bottom quarks, and one charm-quark jet is expected in every second
hadronic \W decay, while gluon jets are only expected if additional 
radiation occurs).
The corrections are derived from simulated events as described 
in~\cite{Abulencia:2005aj}.
Light- and bottom-quark jets are corrected differently, and thus
this last correction can only be applied once a jet is assigned to a 
final parton.
In contrast, the \dzero experiment absorbs these corrections into the 
transfer functions used in the Matrix Element and Ideogram analyses.
For corrections that are identical for both data and simulation, 
the measured top quark mass is not systematically shifted since the
measurement calibration is based on the simulation.
The correction may however lead to an improvement of the statistical 
sensitivity in template-based measurements where no \mtop dependent
likelihood is derived on an event-by-event basis.

\subsubsection{Relative Jet Energy Scale Between Data and Simulation}
\label{detcalib.jets.datamcjes.sec}
In all top quark mass measurements, simulated events are used 
to calibrate the measurement technique.
Thus, if the corrected jet energies in the data systematically do not
reproduce the particle jet energies, and the same effect is present
in Monte Carlo simulated events, the calibration procedure assures
that the top quark mass is still measured correctly.
Consequently, only uncertainties on the 
{\em relative data/Monte Carlo jet energy scale} enter the systematic
error on the top quark mass.

\subsubsection{Jet Energy Resolution}
\label{detcalib.jets.jer.sec}
For a given event sample and analysis technique the 
statistical uncertainty on the top quark mass is dominated by the jet energy
resolution.
To illustrate this, events with a top quark involving a leptonically decaying
\W have been passed through the full simulation of the \dzero
detector, and the effect of the detector resolution on the
reconstructed top quark mass distribution is studied.
Of the three top quark decay products, either for the bottom quark or
the charged lepton the reconstructed momentum vector is taken, while
the true momentum vectors are used for the other two decay products.
The results in the left plot of 
\Fref{mtopresolution-emujet.fig} show that 
the inclusion of the jet resolution has the largest effect on
the distribution of the reconstructed top quark mass.
The tails visible when using the reconstructed muon momentum are due
to the fact that the momentum resolution degrades with increasing
$\pt$; the effect on the reconstructed top quark mass distribution 
is demonstrated in the right plot of \Fref{mtopresolution-emujet.fig}.
\begin{figure}
\begin{center}
\includegraphics[width=0.45\textwidth]{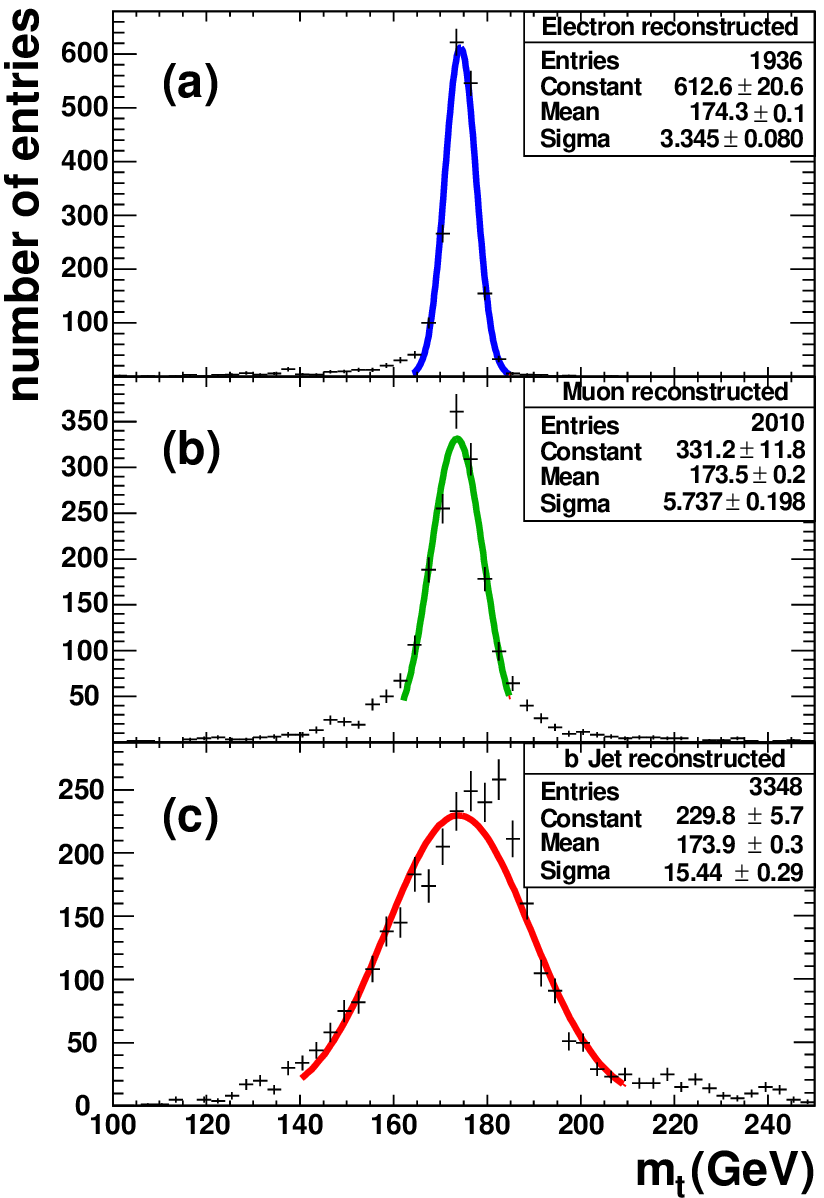}
\includegraphics[width=0.462\textwidth]{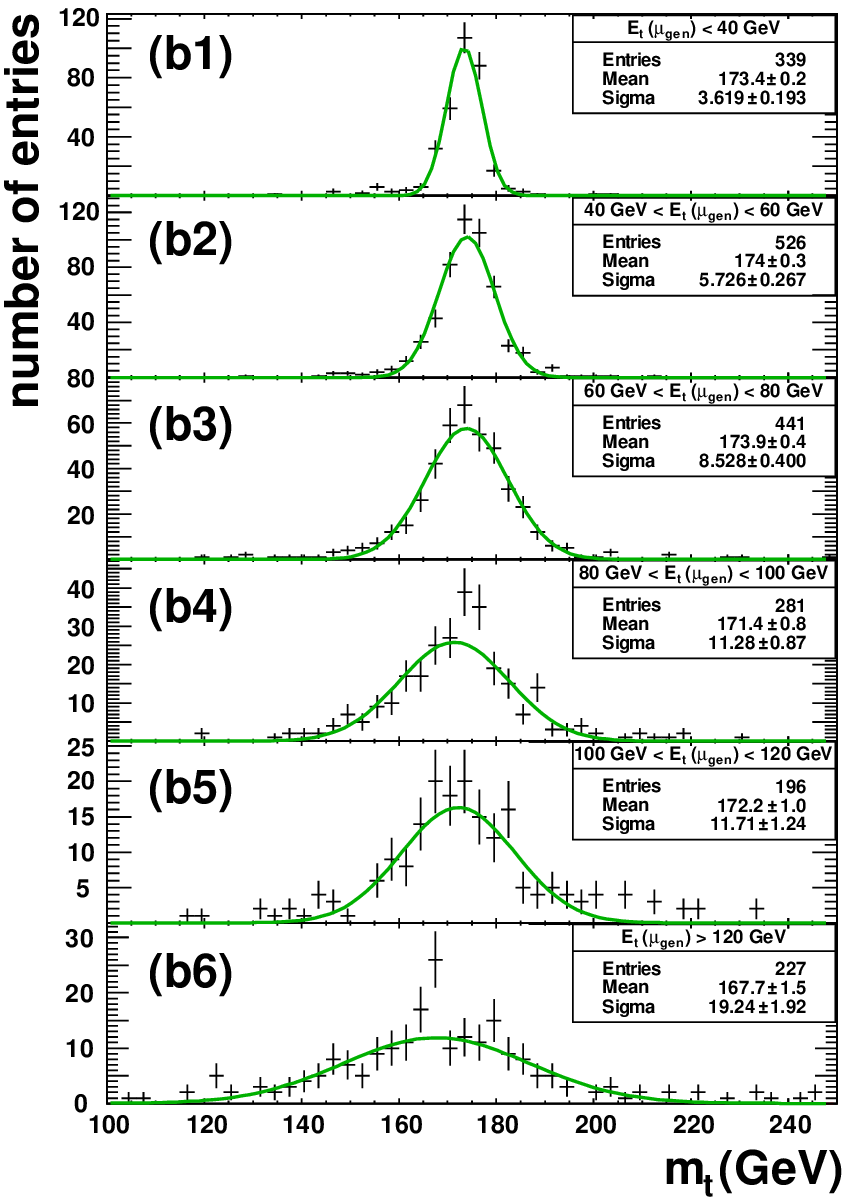}
\caption{\captionfont\label{mtopresolution-emujet.fig}
Simulated top quark mass distributions for top quark decays at
$\mtop=175\,\GeV$ involving a leptonic \W decay,
taking the momentum as reconstructed with the \dzero detector 
for (a) the electron, (b) the muon, or (c) the \bquark jet, and
the true momenta for the other two decay products.  The effect for 
muons of different true
transverse momentum is shown separately in plots (b1) to (b6).
The mean and width of the top quark mass distribution are determined
with a Gaussian fit and given in units of GeV.}
\end{center}
\end{figure}

The CDF experiment has tuned the simulation so that not only 
the mean shower energy in single track data is reproduced (which is
relevant for the overall jet energy scale, see
Section~\ref{detcalib.jets.jes.sec}), but also the parameters
describing the shower shape~\cite{bib-CDFruniijes}.
Consequently, the jet energy resolution is taken from the 
simulation.

At the \dzero experiment, the jet energy resolution is measured
from $\gamma$+jet (below a jet $\Et$ of 50~GeV) and dijet events
(above 50~GeV).
The same measurement is performed in data and Monte Carlo, and 
the resolution of simulated jets is smeared to reproduce the data.
The measured top quark mass depends on
the modeling of the jet resolution because the event selection 
in general requires a minimum jet $\Et$.
Furthermore, an accurate modeling of the resolution allows the
observed statistical error to be compared with expectations from 
the simulation.

\subsection{Efficiency of Bottom-Quark Jet Identification}
\label{detcalib.btagging.sec}
When the efficiency to identify \bquark-quark jets is defined for 
taggable jets (cf.\ the definition of taggability in
Section~\ref{reco.reco.btagging.sec}),
it becomes independent from detector inefficiencies.
The {\em taggability} is related to the efficiency with which tracks
are reconstructed.
To take into account the geometrical acceptance of the silicon detector, the
\dzero experiment measures the taggability of jets in bins
of the quantity 
$\left|z^{\rm pv}\right|\times{\rm sign}\left(z^{\rm pv}\eta^{\rm jet}\right)$
where $z^{\rm pv}$ and $\eta^{\rm jet}$ are the $z$ position of the
primary vertex and the pseudorapidity of the jet,
respectively~\cite{bib-D0ljetsbtagxspaper}.
The measured taggability is parametrized as a function of jet
transverse momentum and pseudorapidity, and the relative taggabilities
of light, charm, and bottom jets are determined from the simulation.
The results of the study are shown in Figure~\ref{taggability.fig}.

\begin{figure}
\begin{center}
\includegraphics[width=0.45\textwidth]{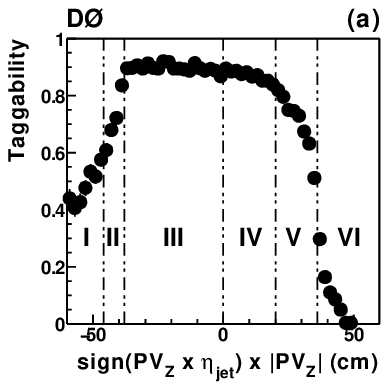}
\includegraphics[width=0.6\textwidth]{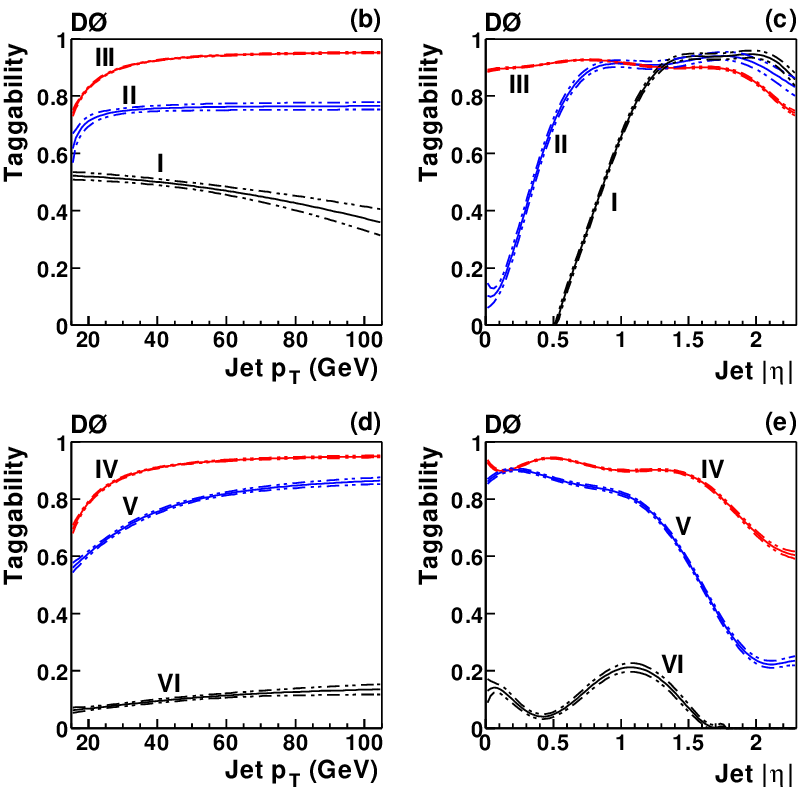}
\includegraphics[width=0.6\textwidth]{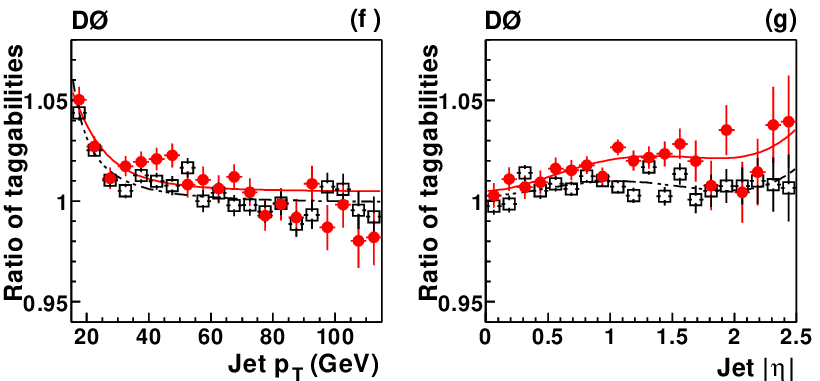}
\caption{\captionfont\label{taggability.fig}Jet taggability measurements by the 
  \dzero experiment~\cite{bib-D0ljetsbtagxspaper}.  (a): Definition of
  taggability regions.  (b)-(e): Taggability measured in the data 
  as a function of jet \pt ((b), (d)) and jet $|\eta|$ ((c), (e)) for 
  jets in the various regions defined in (a).  (f), (g): Taggability
  ratios of \bquark to light (full circles) and \cquark to light (open
  squares) jets determined from simulated events.}
\end{center}
\end{figure}

Both CDF and \dzero determine the {\em $b$-tagging efficiency} for 
bottom-quark jets and the mistag rate for light-flavor jets from 
data, with additional corrections based on the
simulation~\cite{bib-D0ljetsbtagxspaper,bib-CDFljetsbtagxspaper}.
The efficiency for bottom-quark jets is measured on a dijet event
sample whose bottom-quark content is enhanced by requiring the 
presence of an electron (CDF) or a muon (\dzero) within one of the
jets as an indication of a semileptonic heavy hadron decay.

The CDF experiment determines the bottom-quark content of their 
calibration sample by reconstructing $\PDz\to\PKm\Ppip$ decays
or muons in the jet containing the electron, both of which are 
additional signatures for a heavy hadron decay.
The \dzero experiment uses the transverse momentum spectrum of the 
muon relative to the axis of its jet to measure the bottom-quark
content of the calibration sample.

Both CDF and \dzero thus measure the $b$-tagging efficiency for 
bottom-quark jets with a semileptonic decay.
Corrections to obtain the efficiency for inclusive bottom-quark jets
are derived from the simulation.
The CDF experiment also takes the dependence of the $b$-tagging efficiency
on jet energy, pseudorapidity, and track multiplicity from the 
simulation, while the overall normalization is determined from the data
measurement.

Both CDF and \dzero measure the light-flavor tagging rate (mistag
rate) on the data using the rate of jets that contain a secondary
vertex with negative decay length significance
$L_{xy}/\sigma(L_{xy})$ (cf.\ Section~\ref{reco.reco.btagging.sec}).
After correction for the contribution of heavy-flavor jets to such
tags and the presence of long-lived particles in light-flavor jets,
this rate is a measure of the probability that a light-flavor jet
gives a secondary vertex tag with positive $L_{xy}/\sigma(L_{xy})$.
The $b$-tagging efficiency for charm-quark jets cannot easily be determined
from data, and thus the ratio of efficiencies for charm- and
bottom-quark jets is taken from the simulation.

The $b$-tagging efficiencies of the CDF and \dzero secondary vertex
tagging algorithms for taggable jets are shown in
\Fref{btagefficiency.fig}.
\begin{figure}
\begin{center}
\includegraphics[width=0.55\textwidth]{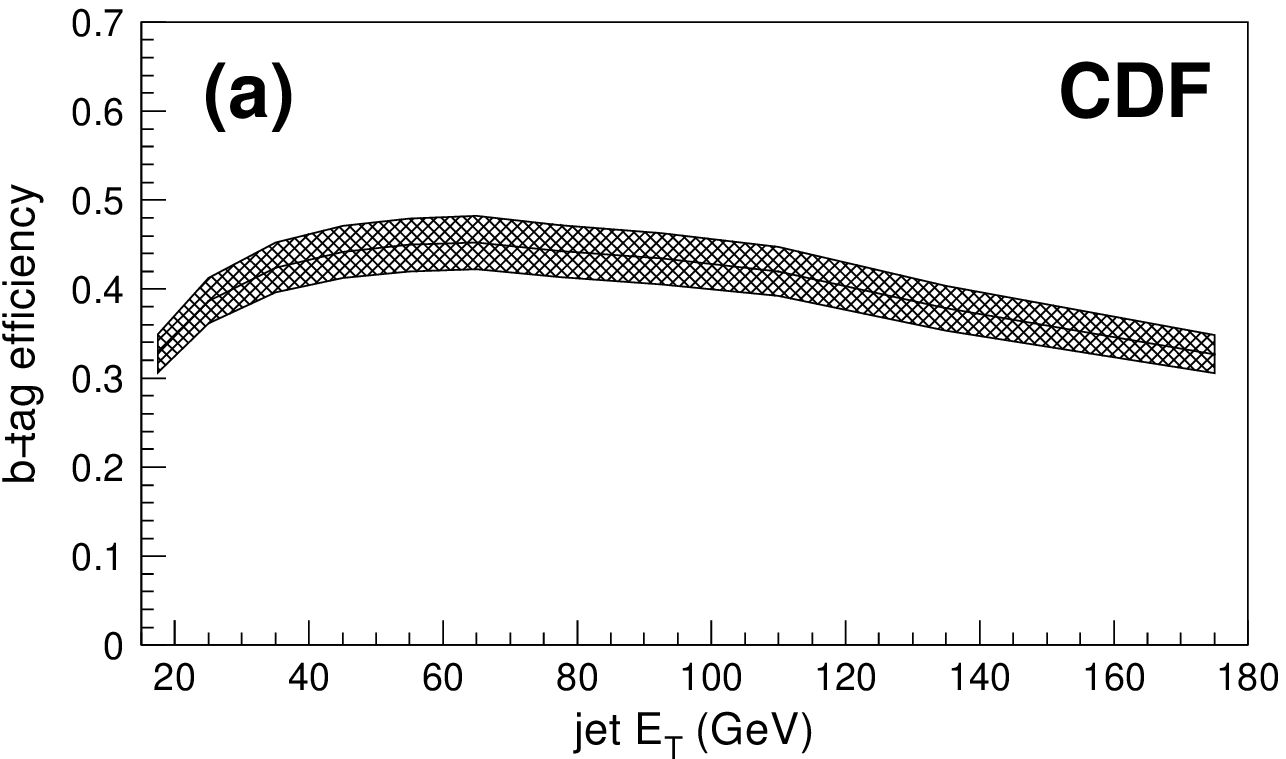}
\includegraphics[width=0.6\textwidth]{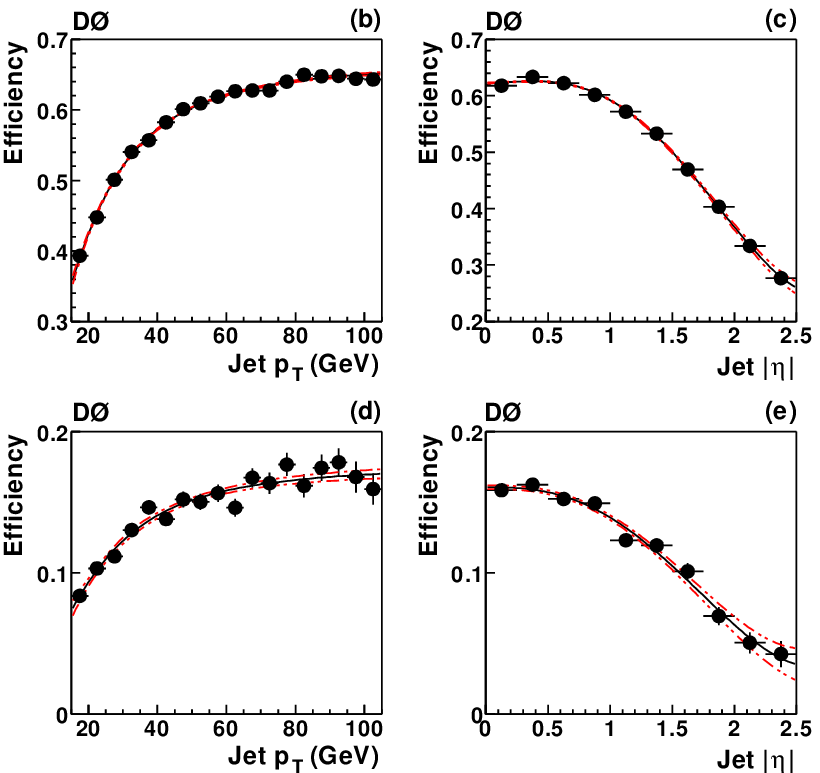}
\caption{\captionfont\label{btagefficiency.fig}Secondary vertex $b$-tagging efficiencies for taggable jets.
  (a): CDF experiment~\cite{Abulencia:2005aj}.  Efficiency for bottom-quark jets with $|\eta|<1$
  as a function of jet $\Et$ together with the $\pm1\sigma$
  uncertainty range.
  (b)-(e): \dzero experiment~\cite{bib-D0ljetsbtagxspaper}.
  Measured efficiency for bottom-quark jets as a function of (b) jet
  \pt and (c) jet $|\eta|$ and for charm-quark jets as a function of 
  (d) jet \pt and (e) jet $|\eta|$; the $\pm1\sigma$ uncertainties of
  the fits are
  indicated by the red dotted lines.}
\end{center}
\end{figure}

\clearpage
\section{Methods for Top Quark Mass Measurements}
\label{methods.sec}
\begin{center}
\begin{tabular}{p{15cm}}
{\it So far, the report described the selection of top-quark events
  and the calibration of the detectors.
  This section now introduces a classification of the methods to determine the
  top quark mass from the events selected.
  Following this classification, the subsequent
  sections then give details about each of the methods
  together with concrete examples.}
\end{tabular}
\end{center}

Experimental results for the top quark mass can be grouped
according to the \ttbar decay channel analysed, 
cf.\ Section~\ref{toprodec.topeventtopologies.sec}.
A comparison between the measurements in different channels
allows to search for an indication of differences and 
thus for new physics effects beyond the Standard Model.

In this section, another classification is introduced according to the 
measurement technique applied.
Even though top quarks decay before hadronization, the information on
the top quark mass is still diluted
in the events measured in the detector by physics effects
(initial- and final-state radiation and hadronization) and the 
detector resolution.
In broad terms, the following different approaches have been followed by the
Tevatron experiments to deal with this complication:
\paragraph{\bf Template Method:}
\label{methods.templatemethod.sec}
A measurement quantity per event called {\em estimator} (mostly a
single number, but in some measurements a vector of numbers) that is 
correlated with the top quark mass is computed per event. 
Any measured quantity in the event that is correlated with the mass 
of the decaying top quark can be used as estimator in the analysis.
In all cases, it is mandatory to understand the exact top quark mass 
dependence of the distribution of this quantity.

In lepton+jets and \alljets events where enough decay products are 
reconstructed, the smallest {\em statistical} error is obtained if 
the invariant masses of the two top quarks are explicitly
reconstructed to obtain the estimator; this is however not necessary
(and not possible in dilepton events without additional assumptions 
because of the two neutrinos in the final state).
In addition, alternative techniques have been developed for
lepton+jets events to reduce the sensitivity
to {\em systematic} errors by a careful selection of the estimator.
These are already being explored at the Tevatron and will become
much more important at the LHC.

The distribution of the estimator
for the set of selected data events is compared with the 
expected distribution for various assumed values of the top quark
mass.  
This so-called {\em template}
distribution is generated using simulated signal and background
events, taking efficiencies and the 
relevant cross sections into
account.  
The values of the estimator in the data events are then compared
to the template distributions in a fit to determine the
top quark mass.
To increase the statistical power of the method, the event sample is
often divided into subsamples with different signal purity, for
example according to the number of \bquark-tagged jets per event.
\paragraph{\bf Matrix Element Method:}
\label{methods.matrixelementmethod.sec}
For each selected event, the likelihood to observe it is 
calculated as a function of the assumed top quark mass.  
To this end, all possible reactions yielding final states that could
have led to the observed event are considered.
An integration is performed
over all possible momentum configurations of the final state
particles for all relevant reactions.
In this integration, the probability of the colliding partons to
have a given momentum fraction of the proton or antiproton is taken into
account using the appropriate PDFs.
Similarly, the likelihood to obtain the
detector measurement for an assumed final state is accounted for by a
transfer function that relates an assumed final-state
momentum configuration to the measured quantities in the detector.
Here, a choice can be made which detector measurements to use in the 
analysis; for example, the measured missing transverse momentum is not
explicitly used in the Matrix Element measurements in the lepton+jets
channel.

The Matrix Element method accounts for the fact that the accuracy of
the information on the
top quark mass contained in different events is in general different:
\begin{list}{$\bullet$}{\setlength{\itemsep}{0.5ex}
                        \setlength{\parsep}{0ex}
                        \setlength{\topsep}{0ex}}
\item
Depending on the event kinematics and characteristics like the quality
of a \bquark tag, some selected events have a higher likelihood of being
a \ttbar event for a certain top quark mass than others.
\item
Depending on the energies and directions of the final-state particles,
the resolutions of the measured momenta and thus of the
top quark mass generally differ between events.
\end{list}
Because a likelihood as a function of assumed top quark mass is calculated
separately for each event, and the likelihood for the entire event 
sample is obtained as the product of the individual event
likelihoods, each event contributes to the measurement with its
appropriate weight, and the Matrix Element method minimizes the
statistical uncertainty of the measurement.

However, the integration over final-state momenta is complex, and it is
impossible in practice to use full detector simulation to evaluate the
transfer function during the integration.
Simplifying assumptions are thus made in the integration, and the
measurement is then calibrated using fully simulated events.
Still, a Matrix Element measurement requires significantly 
more computation time than a template analysis.
The Matrix Element method was first used by \dzero for Tevatron \runi
data~\cite{bib-nature}, 
where it yielded the single most precise measurement, and it
also currently yields the single most precise measurement at
\runii~\cite{bib-CDFljetsme}.
\paragraph{Ideogram Method:}
\label{methods.ideogrammethod.sec}
The Ideogram method can be regarded as an approximation to the 
Matrix Element method.
It does not make use of the full kinematic characteristics of each
selected event, but only relies on information about the invariant
masses of the top and antitop quarks and \W bosons.
The description of wrong jet-parton assignments and of background
events is even further simplified.

The statistical sensitivity is not substantially reduced relative to
that of the Matrix Element method if the 
signal to background ratio is large.
In particular, the method retains the benefits of using a per-event
likelihood as a function of assumed top quark mass.
The computations are however simpler than for the 
Matrix Element method, making the Ideogram method a candidate for 
future analysis of large-statistics data samples.

\paragraph{}

If the measured jet energies are used in the computation of the
estimator or in the transfer function in order to minimize the 
statistical uncertainty, the dominant systematic uncertainty is due
to the hadronic jet energy scale.
The dependence of the top quark mass on external jet energy scale
measurements and thus the associated systematic uncertainty is
significantly reduced if an overall jet energy scale
factor can be fitted simultaneously with the top quark mass using the
same \ttbar events.
The information on the jet energy scale comes mainly from hadronic
$\W\to\qqbarprime$ decays where a constraint to the 
known\footnote{The \W mass is known from LEP and the
Tevatron~\cite{bib-pdg} with a precision far beyond what is needed
for the determination of the hadronic energy scale.} \W mass can be
used.
This in situ calibration also reduces the
systematic correlation between measurements on different event
samples, which further improves the combination of Tevatron results.

\paragraph{}

In Section~\ref{templatemeasurements.sec} a description of the
estimators used in the various template measurements at the Tevatron
is given.
Section~\ref{memeasurements.sec} gives details about the 
computation of the mass-dependent event likelihood in the
measurements using the Matrix Element method, and
Section~\ref{idmeasurements.sec} outlines the analyses using the
Ideogram method.
The final top quark mass determination from an event sample is described in
Section~\ref{massfit.sec}, and systematic uncertainties 
are discussed in Section~\ref{systuncs.sec}.
Since there is no no a priori fundamental difference between 
the CDF and \dzero experiments on the level of reconstructed \ttbar
decay products, 
the following sections describe the measurement
techniques based on individual analyses as examples.
A full account of all results is given in Section~\ref{resinterp.sec}.

\clearpage
\section{The Template Measurement Method}
\label{templatemeasurements.sec}
\begin{center}
\begin{tabular}{p{15cm}}
{\it The template method is the ``standard'' technique used at the
  Tevatron to determine the top quark mass from the selected \ttbar
  events.
  This section introduces the key concept of the template method,
  the per-event estimator of the top quark mass.
  Based on recent Tevatron measurements as examples, the calculation
  of estimators for the various \ttbar event topologies is
  described.}
\end{tabular}
\end{center}

In this section, template-based measurements of the top quark mass
are reviewed.
The general description of the analyses is complemented with
concrete examples from recent Tevatron measurements.

Measurements using the template method are based on the determination
of an estimator for each event, which is a quantity that captures the
information about the top quark mass contained in the event.
Depending on the event topology analyzed, but also depending on the
relative importance of statistical and systematic errors, various
choices of estimators are possible.
In addition, a second estimator is introduced when the jet energy
scale is measured simultaneously with the top quark mass.
The different choices of estimators used in Tevatron analyses and 
their computation are described in this section.

The values of the estimator
for the selected data events are compared with the 
expected distribution as a function of the true top quark
mass.  
These expected estimator
distributions (templates) are generated using simulated signal and background
events for a discrete number of true values of the top quark
mass.  
Trigger and selection efficiencies and the 
appropriate cross sections (as a function of the top quark mass) are
taken into account.  
The comparison of the sum of signal and background templates for various
top quark mass hypotheses with the observed estimator
distribution in the data then yields the likelihood to observe this
event sample as a function of assumed top quark mass.
The top quark mass is usually extracted in a fit of the measured
events to the generated template distributions, for which
a continuous parametrization of the templates as a
function of true top quark mass is obtained.
This parametrization is also
described in this section.
The fitting procedure
to determine the top quark mass is then outlined in 
Section~\ref{massfit.sec}.

To the extent that the simulation used to derive the template 
distributions is accurate, the fit yields a measurement of the 
true top quark mass.
Possible deficiencies of the simulation have to be studied and 
corresponding systematic uncertainties assigned, as described in 
general in Section~\ref{systuncs.sec}.

To extract mass information from the top quark decay products,
any measured quantity in the event that is correlated with the mass 
of the decaying top quark can be used as estimator.
In \ljets and \alljets events, taking the explicitly reconstructed
invariant top quark mass as estimator allows to extract
the largest statistical information per event.
An example for the \ljets channel is described in detail
in Section~\ref{templatemeasurements.kinrec.sec}.
Alternative estimators have been developed to reduce the sensitivity
to systematic uncertainties by a careful selection of the quantity
that is used to determine the top quark mass.
For an example of a Tevatron study, see
Section~\ref{templatemeasurements.nocalo.sec}.
Such techniques will become
more important at the LHC with its much larger \ttbar data sets.
The \alljets decay channel is described in
Section~\ref{templatemeasurements.allhad.sec}.
Dilepton events do not allow a full reconstruction of the event 
kinematics if the top quark mass is assumed to be unknown.
Therefore, techniques have been developed to determine a likely
kinematic configuration by making additional assumptions, and the
top quark mass reconstructed for this configuration is then used 
as the estimator, see 
Section~\ref{templatemeasurements.dilepton.sec}.

\subsection{Full Kinematic Reconstruction of Lepton+Jets Events}
\label{templatemeasurements.kinrec.sec}

The full kinematic reconstruction of \ljets events with a kinematic
fit that assumes a \ttbar event configuration yields a fitted top quark 
mass per event that can be used as estimator of the true top quark mass.
This technique is most commonly used in template based measurements
in the \ljets decay channel at the Tevatron.

The top quark mass information is then mostly based on the measured
jets.
Both CDF and \dzero determine a reference jet energy scale, including $\eta$
and $\Et$ dependence, as described in Section~\ref{detcalib.jets.sec},
and correct jet energies and missing transverse momentum according 
to this scale.
With the current Tevatron statistics, measurements in the \ljets
channel would be systematically limited by the uncertainty on the overall 
jet energy scale factor \JES, unless this factor is determined 
in situ from the same data as well.
The reconstructed hadronic \W mass is highly correlated with the jet energy
scale, but not with the reconstructed top quark mass.
It can therefore be used as an additional
estimator to measure an overall deviation from the reference jet energy
scale.

As an example, the CDF template 
measurement~\cite{Abulencia:2005aj,bib-CDFmtop_ljetstemplate_prel}
is described in detail here.
\paragraph{Event Selection:}
\label{templatemeasurements.kinrec.evtsel.sec}
The \ttbar candidate events are selected by requiring the presence of 
one electron or muon, missing transverse energy, and four or more
jets, as outlined in Section~\ref{reco.reco.sec}.
The electron (or muon) must have a transverse energy (or momentum)
larger than $20\,\GeV$ and pass quality and isolation criteria,
and missing transverse energy of at least $20\,\GeV$ is required.
The transverse energy requirements on the jets (reconstructed with a 
cone algorithm with radius $\DeltaR=0.4$ as described in 
Section~\ref{reco.reco.jetsel.sec}) depend on the event category
described below.

In hadronic \W decays almost exclusively light-quark jets 
(including charm jets)
are produced; thus $b$-tagged jets are likely to be direct top-quark 
decay products.
This analysis does not consider $b$-tagged jets as \W decay products.
Thus, the number of jet-parton assignments is reduced 
in events with $b$-tagged jets.
As it is then more likely that the 
kinematic fit selects the correct assignment, the reconstructed
mass distribution becomes sharper, and these events contribute
more mass information.
Furthermore, the signal to background ratio is larger in events with
$b$ tags.
Finally, the jets in background events have mostly low transverse
energies.
Consequently, the events are grouped into four categories with 
different jet $\Et$ cuts to improve
the statistical error of the measurement.
The criteria for the classification are summarized in
\Tref{CDFljetstemplatecategories.table}.

\begin{table}
\begin{center}
\begin{tabular}{lcccc}
\hline\hline
 Category & 2-tag & 1-tag (T) & 1-tag (L) & 0-tag \\
\hline\hline
 $b$ tagged jets                  & $n_b\ge2$                  
                                  & $n_b=1$
                                  & $n_b=1$
                                  & $n_b=0$ \\
 $\Et$ of three leading jets      & $\Et\!>\!15\,\GeV$
                                  & $\Et\!>\!15\,\GeV$
                                  & $\Et\!>\!15\,\GeV$
                                  & $\Et\!>\!21\,\GeV$ \\
 $\Et$ of fourth jet              & $\Et\!>\!\enspace8\,\GeV$
                                  & $\Et\!>\!15\,\GeV$ 
                                  & $\Et\!>\!\enspace8\,\GeV$
                                  & $\Et\!>\!21\,\GeV$ \\
\hline\hline
 expected $S:B$ ratio             & $10.6\,:\,1$
                                  & $3.7\,:\,1$
                                  & $1.1\,:\,1$
                                  & no constraint used \\
\hline\hline
\end{tabular}
\caption{\captionfont\label{CDFljetstemplatecategories.table}CDF lepton+jets template measurement~\cite{Abulencia:2005aj}: 
  Event selection requirements for the four categories 
  together with
  the expected ratios of signal and background events used as
  constraints in the measurement.
  Events in the 1-tag (T) category are excluded from the 1-tag (L)
  category.}
\end{center}
\end{table}

\paragraph{Estimator of the Top Quark Mass:}
\label{templatemeasurements.kinrec.mtopestimator.sec}
Every selected event is subjected to a kinematic fit, and the 
jet-parton assignment that yields the best $\chi^2$ for a
\ttbar hypothesis is chosen to compute the estimator of the top quark
mass.
The fit uses the reconstructed charged
lepton, the four highest $\Et$ jets, and the unclustered 
momentum\footnote{The presence of an energetic neutrino is signaled
by large missing transverse energy.  However, the missing 
transverse momentum vector is 
derived experimentally as the vector sum of the momenta of all
energy deposits in the calorimeter, and thus the uncertainty on the 
missing transverse momentum depends on the jet activity in the event.
The uncertainty on the unclustered momentum is however approximately
independent of the rest of the event.  Therefore the unclustered
momentum is used in the kinematic reconstruction.} as
inputs.
Of the 24 possible assignments of four jets to four final-state
partons, 12 need not be considered as they correspond to an
interchange of the two jets assumed to come from the hadronic \W
decay, yielding identical reconstructed invariant masses and thus
no change to the $\chi^2$ value described below.
For each of the remaining 12 assignments, two possible
solutions for the $z$ component of the neutrino momentum exist
for a given value of the \W mass.
Therefore, 24 different combinations need to be considered
in the 0-tag
event category; there are 6 combinations for events with one 
$b$-tagged jet; for 2-tag events, two combinations exist.

The following $\chi^2$ is minimized for all different
combinations:
\begin{eqnarray}
  \nonumber
    \chi^2
  & = &
      \mathop{\sum_{i=\rm lepton,}^{}}_{\rm 4 jets}
        \frac{\left( \pt^{i,\ {\rm fit}} - \pt^{i,\ {\rm meas}} \right)^2}
             {\sigma_i^2}
    + \sum_{j=x,\,y}
        \frac{\left( {\ptuncl}^{j,\ {\rm fit}} - {\ptuncl}^{j,\ {\rm meas}} \right)^2}
             {\sigma_j^2} \\
  & &
    + \frac{\left( m_{\ell\nu}^{\rm fit} - \MW \right)^2}
           {\GW^2}
    + \frac{\left( m_{jj}^{\rm fit} - \MW \right)^2}
           {\GW^2}
    + \frac{\left( m_{b\ell\nu}^{\rm fit} - \mtop^{\rm reco} \right)^2}
           {\Gtop^2}
    + \frac{\left( m_{bjj}^{\rm fit} - \mtop^{\rm reco} \right)^2}
           {\Gtop^2}
  \ .
  \label{CDFljetstemplatechisq.eqn}
\end{eqnarray}
Here, the symbol $\pt^{i,\ {\rm meas}}$ denotes the transverse momenta
of the charged lepton and four highest $\Et$ jets as measured in the
detector, and the $\sigma_i$ are the corresponding uncertainties.
The jet and lepton angles are assumed to be well-measured and are not 
varied in the fit.
The transverse momentum varied in the fit is called $\pt^{i,\ {\rm fit}}$.
Similarly, the measured unclustered momentum components along $x$ and
$y$, their uncertainties, and the fitted unclustered momentum
enter the $\chi^2$ in the second term.
The masses of the two \W bosons and the two top quarks in an event
need not be equal, but can vary around the \W boson mass,
$\MW=80.42\,\GeV$, and the reconstructed top quark mass, 
$\mtop^{\rm reco}$, according to 
the decay widths.
The Breit-Wigner resonances are approximated with Gaussians with 
widths $\GW$ and $\Gtop$.
In each step of the 
minimization procedure the calculated masses $m^{\rm fit}$ 
are compared with $\MW$ and $\mtop^{\rm reco}$.
Note that the quantity $\mtop^{\rm reco}$
extracted from each event is not a direct measurement of 
the top quark mass; rather, this quantity is the 
estimator whose distribution for all data events is then compared
with the corresponding distribution in simulated samples.
To obtain different values for the two neutrino solutions, the two
possible values for the neutrino $z$ momentum are computed assuming
the nominal \W mass and are used to initialize the fit.

For each event, the estimator $\mtop^{\rm reco}$ is taken to be
the value obtained from the combination that yields the smallest
$\chi^2$.
If this $\chi^2$ is greater than 9, the event is rejected altogether
from the top quark mass fit.
The $\mtop^{\rm reco}$ distributions for \ttbar
signal events with $\mtop=178\,\GeV$ and the default jet energy scale
are shown
in Figure~\ref{CDFsignalmtoptemplates.fig} for all four event
categories.
It is evident that events in the 2-tag category have the best
$\mtop^{\rm reco}$ resolution.

\begin{figure}[h]
\begin{center}
\includegraphics[width=0.8\textwidth]{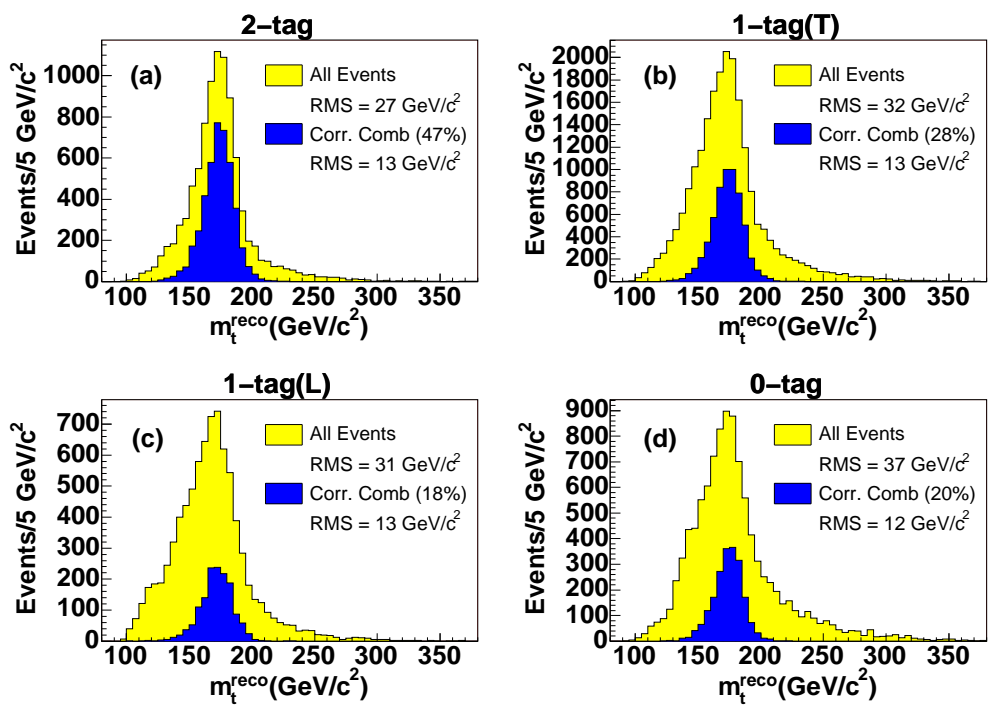}
\caption{\captionfont\label{CDFsignalmtoptemplates.fig}CDF lepton+jets template measurement~\cite{Abulencia:2005aj}: 
Template $\mtop^{\rm reco}$ distributions for simulated signal events 
with $\mtop=178\,\GeV$ and the default jet energy scale in the 
2-tag (a), 1-tag(T) (b), 1-tag(L) (c), and 0-tag (d) categories
defined in Table~\ref{CDFljetstemplatecategories.table}.
The dark blue areas show the distributions for events where the
correct jet-parton assignment has been chosen.}
\end{center}
\end{figure}

\paragraph{Estimator of the Jet Energy Scale:}
\label{templatemeasurements.kinrec.jesestimator.sec}
To determine an estimator of the jet energy scale, no kinematic
fit is applied.
The masses $m_{jj}$ of all dijet combinations that do not
involve a $b$-tagged jet are considered.
There are between one (events in the 2-tag category) and 6 (0-tag
category) such combinations per event.
The $m_{jj}$ distributions for \ttbar
signal events with $\mtop=178\,\GeV$ and the default jet energy scale 
are shown 
in Figure~\ref{CDFsignalmwtemplates.fig}.

\begin{figure}
\begin{center}
\includegraphics[width=0.8\textwidth]{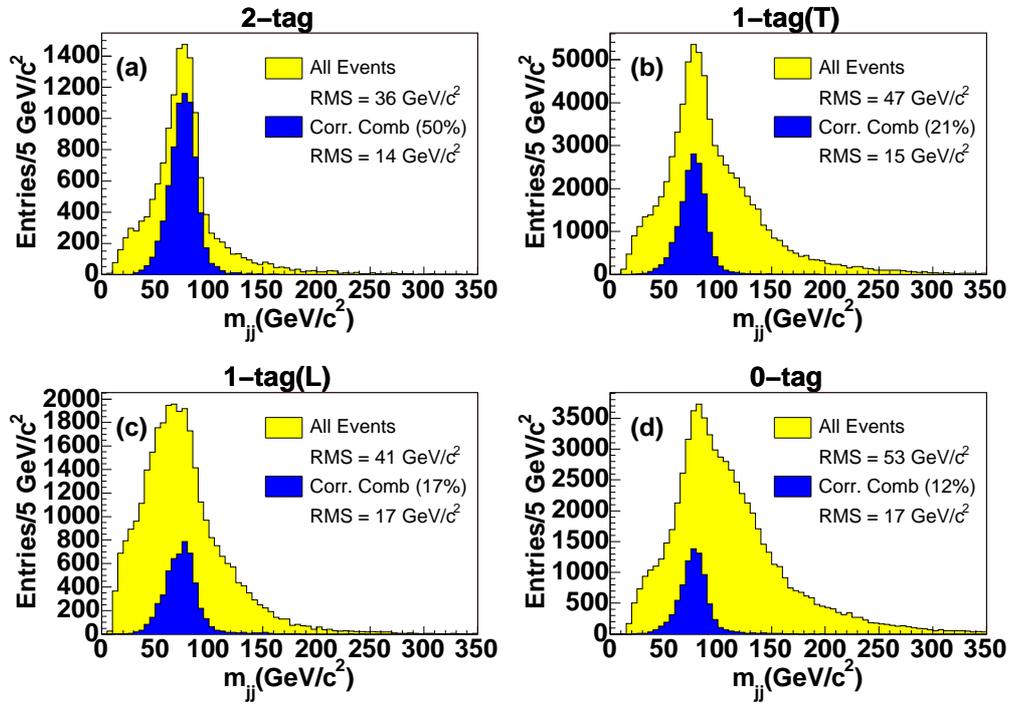}
\caption{\captionfont\label{CDFsignalmwtemplates.fig}CDF lepton+jets template measurement~\cite{Abulencia:2005aj}: 
Template $\m_{jj}$ distributions for simulated signal events 
with $\mtop=178\,\GeV$ and the default jet energy scale in the 
2-tag (a), 1-tag(T) (b), 1-tag(L) (c), and 0-tag (d) categories.
The dark blue areas show the distributions for dijet pairs
that correspond to the \W decay products.}
\end{center}
\end{figure}

\paragraph{Template Parametrization:}
\label{templatemeasurements.kinrec.templateparametrization.sec}
The measurements of the top quark mass and the jet energy scale are
obtained from an unbinned likelihood fit of the $\mtop^{\rm reco}$ and
$m_{jj}$ values computed in the data events to the 
predictions from the simulation.
However, samples of Monte Carlo simulated events are only available for 
discrete values of the top quark mass and jet energy scale.
The solution is to describe the $\mtop^{\rm reco}$ and
$m_{jj}$ template distributions with functions whose
parameters depend on the true values of the top quark mass and jet
energy scale.
It is then possible to continuously vary the values of the top quark
mass and jet energy scale in the fit of the $\mtop^{\rm reco}$ and
$m_{jj}$ values from the data sample.
The convention adopted in this measurement\footnote{
In the Matrix Element analyses, on the other hand, the relative
deviation \JES from the reference jet energy scale is measured, as 
described in Section~\ref{memeasurements.tf.jt.sec}.}
is to consider deviations $\DeltaJES$ from the reference jet energy scale
in units of its uncertainty $\sigma_c$ as a function of jet $\Et$ 
and $\eta$, cf.\ Figure~\ref{CDFjesunc.fig}.

Examples of Monte Carlo template distributions of $\mtop^{\rm reco}$ and
$m_{jj}$ for signal \ttbar events are shown in 
Figure~\ref{CDFparametrizedsignaltemplates.fig}
for various values of the true top quark mass and jet energy scale.
The same functional form is used to describe the $\mtop^{\rm reco}$ and
$m_{jj}$ templates.
It is a linear combination of two Gaussians with independent
parameters (to account for those cases where the \W or top quark masses
are well-reconstructed, i.e.\ where the correct combination has been
chosen) and a gamma distribution (to describe incorrect combinations).
The nine parameters describing a template for a given pair of true
(\mtop, \DeltaJES) values are themselves assumed to depend linearly
on both of these true values.
This assumption is justified because in the measurement, the values of 
\mtop and \DeltaJES need only be varied in a relatively small range
since they are already known a priori to a certain precision.
In Figure~\ref{CDFparametrizedsignaltemplates.fig}, the parametrizations of the template
distributions are overlaid.

\begin{figure}
\begin{center}
\includegraphics[width=0.9\textwidth]{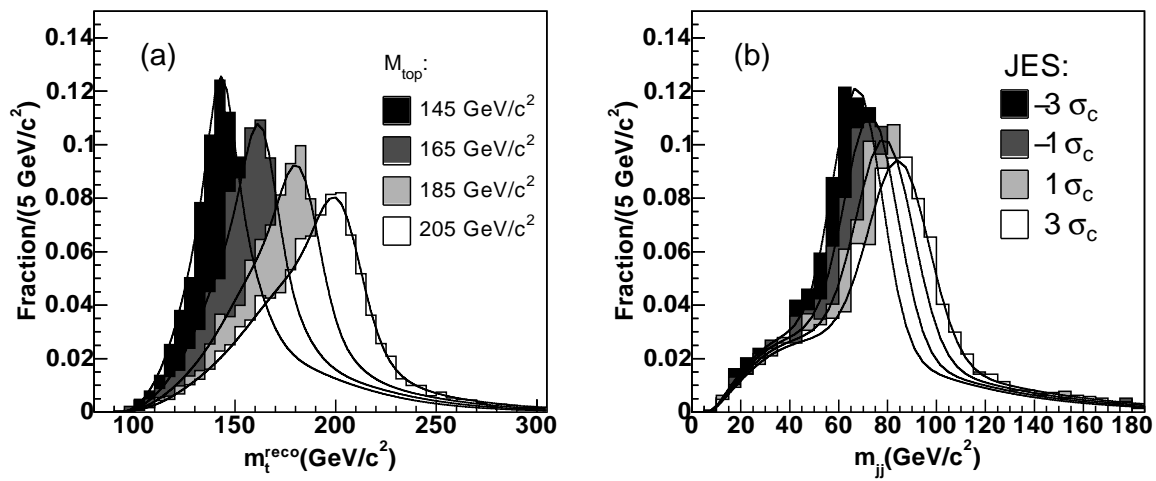}
\caption{\captionfont\label{CDFparametrizedsignaltemplates.fig}CDF lepton+jets template measurement~\cite{Abulencia:2005aj}: 
Template distributions for simulated signal \ttbar events in the 1-tag(T)
category;
(a) $\mtop^{\rm reco}$ templates for various true top 
quark masses at $\DeltaJES=0$ and 
(b) $m_{jj}$ templates for various true values of 
\DeltaJES at $\mtop=180\,\GeV$.
The parametrizations of the template distributions are overlaid.}
\end{center}
\end{figure}

The $\mtop^{\rm reco}$ and $m_{jj}$
template distributions for background events do not depend on
the true value of the top quark mass.
In principle, there is a dependence on the jet energy scale.
However, it has been verified that a variation of the jet energy
scale affects mostly the overall number of background events while
leaving the shape of the template distributions essentially unchanged.
The relative contributions of backgrounds from different sources
are kept constant, only the overall background normalization is
allowed to vary (constrained by the expectation except in the 0-tag
category).
Therefore, a single $\mtop^{\rm reco}$ and $m_{jj}$
template distribution is used to describe
the background in each event category.
The background templates are also described with functions, in this case
not to obtain a continuous parametrization but to become insensitive
to statistical fluctuations due to the limited size of simulated
background samples.
As an example, the $\mtop^{\rm reco}$ and $m_{jj}$
template distributions for background in the 
1-tag(T) category are shown in 
Figure~\ref{CDFparametrizedbackgroundtemplates.fig} together with 
their parametrizations.

\begin{figure}
\begin{center}
\includegraphics[width=0.9\textwidth]{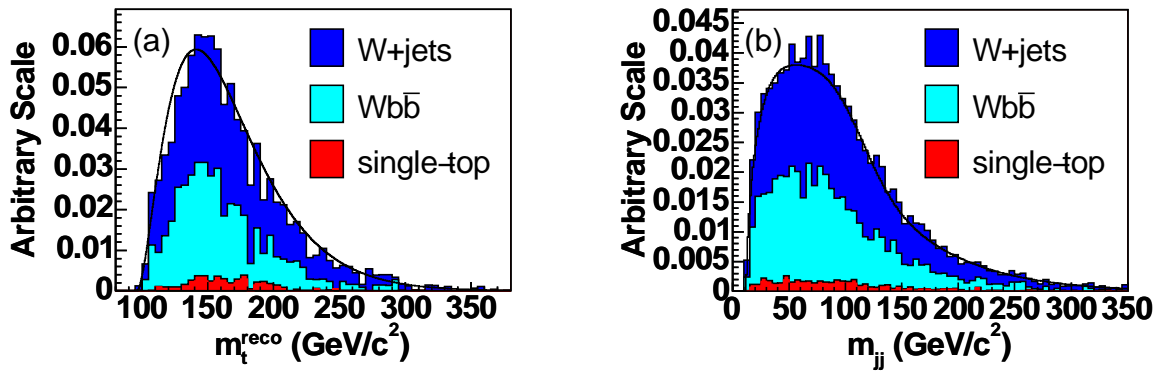}
\caption{\captionfont\label{CDFparametrizedbackgroundtemplates.fig}CDF lepton+jets template measurement~\cite{Abulencia:2005aj}: 
Template distributions for simulated background events in the 1-tag(T)
category with the contributions from individual background sources;
(a) $\mtop^{\rm reco}$ templates and 
(b) $m_{jj}$ templates.
The parametrizations of the template distributions are overlaid.}
\end{center}
\end{figure}

\subsection{Estimators Independent of the Jet Energy Scale}
\label{templatemeasurements.nocalo.sec}

The full kinematic reconstruction of \ttbar events in the \ljets
channel relies on measurements of four jet energies and thus 
depends on the determination of the jet energy scale, as explained
in the previous section.
Measurements that are completely
independent of the jet energy scale will be increasingly
important for the overall top quark mass combination with decreasing
statistical uncertainties.
The energy of $b$ quarks from top quark decay is correlated with the 
top quark mass; the energy of the bottom hadron and consequently its
decay length (distance between production and decay vertex) are then
also correlated with the top quark mass.
The CDF collaboration has used the decay length in the plane
perpendicular to the beam axis as an estimator in a top quark mass
measurement~\cite{bib-CDFLxy}.
This measurement is independent of the jet energy scale (except for
the fact that secondary vertices are only looked for in reconstructed
jets), but has in principle a larger dependence on modeling of 
$b$-quark fragmentation than measurements with full reconstruction of
the \ttbar event kinematics.

The measurement is based on an event sample that is triggered with an
inclusive lepton trigger and selected requiring the presence of 
an isolated charged lepton with $\et>20\,\GeV$, missing transverse
energy $\etmiss>20\,\GeV$, and at least three jets with 
$\et>15\,\GeV$, at least
one of which is $b$-tagged by the presence of a secondary vertex, as
explained in more detail in Section~\ref{reco.reco.sec}.
Note that it is not necessary to require the presence of four
reconstructed jets, as a full reconstruction of the final state is not
attempted. 
The signed decay length $\Lxy$ corresponding to a secondary vertex
is calculated as the vector from the primary
to the secondary vertex, projected first onto the jet momentum vector
and then into the $xy$ plane perpendicular to the beam axis.

The measurement is based on the mean decay length
$\langle\Lxy\rangle$ in all jets with a secondary vertex with $\Lxy>0$
in the selected sample.
To check the modeling of the decay length,
the $\Lxy$ distributions in 
doubly $b$-tagged dijet events (which contain mostly $b$-quark jets)
and events that contain at most two
jets but otherwise pass the \ttbar event selection (which are
depleted in $b$-quark content) are compared between data and
simulation.
Decay length values in the simulation are scaled by the ratio
between the mean values $\langle\Lxy\rangle$ found in the data and
simulation.
In Figure~\ref{CDFLxy.fig}, the distribution of positive $\Lxy$ values
measured in the data
is shown together with the expected contributions from signal and
background events after the scaling procedure.
\begin{figure}
\begin{center}
\includegraphics[width=0.9\textwidth]{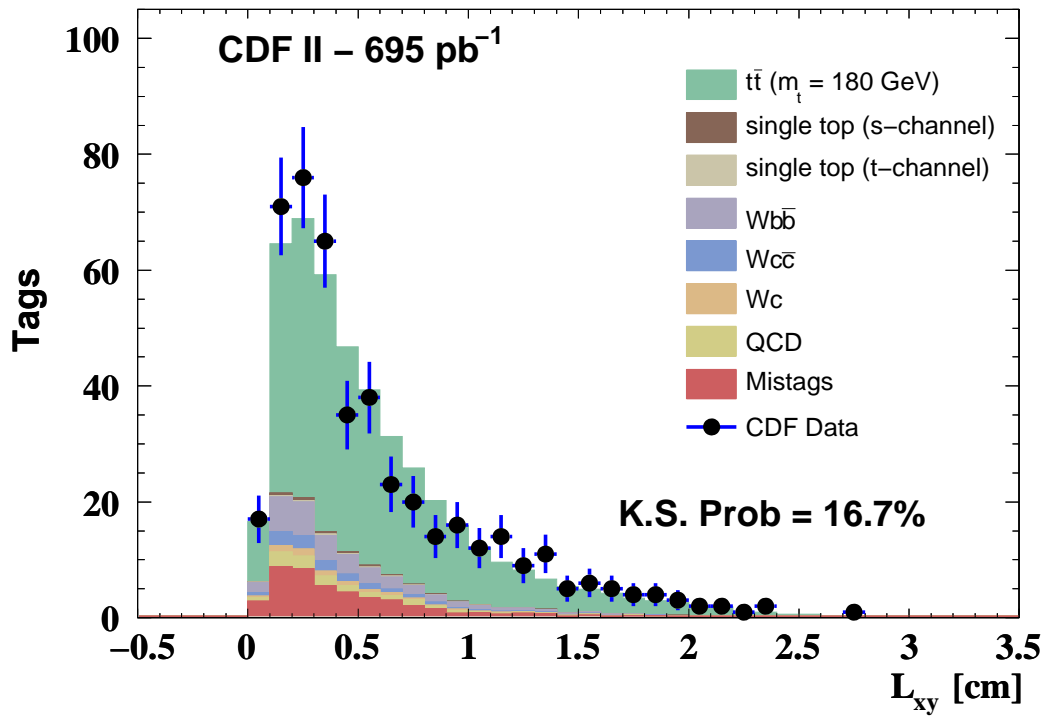}
\caption{\captionfont\label{CDFLxy.fig}CDF lepton+jets measurement using the decay length 
technique~\cite{bib-CDFLxy}:
$\Lxy$ distribution (from jets with $\Lxy>0$) in the data (points with
error bars) together with the expected contributions from signal and
background processes as explained in the plot.  The simulated
distributions have been normalized to the observed number of events.
The result of a Kolmogorov-Smirnov test is indicated
in the plot.}
\end{center}
\end{figure}

The expected distribution of the mean decay length,
$\langle\Lxy\rangle$, is then obtained from pseudo-experiments using
simulated events for true values of the top quark mass between 130
and 230\,\GeV.
The $\langle\Lxy\rangle$ values are fitted with a third degree polynomial
as a function of \mtop, and this parametrization is used to 
obtain the measurement of \mtop from the value of $\langle\Lxy\rangle$
observed in the data sample.

\subsection{Estimators in the All-Jets Channel}
\label{templatemeasurements.allhad.sec}

As the \ljets channel, the \alljets channel also offers the
possibility of full reconstruction of the event kinematics; the 
challenge here is the large background from QCD multijet events, and
also the combinatorial background.

\paragraph{Event Selection:}
\label{templatemeasurements.allhad.evtsel.sec}
In the CDF template analysis in the \alljets
channel~\cite{bib-CDFalljetsxspaper,bib-CDFallhadtemplate}, events are used if they pass a
multi-jet trigger, contain no isolated energetic
leptons and no significant missing transverse energy, and between 6
and 8 jets with $\et>15\,\GeV$, $|\eta|<2.0$, and $\DeltaR>0.5$
between the jets.
The output of an artificial neural network trained to identify \ttbar
events is used to further reduce the contribution from QCD multijet
background.
It is calculated from the following inputs:
\begin{list}{$\bullet$}{\setlength{\itemsep}{0.5ex}
                        \setlength{\parsep}{0ex}
                        \setlength{\topsep}{0ex}}
\item
The scalar sum \Ht of transverse energies of all jets in the event, and
the scalar sum of transverse energies of all but the two highest-$\et$
jets;
\item
the centrality ${\cal C}=\Ht/\sqrt{\hat{s}}$, where $\sqrt{\hat{s}}$ is
the invariant mass of the event calculated from the reconstructed
jets, and the aplanarity 
${\cal A}=\frac{3}{2}\lambda_1$; the symbol $\lambda_1$ denotes the smallest of
the three eigenvalues of the normalized momentum tensor
\begin{equation}
  \label{normalizedmomentumtensor.eqn}
    {\cal M}_{ij}
  =
    \frac{\displaystyle\sum_a p_{a,i} p_{a,j}}
         {\displaystyle\sum_a \left(\vec{p}_a \right)^2}
  \ ,
\end{equation}
where $\vec{p}_{a}$ is the reconstructed momentum vector of jet $a$, and $i$ and
$j$ are Cartesian coordinates;
\item
the minimum and maximum dijet and trijet masses;
\item
the quantity $\et^{1*} = \et^1 \sin^2\theta_1^*$, where $\et^1$ is the
transverse energy of the highest-$\et$ jet in the event, and 
$\theta_1^*$ denotes its polar angle in the all-jets rest frame, and 
the corresponding quantity for the second-highest-$\et$ jet; and
\item
the geometric average over the $\et^{*}$ values for all but the two
highest-$\et$ jets in the event.
\end{list}
The distribution of neural network output values, $NN$, is shown in 
Figure~\ref{CDFallhadtemplateNN.fig} for the data and the expected
\ttbar contribution.
\begin{figure}
\begin{center}
\includegraphics[width=0.9\textwidth]{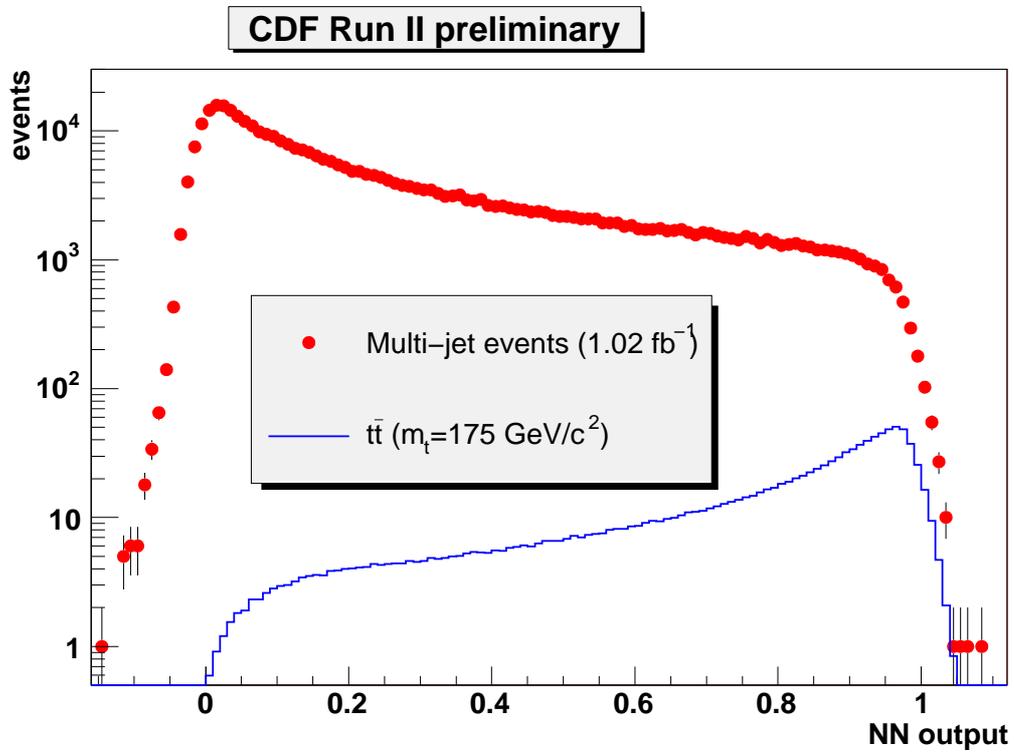}
\caption{\captionfont\label{CDFallhadtemplateNN.fig}CDF \alljets template measurement~\cite{bib-CDFallhadtemplate}:
Distribution of neural network output values $NN$ for data events
(points with error bars) and the expected contribution from \ttbar
events (solid histogram).}
\end{center}
\end{figure}
Events are selected if they satisfy $NN>0.91$.
The background is further reduced in the subsequent analysis because
only events with at least one $b$-tagged jet are used.

\paragraph{Estimator of the Top Quark Mass:}
\label{templatemeasurements.allhad.estimator.sec}
In every selected event, the 6 highest-$\et$ jets are assumed to be 
\ttbar decay products and used to reconstruct the event in a 
kinematic fit to the \ttbar hypothesis similar to the one described
in
Section~\ref{templatemeasurements.kinrec.mtopestimator.sec}
with a $\chi^2$ given by
\begin{eqnarray}
  \nonumber
    \!\!\!\!\!\! \chi^2 \!\!
  & \!\! = \!\! &
    \!\!\!
      \sum_{i=\rm 6\, jets}
        \frac{\left( \pt^{i,\ {\rm fit}} - \pt^{i,\ {\rm meas}} \right)^2}
             {\sigma_i^2}
  \\
  & \!\! & \!\!\!
    + \frac{\left( m_{\rm dijet\,1}^{\rm fit} - \MW \right)^2}
           {\GW^2}
    + \frac{\left( m_{\rm dijet\,2}^{\rm fit} - \MW \right)^2}
           {\GW^2}
    + \frac{\left( m_{\rm trijet\,1}^{\rm fit} - \mtop^{\rm reco} \right)^2}
           {\Gtop^2}
    + \frac{\left( m_{\rm trijet\,2}^{\rm fit} - \mtop^{\rm reco} \right)^2}
           {\Gtop^2}
  \, .\
  \label{CDFallhadtemplatechisq.eqn}
\end{eqnarray}
Each $b$-tagged jet is considered in turn as a $b$ jet (with specific
$b$-jet energy corrections applied), jet-parton combinations
that assign it to a \W decay product are rejected, and the
reconstructed top quark mass $\mtop^{\rm reco}$ corresponding to the combination
that yields the best $\chi^2$ is taken as estimator.
Combinations with $\chi^2>16$ are rejected.
There are thus up to $\Ntag$ estimators in an event with $\Ntag$ tagged
jets.

\paragraph{Template Parametrization:}
\label{templatemeasurements.allhad.templateparametrization.sec}
Similar to the procedure described in
Section~\ref{templatemeasurements.kinrec.templateparametrization.sec}, 
the $\mtop^{\rm reco}$ template distributions for various input
top quark masses are fitted with functions whose
parameters depend on the true value of the top quark mass.
Examples of Monte Carlo template $\mtop^{\rm reco}$ distributions 
for signal \ttbar events are shown in 
Figure~\ref{CDFallhadparametrizedtemplates.fig}(a)
for various values of the true top quark mass.
The same functional form as in
Section~\ref{templatemeasurements.kinrec.templateparametrization.sec}
is used to describe the templates, with linear dependence of the
parameters on the true top quark mass.

The background template is derived from the data.
The jet tagging probability is obtained from a signal-depleted sample
of events with exactly four jets.
The background shape is then determined from the \ttbar candidate event sample by
weighting each jet in turn by its tagging probability, rather than
imposing an actual $b$-tagging requirement.
The signal contribution to this background template estimate is 
obtained using the simulation and subtracted.
The procedure for determining the background shape is validated using
events with low neural network output $NN$.
The background template, shown in
Figure~\ref{CDFallhadparametrizedtemplates.fig}(b), is
parametrized with two gamma functions plus one Gaussian. 
\begin{figure}
\begin{center}
\includegraphics[width=0.45\textwidth]{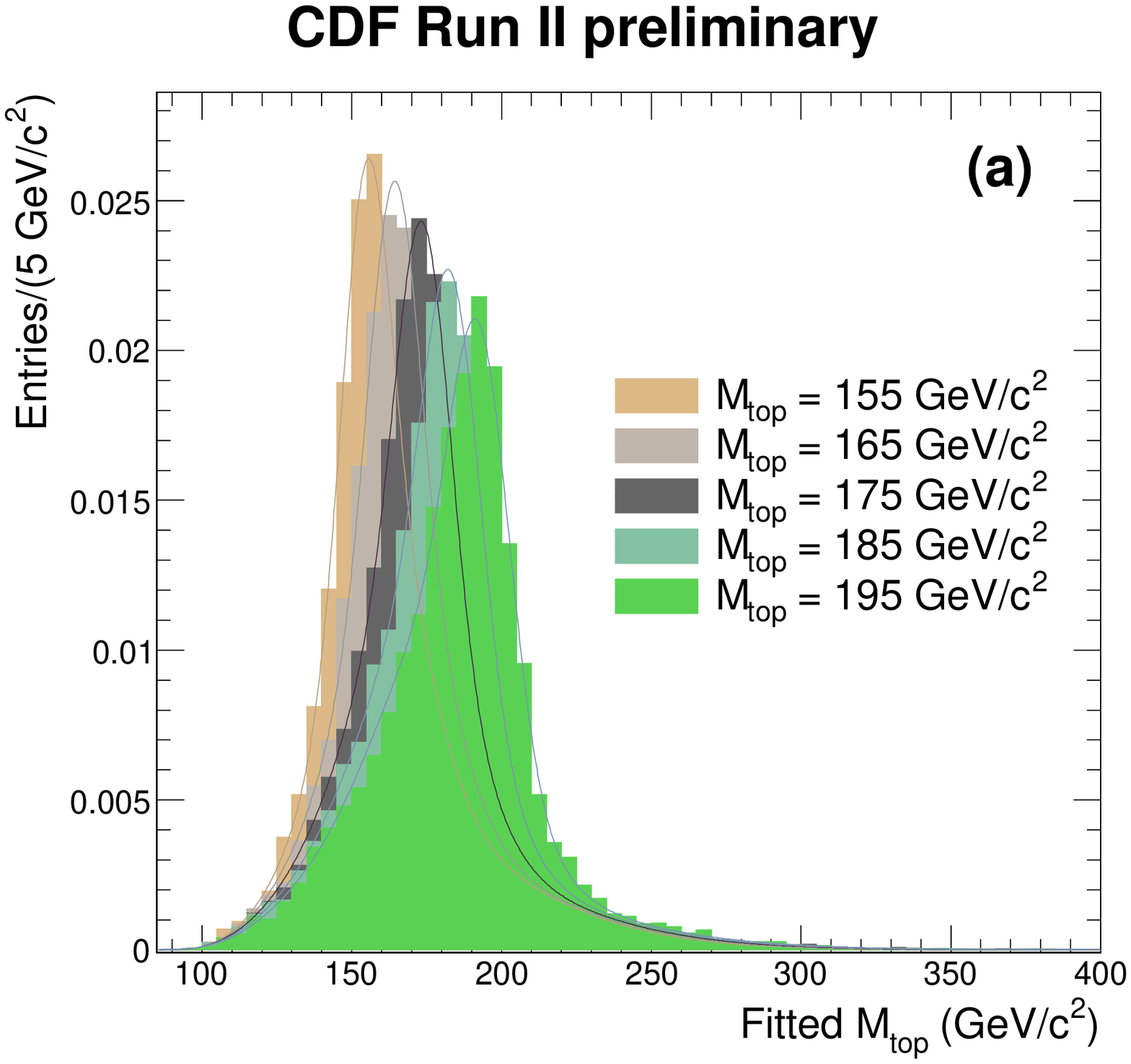}
\includegraphics[width=0.45\textwidth]{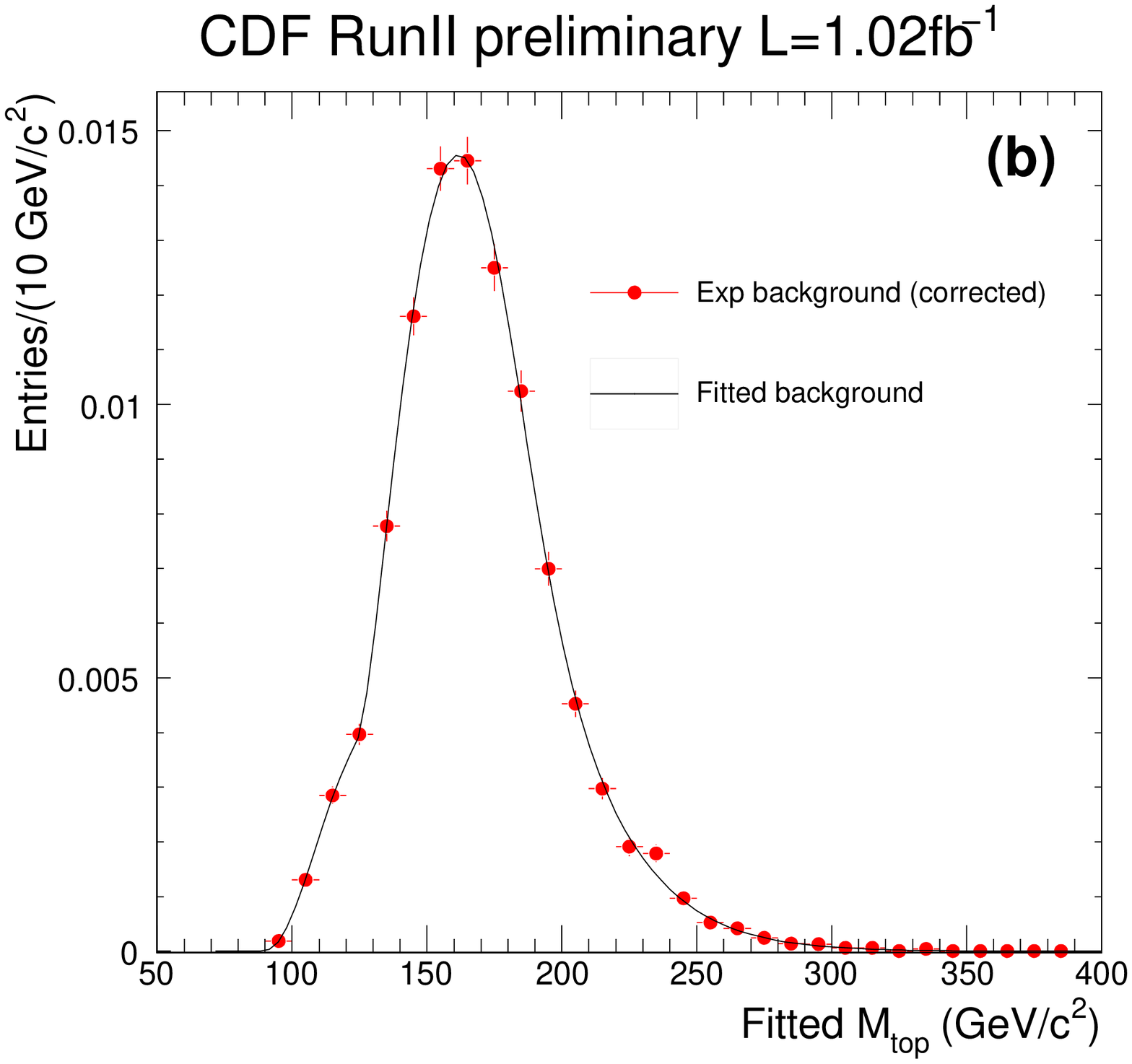}
\caption{\captionfont\label{CDFallhadparametrizedtemplates.fig}CDF \alljets template measurement~\cite{bib-CDFallhadtemplate}: 
Template $\mtop^{\rm reco}$ distributions for (a) signal \ttbar events 
for various true top quark masses and (b) background events.
The parametrizations of the template distributions are overlaid.}
\end{center}
\end{figure}

\subsection{Estimators in the Dilepton Channel}
\label{templatemeasurements.dilepton.sec}

As explained in
Section~\ref{toprodec.topeventtopologies.dileptoneventopology.sec},
dilepton events are kinematically underconstrained if the top quark
mass is not assumed to be known.
It is therefore not possible to use full
reconstruction of the event kinematics to obtain an estimator
$\mtop^{\rm reco}$ as in the \ljets or \alljets channels, 
cf.\ Sections~\ref{templatemeasurements.kinrec.sec}
and~\ref{templatemeasurements.allhad.sec}.
Nevertheless, for a given selected dilepton event, some top quark
mass hypotheses are still more likely than others, and this allows
a top quark mass measurement.
To determine the relative likelihoods of
different top quark mass assumptions, an integration is performed over
undetermined kinematic quantities of the event.
Various methods have been developed for this integration and for
obtaining a top quark mass estimator; in the following, 
these methods are described in turn.
The event selection criteria for all analyses are based on the 
general topology of a dilepton \ttbar event as described in
Section~\ref{reco.backgrounds.dilepton.sec}:
Typically, two oppositely-charged isolated energetic leptons
inconsistent with the $\Z\to\lplm$ hypothesis, at least
two energetic jets, and significant missing transverse energy are
required, and the two highest-\et jets are assumed to be \ttbar decay
products. 

\paragraph{Neutrino Weighting Method:}
\label{templatemeasurements.dilepton.neutrinoweighting.sec}
This method has been used by both CDF and \dzero in 
\runi~\cite{bib-runineutrinoweighting} and 
\runii~\cite{bib-CDFcombineddileptontemplatepaper,bib-Dzerodileptonneutrinoweighting}.
Since the $\bquark$ and $\antibquark$ jets are not distinguished in
the reconstruction, there are two possible jet-parton assignments, 
corresponding to two possible jet-lepton pairings per event.
For each pairing, a scan over assumptions for \mtop and the 
(anti-)neutrino pseudorapidities $\eta_{\nu}$ and
$\eta_{\overline{\nu}}$ is performed.
Disregarding the measured missing momentum in the event, 
the event kinematics are reconstructed for each assumption
by imposing a \ttbar event hypothesis.
This leads to four solutions per 
$\left(\mtop,\ \eta_{\nu},\ \eta_{\overline{\nu}}\right)$ assumption.
According to the measured missing transverse momentum \ptmissvec,
each solution $i$ is assigned a weight $w_i$ of
\begin{equation}
  \label{weightforonesolution.eqn}
    w_i 
  = 
    \exp\left( - \frac{ \left( \ptmissx - \pnux - \pnubarx \right)^2 }
                      { 2 \sigma_x^2 } \right)
    \exp\left( - \frac{ \left( \ptmissy - \pnuy - \pnubary \right)^2 }
                      { 2 \sigma_y^2 } \right)
  \ ,
\end{equation}
where $\left(\sigma_x,\ \sigma_y\right)$ denotes the missing
transverse momentum resolution,
and $\vec{p}_{\nu}$ and $\vec{p}_{\overline{\nu}}$ are the neutrino momenta obtained for the 
given solution.
For a given assumption, the four solutions have equal a priori
probability; the assumption can thus be assigned a weight of
\begin{equation}
  \label{weightforoneassumption.eqn}
    w\left(\mtop,\ \eta_{\nu},\ \eta_{\overline{\nu}},\ {\rm pairing}\right)
  =
    \sum_{i=1}^4 w_i
  \ .
\end{equation}

The a priori probabilities of the neutrino and antineutrino
pseudorapidities are determined from simulated \ttbar events.
They are uncorrelated and can be described by a Gaussian centered
around zero whose width is nearly independent of the true top quark
mass (cf.\ Figure~\ref{CDFdileptonneutrinoetadistribution.fig}).
To obtain the weight as a function of 
the top quark mass alone, a scan over the unknown values of 
$\eta_{\nu}$ and $\eta_{\overline{\nu}}$ is performed, and the 
corresponding weights are multiplied with the a priori probabilities
$P\left(\eta_{\nu},\ \eta_{\overline{\nu}}\right)$ of 
the neutrino and antineutrino pseudorapidities. 
Finally, as both jet-lepton pairings also have the same a priori
probability, one obtains
\begin{equation}
  \label{weightasfunctionofmtop.eqn}
    w(\mtop)
  =
    \sum_{\rm pairings\ } 
       \sum_{\,\eta_{\nu},\, \eta_{\overline{\nu}}}
         P\left(\eta_{\nu},\ \eta_{\overline{\nu}}\right)
         w\left(\mtop,\ \eta_{\nu},\ \eta_{\overline{\nu}},\ {\rm pairing} \right)
  \ .
\end{equation}
While the weight distribution $w(\mtop)$ for a single event can have
more than one relative maximum, the average weight distribution for
many simulated events has one maximum close to the true top quark
mass, as shown in Figure~\ref{neutrinoweightingweightdistribution.fig}.

\begin{figure}[h]
\begin{center}
\includegraphics[width=0.6\textwidth]{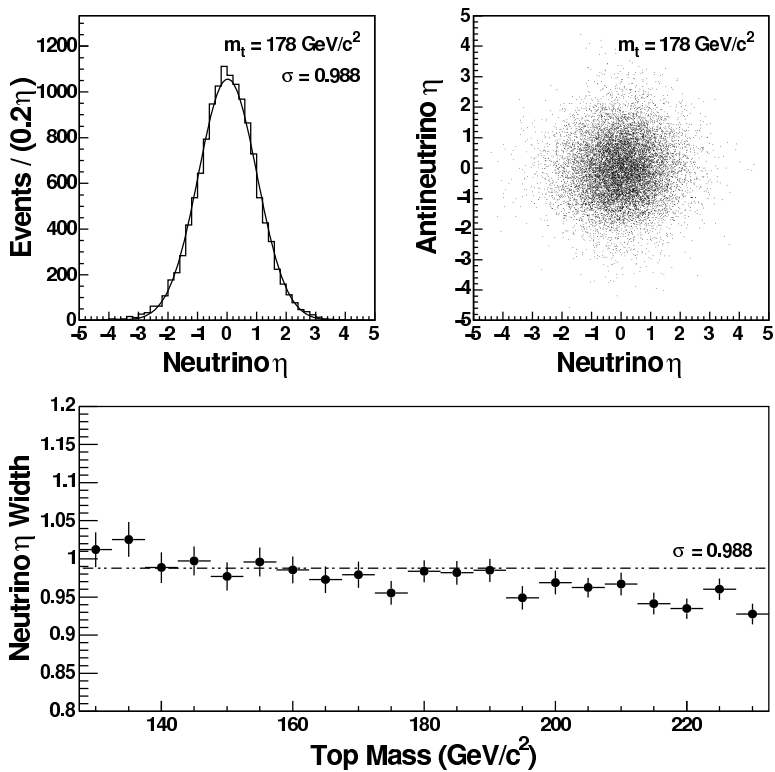}
\caption{\captionfont\label{CDFdileptonneutrinoetadistribution.fig}CDF dilepton neutrino weighting template 
measurement~\cite{bib-CDFcombineddileptontemplatepaper}:
The distribution of neutrino pseudorapidities $\eta$ with a Gaussian
fit (upper left) and the correlation with the antineutrino
pseudorapidity (upper right), determined from simulated \ttbar events
with $\mtop=178\,\GeV$ using \herwig Monte Carlo.  The width of the
fitted Gaussian as a function of \mtop is shown in the lower plot;
here, the line indicates the width for $\mtop=178\,\GeV$.}
\end{center}
\end{figure}

\begin{figure}[ht]
\begin{center}
\includegraphics[width=0.45\textwidth]{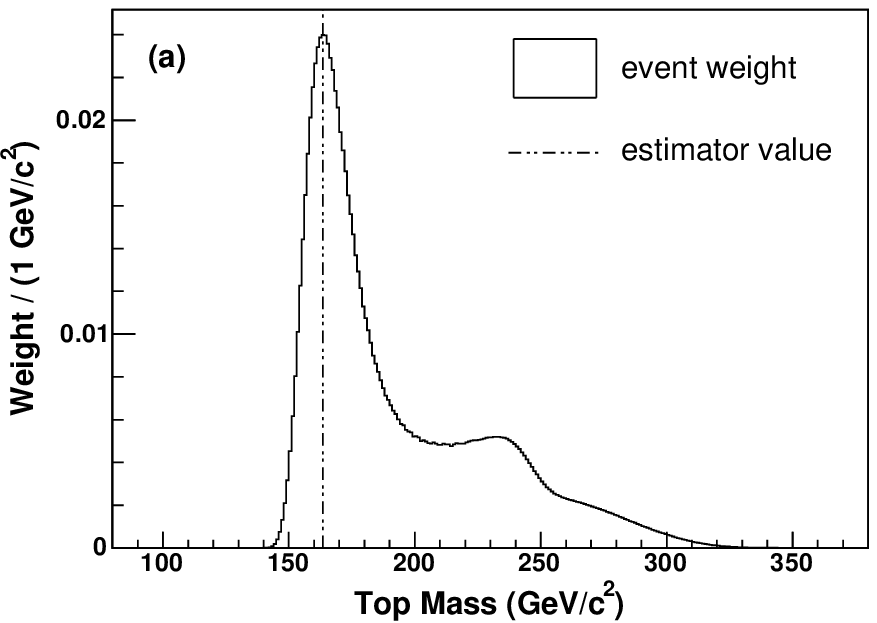}
\includegraphics[width=0.46\textwidth]{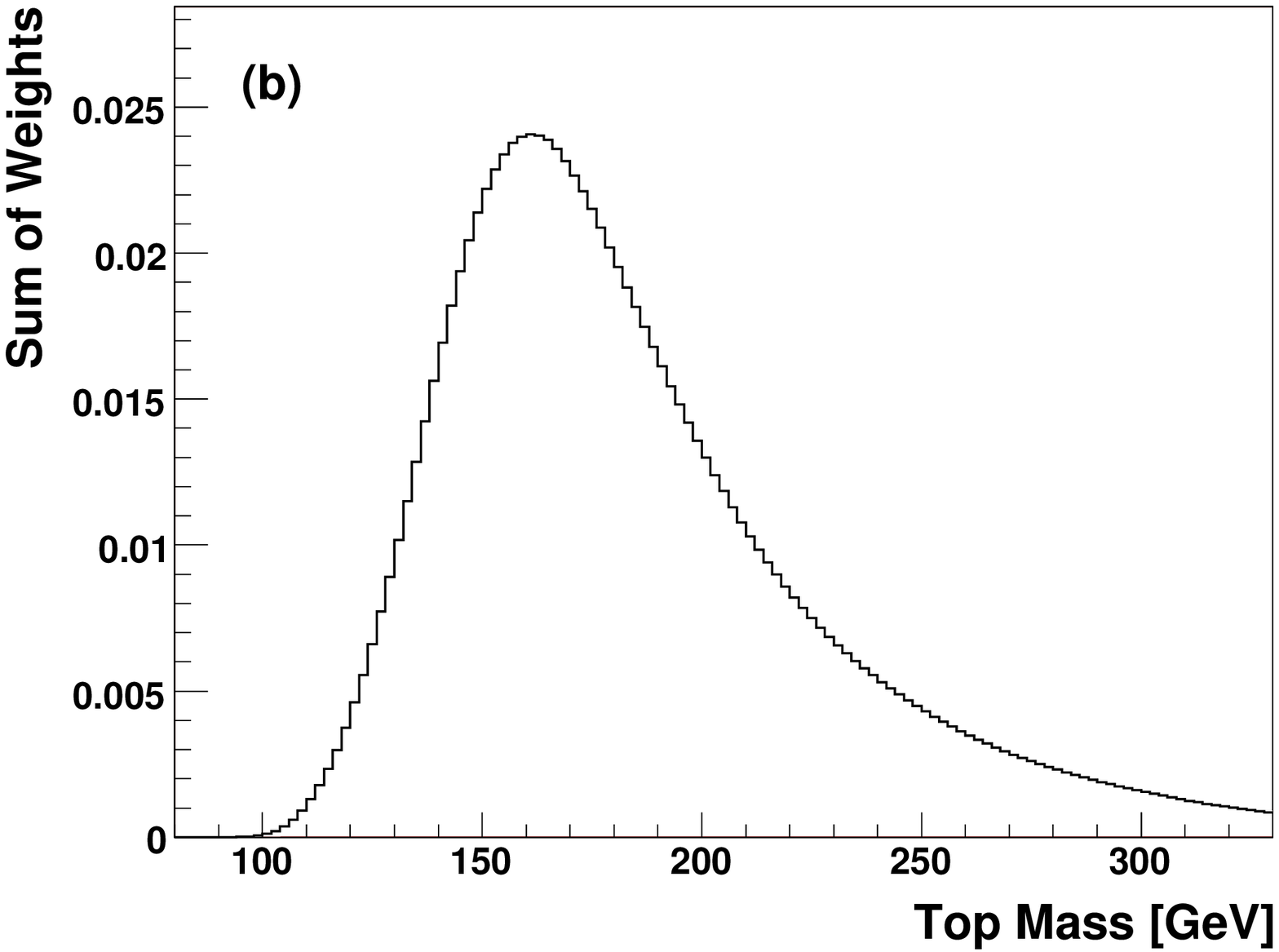}
\caption{\captionfont\label{neutrinoweightingweightdistribution.fig}Dilepton neutrino weighting template measurements:
(a) The weight distribution as a function of assumed top quark mass 
for one \ttbar event with $\mtop=170\,\GeV$ simulated using \herwig,
reconstructed at CDF~\cite{bib-CDFcombineddileptontemplatepaper}.
The top quark mass value taken as estimator for this event is
indicated by the vertical line.
(b) The average weight distribution (obtained by summing weights for
many simulated events) for \ttbar events with
$\mtop=175\,\GeV$ generated with \pythia, reconstructed at 
\dzero~\cite{bib-Dzerodileptonneutrinoweighting}.}
\end{center}
\end{figure}

Both
CDF~\cite{bib-CDFcombineddileptontemplatepaper}
and \dzero~\cite{bib-Dzerodileptonneutrinoweighting}
have performed measurements where the \mtop value that maximizes the
weight distribution is taken as the top quark mass estimator for a given
event.
These analyses then proceed as described above in 
Section~\ref{templatemeasurements.kinrec.templateparametrization.sec}:
The templates are fitted as a function of the most likely top quark mass,
and for the signal templates, the dependence of the fit parameters on
the true top quark mass is parametrized as
well~\cite{bib-CDFcombineddileptontemplatepaper}
or included in a two-dimensional
fit of the templates as a function of the estimator value and the true
input top quark mass~\cite{bib-Dzerodileptonneutrinoweighting}.

This procedure where the \mtop value that maximizes the weight is
taken as estimator does not take into account that events with a broad
maximum contain less \mtop information than events where the maximum
is strongly peaked.
\dzero has therefore made measurements that use a vector of
estimators.
This vector either contains the weight histogram integrated in a few
coarse bins
or the mean and RMS of the weight
distribution~\cite{bib-Dzerodileptonneutrinoweighting}.
These multidimensional estimators cannot be fitted like in the
one-dimensional case; therefore, the signal and background probability
densities for a given vector of estimators are determined from the
density and weights of simulated events with estimators that have 
nearby values in estimator-space.
A parametrization of probability densities as a continuous function of
assumed top quark mass is not performed, either.
The gain from these methods is modest: \dzero finds that 
relative to the analysis that uses as estimator the \mtop value that 
maximizes the weight, the 
expected uncertainty decreases by 7\% when using five bins, but even
increases by 5\% when using the mean and
RMS~\cite{bib-Dzerodileptonneutrinoweighting}.
Rather than introducing multidimensional estimators for the same
quantity, a more natural approach to extracting more information from
the events is to go beyond template methods
altogether, as described for example in Section~\ref{memeasurements.sec}.

\paragraph{Neutrino {\boldmath$\phi$\unboldmath} Weighting Method:}
\label{templatemeasurements.dilepton.neutrinophiweighting.sec}
Instead of assuming the pseudorapidities $\eta$ for both neutrino momenta,
one can assume values of their azimuthal angles.
This procedure is used by
CDF~\cite{bib-CDFcombineddileptontemplatepaper}.
A top quark mass estimator for a selected event is obtained via the
following steps:
\begin{list}{$\bullet$}{\setlength{\itemsep}{0.5ex}
                        \setlength{\parsep}{0ex}
                        \setlength{\topsep}{0ex}}
\item
For each assumed pair of neutrino $\phi$ values, the event 
kinematics is reconstructed in a kinematic fit assuming a \ttbar event
that constrains the lepton-neutrino pairs to the \W boson mass and
constrains the two top quarks to have equal masses within the top
quark width.
For each pair, there are 8 solutions arising from the two-fold
ambiguity in solving for the neutrino longitudinal momentum and from 
the two lepton-jet pairings, and the fitted top quark mass from the 
solution with the smallest $\chi^2$ is taken.
\item
A $12\times12$ grid of assumed neutrino $\phi$ values is
tested.
The top quark mass calculated for each given
pair of assumed neutrino $\phi$ values is weighted by its
$\chi^2$ probability.
The weighted average of the top quark masses is then taken as
estimator for the event, where only assumptions with a weight
of at least 30\% of the maximum weight in the event are considered.
\end{list}

\paragraph{Full Kinematic Analysis:}
\label{templatemeasurements.dilepton.fullkinematicanalysis.sec}
A third method employed by CDF uses an assumed value $\pz(\ttbar)$ of the
longitudinal momentum component of the \ttbar system to solve the 
event kinematics~\cite{bib-CDFcombineddileptontemplatepaper}.
The $\pz(\ttbar)$ distribution is expected to be centered
around zero with a width of $180\,\GeV$, where the width changes by
$10\%$ when varying the top quark mass between $140$ and $200\,\GeV$.
There are two possible jet-lepton pairings, 
each with up to four different 
solutions to the kinematic equations
for a given assumed value of $\pz(\ttbar)$.
The up to eight possible solutions
for the top quark mass under the assumption of a \ttbar event are
calculated.
This procedure is repeated 10000 times, with $\pz(\ttbar)$ values drawn
from the expected distribution, and with the measured jet energies
and the missing momentum varied within their resolutions.
Of the four most probable values corresponding to the four solutions
for a given jet-lepton pairing, the 
one yielding the smallest \ttbar invariant mass is retained, if any
solution was found.

From the resulting distributions of top quark masses for each of the
two jet-lepton pairings,
the one pairing is chosen for which the number of trials
that yielded no solution is smaller.
The most probable value of the corresponding distribution of top quark mass
solutions is taken as the estimator.
The procedure slightly favors lower top quark masses; however
this is valid since only an 
{\em estimator} for the top quark mass is desired, and the bias introduced
is corrected for when the method is calibrated using simulated events.

\paragraph{Matrix Weighting Method:}
\label{templatemeasurements.dilepton.matrixweighting.sec}
In the Matrix Weighting technique employed by 
\dzero~\cite{bib-Dzerodileptonmatrixweighting} the kinematics of the \ttbar candidate
event is solved by assuming a value for the top and antitop quark
masses.
There are four solutions to the 
kinematic equations for a given jet-lepton pairing.
The weight for a given solution is computed from the proton and
antiproton parton
distribution functions \fPDF and \fbarPDF and
the probability $p\left(E_\ell^*|\mtop\right)$ of a charged lepton $\ell$
to have energy $E_\ell^*$ in the top quark rest frame as
\begin{equation}
  \label{matrixweightingweight.eqn}
    w 
  = 
    \fPDF(x) 
    \fbarPDF(\overline{x}) 
    p\left(E_\ell^*|\mtop\right)
    p\left(E_{\overline{\ell}}^*|\mtop\right)
  \, ,
\end{equation}
where $x$ and $\overline{x}$ denote the momentum fractions of the 
colliding partons in the proton and antiproton.
To compute the total weight for a given \mtop assumption, the weights
for both jet-parton assignments and all solutions are summed.
Using resolution sampling, the above procedure is repeated many times
with reconstructed energies/momenta and the missing transverse
momentum drawn from distributions according to the detector
resolution, and for each assumed top quark mass, the mean total weight
is determined.
The value of the assumed top quark mass where the mean total weight
reaches its maximum is then taken as the estimator for the
measurement.

\clearpage
\section{The Matrix Element Measurement Method}
\label{memeasurements.sec}
\newcommand{\newlineonlyintwocol}{}
\newcommand{\nonumberonlyintwocol}{}
\newcommand{\figMEcalibwidth}{0.49}
\newcommand{\figbtagwidth}{0.45}

\begin{center}
\begin{tabular}{p{15cm}}
{\it The previous section described different possibilities for
  computing an estimator of the top quark mass in each \ttbar
  candidate event, and how the expected estimator distribution as a
  function of the assumed top quark mass can be used in a top quark
  mass measurement.  In this section, a different measurement 
  strategy is described, where for each selected event a
  likelihood as a function of the assumed top quark mass is
  calculated.  The section starts with the definition of this
  likelihood, then describes the parametrization of the detector
  resolution needed to compute it, and continues with an in-depth
  discussion of how the likelihood is calculated for the various
  (signal and background) processes via which a candidate event may
  have been produced.}
\end{tabular}
\end{center}

The Matrix Element method is based on the likelihood to observe a
given event in the detector, calculated as a function of 
assumed top quark mass.
The Matrix Element method was first used by the \dzero collaboration 
for the measurement in the \ljets channel at Tevatron
\runi~\cite{bib-nature}, where it yielded the single most precise
measurement of the top quark mass.
The method has been applied to the measurement in the \ljets channel
at \runii by both CDF~\cite{bib-CDFljetsme} and \dzero~\cite{bib-me},
and CDF has also used it in the dilepton
channel~\cite{bib-CDFdileptonMEpaper,bib-CDFdileptonme,bib-CDFdileptonmewithbtagging}. 
The Dynamical Likelihood method follows a similar concept.
It has been used by CDF in the \ljets~\cite{bib-CDFljetsDLMmass} 
and dilepton~\cite{bib-CDFdileptonDLMmass} channels at \runii and is
described in this section together with the Matrix Element method.

The selection of events used in the analyses is briefly
described in Section~\ref{memeasurements.evtsel.sec}.
An overview of the calculation of the event likelihood is given in 
Section~\ref{memeasurements.pevt.sec}, and the calculation of the 
likelihood for a given process is explained in
Section~\ref{memeasurements.pprc.sec}.
Section~\ref{memeasurements.tf.sec} discusses the parametrization of the
detector response.
This includes a description of how $b$-tagging information
can be used in the analysis.
Technical details on the computation of the signal and background
likelihoods are given in 
Sections~\ref{memeasurements.psgn.sec} 
and~\ref{memeasurements.pbkg.sec} with an emphasis on the \dzero \runii
analysis in the \ljets channel which serves as an
example.

\subsection{Event Selection}
\label{memeasurements.evtsel.sec}
The selection of events for the measurements with
the Matrix Element method is very similar to the general criteria
described in Sections~\ref{reco.backgrounds.dilepton.sec}
and~\ref{reco.backgrounds.ljets.sec} and to the selections used in 
measurements based on the template method, see
Section~\ref{templatemeasurements.sec}.

There is, however, one aspect that deserves special attention.
Leading-order matrix element calculations are used to evaluate
likelihoods with which the selected events are produced in the
signal and background processes.
Initial- and final-state gluon radiation are therefore not accounted
for in these likelihoods.
This is not a problem since the calibration of the measurement is
based on fully simulated events which do include gluon radiation,
cf.\ Section~\ref{massfit.validationandcalibration.sec}.
Nevertheless, both CDF and \dzero select only events with {\em exactly} four
jets for their Matrix Element measurements in the \ljets channel.
While not completely removing events with significant gluon radiation,
this cut still reduces their contribution to the event sample.
Also, the complication of selecting the jet from radiation and
assigning it either to initial-state radiation or to final-state radiation 
off one of the \ttbar decay products is avoided.
On the other hand, the measurements in the dilepton channel are more
severely limited by statistics.
Here the event selection requires two
{\em or more} jets, as for the template analyses, and any jets but the two
highest-\et ones are assumed to be due to initial-state radiation.

In the \ljets channel, the \dzero experiment has performed both a
topological and a \bquark-tagging measurement (i.e.\ disregarding/using
\bquark-tagging information); in both analyses, all events
(irrespective of the number of \bquark tags) are used.
Unless noted explicitly, the following description refers to the 
\bquark-tagging analysis.
In contrast, the CDF event selection in the \ljets channel requires at
least one jet to be \bquark-tagged; this selection is used in 
the Matrix Element and also the Dynamical Likelihood measurement.
In the dilepton channel, CDF has performed two measurements based on
the Matrix Element method, one not using \bquark-tagging information
and the other requiring at least one \bquark-tagged jet.
The dilepton measurement with the Dynamical Likelihood technique does
not use \bquark-tagging information.

\subsection{The Event Likelihood}
\label{memeasurements.pevt.sec}
To make maximal use of the kinematic information on the top quark mass
contained in the event sample,
for each selected event 
the likelihood \pevt that this event is observed is calculated 
as a function of the assumed top quark mass.
In analyses where additional parameters like the overall jet energy
scale are to be measured simultaneously with the top quark mass, 
\pevt is also a function of the assumed values of these parameters.
The likelihoods for all events are then combined to
obtain the sample likelihood,
and the measurement of the top quark mass and of the other parameters,
if applicable, is extracted from this sample likelihood.
To make the likelihood calculation tractable, simplifying assumptions
in the description of the physics processes and the detector response
are introduced as described in this section.
Before applying it to the data, the measurement technique is however
calibrated using fully simulated events, and 
the assumptions the full simulation makes to describe the physics 
processes are accounted for by systematic uncertainties.

It is assumed that the physics processes that can lead to an observed
event do not interfere.
The likelihood \pevt then in principle has to be composed from
likelihoods for all these processes as
\begin{equation}
  \label{basicPevt.eqn}
  \pevt = \sum_{{\rm processes}\ P} f_P  L_P
  \ ,
\end{equation}
where $L_P$ is the likelihood for the event to be created via
a given process $P$, and $f_P$
denotes the fraction of events from that process in the event sample.
In practice, not all possible processes can be accounted for 
explicitly, and 
simplifying assumptions are made in the calculation of the
likelihoods for the individual processes.
The measurement result therefore has to be corrected accordingly.

A likelihood \psgn is calculated for the event
to be produced in the signal \ttbar reaction; this likelihood will 
depend on the assumed top quark mass.
The \ttbar production processes taken into account are listed in
Table~\ref{pevt.table}.
In their measurements using the Matrix Element method, 
CDF and \dzero have also included
likelihoods for the event to be produced via the dominant background
processes; this maximizes the separation
between signal and background events and keeps corrections to the
final result small.
In contrast, the CDF analyses
using the Dynamical Likelihood technique omit an explicit treatment
of background at this stage and apply a correction for all 
backgrounds to the final result.
When applying the Matrix Element method to the \ljets channel, 
CDF and \dzero choose to determine the jet energy scale and the signal
fraction together with the top quark mass.

The event likelihood \pevt can thus be
expressed as
\begin{equation}
  \label{eq:MEpevt}
    \pevt\left(x;\,\mtop,JES,\ftop\right) 
  = 
                     \ftop \psgn\left(x;\,\mtop,JES\right) 
    + \left(1-\ftop\right) \pbkg\left(x;\,JES\right)
  \ ,
\end{equation}
where \pbkg is a weighted sum of likelihoods for all background
processes according to Equation~(\ref{basicPevt.eqn}).
The symbol $x$ denotes the kinematic variables of the event, \ftop is the
signal fraction of the event sample, and \psgn and \pbkg are the 
likelihoods for observing the event if it is produced via
the \ttbar or any of the background processes, respectively.
The values of \ftop and $JES$ are fixed if applicable, depending 
on the details of the analysis.
An overview of the event likelihood calculation in the analyses
described here is given in
Table~\ref{pevt.table}.
\begin{table}[htbp]
\begin{center}
\begin{tabular}{ccccccc}
\hline
\hline
    channel 
  & method 
  & exp. 
  & \multicolumn{1}{@{}c@{}}{parameters }
  & \begin{tabular}{@{}c@{}}
      signal\\ 
      processes
    \end{tabular}
  & \begin{tabular}{@{}c@{}}
      background\\
      processes
    \end{tabular}
  & reference
\\
\hline
  \ljets   & ME & \dzero & \mtop, $JES$, \ftop 
                         & $\qqbar\to\ttbar$ 
                         & \wfourpshort
                         & \cite{bib-me} \\
  \ljets   & ME & CDF    & \mtop, $JES$, \ftop 
                         & $\qqbar\to\ttbar$ 
                         & \wfourpshort
                         & \cite{bib-CDFljetsme} \\
  \ljets   & DL & CDF    & \mtop               
                         & $\qqbar\to\ttbar$, $gg\to\ttbar$ 
                         & --- 
                         & \cite{bib-CDFljetsDLMmass} \\
\hline
  dilepton & ME & CDF    & \mtop
                         & $\qqbar\to\ttbar$ 
                         & \begin{tabular}{@{}c@{}}
                             \Zgammastartwopshort,\\
                             \WWtwopshort,\\
                             \Wthreepshort
                           \end{tabular}
                         & \cite{bib-CDFdileptonMEpaper,bib-CDFdileptonme} \\
  dilepton & DL & CDF    & \mtop               
                         & $\qqbar\to\ttbar$, $gg\to\ttbar$ 
                         & --- 
                         & \cite{bib-CDFdileptonDLMmass} \\
\hline
\hline
\end{tabular}
\caption{\captionfont\label{pevt.table}Overview of the \pevt calculation in the 
\mtop measurements using the Matrix Element (ME) and Dynamical
Likelihood (DL) methods.  The column entitled ``parameters'' lists the 
quantities that are measured in the analysis, and the signal and
background processes taken into account in the event likelihood
are listed in the following columns.  The symbol ``p'' refers to any
light parton, i.e.\ a \uquark, \dquark, \squark, or \cquark quark (or
antiquark) or a gluon.  The lines describing the Matrix Element 
measurements by \dzero 
in the \ljets channel and by CDF in the dilepton channel refer to both 
the topological and the \bquark-tagging analyses (the \WWtwopshort
background is negligible and not considered in the dilepton
measurement using \bquark tagging).}
\end{center}
\end{table}

To extract the top quark mass from a set of $N$ measured
events $x_1,..,x_N$, a likelihood function for the event sample
is built from the
individual event likelihoods calculated according to
Equation~(\ref{eq:MEpevt}) as
\begin{equation}
  \label{eq:MElhood-fnc}
    L(x_1,..,x_N;\,\mtop,JES,\ftop) 
  = 
    \prod_{i=1}^{N}\pevt(x_i;\,\mtop,JES,\ftop)
  \, .
\end{equation}
This likelihood is maximized to determine the 
top quark mass (and additional parameters, if applicable).

\subsection{The Likelihood for one Process}
\label{memeasurements.pprc.sec}
To evaluate the likelihood for an observed event to be produced via
a given process $P$, all possible configurations $y$ of the
four-momenta of the final-state
particles that could have led to the observed
event $x$ are considered.
In practice, also a sum over different non-interfering processes is
performed; thus only one likelihood is computed for different color
or flavor configurations if the differential cross section is
identical.
The likelihood for a final state with $n_f$ partons and 
given four-momenta $y$ to be produced in the hard-scattering process is 
proportional to the differential cross section ${\rm d}\sigma_P$ of the 
corresponding process, given by
\begin{eqnarray}
  \label{eq:MEdsigmahs}
    \dsigmaP(a_1 a_2 \to y)
   = 
    \frac{(2 \pi)^{4}\! \left|\mathscr{M}_P\left(a_1 a_2 \to y\right)\right|^{2}}
         {\qone \qtwo s}
    {\rm d}\Phi_{n_f}
   \ , \nonumberonlyintwocol \newlineonlyintwocol
\end{eqnarray}
where 
$a_1 a_2$ and $y$ stand for the kinematic variables of the partonic
initial and final state, respectively.
The symbol $\mathscr{M}_P$ denotes the matrix element for this process,
$s$ is the center-of-mass energy squared of the collider, $\qone$ and $\qtwo$ are the
momentum fractions of the colliding partons $a_1$ and $a_2$
(which are assumed to be 
massless) within the colliding proton and antiproton\footnote{
The following discussion is based on situation at the Tevatron \ppbar collider as a
concrete example but is equally valid for the LHC when the antiproton
is replaced with a proton and the appropriate PDF is used.}, and
${\rm d}\Phi_{n_f}$ is an element of $n_f$-body phase space.

To obtain the differential cross section 
$\dsigmaP(\ppbar\to y)$
in \ppbar collisions, the differential cross section
from equation \Eref{eq:MEdsigmahs} is convoluted 
with the parton density functions (PDF) and summed over all possible flavor
compositions of the colliding partons,
\begin{equation}
  \label{eq:MEdPpp}
    \dsigmaP(\ppbar\to y)
  = 
    \int\limits_{\qone, \qtwo} \sum_{a_1, a_2}
    {\rm d}\qone {\rm d}\qtwo\ 
    f_{\rm PDF}^{a_1} (\qone)\ 
    \bar{f}_{\rm PDF}^{a_2} (\qtwo)\ 
    \dsigmaP(a_1 a_2 \to y)
  \ ,
\end{equation}
where $f_{\rm PDF}^a (\qgeneral)$ and 
$\bar{f}_{\rm PDF}^a (\qgeneral)$ denote
the probability densities to find a parton of
given flavor $a$ and momentum fraction $\qgeneral$ in the proton or
antiproton, respectively.
This equation directly reflects QCD factorization as introduced in
Section~\ref{toprodec.topproduction.sec} and corresponds to 
Equation~(\ref{factorization.eqn}).

The finite detector resolution is taken into account via a
convolution with
a transfer function $W(x,y;\,JES)$ that describes the probability to
reconstruct a partonic final state $y$ as $x$ in the detector.
The differential cross section to observe a given reconstructed 
event then becomes
\begin{eqnarray}
  \label{eq:MEdsigmapp}
    \dsigmaP(\ppbar\to x)
  & = &
    \int\limits_{y} \dsigmaP(\ppbar\to y)\
    W(x,y;\,JES)
  \ .
\end{eqnarray}

Only events that are inside the 
detector acceptance and that pass the trigger conditions and offline
event selection are used in the measurement.
Because of the selection cuts, the 
corresponding overall detector efficiency depends both on \mtop
and on the jet energy scale.
This is taken into account in the cross section of
events observed in the detector:
\begin{equation}
  \label{eq:MEsigmaobs}
    \sigmaPobs
  =
    \int\limits_{x,y} {\rm d}\sigmaP(\ppbar\to y)
    W(x,y;\,JES)
    f_{\rm acc}(x)
    {\rm d}x
  \ ,
\end{equation}
where $f_{\rm acc}=1$ for selected events and $f_{\rm acc}=0$ otherwise.

For example, the likelihood to observe a \ttbar event as 
$x$ in the detector is given by
\begin{eqnarray}
  \nonumber
    \psgn(x;\,\mtop,JES)
  & = &
    \frac{\dsigma_{\ttbar}(\ppbar\to x;\,\mtop,JES)}
         {\sigmaobs_{\ttbar}(\mtop,JES)}
  \\
  \label{eq:MEpsgn}
  & = &
    \frac{1}{\sigmaobs_{\ttbar}(\mtop,JES)}
    \int\limits_{\qone,\qtwo,y} 
    \sum_{a_1,\,a_2}
    {\rm d}\qone {\rm d}\qtwo\,
    f_{\rm PDF}^{a_1} (\qone)\, 
    \bar{f}_{\rm PDF}^{a_2} (\qtwo)\, 
    \times
  \\
  \nonumber
  & & 
    \frac{(2 \pi)^{4} \left|\mathscr{M}_{\ttbar}(a_1 a_2\to y)\right|^{2}}
         {\qone \qtwo s}
    {\rm d}\Phi_{6}
    W(x,y;\,JES)
  \ .
\end{eqnarray}
The contributions to the likelihood are visualized schematically
in Figure~\ref{psgn_schematic.fig}.
A similar formula holds for the likelihood to observe
a background event as $x$, except this likelihood does not depend on the 
top quark mass.
\begin{figure}
\begin{center}
\includegraphics[width=0.8\textwidth]{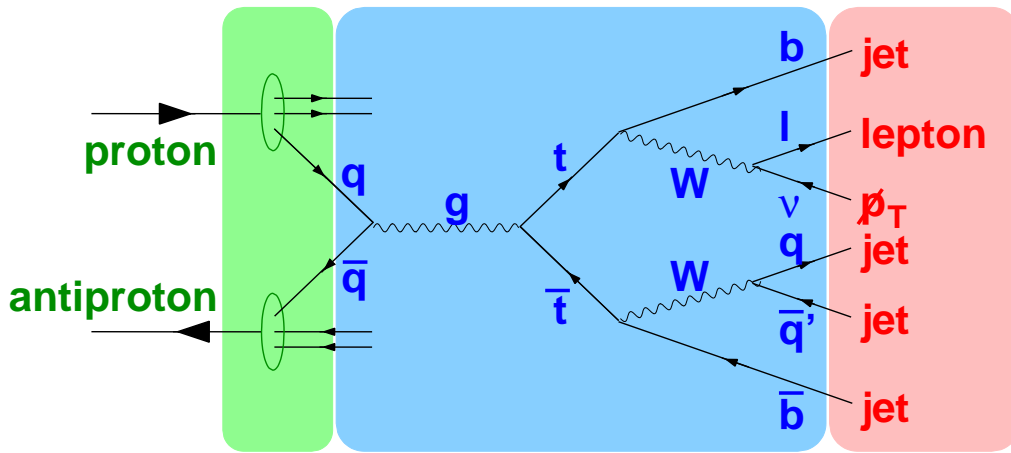}
\caption{\captionfont\label{psgn_schematic.fig}Schematic representation 
  of the calculation of the
  likelihood for \ttbar production to lead to a given observed
  \ljets event.  The observed event $x$, shown in red at the right,
  is fixed while integrating over all possible momentum configurations 
  $y$ of final-state particles (shown in blue).  All possible
  assignments of final-state partons arising from the process shown by
  the Feynman diagram to the measured jets in the detector are
  considered (only one possibility is shown here).  
  The differential cross section for the process shown by
  the diagram is convoluted with the probability for the final-state
  partons to yield the observed event (transfer function), and with
  the probability to find initial-state partons, shown in green to the
  left, of the given flavor
  and momenta inside the colliding proton and antiproton (parton
  distribution function).  For each
  partonic final state under consideration, the initial
  state parton momenta are known by energy and momentum conservation.}
\end{center}
\end{figure}

Details of the parametrization of the detector response are 
given in Section~\ref{memeasurements.tf.sec}.
The parametrization of the matrix element and the computation of 
\psgn are described in Section~\ref{memeasurements.psgn.sec}.
The determination of \pbkg is discussed in 
Section~\ref{memeasurements.pbkg.sec}.

\subsection{Description of the Detector Response}
\label{memeasurements.tf.sec}
The transfer function $W(x,y;\,JES)$ relates the characteristics $y$
of the final-state partons to the 
measurements $x$ in the detector.
The symbol $x$ denotes measurements of the jet and charged lepton
energies or momenta and directions as well as $b$-tagging information
for the jets.
A parametrization of the detector resolution is used in the
likelihood calculation because the full \geant-based simulation would
be too slow.
The full simulation is however used to generate the simulated events
with which the method is calibrated.
In this section, the general form of the transfer function is first
described, followed by a discussion of the individual factors.

\subsubsection{General Form of the Transfer Function}
\label{memeasurements.tf.general.sec}
The transfer function $W(x,y;\,JES)$ describes the probability density ${\rm d}P$
to reconstruct a given assumed partonic final state $y$ as $x$ in the 
detector:
\begin{equation}
  \label{tfdef.eqn}
    {\rm d}P
  = 
    W(x,y;\,JES) {\rm d}x
  \ .
\end{equation}
Because the final-state partons are assumed to give rise to some
measured event $x$, the normalization condition
\begin{equation}
  \label{tfnorm.eqn}
    \int_x W(x,y) {\rm d}x
  =
    1
\end{equation}
holds, where the integral is over all possible observed events $x$.

The transfer function is assumed to factorize into contributions from 
each measured final-state particle.
Aspects to be considered in the transfer function are in principle the
measurement of the momentum of a particle (both of its energy and of its
direction) as well as its identification.

The relative importance of the energy resolution for electrons, muons, 
and quarks for the top quark mass reconstruction has been studied 
qualitatively in simulated top quark decays with a leptonic \W
decay, reconstructed with the \dzero detector.
Of the three top quark decay products, the reconstructed momentum from
either the charged lepton or the \bquark jet is taken, while the true
values are taken from the simulation for the other two particles.
The three plots in Figures~\ref{mtopresolution-emujet.fig}(a), (b), and
(c) show qualitatively that the resolution of the top quark mass is
dominated by the jet energy resolution, while the effect of the
electron resolution is comparatively small.
The effect of the muon resolution at high muon transverse momentum is 
of the same order as that of the jet resolution, while muons are 
comparatively well-measured at low $\pt$, as shown in
Figures~\ref{mtopresolution-emujet.fig}(b1)-(b6).

Consequently, the following assumptions are made about how final-state
particles are 
measured in the detector, which allow reducing the dimensionality
of the integration over 6-particle phase space described in 
Section~\ref{memeasurements.pprc.sec}:
\begin{list}{$\bullet$}{\setlength{\itemsep}{0.5ex}
                        \setlength{\parsep}{0ex}
                        \setlength{\topsep}{0ex}}
\item{\bf Electrons:}
  Apart from efficiency losses (which are not described by the
  transfer function), electrons are assumed to be
  unambiguously identified (i.e.\ an electron is not reconstructed as
  a muon or a hadronic jet).
  The electron direction and energy are both assumed to be
  well-measured, i.e.\ during integration, the final-state
  electron is assumed to be identical to the measured particle.
\item{\bf Muons:}
  As for electrons, muons are assumed to be unambiguously
  identified, and their direction to be precisely measured.
  While CDF also considers the muon $\pt$ to be well-measured, \dzero
  introduces a transfer function that allows for a finite resolution.
  This has primarly an effect for muons with large $\pt$.
\item{\bf Energetic quarks and gluons:}
  Energetic quarks and gluons are almost always
  reconstructed as a jet in the detector.
  There is however a small probability that they are reconstructed as
  an isolated {\em fake} lepton, and it depends on the process
  considered whether or not this possibility has to be taken into
  account:
  If the number of energetic leptons required in the event selection
  is already present in the final state, the possiblity that this
  lepton is not reconstructed {\em and} a quark/gluon fakes a
  lepton that is selected can be ignored.
  If fewer energetic leptons are produced than required in the selection,
  it is assumed that the signatures of the remaining
  leptons in the detector have arisen from quarks or gluons.
  This effect becomes important in the case of dilepton events, where
  background from \Wthreepshort production is explicitly taken into account.

  Both \dzero and CDF assume the direction of the final-state quarks
  and gluons to be well-measured by the jet directions.
  Transfer functions are introduced for the jet energy (or transverse
  energy).

  The ability of the detector to distinguish quarks and gluons and to 
  identify the flavor of final-state quarks is limited.
  The probability to obtain a $b$-tagged jet is largest for \bquark
  quarks, and it is still larger for \cquark quarks than for light
  quarks or gluons.
  While the \dzero measurement takes into account the possibility that a
  $b$ tag is faked by a \cquark or light quark or gluon, CDF makes the 
  approximation that this probability is zero for signal events where 
  enough \bquark quarks are present in the final state.
  This has consequences for the assignment of final-state quarks to
  measured jets in \ljets \ttbar events, where there are two \bquark
  quarks and two light (or charm) quarks in the final state.
  In dilepton \ttbar events with only two \bquark quarks in the final
  state, this complication does not arise.
\item{\bf Neutrinos:}
  Neutrinos are not measured in the detector.
  The presence of energetic neutrinos can be inferred from the transverse
  momentum imbalance ($\ptmissvec$); however this quantity depends on the 
  momenta of the other objects measured in the detector.
  The observed missing transverse momentum is due to the neutrinos as
  well as to mismeasurement of jet energies and of the energy \etrecoil
  from other objects in the detector against which the \ttbar
  system recoils.

  In the Matrix Element analyses, the \ttbar transverse momentum is 
  assumed to be zero, and the transverse components of the sum of 
  neutrino momenta is obtained as the negative vector sum of all 
  other assumed final-state particle transverse momenta.
  The dilepton measurement includes a transfer function factor that 
  describes the likelihood with which the measured value of
  \etrecoil is obtained if the assumed value is zero.
  The Matrix Element measurements in the \ljets channel do not include
  this factor.

  In the Dynamical Likelihood measurements, the transverse components
  of the sum of neutrino momenta are taken as the negative vector sum of all 
  other assumed final-state particle transverse momenta and the
  unclustered transverse momentum in the calorimeter; consequently the
  \ttbar transverse momentum is assumed to be minus the unclustered
  transverse momentum and not necessarily zero.
  No transfer function factor for the unclustered transverse momentum 
  is included.
\end{list}

In addition to the energy resolution, 
one has to take into account the fact that the jets in the
detector cannot be assigned unambiguously
to a specific final-state parton.
(Similarly, it is not known which reconstructed electron was faked by
which final-state parton, if applicable.)
Consequently, all possibilities must be considered in principle, and
their contributions to the transfer function summed.

If no $b$-tagging information is used, the transfer function 
$W(x,y;\,JES)$ is given by
\begin{eqnarray}
  \label{eq:tfdefinition-topo}
    W(x,y;\,JES)
  & = &
    \prod_{e=1}^{n_e} \delta^{(3)} \! \left(   \vec{p}_e^{\,\rm rec}
                                             - \vec{p}_e^{\,\rm ass} \right)
  \times
  \\
  \nonumber
  & &
    \prod_{m=1}^{n_\mu} \delta^{(2)} \! \left(   \vec{u}_\mu^{\,\rm rec}
                                               - \vec{u}_\mu^{\,\rm ass} \right)
    W_{\mu}\left( \qoverptmurec,\qoverptmugen \right)
  \times
  \\
  \nonumber
  & &
    \frac{1}{n_{\rm comb}} 
    \sum_{i=1}^{n_{\rm comb}}
    \prod_{j=1}^{n_j} 
    \delta^{(2)} \! \left(   \vec{u}_{{\rm jet}\,j}^{\,\rm rec}
                           - \vec{u}_{{\rm parton}\,k}^{\,\rm ass} \right)
    W_{\rm jet} \left( E_{{\rm jet}\,j}^{\,\rm rec},\ 
                       E_{{\rm parton}\,k}^{\,\rm ass};\ 
                       JES \right)
  \times
  \\
  \nonumber
  & &
    \prod_{\xi=x,y}
    W_{\rm recoil} \left(   \left( \vec{p}_{\rm recoil}^{\,\rm rec}\right)_\xi 
                          - \left(-\vec{p}_{\ttbar}^{\,\rm ass}\right)_\xi \right)
  \ ,
\end{eqnarray}
where the four lines represent the contributions from electrons,
muons, jets, and the recoil energy of the \ttbar system, respectively.
It is understood that a term only appears if the corresponding
particle appears in the final state under consideration.
The symbols $n_e$, $n_\mu$, and $n_j$ stand for the numbers of electrons, 
muons, and jets in the final state, and $e$, $m$, and $j$ stand for a
specific reconstructed particle.
The number of possible assignments of jets $j$ to final-state partons
$k$ is denoted by $n_{\rm comb}$, and $i$ stands for one specific
permutation.
The summation over reconstructed jets $j$ implies a sum over
final-state partons $k$.
The reconstructed (rec) and assumed (ass) values of the
energy $E$ and momentum vector $\vec{p}$, the unit vector $\vec{u}$
along the direction of the momentum, and the charge $q$ of a particle
enter the transfer function.
The terms describing the muon and jet resolution are parametrized
in \qoverptmu and the jet/parton energy, respectively, since these
are the quantities measured in the detector.
For the \ttbar recoil, the $x$ and $y$ components are assumed to be 
independent.
An additional term is added corresponding to a sum over all
possibilities for final-state partons faking an electron, if applicable.
In Equation~(\ref{eq:tfdefinition-topo}) it is assumed that
reconstructed charged leptons can be unambiguously assigned to 
final-state leptons; this is justified even in the
dilepton channel because both jet-parton assignments are considered.

If no $b$-tagging information is used, the information from the reconstructed 
jet momentum vectors determines the relative weight of different
jet-parton assignments for a given partonic final state.
The inclusion of $b$-tagging information allows for an
improved identification of the correct jet-parton assignment
in final states like \ljets \ttbar events that contain 
\bquark quarks as well as light partons.
This can be encoded in the jet transfer function by an additional 
factor $W_b$ for each jet:
\begin{eqnarray}
  \label{eq:tfdefinition-btag}
    W(x,y;\,JES)
  & = &
    \prod_{e=1}^{n_e} \delta^{(3)} \! \left(   \vec{p}_e^{\,\rm rec}
                                             - \vec{p}_e^{\,\rm ass} \right)
  \times
  \\
  \nonumber
  & &
    \prod_{m=1}^{n_\mu} \delta^{(2)} \! \left(   \vec{u}_\mu^{\,\rm rec}
                                               - \vec{u}_\mu^{\,\rm ass} \right)
    W_{\mu}\left( \qoverptmurec,\qoverptmugen \right)
  \times
  \\
  \nonumber
  & &
    \frac{1}{n_{\rm comb}} 
    \sum_{i=1}^{n_{\rm comb}}
    \left(
    \prod_{j=1}^{n_j} 
    \delta^{(2)} \! \left(   \vec{u}_{{\rm jet}\,j}^{\,\rm rec}
                           - \vec{u}_{{\rm parton}\,k}^{\,\rm ass} \right)
    W_{\rm jet} \left( E_{{\rm jet}\,j}^{\,\rm rec},\ 
                       E_{{\rm parton}\,k}^{\,\rm ass};\ 
                       JES \right)
    \right.
  \times
  \\
  \nonumber
  & &
    \left.
    \phantom{
    \frac{1}{n_{\rm comb}} 
    \sum_{i=1}^{n_{\rm comb}}
    \left(
    \prod_{j=1}^{n_j} 
    \right.
    }
    W_b \left( {\cal B}_{{\rm jet}\,j}^{\,\rm rec},\ 
               \phi_{{\rm parton}\,k}^{\,\rm ass} \right)
    \right)
  \times
  \\
  \nonumber
  & &
    \prod_{\xi=x,y}
    W_{\rm recoil} \left(   \left( \vec{p}_{\rm recoil}^{\,\rm rec}\right)_\xi 
                          - \left(-\vec{p}_{\ttbar}^{\,\rm ass}\right)_\xi \right)
  \ .
\end{eqnarray}
The quantity $W_b$ describes the probability for parton $k$ with given assumed
flavor $\phi_{{\rm parton}\,k}^{\,\rm ass}$ to be reconstructed with 
$b$-tagging information ${\cal B}_{{\rm jet}\,j}^{\,\rm rec}$.
If $b$ tagging is used as a binary decision (of a jet to be $b$-tagged
or not to be $b$-tagged) as is the case in the analyses described
here, then one simply has
\begin{equation}
  \label{binarybtaggingweight.eqn}
    W_b \left( {\cal B}_{{\rm jet}\,j}^{\,\rm rec},\ 
               \phi_{{\rm parton}\,k}^{\,\rm ass} \right)
  = 
    \left\{
    \begin{array}{r @{\ \ } l @{}}
          \epsilon_b \left( \phi_{{\rm parton}\,k}^{\,\rm ass} \right)
      &
        {\rm if\ the\ jet\ }j\ {\rm is\ }b{\text -}{\rm tagged\ and}
      \vspace{1.5ex}\\
        1-\epsilon_b \left( \phi_{{\rm parton}\,k}^{\,\rm ass} \right)
      &
        {\rm otherwise,}
    \end{array}
    \right.
\end{equation}
where $\epsilon_b \left( \phi \right)$ is the \bquark-tagging
efficiency for a jet from a parton of given flavor $\phi$.

In principle, the transfer function can still depend on the top quark
mass.
This comment applies in particular to the term describing the jet
transfer function: 
The event topology and thus the angular
separation between the jets depends on the top quark mass.
For example, the probability of misassignment of particles to the
wrong jet can therefore slightly depend on the top quark mass.
A study of the \mtop dependence of the transfer function 
is described in~\cite{bib-CDFljetsDLMmass}.
None of the analyses described here parametrize the \mtop dependence
explicitly in the transfer function; instead, the analyses rely on the 
calibration with fully simulated events to correctly account for any
such dependence in the measurement on data.

\subsubsection{Simplifying Assumptions}
\label{memeasurements.tf.simplif.sec}
In general, the analyses do not use the full transfer function given
in Equation~(\ref{eq:tfdefinition-btag}) but introduce further
simplifications.
These are discussed in this section.

\paragraph{\bf Muon Transfer Function:}
\label{memeasurements.tf.simplif.mu.sec}
The CDF analyses treat the muon transverse momentum as a well-measured
quantity, similar to the electron energy.
This is justified since muons at very high \pt such that they have a
sizeable effect on the top quark mass resolution consistute a small
fraction of the sample 
(cf.\ Figures~\ref{mtopresolution-emujet.fig}(b1)-(b6)). 
This assumption will lead to an increased pull width (since the 
uncertainty on the muon \pt is set to zero) to be
accounted for in the calibration, and to a slightly increased
measurement uncertainty (since the relative weight of events with 
and without a high-\pt muon is non-optimal).

\paragraph{\bf Jet Transfer Function:}
\label{memeasurements.tf.simplif.jt.sec}
The hadronization process depends on what kind of parton 
initiates a jet.
The analyses use the same transfer function to describe light-quark
(\uquark, \dquark, \squark, and \cquark) 
and gluon jets; an independent transfer function is 
used for \bquark jets.
The \dzero experiment further distinguishes between \bquark jets that
contain a reconstructed (soft) muon and other \bquark jets:
The muon is taken as an indication for a semimuonic bottom- or 
charm-hadron decay, and the special transfer function allows to account on
average for the energy carried by the unreconstructed neutrino.
Semielectronic decays or semimuonic decays where the muon is not
identified are not treated explicitly and still have to be accounted
for on average by the generic \bquark-quark transfer function.

\paragraph{\bf Treatment of {\boldmath\bquark}-Tagged Jets:}
\label{memeasurements.tf.simplif.btagging.sec}
The $b$-tagging efficiency is much larger for \bquark-quark jets 
than for jets from light quarks or gluons.
The CDF collaboration therefore makes the assumption for
the calculation of the 
likelihood in their \ljets
analyses that a $b$-tagged jet {\em always} corresponds to a 
\bquark quark, if \bquark quarks are present in the final state.
Since CDF requires at least one $b$-tagged jet in the event,
the computation time is significantly reduced
since fewer
jet-parton assignments remain to be considered.
This assumption corresponds to the approximation that the $b$-tagging
efficiency for jets without a \bquark quark is zero, and
Equation~(\ref{binarybtaggingweight.eqn}) becomes
\begin{equation}
  \label{binarybtaggingweightpruning.eqn}
    W_b \left( {\cal B}_{{\rm jet}\,j}^{\,\rm rec},\ 
               \phi_{{\rm parton}\,k}^{\,\rm ass} \right)
  = 
    \left\{
    \begin{array}{c @{\ \ } l @{}}
            0
          &
            {\rm for\ }b{\rm{\text -}tagged\ light{\text -}quark / gluon\ jets,}
          \\
            \epsilon_b(b)
          &
            {\rm for\ }b{\rm{\text -}tagged\ }\bquark{\text -}{\rm quark\ jets,}
          \vspace{1.5ex}\\
            1
          &
            {\rm for\ untagged\ light{\text -}quark / gluon\ jets,\ and}
          \\
            1-\epsilon_b(b)
          &
            {\rm for\ untagged\ }\bquark{\text -}{\rm quark\ jets.}
    \end{array}
    \right.
\end{equation}
For the remaining jet-parton assignments, the 
overall factor $\prod_{j=1}^{4} W_b\left( {\cal B}_{{\rm jet}\,j}^{\,\rm rec},\ 
                \phi_{{\rm parton}\,k}^{\,\rm ass} \right)$
in the signal likelihood is then
\begin{equation}
  \label{binarybtaggingweightproductpruning.eqn}
    \!\!\!\prod_{j=1}^{4} W_b\left( {\cal B}_{{\rm jet}\,j}^{\,\rm rec},\ 
                              \phi_{{\rm parton}\,k}^{\,\rm ass}
                              \right)
  =
    \left\{\!\!
    \begin{array}{c @{\times} c @{\times} c @{\times} c @{\ \ } l @{}}
        \epsilon_b(b) & \left( 1-\epsilon_b(b) \right) & 1 & 1
      &
        {\rm for\ events\ with\ one\ }b{\rm{\text -}tagged\ jet\ and}
      \\
        \epsilon_b(b) & \epsilon_b(b) & 1 & 1
      &
        {\rm for\ events\ with\ two\ }b{\rm{\text -}tagged\ jets,}
    \end{array}
    \right.
\end{equation}
where the two factors for the two jets that are assumed to originate from
\bquark quarks are given first.
For both single- and double-tagged events, this product is almost identical
for all jet-parton assignments considered.
In the measurement using the Dynamical Likelihood method, in which
only the signal likelihood is considered, this yields one
multiplicative scale factor for each event likelihood.
Such a scale factor is irrelevant for the \mtop fit and can thus be
neglected.

For the calculation of the background likelihood in the Matrix
Element measurements in the \ljets channel, only processes without
\bquark quarks in the final state are considered, and the 
overall factor $\prod_{j=1}^{4} W_b\left( {\cal B}_{{\rm jet}\,j}^{\,\rm rec},\ 
               \phi_{{\rm parton}\,k}^{\,\rm ass} \right)$
can again be considered identical to a good approximation 
for all jet-parton assignments, such that it only depends on the
number of $b$-tagged jets in the event.

To proceed, one can divide the event sample into events with zero,
one, and two $b$ tags and determine the relative normalization
of the signal and background likelihoods separately for the three subsamples.
This approach was chosen in the \dzero analysis and is described in
the following.

In contrast to the approximation made by CDF that $b$-tagged jets
always correspond to \bquark quarks, the \dzero \ljets
analysis considers all possible jet-parton assignments even
if $b$-tagged jets are present.
For the \ttbar likelihood, 
the factor $W_b\left( {\cal B}_{{\rm jet}\,j}^{\,\rm rec},\ 
               \phi_{{\rm parton}\,k}^{\,\rm ass} \right)$
is taken from Equation~(\ref{binarybtaggingweight.eqn}) with the only
modification that it is approximated to 1.0 for all jets if none of
the four jets is $b$-tagged.
Nevertheless, 
assumptions on the jet flavors are introduced for the calculation of
$W_b\left( {\cal B}_{{\rm jet}\,j}^{\,\rm rec},\ 
           \phi_{{\rm parton}\,k}^{\,\rm ass} \right)$
such that the likelihoods for \ttbar events with $W$ decays to 
$u\bar{d}'$ and $c\bar{s}'$ final states
need not be calculated separately,
allowing for a reduction of the computation time.
If an event contains exactly one $b$-tagged jet, the quarks from the 
hadronic $W$ decay are both assumed to be light quarks ($u$, $d$, or $s$).
This is justified since the tagging efficiencies for $b$ jets are 
much larger than those for other flavors, and there are two $b$ jets
per event.
For events with two or more $b$-tagged jets, a charm jet from the 
hadronic $W$ decay is tagged in a non-negligible fraction of cases.
Consequently, the quarks from the hadronic $W$ decay are assumed to be charm
quarks if the corresponding jet has been tagged, and light quarks
otherwise.

The improvement from the inclusion of 
jet-parton assignments with a tagged charm jet
in the likelihood calculation can be seen by comparing the signal and
background likelihoods.
Figure~\ref{fig:nopruning}(a) shows the ratio of \ttbar to background
likelihoods in simulated \ljets 
\ttbar events reconstructed at \dzero with two $b$-tagged jets
when only the two jet-parton assignments in which tagged jets
are assigned to $b$ quarks are considered in the signal likelihood 
calculation. 
The hatched histogram shows the correct assignments only, whereas the 
open histogram shows all combinations, including the ones in which a 
charm quark from the \W decay was tagged.
Figure~\ref{fig:nopruning}(b) shows the same ratio when all combinations are
included with their appropriate weights as discussed above.
The tail for low signal to background likelihood ratios in
Figure~\ref{fig:nopruning}(a) arises because the correct jet-parton
assignment is not included in the calculation in events where one of
the tagged jets comes from a charm quark.
\begin{figure}[ht]
\begin{center}
\includegraphics[width=\figMEcalibwidth\textwidth]{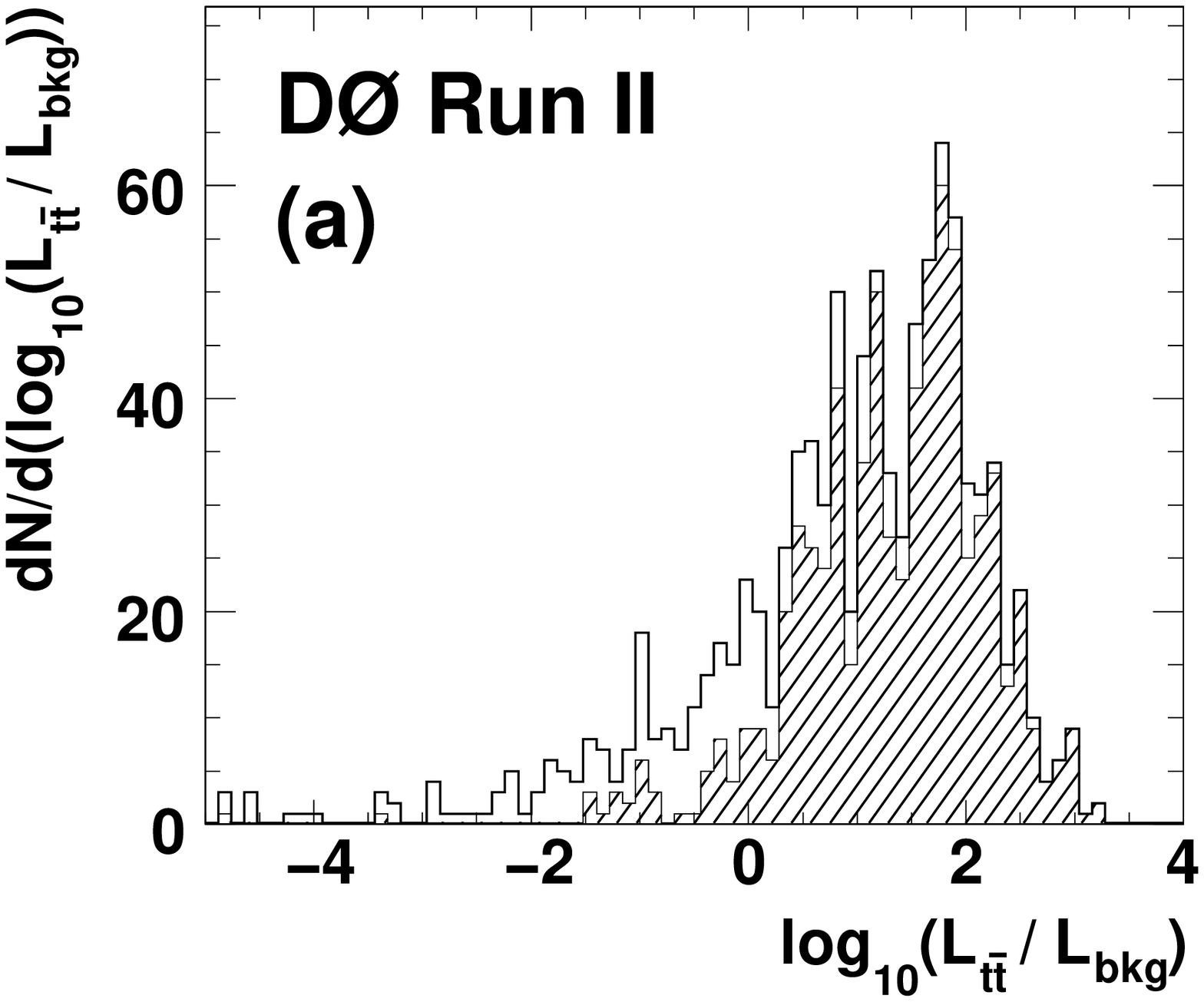}
\hspace{-0.03\textwidth}
\includegraphics[width=\figMEcalibwidth\textwidth]{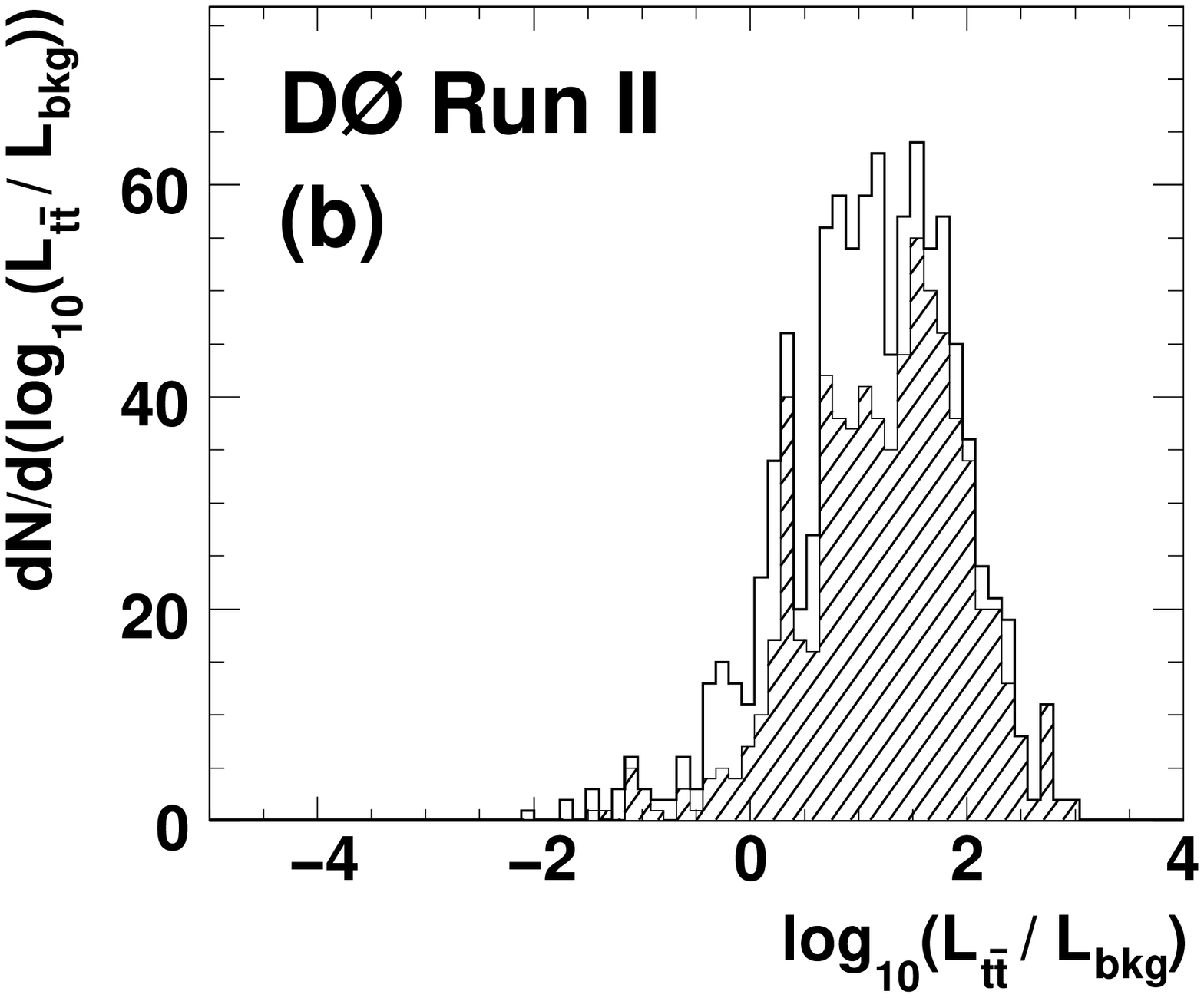}
\caption{\captionfont\label{fig:nopruning}Monte Carlo study of the effect of charm-jet tagging on the
  signal to background likelihood ratio in the \dzero Matrix Element 
  analysis in the \ljets channel~\cite{bib-me}, 
  for \ttbar events generated with $\mtop=175\,\GeV$ 
  reconstructed with the \dzero detector
  that contain two $b$-tagged jets.  
  The \psgn values are calculated for the assumption $\mtop=175\,\GeV$.
  (a) Only the two jet-parton assignments in which tagged jets
  are assigned to $b$ quarks are considered. 
  (b) All weighted jet parton-assignments enter the likelihood calculation.
  In both plots, the hatched histogram corresponds to those cases
  where the two $b$-tagged jets are correctly assigned to $b$ quarks,
  which happens $84\%$ of the time in the double-tag sample.}
\end{center}
\end{figure}

The different 
flavor contributions to the \wjets process are parametrized by the 
\wjets matrix element without heavy flavor quarks in the final state,
which means that the factors 
$W_b\left( {\cal B}_{{\rm jet}\,j}^{\,\rm rec},\ \phi_{{\rm parton}\,k}^{\,\rm ass} \right)$
for the 
background likelihood are all equal for a given event
even if $b$-tagged jets are
present.
Therefore, the factors are omitted altogether from the background
likelihood calculation.
To account for the different amount of background in the event
categories with zero, one, and two (or more) $b$ tags, the relative
normalization of \ttbar and background likelihoods in the three
samples is adjusted accordingly, as suggested by
Equation~(\ref{basicPevt.eqn}).

\paragraph{\bf Transfer Function for the Recoil Energy:}
\label{memeasurements.tf.simplif.recoil.sec}
The transfer function for the recoil energy is so far only included in the
CDF measurement in the dilepton
channel~\cite{bib-CDFdileptonMEpaper}.
In the \ljets analyses, this term is omitted, which leads to a slightly
increased expected statistical measurement uncertainty 
but avoids any
explicit dependence on the modeling of the unclustered transverse
momentum in the 
simulation.

\subsubsection{Parametrization of the Jet Energy Resolution}
\label{memeasurements.tf.jt.sec}
The jet energy transfer function,
    $W_{\rm jet} \left( E_{{\rm jet}\,j}^{\,\rm rec},\ 
                        E_{{\rm parton}\,k}^{\,\rm ass};\ 
                        JES \right)$,
yields the probability density for a measurement 
$E_{{\rm jet}\,j}^{\,\rm rec}$ in the
detector if the true quark energy is $E_{{\rm parton}\,k}^{\,\rm ass}$,
given an overall jet energy scale \jes.
Both CDF and \dzero describe it as a function of the difference 
$\Delta E = E_{{\rm jet}\,j}^{\,\rm rec} - E_{{\rm parton}\,k}^{\,\rm ass}$ 
and the assumed parton energy $E_{{\rm parton}\,k}^{\,\rm ass}$ (the
Dynamical Likelihood analysis uses the transverse energy instead of the
energy).
Different transfer functions are determined for jets from light quarks
and gluons and for \bquark-quark jets, and \dzero also treats \bquark jets with
a soft muon from semimuonic heavy hadron decay separately from other
\bquark jets.
In the \dzero Matrix Element analysis and the CDF measurement with
the Dynamical Likelihood technique, different sets of transfer functions are 
also derived for different $|\eta|$ regions.

In their Matrix Element analyses, both CDF and \dzero 
use a double Gaussian as a function of $\Delta E$
with parameters that depend linearly on the assumed quark energy
to describe the jet energy transfer function\footnote{In the 
Dynamical Likelihood analysis, CDF does not use a
parametrization of the transfer function, but uses random numbers
generated according to the distributions.}.
For the case $JES=1$, it is parametrized as
\begin{eqnarray}
  \label{eq:MEtf}
    W_{\rm jet}(E_{{\rm jet}\,j}^{\,\rm rec}, 
                E_{{\rm parton}\,k}^{\,\rm ass}; 
                JES=1) 
  & = &
    \frac{1}{\sqrt{2\pi}(p_2+p_3p_5)} \times
  \\
  \nonumber
  & &
    \left[\exp{\left(-\frac{(\Delta E-p_1)^2}{2p_2^2}
    \right)} +
    p_3\exp{\left(-\frac{(\Delta E - p_4)^2}{2p_5^2}\right)}\right] 
  \ .
\end{eqnarray}
The parameters $p_i$ are themselves functions of the quark energy, and are
parametrized as linear functions of the quark energy so that
\begin{equation}
p_i = a_i + b_i E_{{\rm parton}\,k}^{\,\rm ass}
\ ,
\end{equation}
with $a_3$ fixed to 0 in the \dzero analysis.
The parameters $a_i$ and $b_i$ are determined in a fit from 
simulated \ttbar events,
after all jet energy corrections have been applied.
The \dzero transfer function for light quarks in the region $|\eta|<0.5$ is 
shown in Figure~\ref{fig:transfer}.

For $JES\neq1$, the jet transfer function is modified as follows:
\begin{equation}
  \label{eq:MEtfjes}
    W_{\rm jet}(E_{{\rm jet}\,j}^{\,\rm rec},
                E_{{\rm parton}\,k}^{\,\rm ass}; 
                JES) 
  =
    \frac{W_{\rm jet}(\frac{E_{{\rm jet}\,j}^{\,\rm rec}}{JES},
                      E_{{\rm parton}\,k}^{\,\rm ass}; 1)}
         {JES} 
  \ ,
\end{equation}
where the factor $JES$ in the denominator ensures the correct
normalization
\begin{equation}
  \label{tfjtnorm.eqn}
     \int_{E_{{\rm jet}\,j}^{\,\rm rec}} 
     W_{\rm jet}(E_{{\rm jet}\,j}^{\,\rm rec},
                 E_{{\rm parton}\,k}^{\,\rm ass}; 
                 JES)
     {\rm d}E_{{\rm jet}\,j}^{\,\rm rec} 
  = 
    1
  \ .
\end{equation}
\begin{figure}[htb]
\begin{center}
\includegraphics[width=0.45\textwidth]{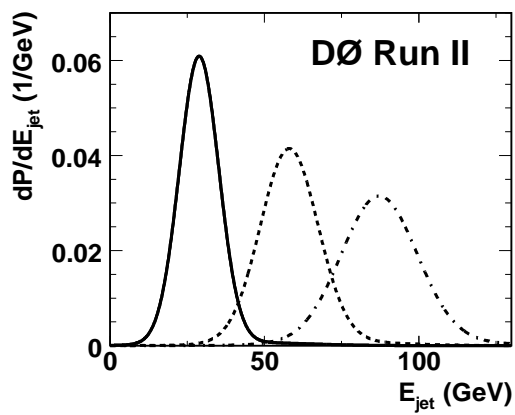}
\caption{\captionfont\label{fig:transfer}Jet energy transfer functions for 
light-quark jets in the \dzero detector in the region
$\left|\eta\right|<0.5$, for parton energies $E_{p}=30\ {\rm GeV}$
(solid), $60\ {\rm GeV}$ (dashed), and $90\ {\rm GeV}$ (dash-dotted
curve).  The parametrization corresponds to the reference jet
energy scale, $JES=1.0$~\cite{bib-me}.}
\end{center}
\end{figure}

\subsubsection{Parametrization of the Muon Momentum Resolution}
\label{memeasurements.tf.mu.sec}
Only the \dzero experiment so far considers the muon resolution 
explicitly in the likelihood calculation.
To describe the resolution of the central tracking chamber, 
the resolution of the charge divided by the transverse momentum of a
particle is considered as a function of pseudorapidity.
The muon transfer function is parametrized as
\begin{equation}
  \label{eq:MEtfmu}
    W_{\mu}\left( \qoverptmurec,\qoverptmugen \right)
  =
    \frac{1}{\sqrt{2\pi}\sigma} 
    \exp\left( -\frac{1}{2} \left( \frac{ \qoverptmurec - \qoverptmugen } 
                                        { \sigma }
                            \right)^2
        \right)
  \ ,
\end{equation}
where $q$ denotes the charge and $\pt$ the transverse momentum assumed
(ass) or reconstructed (rec) for a muon.
The resolution
\begin{equation}
  \label{eq:MEsigmamu}
    \sigma
  =
    \left\{
    \begin{array}{cc}
        \sigma_0
      &
        {\rm for}\ |\eta|\le\eta_0
      \vspace{2ex}\\
        \sqrt{   \sigma_0^2 
               + \left[ c \left( |\eta| - \eta_0 \right) \right]^2 
             }
       &
        {\rm for}\ |\eta|>\eta_0
    \end{array}
    \right.
\end{equation}
is obtained from muon tracks in simulated events.
The muon charge is not used in the calculation of \psgn and \pbkg;
however, for muons with large transverse momentum 
the possibility of reconstruction of a track
bent in the wrong direction is automatically
taken into account in the
transfer function when using this parametrization.

\subsubsection{The Transfer Function for the Unclustered Transverse Momentum}
\label{memeasurements.tf.euncl.sec}
The probability density to observe unclustered transverse momentum in the event
is only used in the CDF dilepton analysis; it is parametrized
as a Gaussian in each of the $x$ and $y$ directions, with no
correlation.

\subsection[The Signal Likelihood \psgn]{The Signal Likelihood {\boldmath\psgn}}
\label{memeasurements.psgn.sec}
When spin correlations between the top and antitop quarks are
neglected, the leading-order matrix element for the process
$\qqbar\to\ttbar$ is given by~\cite{bib-mahlonparke}
\begin{equation}
\label{eq:MEsignalME}
    |\mathscr{M}_{\qqbar\to\ttbar}|^2 
  = 
    \frac{g_s^4}{9} F \overline{F} \left( 2 - \beta^2 s_{qt}^2 \right) 
  \ ,
\end{equation}
where $g_s^2/(4\pi)=\alpha_s$ is the strong coupling constant, 
$\beta$ is the velocity
of the top quarks in the \ttbar rest frame, and $s_{qt}$ denotes the
sine of the angle between the incoming parton and the outgoing top
quark in the \ttbar rest frame.
If the top quark decay products include a leptonically decaying $W$ boson,
while the antitop decay includes a hadronically decaying $W$, one
has
\begin{eqnarray}
\label{eq:MEsignalME_F}
  F
& = &
  \frac{g_w^4}{4}
  \left( \frac{   \mbellnu^2 - \mellnu^2 }
              {   \left( \mbellnu^2 - \mt^2 \right)^2 
                + \left( \mt \Gt \right)^2 } \right)
  \left( \frac{   \mbellnu^2 \left( 1 - \chatbell^2 \right)
                + \mellnu^2 \left( 1 + \chatbell \right)^2 }
              {   \left( \mellnu^2 - \mW^2 \right)^2
                + \left( \mW \GW \right)^2 } \right)
\ , \\
\label{eq:MEsignalME_Fbar}
  \overline{F}
& = &
  \frac{g_w^4}{4}
  \left( \frac{   \mbdubar^2 - \mdubar^2 }
              {   \left( \mbdubar^2 - \mt^2 \right)^2 
                + \left( \mt \Gt \right)^2 } \right)
  \left( \frac{   \mbdubar^2 \left( 1 - \chatbbard^2 \right)
                + \mdubar^2 \left( 1 + \chatbbard \right)^2 }
              {   \left( \mdubar^2 - \mW^2 \right)^2
                + \left( \mW \GW \right)^2 } \right)
\end{eqnarray}
(for the reverse case in \ljets events, replace 
$b \leftrightarrow \overline{b}$, $\ell \leftrightarrow d$, 
and $\nu \leftrightarrow \overline{u}$;
for dilepton events, replace $d$ and $\overline{u}$ by the 
second charged lepton and neutrino, respectively).
Here, $g_w$ denotes the weak charge ($G_{\rm F}/\sqrt{2}=g_w^2/8 m_W^2$),
\mt and \mW are the masses of the top quark (which is to be
measured) and the $W$ boson, and \Gt and \GW are their widths. 
Invariant top and $W$ masses in a particular event are denoted by
$m_{xyz}$ and $m_{yz}$, respectively, where $x$, $y$, and $z$ are the
decay products.
The cosine of the angle between particles $x$ and $y$ in the $W$ rest
frame is denoted by ${\hat c}_{xy}$.
Here and in the following, the symbols $d$ and $\overline{u}$ stand
for all possible decay products in a hadronic \W decay.
The top quark width is given as a function of the top quark 
mass as~\cite{bib-pdg}
\begin{equation}
  \label{topwidth.eqn}
    \Gt
  =
    \frac{G_{\rm F} \mtop^3}
         {8 \pi \sqrt{2}}
    \left( 1 - \frac{\mW^2}{\mtop^2} \right)^2
    \left( 1 + 2\frac{\mW^2}{\mtop^2} \right)
    \left[ 1 - \frac{2\alpha_s}{3\pi}
               \left( \frac{2\pi^2}{3} - \frac{5}{2} \right) \right]
  \ .
\end{equation}

The correct association of reconstructed jets with the final-state
quarks in Equations~(\ref{eq:MEsignalME_F}) and (\ref{eq:MEsignalME_Fbar})
is not known.
Therefore, the transfer function takes into account all possible jet-parton
assignments as described in Section~\ref{memeasurements.tf.sec}.
However, in the case of the signal likelihood for \ljets events, 
the mean value of the two assignments with the 4-momenta of the
quarks from the hadronic $W$ decay interchanged may be computed explicitly
by using the symmetrized formula
\begin{eqnarray}
\label{eq:MEsignalME_Fbarsymm}
  \overline{F}
 & = &
  \frac{g_w^4}{4}
  \left( \frac{   \mbdubar^2 - \mdubar^2 }
              {   \left( \mbdubar^2 - \mt^2 \right)^2 
                + \left( \mt \Gt \right)^2 } \right)
  \left( \frac{   \mbdubar^2 \left( 1 - \chatbbard^2 \right)
                + \mdubar^2 \left( 1 + \chatbbard^2 \right) }
              {   \left( \mdubar^2 - \mW^2 \right)^2
                + \left( \mW \GW \right)^2 } \right)
\end{eqnarray}
instead of~\Eref{eq:MEsignalME_Fbar}, where only the terms containing 
$\chatbbard$ are affected. Consequently, only a summation 
over half the jet-quark assignments remains to be evaluated.

The leading-order matrix element for the process $\glueglue\to\ttbar$ 
is~\cite{bib-CDFljetsDLMmass}
\begin{equation}
\label{eq:MEggME}
    |\mathscr{M}_{\glueglue\to\ttbar}|^2 
  = 
    g_s^4 F \overline{F} 
    \left( \frac{1}{6 \tau_1 \tau_2} - \frac{3}{8} \right)
    \left( \tau_1^2 + \tau_2^2 + \rho - \frac{\rho^2}{4 \tau_1 \tau_2} \right)
  \ ,
\end{equation}
with 
\begin{equation}
\label{eq:MEggMEsymbols}
    \tau_i = \frac{ m_{g_i b\ell\nu}^2 - m_{b\ell\nu}^2 }
                  { m_{b\ell\nu\overline{b}d\overline{u}}^2 }
  \
  {\rm and}\
    \rho = \frac{ 4\mt^2 }{ m_{b\ell\nu\overline{b}d\overline{u}}^2 }
  \ ,
\end{equation}
where $g_i$, $i=1,2$ denotes the two incoming gluons.
Here, again \ttbar spin correlations have been neglected.
This process is only taken into account explicitly in the CDF
measurement based on the Dynamical Likelihood method.
In the Matrix Element measurements it is not computed because the 
top and \W propagator and decay parts of the matrix element, which contain most
of the information on the top quark mass and the separation of signal
and background events, are identical.

The computation of the signal likelihood \psgn 
involves an integral over the momenta of the colliding partons and
over 6-body phase space to cover all
possible partonic final states, cf.~Equation~(\ref{eq:MEpsgn}).
The number of dimensions of the integration is reduced by the 
following conditions:
\begin{itemize}
\item
  The transverse momentum of the colliding partons is assumed to be
  zero, or to be consistent with the observed unclustered transverse energy (in
  the Dynamical Likelihood measurement).
  The transverse momentum of the
  \ttbar system then follows from conservation of 4-momentum
  because the leading-order matrix element is used to 
  describe \ttbar production.  Also, the $z$ momentum and energy of
  the \ttbar system are known from the momenta of the colliding partons.
\item
  The directions of the quarks and the charged lepton in the final
  state are assumed to be exactly measured. 
\item
  The energy of electrons from $W$ decay is assumed to be perfectly
  measured. The corresponding statement is not
  necessarily true for high momentum muons, and an integration over
  the muon momentum is performed in the \dzero analysis.
\end{itemize}

Even after these considerations, a multi-dimensional integral
remains to be calculated.
In the Matrix Element analyses, this calculation
is performed numerically with the Monte Carlo program
\vegas~\cite{Lepage:1977sw,Lepage:1980dq}.

\subsection{The Background Likelihood}
\label{memeasurements.pbkg.sec}
There are in general many background processes that can lead to an
observed event.
It is not problematic per se to not fully account for all backgrounds 
in the event likelihood; in fact, the Dynamical Likelihood measurements by 
CDF omit any explicit treatment of background in the likelihood.
Because of the assumptions made in the Matrix Element technique,
it is always necessary to calibrate the measurement technique with 
pseudo-experiments with varying input top quark masses, jet energy scales,
and sample compositions as described in
Section~\ref{massfit.validationandcalibration.sec}.
An incomplete background likelihood will lead to a shift of the 
measured top quark mass value; this shift will in general depend 
on the top quark mass itself and on the fraction of events in the 
sample that are not accounted for in the overall likelihood.
The shift is determined in the calibration procedure.
When a background term is omitted in the event likelihood, the 
situation will thus be quantitatively, but not qualitatively
different from that in an analysis that includes this term in the 
likelihood.

If several different background
processes have similar kinematic characteristics, it is also
possible to approximately describe the total background by
the likelihood for only one of the background processes, multiplied
by the total background fraction, cf.\ Equation~(\ref{eq:MEpevt}).
This technique has been applied by both CDF and
\dzero in the Matrix Element analyses in the \ljets channel, where a
likelihood for QCD multijet production is not explicitly calculated.
While this is a better approximation than not accounting for multijet 
background at all, it still has to be studied with pseudo-experiments
and taken into account in the calibration.
It should be noted that independently of the definition of the
background likelihood used,
any uncertainty in the characteristics of a background process has to 
be evaluated with pseudo-experiments and
accounted for by a systematic error on the final measurement value, 
see Section~\ref{systuncs.physicsmodeling.background.sec}.

Even if only leading-order background processes and only the most important among
them are considered, it is not practical to explicitly evaluate all
individual diagrams.
Instead, routines from existing Monte Carlo generators are used to 
compute the likelihood for generic processes.
They take into account 
the relative importance of the various subprocesses that contribute
and perform a statistical sampling of all possible spin, flavor, and color
configurations.
Because the background likelihood does not depend on the top quark
mass, it does not have to be computed for as many different
assumptions as the signal likelihood and it is possible to evaluate
the matrix elements without a dedicated routine optimized for speed.

The generic background process taken into account by both CDF and
\dzero for the Matrix Element analyses in the \ljets channel is
the production of a leptonically decaying \W boson in association with
four additional light partons, \wfourpshort.
Events with a leptonically decaying \W boson
and four partons that include heavy-flavor
quarks are not considered separately because their kinematic
characteristics are very similar to those of \wfourpshort events.
QCD multijet production, the second-largest background source, 
is not taken into account explicitly in the event likelihood.

The modeling of the \wfourpshort process in the
\vecbos~\cite{bib-vecbos}
generator is used to calculate the background likelihood \pbkg.
The jet
directions and the charged lepton are taken as well-measured, also
for muons in the \dzero analysis.
The integral over the quark energies in Equation~(\ref{eq:MEdsigmapp}) is
performed by generating Monte Carlo events with parton energies
distributed according to the jet transfer function.
In these Monte Carlo events, the neutrino transverse momentum is 
given by the condition that the transverse momentum of the \wjets
system be zero, while the invariant mass of the charged lepton and
neutrino is assumed to be equal to the \W mass to obtain the 
neutrino $z$ momentum (both solutions are considered).
The mean result from all 24 possible assignments of jets to quarks in the 
matrix element is calculated, and the mean over a number of Monte
Carlo events is taken to be the \pbkg value.
The calibration described in
Section~\ref{massfit.validationandcalibration.sec} supports that it is
not necessary to compute \pbkg for different \jes values; only the
value $\pbkg\left(\jes=1\right)$ is used.

The CDF Matrix Element measurement in the dilepton channel considers
the following backgrounds explicitly in the event likelihood:
\begin{list}{$\bullet$}{\setlength{\itemsep}{0.5ex}
                        \setlength{\parsep}{0ex}
                        \setlength{\topsep}{0ex}}
\item
a leptonically decaying \Z boson in association with two partons,
\Zgammastartwopshort,
\item
two leptonically decaying \W bosons in association with two partons,
\WWtwopshort (this contribution is negligible if a \bquark-tagged jet
is required and thus only considered in the topological analysis), and
\item
one leptonically decaying \W boson in association with three partons,
one of which yields a jet that fakes an isolated charged lepton in the
detector, \Wthreepshort.
\end{list}
Routines from the {\sc alpgen}~\cite{bib-alpgen}
generator are used to perform the statistical sampling to 
average the differential cross section.
In the case of the \Zgammastartwopshort process, in which no energetic
neutrino occurs, the assumption of zero transverse momentum of the 
\Zgammastartwopshort system is relaxed, and an integration over all 
possible values of \pt is performed.
For the \Wthreepshort process, it is assumed that the isolated lepton
originating from the misidentified jet carries most of the jet energy
(otherwise it would not appear isolated in the detector), and the 
jet energy transfer function is taken to relate it with the parton
energy.

\subsection{Normalization of the Likelihood for one Process}
\label{memeasurements.normpprc.sec}
The likelihood for a process has to be normalized by the 
cross section $\sigma^{\rm obs}$ for {\em observed} events in the detector, as
described in Equation~(\ref{eq:MEpsgn}).
The cross section for observed events depends not only on the top
quark mass (in the case of \psgn), but via the jet
$\et$ requirements in the event selection
also on the assumed value of the $JES$ parameter.

To normalize the signal likelihood in the \dzero Matrix Element
analysis, the integral 
$  \sigmaobs_{\ttbar} 
 = \int \dsigma_{\ttbar}(\ppbar\to x;\,\mtop,JES) f_{\rm acc}(x) {\rm d}x$
has been computed
as a function of \mtop and $JES$ as described in
Equation~(\ref{eq:MEsigmaobs}). 
The results are shown in Figure~\ref{fig:MEnrmpsgn} for \ejets and
\mujets events as a function of \mtop for various choices of the
$JES$ scale factor.

\begin{figure}
\begin{center}
\includegraphics[width=0.45\textwidth]{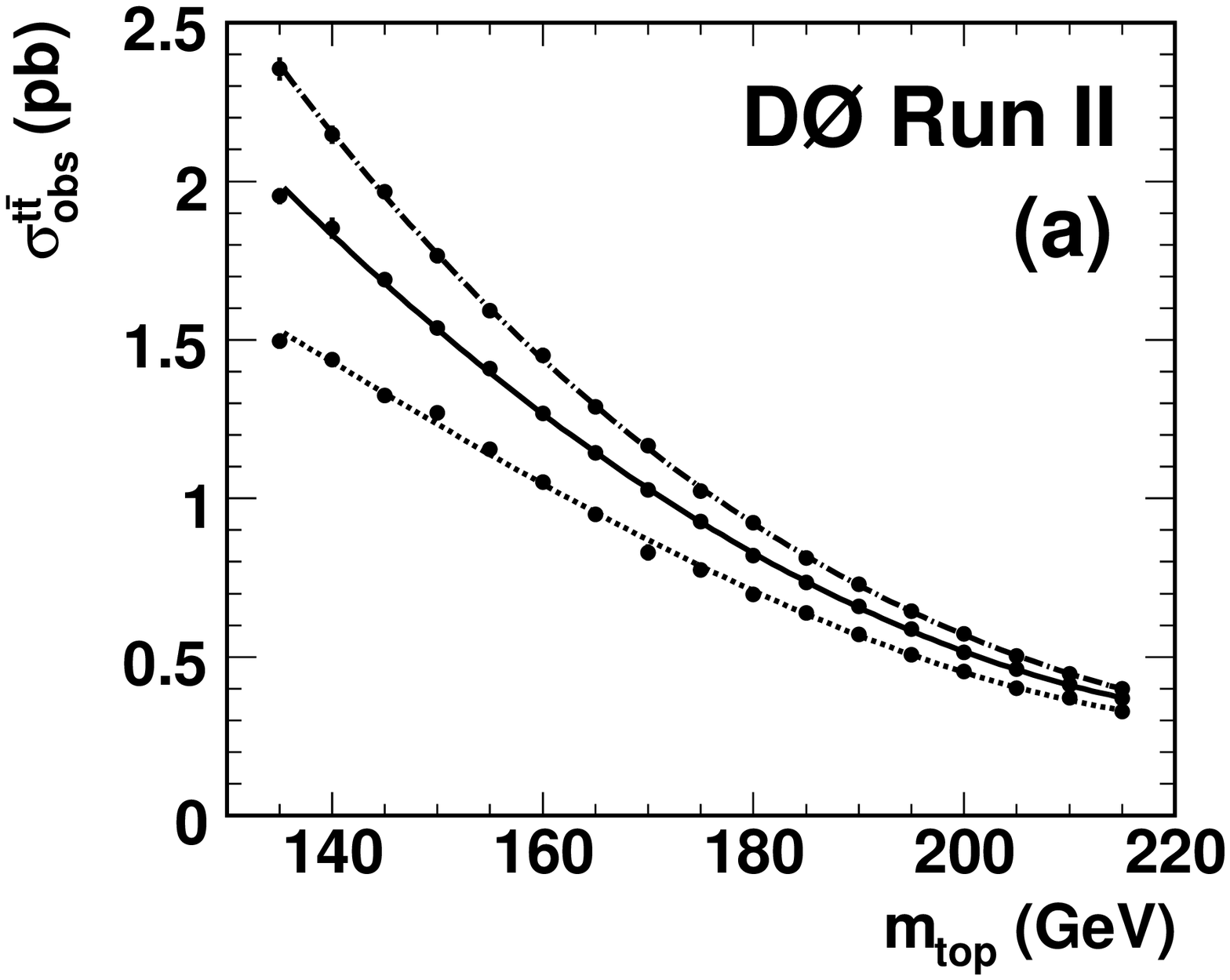}
\includegraphics[width=0.45\textwidth]{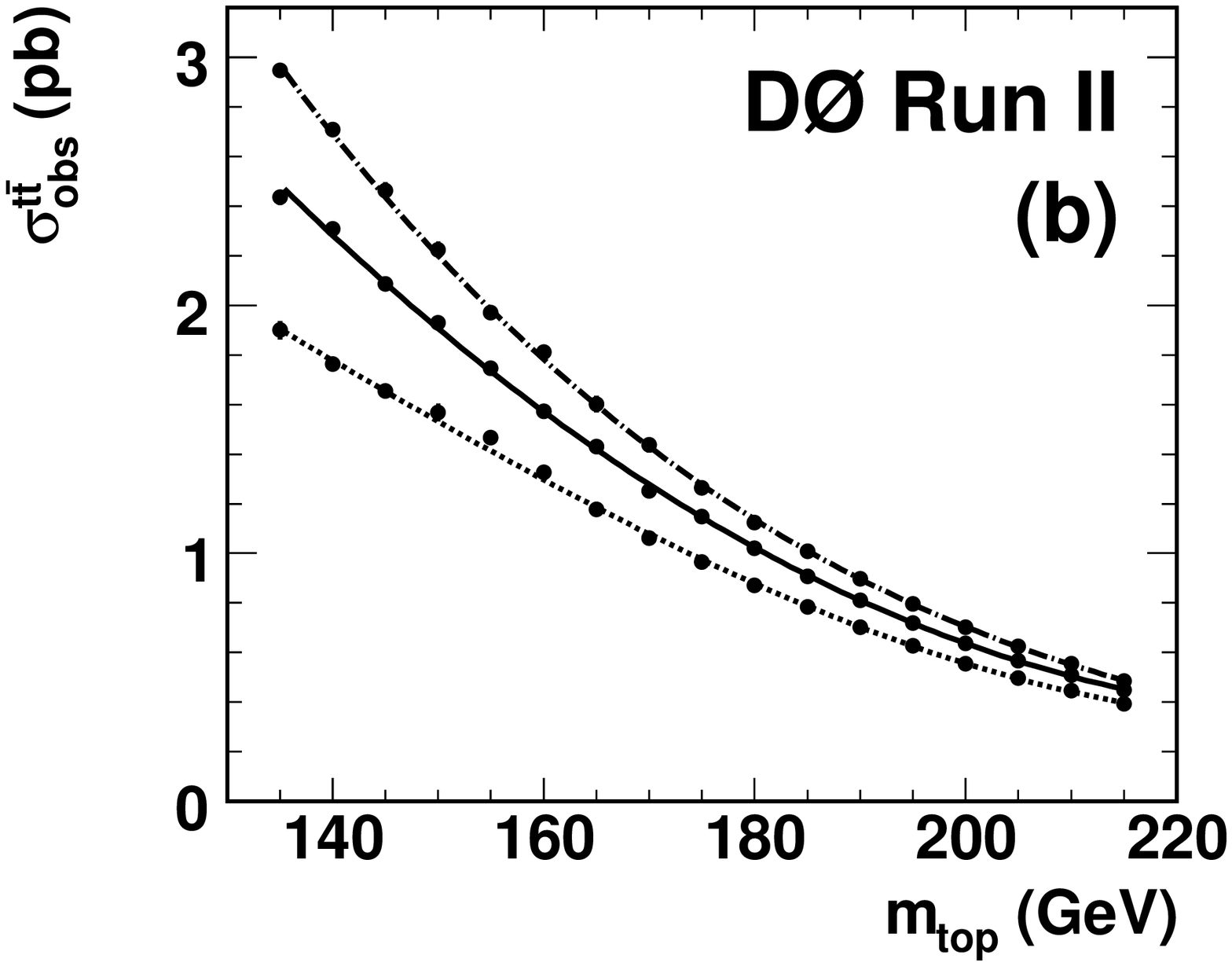}
\caption{\captionfont\label{fig:MEnrmpsgn}Cross section of observed \ttbar events 
  in the \dzero detector~\cite{bib-me}
  computed with the leading-order
  matrix element for (a) \ejets and (b) \mujets
  events as a function of the top quark mass \mtop for different
  choices of the $JES$ scale factor: $JES=1.12$ (dash-dotted), 
  $JES=1.0$ (solid), and $JES=0.88$ (dotted lines).
  The branching fraction $\ttbar\to\bbbar\ell\nu\qqbarprime$ is not
  included, as such a constant overall scale factor is irrelevant
  for the analysis.}
\end{center}
\end{figure}

The normalization of the background likelihoods can in principle be
determined in the same way.
The computation of the integral in
Equation~(\ref{eq:MEsigmaobs}) would be very computing intensive even
though the dependence of \pbkg on the $JES$ parameter does not have to 
be taken into account as shown in the \dzero analysis.
The \dzero experiment has therefore used a different method to compute
the relative normalization of signal and background likelihoods
(an overall scale factor is irrelevant in the analysis), 
assuming the relative contributions of the
individual background subprocesses to the total background likelihood
are known.
This approach makes use of the fact that the fitted signal fraction 
\ftop of the sample will be underestimated if the background
likelihood \pbkg is too large and vice versa.
The relative normalization can therefore be adjusted until the signal
fraction is determined correctly in pseudo-experiments of simulated events.
Note that this technique can only be used when the \ftop parameter
is left free in the fit (i.e.\ when no constraint from the \ttbar
and background cross section is used), as is the case for the CDF and \dzero
Matrix Element measurements in the \ljets channel.
The calibration of the \ftop fit result is further discussed in
Section~\ref{massfit.validationandcalibration.sec}.

\clearpage
\section{The Ideogram Measurement Method}
\label{idmeasurements.sec}

\begin{center}
\begin{tabular}{p{15cm}}
{\it This section describes the Ideogram method and its application in
  top quark mass measurements.  Like the Matrix
  Element and Dynamical Likelihood techniques discussed in the
  previous section, this method is based on a per-event
  likelihood that depends on the top quark mass.
  The signal and background likelihoods are
  however only based on the reconstructed top quark and \W boson masses
  in each event and do not make use of the full
  kinematic information.  This means that the
  amount of computation time needed for the analysis is reduced
  significantly.}
\end{tabular}
\end{center}

The Ideogram method has previously been used to measure the 
\W boson mass at the DELPHI experiment at LEP~\cite{Abreu:1997ic}.
It is now also applied by the \dzero and CDF experiments to measure
the top quark mass using \ttbar events in the \ljets~\cite{bib-ID}
and \alljets channels~\cite{bib-CDFallhadID}, respectively.
In Section~\ref{idmeasurements.evtselkinfit.sec}, the 
event selection and reconstruction using a kinematic fit
are summarized for these two analyses.
The definition of the event likelihood in the Ideogram method is then
discussed in Section~\ref{idmeasurements.pevt.sec} and compared with 
the approach in the Matrix Element and Dynamical Likelihood methods.

\subsection{Event Selection and Kinematic Reconstruction}
\label{idmeasurements.evtselkinfit.sec}
The event selection in the \dzero \ljets analysis is identical to the 
one used in the Matrix Element measurement described in
Section~\ref{memeasurements.evtsel.sec}, except that events are also
used if more than four jets are reconstructed (only the four
highest-$\et$ jets are used to measure the top quark mass).
There is an additional cut on the $\chi^2$ obtained from a kinematic
fit as described below.

The kinematic requirements on the events in the CDF measurement in the 
\alljets channel are similar to those described in
Section~\ref{templatemeasurements.allhad.evtsel.sec}.
The most important requirements are:
\begin{list}{$\bullet$}{\setlength{\itemsep}{0.5ex}
                        \setlength{\parsep}{0ex}
                        \setlength{\topsep}{0ex}}
\item
no significant missing transverse energy,
\item
removal of events with a charged lepton with high \pt,
\item
events must contain between 6 and 8 jets within $|\eta|<2.0$
with $\et>15\,\GeV$, and
\item
the sum of jet transverse energies must satisfy 
$\sum_{\rm jets} \et > 280\,\GeV$.
\end{list}
There are additional event quality cuts and requirements on the event
shape using the aplanarity and centrality.

The events are then subjected to a kinematic fit constraining them to
the \ttbar hypothesis.
In the \dzero \ljets analysis, the kinematic fit is similar to the one described
in Section~\ref{templatemeasurements.kinrec.mtopestimator.sec},
yielding one fitted top quark mass $\mtop^i$, the corresponding
uncertainty $\sigma_{\mtop}^i$ and the best $\chi^2_i$
for each of the 12 different jet-parton assignments (an
interchange of the two jets assumed to come from the hadronic \W decay
does not change the kinematic fit) and for each of the two possible
solutions for the longitudinal neutrino momentum component $\pznu$.
The index $i$ thus runs over 24 different possibilities.
All of these values depend on the assumed value $JES$ of the jet
energy scale.

The kinematic fit in the \alljets case is identical to the one
discussed in Section~\ref{templatemeasurements.allhad.estimator.sec}
except that in this analysis, the masses of the two decaying top quarks per
event are treated as independent fit parameters.
Thus, for each of the 90 jet-parton assignments $i$ that have to be
distinguished in an event, 
the two fitted top quark masses $\mtop^{i,\,1/2}$ and their
uncertainties $\sigma_{\mtop}^{i,\,1/2}$ are determined together with
the minimum $\chi^2_i$.
In the analysis in the \alljets channel, no in situ calibration of
the jet energy scale is performed so far.

\subsection{The Event Likelihood}
\label{idmeasurements.pevt.sec}
The definition of the likelihood \pevt to observe a given selected event is
identical to that used in the Matrix Element method, 
cf.\ Section~\ref{memeasurements.pevt.sec}:
\begin{equation}
  \label{eq:IDpevt}
    \pevt\left(x;\,\mtop,JES,\ftop\right) 
  = 
                     \ftop \psgn\left(x;\,\mtop,JES\right) 
    + \left(1-\ftop\right) \pbkg\left(x;\,JES\right)
  \ ,
\end{equation}
where \psgn and \pbkg are the likelihoods to observe the event if it
was produced via the signal or any of the background processes, respectively,
\ftop is the overall fraction of signal events in the selected event
sample, $x$ denotes the event observables, and \mtop and \jes are the 
assumed values of the top quark mass and jet energy scale which are to
be measured
(in the \alljets analysis, the parameter $JES$ is fixed to 1.0).
The evaluation of the signal and background likelihoods however
differs from that in the Matrix Element method.

The event observables can be classified into the kinematic information 
$x_{\rm kin}$ used in
the kinematic fit to reconstruct the top quark mass and other
variables $x_{{\rm topo/}b}$ (describing the event topology and the
$b$-tagging information) that are uncorrelated with the
top quark mass and used to improve the separation of signal and background
events.
The likelihood for the event to be produced via process $i$
can then in general be written as the product
\begin{equation}
  \label{eq:IDpsgn}
    L_P\left(x;\,\mtop,\jes\right)
  =
    L_P^{\rm kin}\left( x_{\rm kin};\,\mtop,\jes \right)
    L_P^{{\rm topo/}b}\left( x_{{\rm topo/}b} \right)
  \ ,
\end{equation}
where the dependence on \mtop only enters for the signal process,
$P=\ttbar$.
The second term is only included in the \dzero analysis.
It recovers some of the topological information of the event (like the
relative angles between the decay products) that can otherwise only be 
used in the Matrix Element method, while the $L_P^{\rm kin}$ term
in the Ideogram method extracts
information on \mtop only from invariant mass information obtained in the
kinematic fit (which in turn is also insensitive to angular information).
In addition, event quality and $b$-tagging information can be included
in $L_P^{{\rm topo/}b}$.
The kinematic and topological terms in the likelihood are discussed
in turn in the following sections.

\subsubsection{The Kinematic Likelihood for a Process}
\label{idmeasurements.pevt.pkin.sec}
The kinematic part of the signal or background 
likelihood is calculated as a sum over all
jet-parton assignments (and neutrino solutions, in the case of the
\ljets analysis).
The relative likelihood $w_i$ of assignment/solution $i$ to be
correct is obtained from the minimum $\chi^2_i$ of the 
corresponding fit and from $b$-tagging information as
\begin{equation}
  \label{eq:IDwi}
    w_i\left(x_{\rm kin}^i; \mtop, \jes\right)
  =
    \exp\left( -\frac{1}{2} \left(\chi^2\right)^i \right)
    \prod_{j=1}^{n_j} 
    W_b \left( {\cal B}_{{\rm jet}\,j}^{\,\rm rec},\ 
               \phi_{{\rm parton}\,k}^{\,\rm ass} \right)
  \ ,
\end{equation}
where the product runs over all $n_j$ jets in the event, and 
$W_b$ is given by the $b$-tagging efficiencies for light and $b$-quark
jets as defined in Equation~(\ref{binarybtaggingweight.eqn}).
The weights depend on the top quark mass and 
jet energy scale because the results
of the kinematic fit do (including the minimum $\chi^2$).

The kinematic term in the signal likelihood $\psgn^{\rm kin}$ 
describes the correct jet-parton assignment
(``\ca'') and all other assignments (``\wa'') separately and
can be written as
\begin{eqnarray}
  \nonumber
    \psgn^{\rm kin}\left(x_{\rm kin};\,\mtop,\jes\right)
  =
    \sum_i
    w_i\left(x_{\rm kin}^i; \mtop, \jes\right)
    [ \phantom{\ +}
  \!\!&\!\!
        \fca                    
  \!\!&\!\!
        \Sca\left(x_{\rm kin}^i; \mtop, \jes\right)
  \\
  \label{eq:IDpsgnkin}
        \ +
  \!\!&\!\!
        \left( 1 - \fca \right) 
  \!\!&\!\!
        \Swa\left(x_{\rm kin}^i; \mtop, \jes\right) ]
  \ ,
\end{eqnarray}
where $i$ runs over all 24 assignments/solutions, and 
\fca corresponds to the relative weight given to the correct
assignment by the weights $w_i$.
In the \dzero analysis in the \ljets channel, the 
value of \fca is determined from the simulation as 
the average fraction of weights $w_{\rm ca}/\left(\sum_i w_i \right)$
given to the correct assignment; the dependence of \fca
on the total number of reconstructed jets and the number of $b$-tagged
jets is taken into account.

For the correct assignment, the likelihood to observe the 
fitted top quark mass $\mtop^{\rm fit}$ takes into account both the
natural width \Gtop of the top quark and the experimental resolution
$\sigma_{\mtop}^{\rm fit}$, which is assumed to be Gaussian and
determined on an event-by-event basis in the kinematic fit.
The likelihood is given by their convolution
\begin{equation}
  \label{eq:IDpsgnca}
    \Sca\left(x_{\rm kin}^i; \mtop, \jes\right)
  =
    \int_{m'} G\left(\mtop^{{\rm fit},\,i}, m', \sigma_{\mtop}^{{\rm fit},\,i}\right) 
              BW(m', \mtop) {\rm d}m'
  \ ,
\end{equation}
where the integration is over the true mass $m'$ of the top quark in the
given event.
The Gaussian resolution $G$ and the relativistic Breit-Wigner $BW$
can be expressed as
\begin{eqnarray}
  \label{eq:IDpsgncagauss}
    G\left(\mtop^{{\rm fit},\,i}, m', \sigma_{\mtop}^{{\rm fit},\,i}\right) 
  & = &
    \frac{1}{ \sqrt{2\pi}\ \sigma_{\mtop}^{{\rm fit},\,i} }
    \exp\left( -\frac{1}{2} 
                \left( \frac{ \mtop^{{\rm fit},\,i} - m' }
                            { \sigma_{\mtop}^{{\rm fit},\,i} } \right)^2 \right)
  \ {\rm and}
  \\
  \label{eq:IDpsgncabw}
    BW(m', \mtop)
  & = &
    \frac{1}{\pi}
    \frac{ \mtop \Gtop }
         { \left( m'^2 - \mtop^2 \right)^2 + \mtop^2 \Gtop^2 }
  \ ,
\end{eqnarray}
respectively.
The likelihood is sensitive to the jet energy scale via the
$\chi^2$ obtained in the kinematic fit since a
constraint to the known \W boson mass is applied.
In the \ljets analysis, with only one fitted mass $\mtop^{{\rm fit},\,i}$
per jet-parton assignment $i$, $m'$ can be interpreted as the average of the top and
antitop quark masses.
In the analysis in the \alljets channel, the term \Sca contains
one integral as given in Equation~(\ref{eq:IDpsgnca}) for each of the
two fitted masses.

Wrong jet-parton assignments in signal \ttbar events 
cannot easily be described as a similar convolution.
Therefore, the corresponding term $\Swa\left(x_{\rm kin}^i; \mtop, \jes\right)$ 
is given by the distribution 
of fitted masses $\mtop^{\rm fit}$ in simulated \ttbar events,
where the two neutrino solutions for the correct jet-parton
assignment are excluded and all other assignments/solutions are 
weighted with $w_i$.
Even though it describes wrong assignments, \Swa still depends
on the top quark mass.
The fitted uncertainty $\sigma_{\mtop}^{\rm fit}$ is not used.
In some simulated events, the correct jet-parton assignment cannot 
be unambiguously identified.
These events are excluded when determining the shape of \Swa; the
calibration of the measurement technique is however performed using
the full simulation including these events, as described in general
in Section~\ref{massfit.validationandcalibration.sec}, so that the
final measurement result is unbiased.

In both the \ljets and \alljets channels, background is described
with one likelihood \pbkg.
The kinematic term of the background likelihood is given by
\begin{equation}
  \label{eq:IDpbkg}
    \pbkg^{\rm kin}(x_{\rm kin};\,\jes)
  = 
    \sum_i w_i B(x_{\rm kin}^i;\,\jes)
\end{equation}
with a weight $w_i$ per jet-parton assignment/solution $i$ as
defined above in Equation~(\ref{eq:IDwi}).
In the \ljets analysis, $B$ is the shape of the mass spectrum obtained
in simulated \wfourpshort events, where each assignment $i$ enters with its
weight $w_i$ as in the likelihood.
The shape $B$ of the background spectrum does not depend strongly on 
\jes (the number of background events does, but this is not relevant since
the signal fraction \ftop is a free parameter in the measurement), and 
$\pbkg^{\rm kin}$ is always evaluated at $\jes=1$ like in the Matrix
Element analyses, see Section~\ref{memeasurements.pbkg.sec}.

In the \alljets channel, background is described by a mixture of
\bbbarfourpshort events simulated with \alpgen and \sixp events obtained
from the data.
Here, $B$ is a two-dimensional function of the two fitted masses.
As above, it is obtained as the weighted spectrum obtained in the 
background events.
No \jes dependence is included in the likelihood since the \jes
parameter is not fitted.

\subsubsection{The Topological Likelihood for a Process}
\label{idmeasurements.pevt.ptopo.sec}
In the \dzero analysis in the \ljets channel, a term $L_i^{{\rm topo/}b}$ 
is included in
the likelihood that captures the information from the event topology,
the event quality, and the number of $b$-tagged jets in the event.
Note that in the kinematic term of the signal likelihood,
$b$-tagging information is included 
to improve the identification of the correct jet-parton assignment
in signal events, while it is used here to improve the separation
between signal and background.
The inputs used in the calculation are:
\begin{list}{$\bullet$}{\setlength{\itemsep}{0.5ex}
                        \setlength{\parsep}{0ex}
                        \setlength{\topsep}{0ex}}
\item{\bf Topological information:}
Four variables are used.
These are
\begin{list}{$-$}{\setlength{\itemsep}{0.5ex}
                        \setlength{\parsep}{0ex}
                        \setlength{\topsep}{0ex}}
\item
the missing transverse energy \etmiss;
\item
the aplanarity ${\cal A}$ as defined in
Section~\ref{templatemeasurements.allhad.sec}, computed from
the momenta of all jets and the leptonically
decaying \W boson reconstructed in the kinematic fit;
\item
the ratio \Htwoprime of the scalar sum of the jet transverse momenta, 
excluding the highest-\pt jet, and the scalar sum of the longitudinal
momenta of the jets and the reconstructed leptonically decaying \W
boson; and
\item
the quantity
\begin{equation}
  \label{ktminp.eqn}
    \ktminp
  =
    \frac{ \min\left(\DeltaR_{ij}\right) \min\left(E_{T,i}, E_{T,j}\right) }
         { E_{T, W} }
  \ ,
\end{equation}
where $\min\left(\DeltaR_{ij}\right)$ is the minimum distance between
any two jets among the four highest-\pt jets.
\end{list}
Although other variables like the scalar sum of all jet transverse
momenta have better separation power between signal and background,
these variables are correlated with the top quark mass, which is why
they are not used in the topological likelihood.
The separation obtained with these topological variables is shown in 
Figures~\ref{idlikelihood.fig}(a) and~(d).
\item{\bf Fraction of track {\boldmath\pt} contained in jets:}
Considering scalar sums of track transverse momenta,
this variable is defined as the fraction of track \pt contained within
the reconstructed jets of the event (i.e.\ within $\DeltaR<0.5$ of the
calorimeter jet axes).
This variable distinguishes clean events from events with poorly
defined jets and is uncorrelated with the topological information
described above.
It provides separation in particular between \ttbar and QCD multijet background
events, as can be seen in Figures~\ref{idlikelihood.fig}(b) and~(e).
\item{\bf\boldmath \bquark-tagging information:}
Finally, the number of $b$-tagged jets in the event is used as an
input to the likelihood.
\end{list}
A likelihood discriminant $D$ is created from all input variables.
The topological/$b$-tagging terms 
$L_{\ttbar}^{{\rm topo/}b}\left( x_{{\rm topo/}b} \right)$
and
$L_{\rm bkg}^{{\rm topo/}b}\left( x_{{\rm topo/}b} \right)$
of the likelihood are then
given by the fraction of signal or background events at the value
of $D$ reconstructed for a particular event.
These are shown in Figures~\ref{idlikelihood.fig}(c) and~(f).
Note that while topological and $b$-tagging information are
also used in the Matrix Element analysis, the $\pt$ fraction variable
is unique to the Ideogram analysis.
\begin{figure}[ht]
\begin{center}
\includegraphics[width=\textwidth]{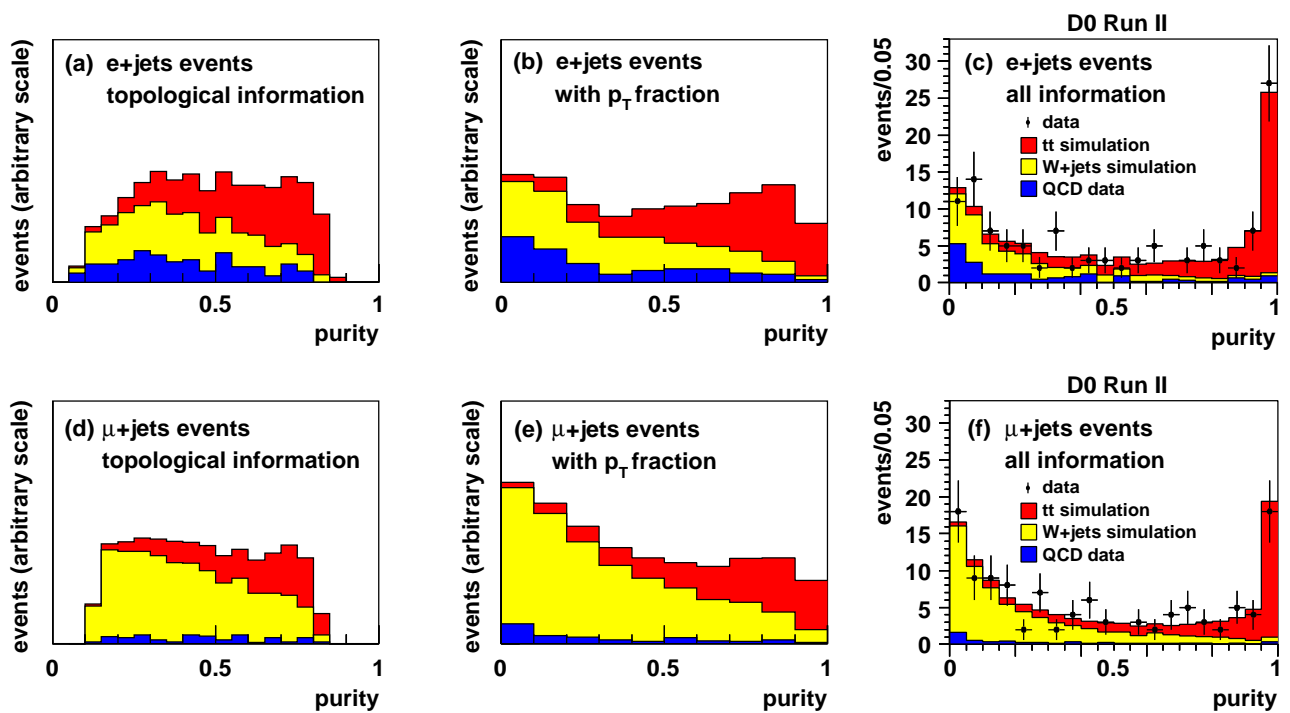}
\caption{\captionfont\label{idlikelihood.fig}\dzero \ljets Ideogram 
  measurement: Topological likelihood for \ejets (upper
  plots) and \mujets (lower plots) events.  Plots (a) and (d) show the
  separation between \ttbar signal (red), \wjets background (yellow),
  and QCD multijet background (blue) when only topological information
  is used.  In plots (b) and (e), information from the track \pt
  fraction contained in jets is included.  
  Plots (a), (b), (d), and (e) show the expected distributions
  on an arbitrary linear vertical scale.
  Plots (c) and (f)
  show the final distributions used to compute the topological
  likelihood, which also include \bquark-tagging
  information.  In these plots, the expectations are scaled to the 
  results from $425\,\ipb$ of data, which are
  are overlaid~\cite{bib-ID}.  In all plots, the \ttbar and \wjets
  predictions are from simulated events, while the QCD multijet
  distribution has been obtained from data using a signal depleted sample.}
\end{center}
\end{figure}

\clearpage
\section{The Top Quark Mass Fit and its Calibration}
\label{massfit.sec}

\begin{center}
\begin{tabular}{p{15cm}}
{\it With the methods presented in
  Sections~\ref{templatemeasurements.sec}, \ref{memeasurements.sec},
  and \ref{idmeasurements.sec}, a likelihood for a sample of
  selected events to be consistent with a given top quark mass
  hypothesis can be computed.  This section describes how this
  information is used to determine the measurement value of the top
  quark mass and its (statistical) uncertainty.  The calibration
  of the measurement method with simulated experiments is also 
  discussed.}
\end{tabular}
\end{center}

The previous sections describe how
the likelihood as a function of the top quark mass hypothesis 
to obtain the observed data sample can be determined:
via the comparison of the estimator distributions in data and
simulation (template method, 
cf.\ Section~\ref{templatemeasurements.sec}), or using likelihoods
calculated for each individual event with the Matrix Element 
(Section~\ref{memeasurements.sec}) or Ideogram methods
(Section~\ref{idmeasurements.sec}).
Section~\ref{massfit.procedure.sec} describes the step of obtaining
a (raw) measurement value of the top quark mass and its statistical
uncertainty from this information.

This (raw) measurement value is only correct if the assumptions made
to derive it reflect reality.
Uncertainties on these assumptions will translate into systematic 
uncertainties on the measurement, as described in
Section~\ref{systuncs.sec}. 
On the other hand, known deficiencies or approximations in the 
technique used to determine the likelihoods can be corrected for
by calibrating the measurement with fully simulated events.
This step also allows for a test of the uncertainties obtained in the
fitting procedure and a comparison of the measured uncertainty in 
data with expectations.
It is further described in
Section~\ref{massfit.validationandcalibration.sec}.

\subsection{The Fitting Procedure}
\label{massfit.procedure.sec}
The technical details of how a (raw) measurement of the top quark mass
is extracted from the likelihood information varies between the
individual analyses. 
For example, the procedure depends on whether the likelihood
is known for arbitrary top quark masses or only for a discrete set
of values.
Furthermore, some analyses require the simultaneous measurement of the 
top quark mass and jet energy scale.

To cover the techniques applied, the fitting procedures used in the 
CDF template measurement in the \ljets 
channel~\cite{Abulencia:2005aj,bib-CDFupdates} and in the \dzero Matrix
Element measurement in the \ljets channel~\cite{bib-me} are described
as examples in Sections~\ref{massfit.procedure.cdfljetstemplate.sec}
and~\ref{massfit.procedure.dzerome.sec}, respectively.
The fitting procedure does not depend a priori on the \ttbar
event topology.

\subsubsection{Fitting Procedure in the CDF Lepton+Jets Template Analysis}
\label{massfit.procedure.cdfljetstemplate.sec}
This section describes the fit used in the CDF template analysis in
the \ljets channel
to determine the top quark mass \mtop and the
jet energy scale \jes (as well as the signal fraction).

The event selection, estimators, and template parametrizations are 
described in Section~\ref{templatemeasurements.kinrec.sec} and are
briefly recapitulated here:
\begin{list}{$\bullet$}{\setlength{\itemsep}{0.5ex}
                        \setlength{\parsep}{0ex}
                        \setlength{\topsep}{0ex}}
\item
The selected events are grouped into four categories with different
expected signal to background ratio depending on the number and 
transverse energies of \bquark-tagged jets in the event.
\item
The top quark mass $\mtop^{\rm reco}$ obtained in a kinematic fit 
of each measured event to the \ttbar hypothesis is used as estimator
for the top quark mass; the dijet mass $\m_{jj}$ is taken as
estimator for the jet energy scale whose deviation $\Delta_{JES}$ 
from the standard scale is measured in units of its uncertainty
$\sigma_c$.
\item
For signal \ttbar events, the $\mtop^{\rm reco}$ templates are 
parametrized as functions of $\mtop^{\rm reco}$ as well as of the 
true top quark mass \mtop; similarly the $\m_{jj}$ templates are 
parametrized as functions of $\m_{jj}$ and the parameter
$\Delta_{JES}$ to be measured.
The background templates are parametrized as functions of 
$\mtop^{\rm reco}$ and $\m_{jj}$, too.
\end{list}

The $\mtop^{\rm reco}$ and $m_{jj}$ values in the data
events are compared to the signal and background templates in an
unbinned likelihood fit, which determines the top quark mass, 
jet energy scale, and the number of signal and background events
in each of the four event categories.
The likelihood for one event category
is computed as the product of likelihoods for each data
event, which in turn contain four terms each:
\begin{list}{$\bullet$}{\setlength{\itemsep}{0.5ex}
                        \setlength{\parsep}{0ex}
                        \setlength{\topsep}{0ex}}
\item
the likelihood to measure the reconstructed value of 
$\mtop^{\rm reco}$, obtained from the linear combination of 
signal and background $\mtop^{\rm reco}$ templates for given
\mtop and \DeltaJES hypotheses, with relative contributions 
of signal and background also determined
in the fit (mainly sensitive to \mtop);
\item
a similar term based on $m_{jj}$ (mainly sensitive to
\DeltaJES);
\item
a term describing the probability of having certain numbers of 
signal and background events in the data, given the total number
of selected events; and
\item
a constraint on the expected number of background events (not for
the 0-tag event category).
\end{list}
The likelihoods for all four event categories are then multiplied,
and a constraint to the 
a priori knowledge of the jet energy scale is included as
another overall factor.
Since the \mtop and \DeltaJES parameters are the same in all
event categories, a total of ten parameters are determined in the fit.
These parameters are determined simultaneously using
\minuit~\cite{bib-minuit}.

\subsubsection{Fitting Procedure in the \dzero Lepton+Jets Matrix Element Analysis}
\label{massfit.procedure.dzerome.sec}
In an analysis with parametrized templates, it is possible to let the
minimization program (e.g.\ \minuit) decide for which
assumed parameter values to evaluate the overall likelihood.
This is impractical for the Matrix Element and Ideogram methods, where
the calculation of the overall likelihood for one 
hypothesis is a time-consuming process.
In these analyses, a different approach is therefore followed:
\begin{list}{$\bullet$}{\setlength{\itemsep}{0.5ex}
                        \setlength{\parsep}{0ex}
                        \setlength{\topsep}{0ex}}
\item
In a first step, the overall likelihood is calculated for 
each hypothesis in a grid of assumed parameter values.
\item
Second, the dependency of the likelihood on the parameters that 
are to be measured is fitted with a function.
\item
The minimum of this function yields the central measurement value, 
and the statistical uncertainty is given by the $68\%$ confidence 
region around this central value.
\end{list}
As an example, the fitting procedure used in the \dzero Matrix Element
measurement in the \ljets channel is described here.

Also in this analysis a simultaneous measurement of the top quark
mass \mtop, jet energy scale \jes, and signal fraction \ftop is
performed.
For each selected event, the signal likelihood is 
evaluated for a grid of assumed \mtop and \jes values in steps of 
$2.5\,\GeV$ and $0.01$.
The background likelihood is calculated for $\jes=1$ only and is assumed
not to depend on the \jes parameter value.

For any given (\mtop,\,\jes) assumption, the likelihood as a function
of \ftop can then be calculated easily as the linear combination given
in Equation~(\ref{eq:MEpevt}).
The signal fraction \ftopbest that maximizes the overall likelihood
is calculated for each (\mtop,\,\jes) parameter pair, and the likelihood value
corresponding to this value is used in further computations.
The overall result quoted for the fitted signal fraction \ftop is
derived from the
value obtained at the (\mtop,$JES$) point in the grid with the maximum
likelihood value for the event sample.
The uncertainty on \ftop is computed by varying \ftop at fixed \mtop and
$JES$ until $\Delta(-\ln L)=+\frac{1}{2}$.
This uncertainty does not account for correlations between \ftop, \mtop, and $JES$.

The result for the top quark mass is obtained from a projection of the 
two-dimensional grid of likelihood values onto the \mtop axis.
In this projection, correlations are taken into account.
The likelihood for a given \mtop hypothesis is obtained as the integral
over the likelihood as a function of $JES$, using linear interpolation
between the grid points and
Gaussian extrapolation to account for the tails for $JES$ values outside
the range considered in the grid.

The likelihoods as a function of assumed top quark mass are converted
to $-\ln L$ values.
These $-\ln L$ points are then fitted with a fourth order polynomial in the 
region defined by the condition $\Delta\ln L<3$ around the best value.
The \mtop value that maximizes the fitted likelihood is taken to be
the measured value of the top quark mass.
The lower and upper uncertainties on the top quark mass are defined 
such that 68\% of the total likelihood integral is enclosed by the 
corresponding top quark mass values, with equal likelihood values at both
limits of the 68\% confidence level region.
The same projection and fitting procedure is applied to determine the 
value of the $JES$ parameter.

The inclusion of \bquark-tagging information introduces two
significant improvements to the analysis:
Both the separation between signal and background and the
identification of the correct jet-parton assignment (under the signal
hypothesis) are improved.
Since the signal and background likelihoods are evaluated on an
event-by-event basis and the \bquark-tagging information is encoded in
the transfer function (cf.\ Section~\ref{memeasurements.tf.sec}), both
aspects are in principle addressed, and it should not be necessary to
divide the event sample into subsamples of different purity like in
the template analysis described in
Section~\ref{massfit.procedure.cdfljetstemplate.sec}.
Nevertheless, the \dzero experiment has taken a different approach.
The transfer function $W(x,y;\,JES)$ given in
Equation~(\ref{eq:tfdefinition-btag}) is modified to obtain
\begin{equation}
  \label{modifiedtf.eqn}
    \tilde{W}(x,y;\,JES)
  =
    \frac{W(x,y;\,JES)}
         {\displaystyle
          \sum_{i=1}^{n_{\rm comb}}
          \prod_{j=1}^{n_j}
          W_b \left( {\cal B}_{{\rm jet}\,j}^{\,\rm rec},\ 
                     \phi_{{\rm parton}\,k}^{\,\rm ass} \right)
         }
  \ .
\end{equation}
This new transfer function is used in the computation of the signal 
likelihood so that the \psgn values can be compared with those of the
background likelihood, which in turn are computed without taking
\bquark-tagging information into account at all, i.e.\ with the
transfer function given in Equation~(\ref{eq:tfdefinition-topo}).

This method implies that only the identification of the correct
jet-parton assignment in \ttbar events is improved.
To recover the enhanced separation of signal and background events,
the event sample is subdivided into three categories based on the
number of \bquark-tagged jets per event.
Overall values of the top quark mass \mtop, jet energy scale \jes,
and signal fraction \ftop are determined for all three categories
together by relating the sample composition in each category to the
overall signal fraction.
It should be possible in future updates of the
measurement to use the full transfer function for the background
likelihood and thus avoid fitting different subsamples of events.

\subsection{Validation and Calibration of the Measurement}
\label{massfit.validationandcalibration.sec}
If the model used to describe the data is correct, then
the measurement method should yield unbiased results and the
correct statistical uncertainty.
To {\em validate} the measurement technique, this assumption 
can be verified with simulated {\em pseudo-experiments} using
events that have been generated with this model.

However, most analysis techniques involve some simplifications, for example
via the template parametrization or the simplified treatment of
detector resolution and physics processes 
in the Matrix Element and Ideogram methods.
Given these simplifications, it cannot be assumed that 
every aspect of the data is accounted for.
To {\em calibrate} the measurement, it is first essential that 
the agreement between data and the full simulation is verified.
Monte Carlo events generated with the 
full simulation are then used to compose pseudo-experiments for
the calibration.

The following information is obtained from pseudo-experiments:
\begin{list}{$\bullet$}{\setlength{\itemsep}{0.5ex}
                        \setlength{\parsep}{0ex}
                        \setlength{\topsep}{0ex}}
\item
The relation between the expected (mean) raw measurement value
$\langle \mtop^{raw} \rangle$ and the true input value \mtop.  
Because the mass range of interest is limited a priori to a range
around a value $\mtop^0$, it is usually parametrized
as a linear function in \mtop as
\begin{equation}
  \label{calib.eqn}
    \left\langle \mtop^{raw} \right\rangle
  =
    \mtop^0 + s \left( \mtop - \mtop^0 \right) + o
  \, .
\end{equation}
The symbols $s$ and $o$ stand for the slope of the calibration curve
and for the offset at mass $\mtop^0$.
\item
The width $w$ of the pull distribution.
To test that the fitted uncertainties describe the actual measurement
uncertainty, the deviation of the measurement value
from the true value is divided by the fitted measurement uncertainty
in each pseudo-experiment.
The width of this distribution of deviations normalized by the measurement
uncertainty is referred to as {\it pull width}.
\item
The expected distribution of measurement uncertainties.
\end{list}
This information can be determined accordingly for any other parameter
that is measured (\jes and \ftop, if applicable).
In the validation step, values of $s=1$, $o=0$, and $w=1$ are expected.
Because of simplifications in the measurement technique, this is in
general not true for the calibration based on the full simulation.
The values of $s$ and $o$ obtained in the calibration are used to correct
the raw measurement value, and the measurement uncertainty is adjusted
according to the value of $w$.
As an example, the 
results from the validation and calibration of the
\dzero measurement with the Matrix Element method in the \ljets channel
are described in the following paragraphs.

\paragraph{Validation:}
\label{massfit.validationandcalibration.MEvalidation.sec}
To validate the Matrix Element method, the \dzero
collaboration has generated events with leading-order event generators
(\madgraph~\cite{bib-madgraph} for \ttbar events and \alpgen
for \wjets events), i.e.\ not including
initial- or final-state radiation, for various values of the top quark mass and
jet energy scale.
These events have been smeared according to 
the transfer function described in
Section~\ref{memeasurements.tf.sec}.
The events are required to pass a simplified kinematic selection
similar to the actual event selection, and the normalization of the
likelihood is determined for this selection according to 
Equation~(\ref{eq:MEsigmaobs}).

Pseudo-experiments are composed of these events with the number of 
signal and background events as observed in the data, and the
measurement values \mtop and $JES$ are determined for each 
pseudo-experiment.
In a 
test where $b$-tagging information is not used (jets are assumed
not to be $b$-tagged),
the fitted top quark mass and jet energy scale are 
unbiased within statistical uncertainties of the test of $300\,\MeV$
and $0.003$, respectively.
Furthermore, the fitted \mtop value does not depend on the input 
$JES$ value used
in the generation of the pseudo-experiments, and similarly, 
the fitted $JES$ value is
independent of the true input top quark mass.
The pull width is in agreement with $1.0$.
This validation study and its results are described in detail
in~\cite{bib-philipp}.

\paragraph{Calibration:}
\label{massfit.validationandcalibration.MEcalibration.sec}
For the calibration, fully simulated \ttbar and \wjets events are
used to compose pseudo-experiments with the same numbers of events
as measured in the data\footnote{In the pseudo-experiments for their 
  Matrix Element analysis, the \dzero
  experiment chooses to describe the multijet background 
  with additional \wjets events because the 
  kinematic characteristics are similar.  This simplification is 
  accounted for with a systematic uncertainty.}.
As an example, the calibration curves for the top quark mass and 
jet energy scale in the \dzero Matrix 
Element measurement (including $b$-tagging information)
are shown in Figures~\ref{fig:MEcalib-btag} and
\ref{fig:MEpullcalib-btag}.
The deviation between input and fitted jet energy scale arises
from the simplified description of the detector response.
Given this offset, the (anti-)correlation between the top quark mass and jet energy scale
measurements explains the observed shift between true and fitted \mtop.
The widths of the pull distributions are slightly larger than one.

\begin{figure}[htbp]
\begin{center}
\includegraphics[width=0.49\textwidth]{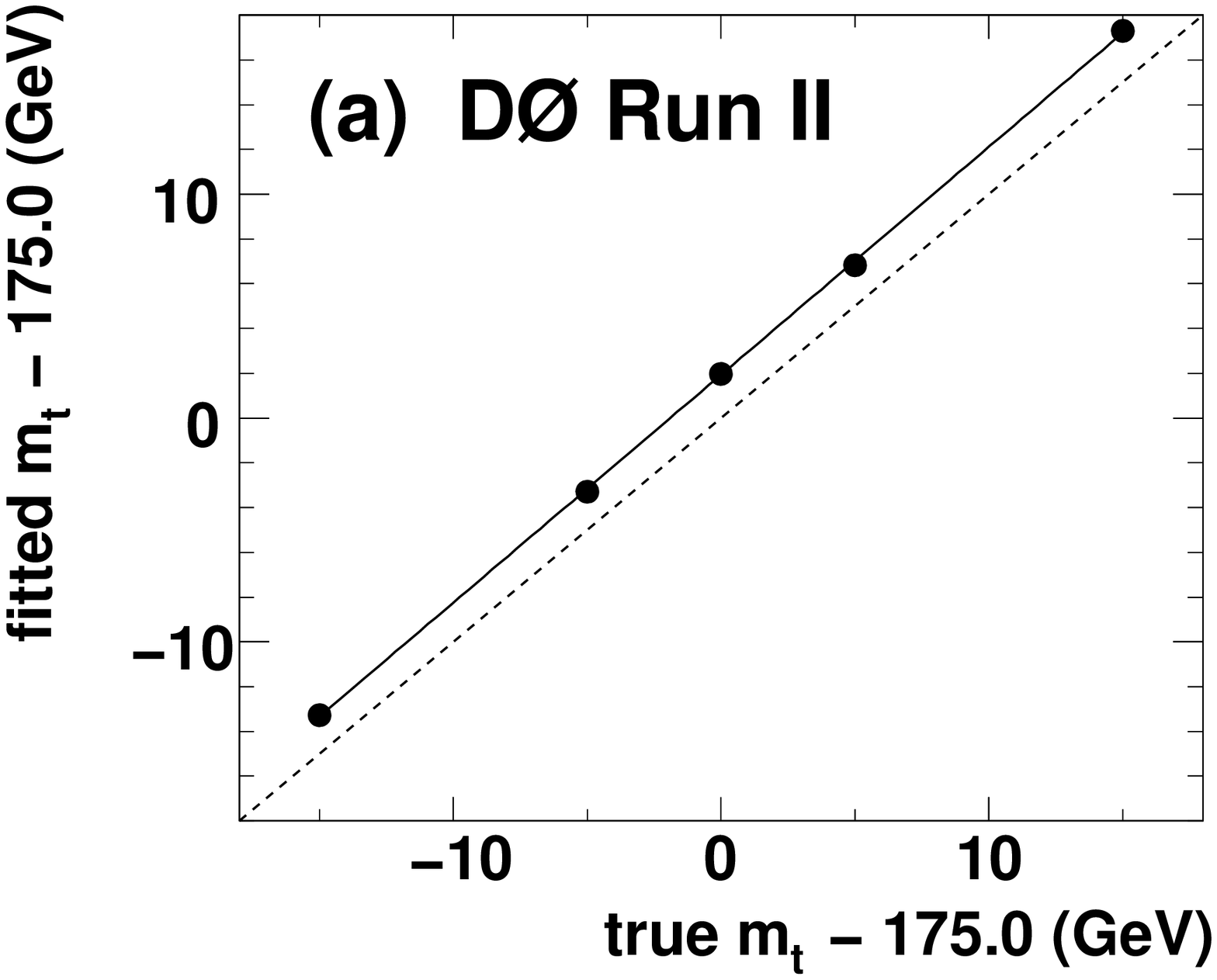}
\hspace{-0.03\textwidth}
\includegraphics[width=0.49\textwidth]{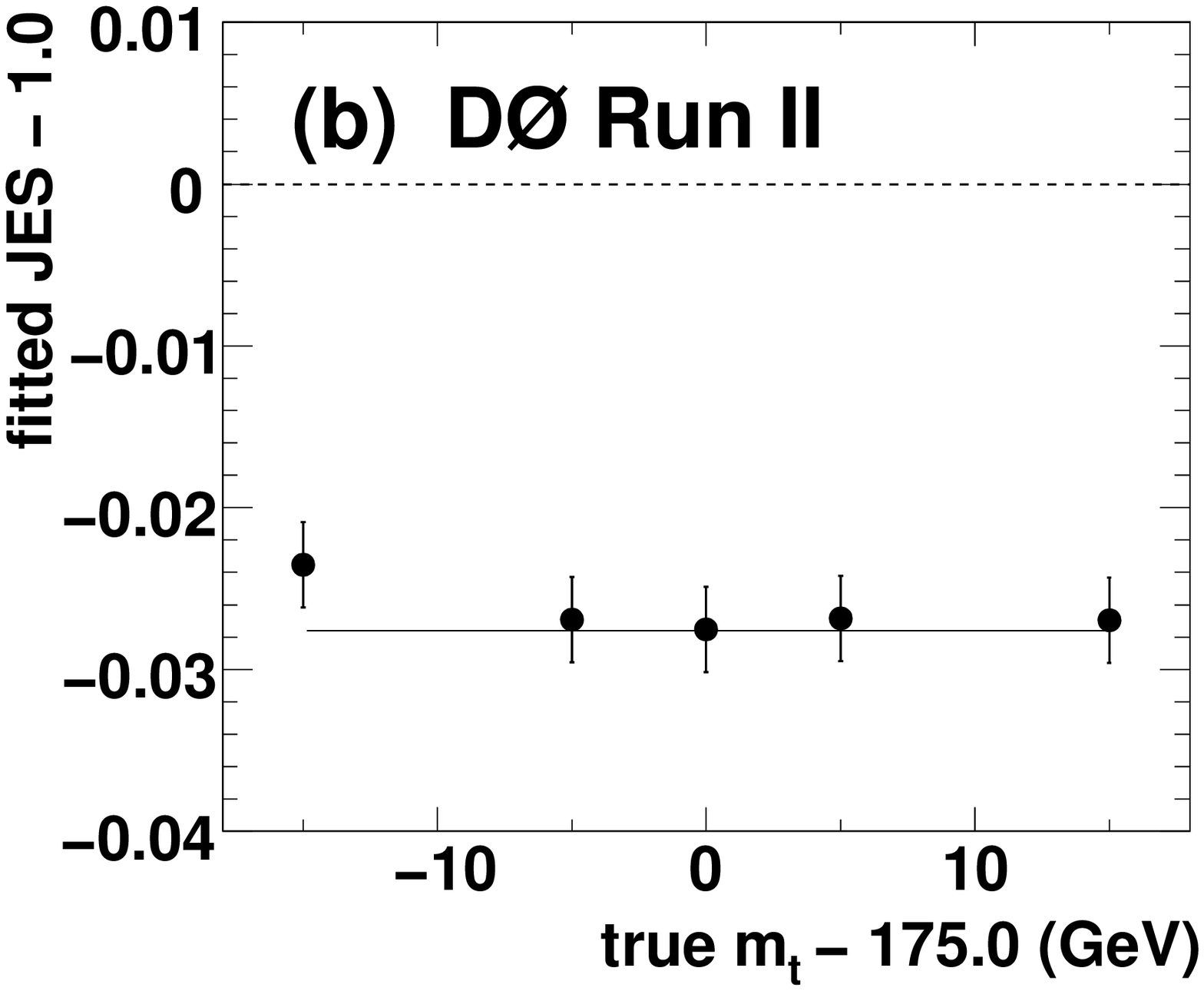}
\includegraphics[width=0.49\textwidth]{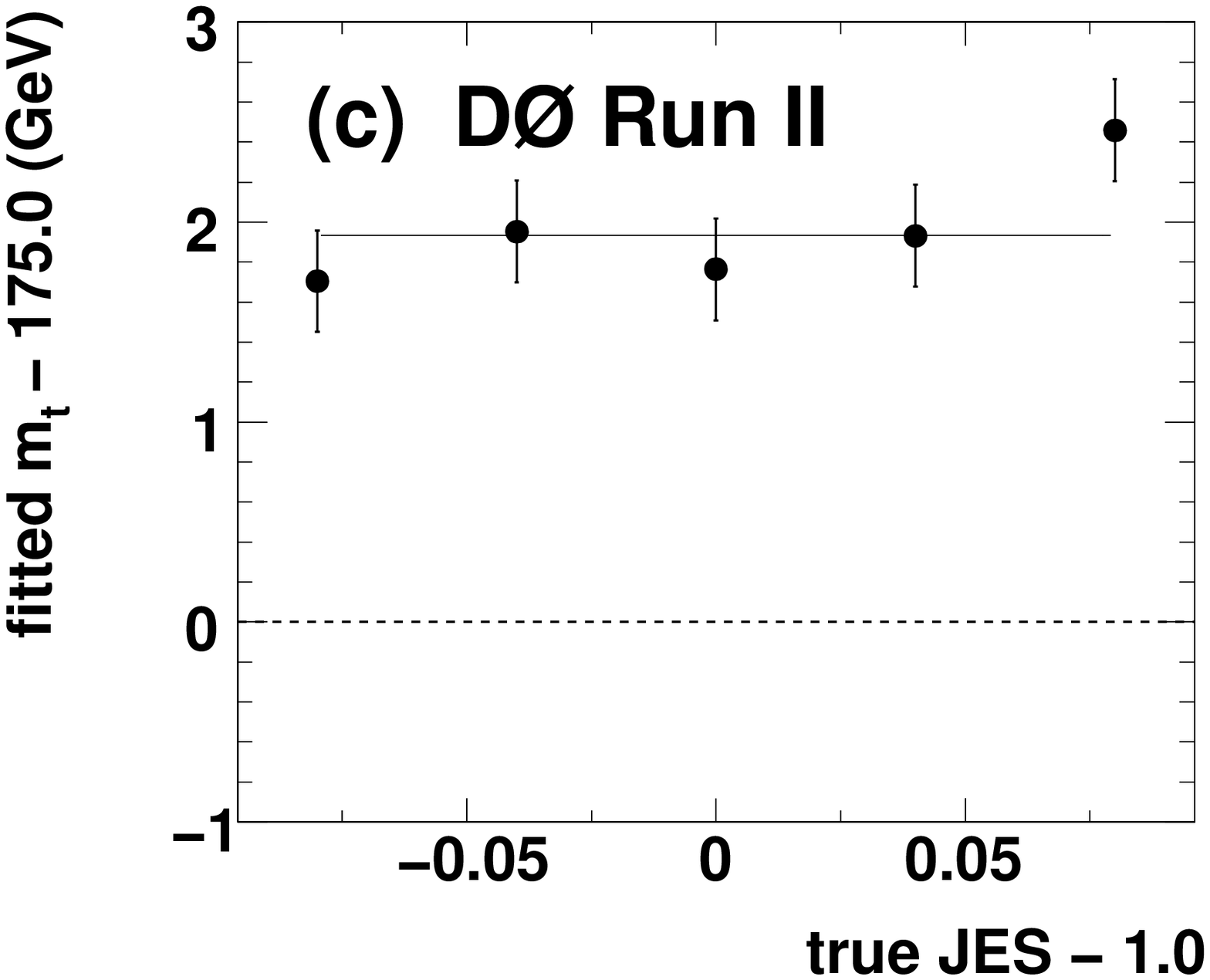}
\hspace{-0.03\textwidth}
\includegraphics[width=0.49\textwidth]{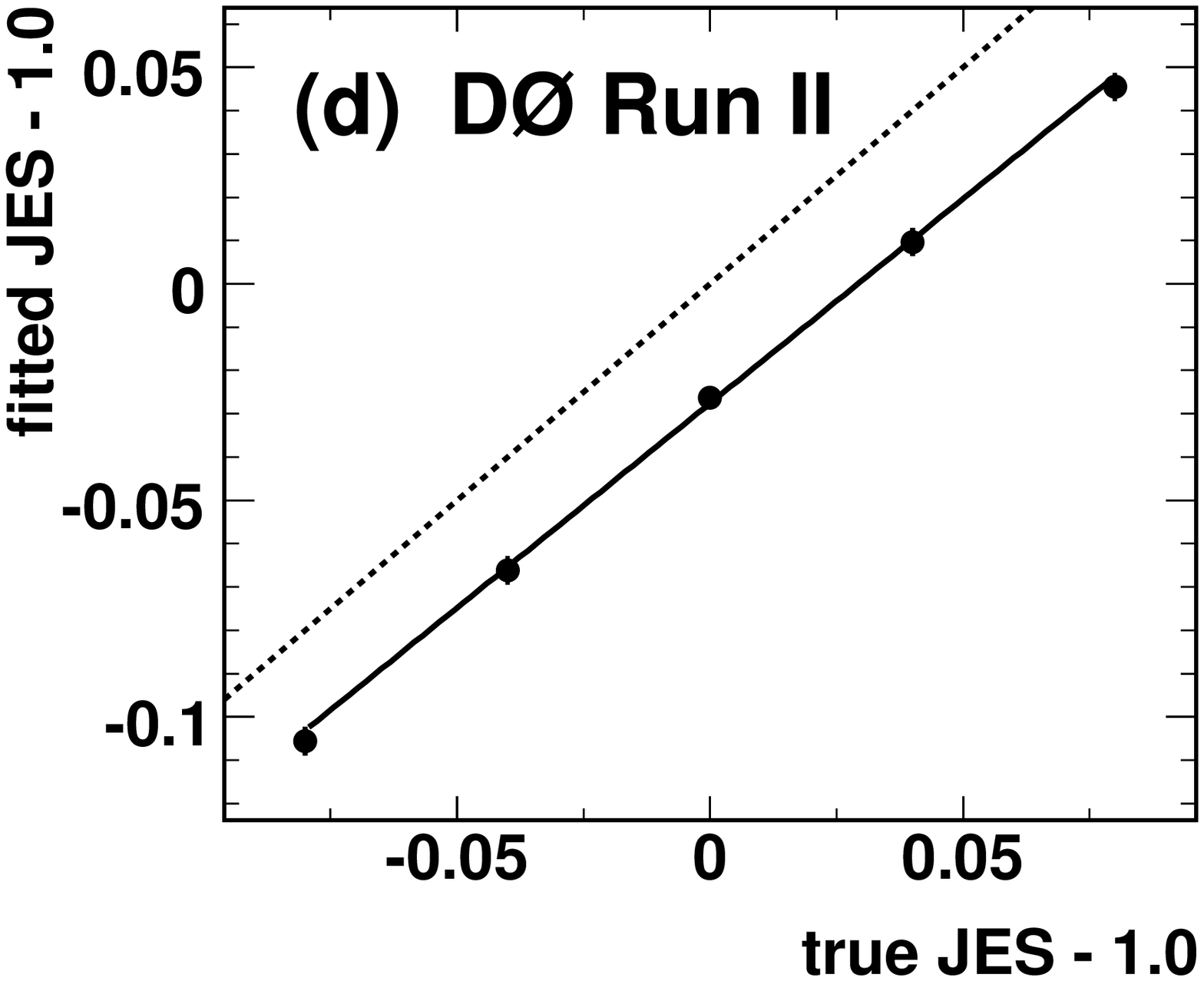}
\caption{\captionfont\label{fig:MEcalib-btag}Calibration of the fitting procedure in the 
\dzero Matrix Element analysis in the \ljets channel~\cite{bib-me}.
The upper plots show the 
reconstructed top quark mass (a) and the measured jet energy scale
(b) as a function of the input top quark mass.
The two lower plots show the reconstructed top quark mass (c) and the 
measured jet energy scale (d) as a function of the
input jet energy scale.
The solid lines show the results of linear fits to the points,
which are used to calibrate the measurement technique.
The dashed lines would be obtained for equal 
fitted and true values of \mtop and $JES$.}
\end{center}
\end{figure}

\begin{figure}[htbp]
\begin{center}
\includegraphics[width=0.49\textwidth]{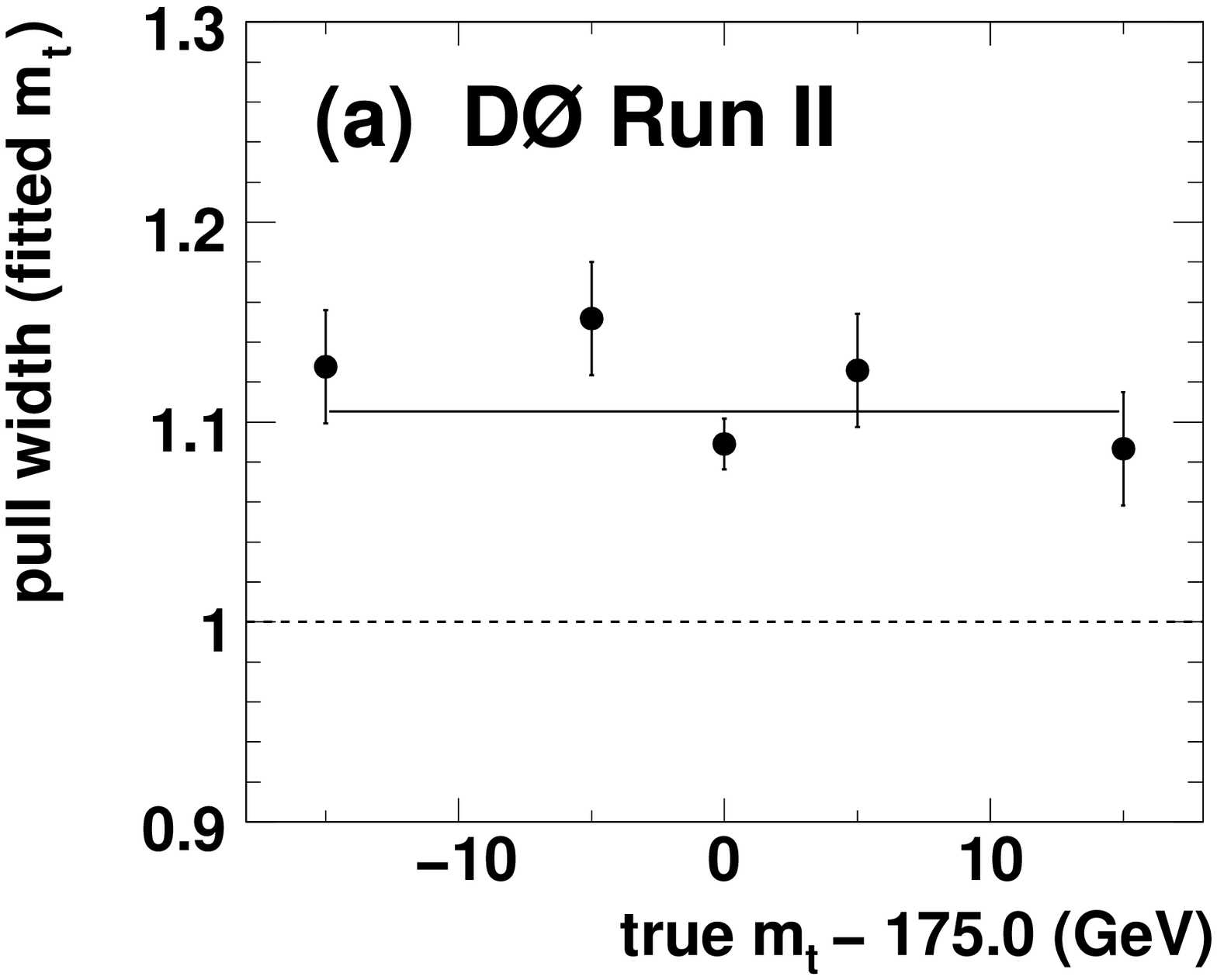}
\hspace{-0.03\textwidth}
\includegraphics[width=0.49\textwidth]{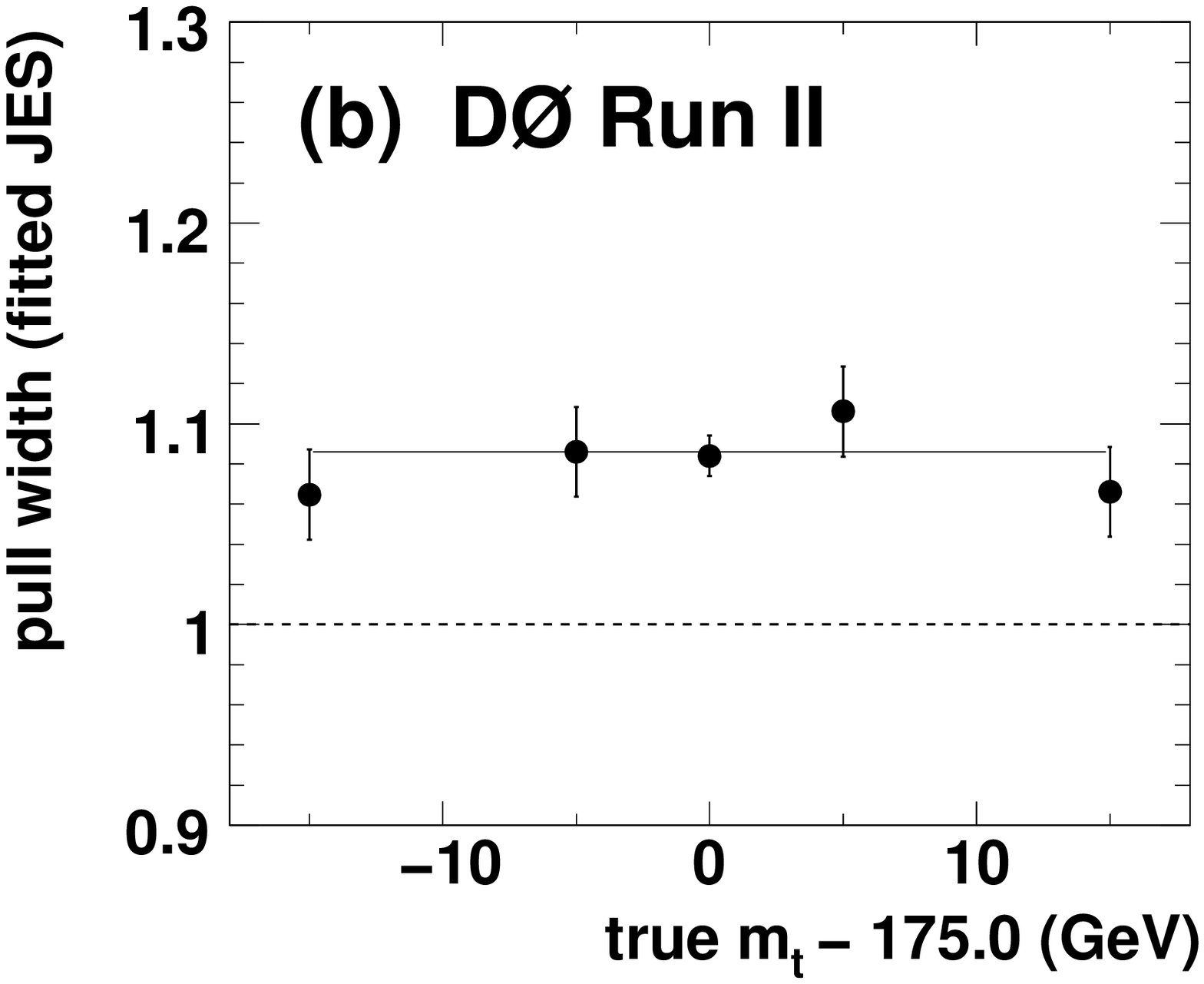}
\includegraphics[width=0.49\textwidth]{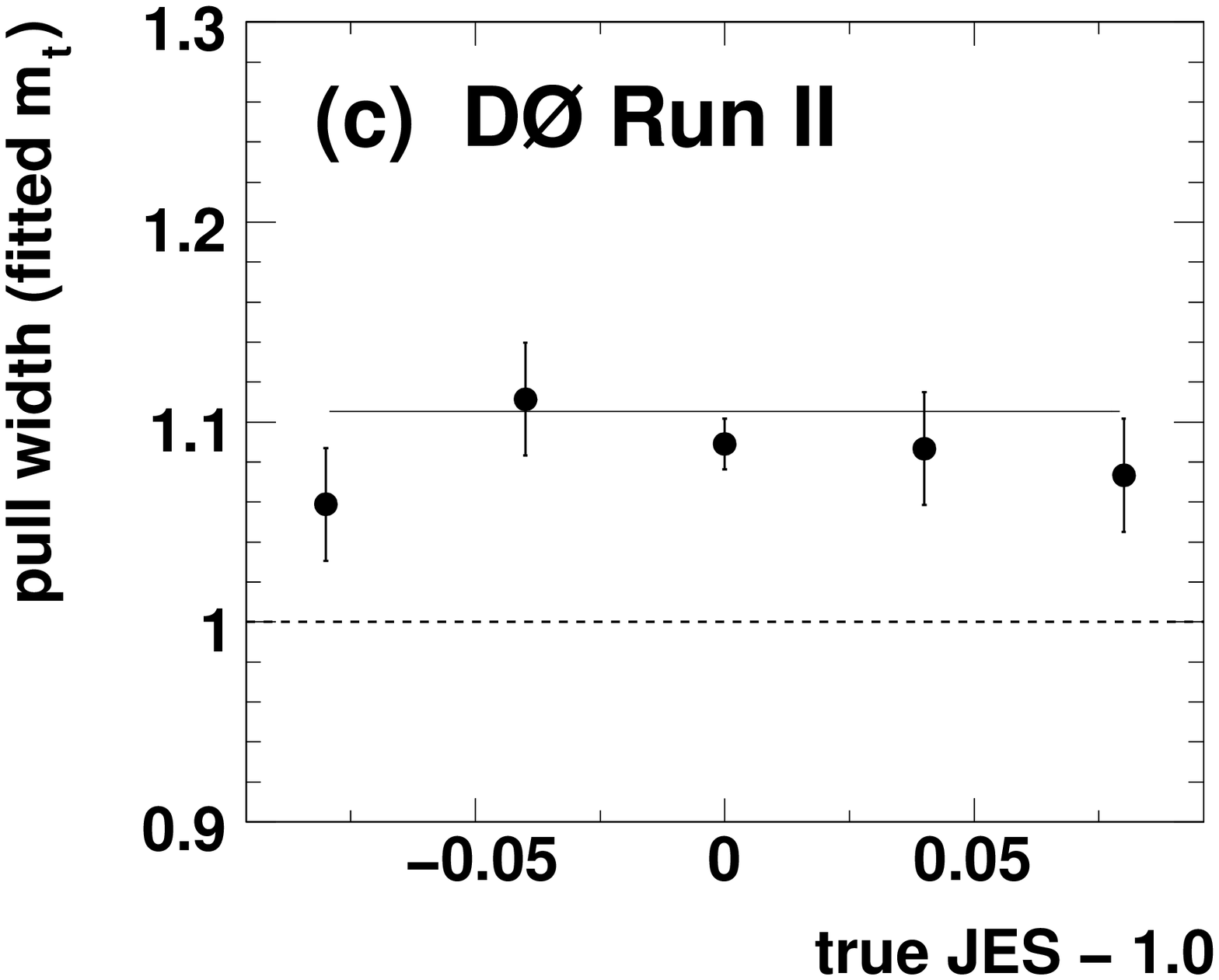}
\hspace{-0.03\textwidth}
\includegraphics[width=0.49\textwidth]{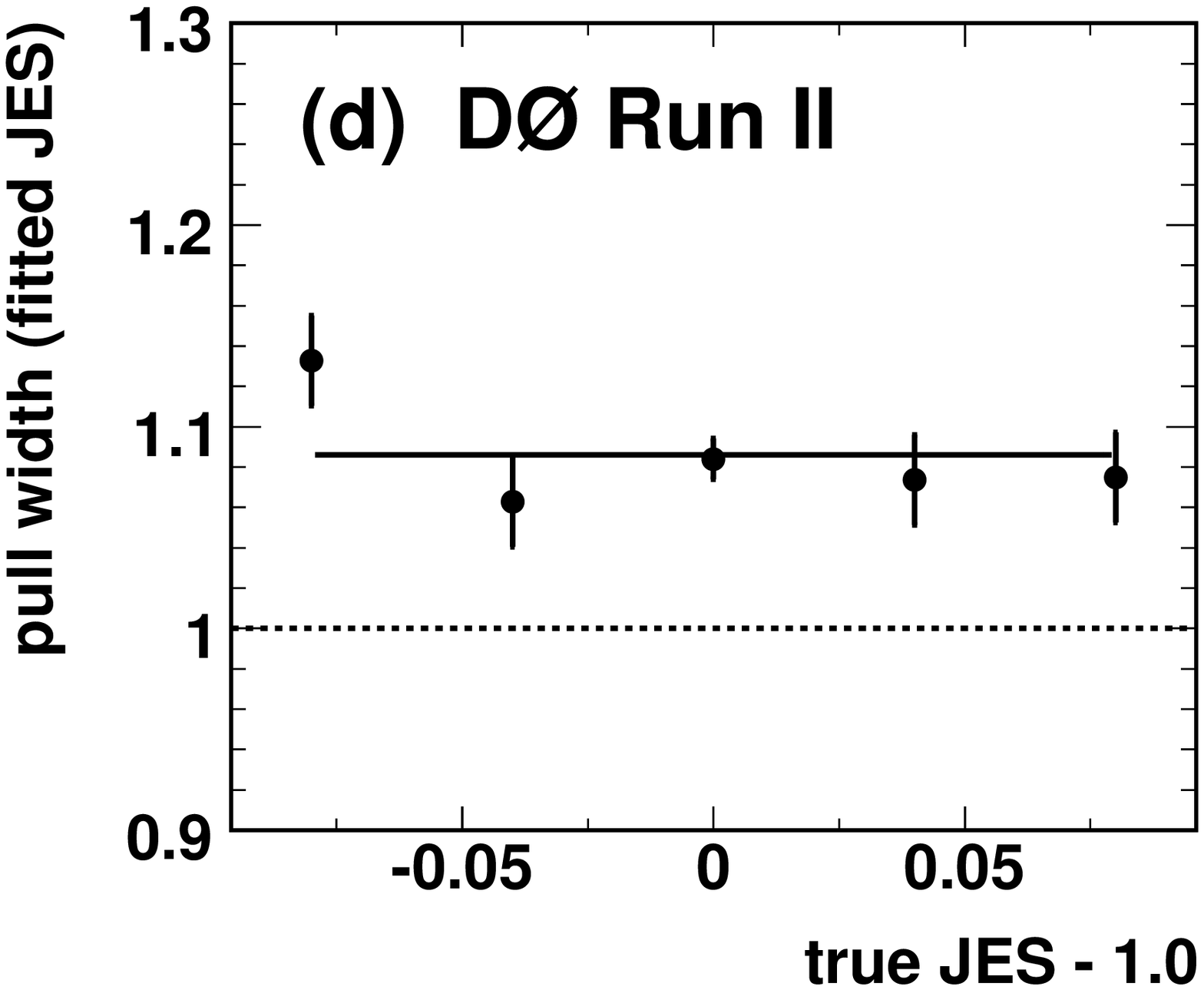}
\caption{\captionfont\label{fig:MEpullcalib-btag}Calibration of the fitting procedure in the 
\dzero Matrix Element analysis in the \ljets channel~\cite{bib-me}.
The upper plots show the 
widths of the pull distributions for the top quark mass
(a) and jet energy scale (b)
as a function of the input top quark mass.
The two lower plots show the widths of the pull distributions for the
top quark mass
(c) and jet energy scale (d) as a function of the
input jet energy scale.
The solid lines show the mean pull width, while
the dashed lines indicate a pull width of $1.0$.}
\end{center}
\end{figure}

For each pseudo-experiment, the statistical uncertainty on the top quark mass 
is multiplied by the pull width, and the resulting distribution of statistical 
uncertainties is shown in Figure~\ref{fig:MEresult-btag-error}.
This allows a comparison with the statistical uncertainty obtained in the data, 
which is also shown in the figure.

The interpretation of such a comparison is not as straightforward as it may seem:
The pseudo-experiments have always been composed with the same expected
numbers of signal and background events.
These numbers have been obtained from the data using a topological
likelihood fit independent of the Matrix Element method; it yields the
\ttbar fraction of the sample with an (absolute) error of 7\% for 
$0.4\,\ifb$ of data (the \ftop result from the Matrix Element method
itself has a similar uncertainty).
If this uncertainty is also taken into account, what at first glance appears to be a 
discrepancy between the fitted top quark mass and jet energy scale 
uncertainties in the data and the expectation becomes
much more consistent.
This is shown in Figure~\ref{fig:MEresult-topo-error-withftopdn}, where
the expected uncertainties for a combined \mtop and \jes fit using
topological information only (i.e.,
no $b$-tagging information) are shown for two types of
pseudo-experiments:
Experiments with a sample composition according to the central
measured value; and experiments with \ftop varied down by
one standard deviation.
Much better agreement between predicted and observed uncertainties
is obtained with the latter class of pseudo-experiments.

\begin{figure}[htbp]                                                              
\begin{center}                                                                  
\includegraphics[width=\figbtagwidth\textwidth]{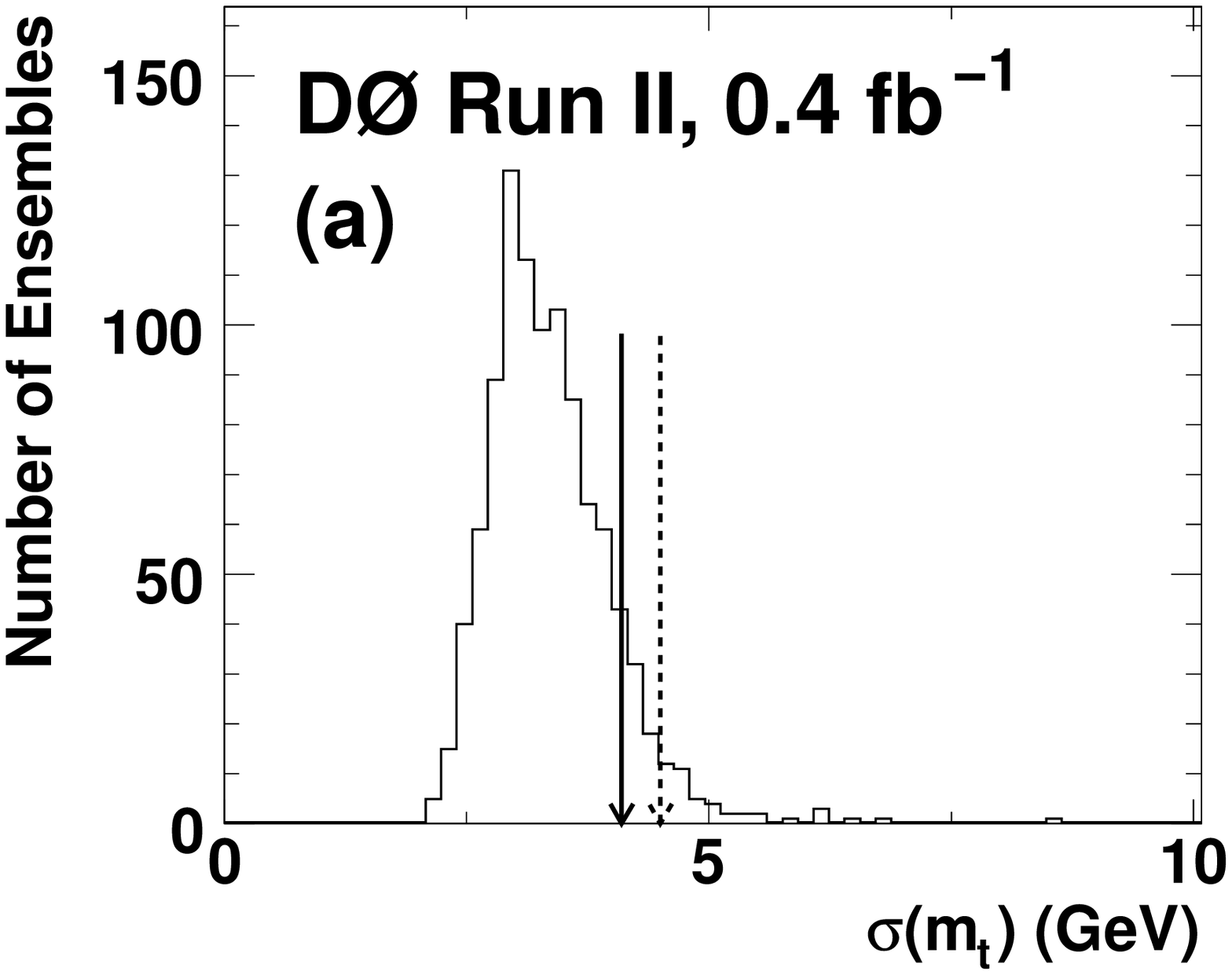}
\includegraphics[width=\figbtagwidth\textwidth]{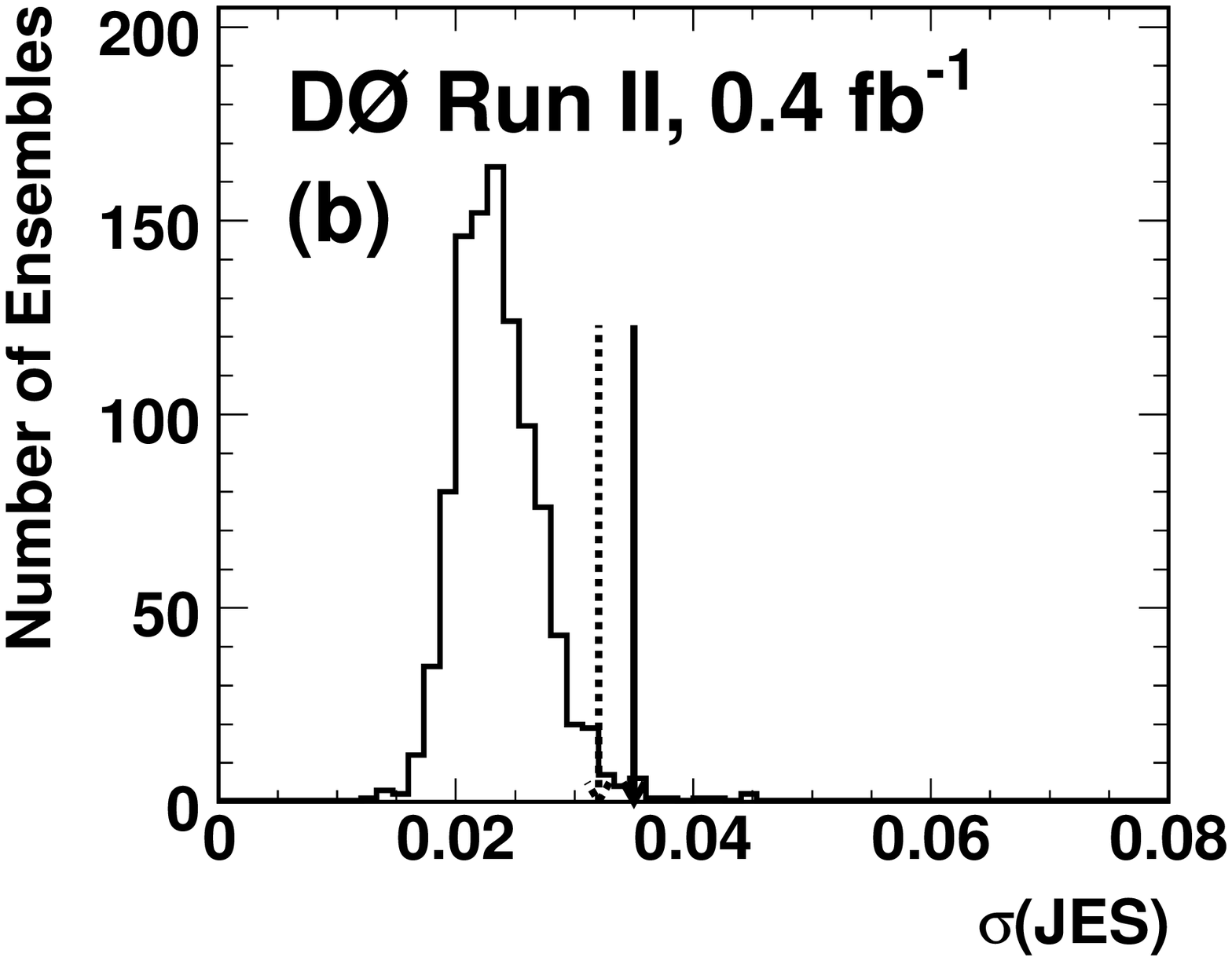}
\caption{\captionfont\label{fig:MEresult-btag-error}Test of the uncertainties on (a) \mtop 
and (b) $JES$ obtained in the 
\dzero Matrix Element analysis~\cite{bib-me}.
The distributions of fitted uncertainties obtained          
from pseudo-experiments are shown by the histograms.  
The histograms show the combined distributions of upper and lower
uncertainties as the individual distributions are very similar.
The upper and lower uncertainties observed 
in the data are indicated by the solid 
and dashed arrows, respectively.}
\end{center}                                                                    
\end{figure}

\begin{figure}[htbp]                                                              
\begin{center}                                                                  
\includegraphics[width=\figbtagwidth\textwidth]{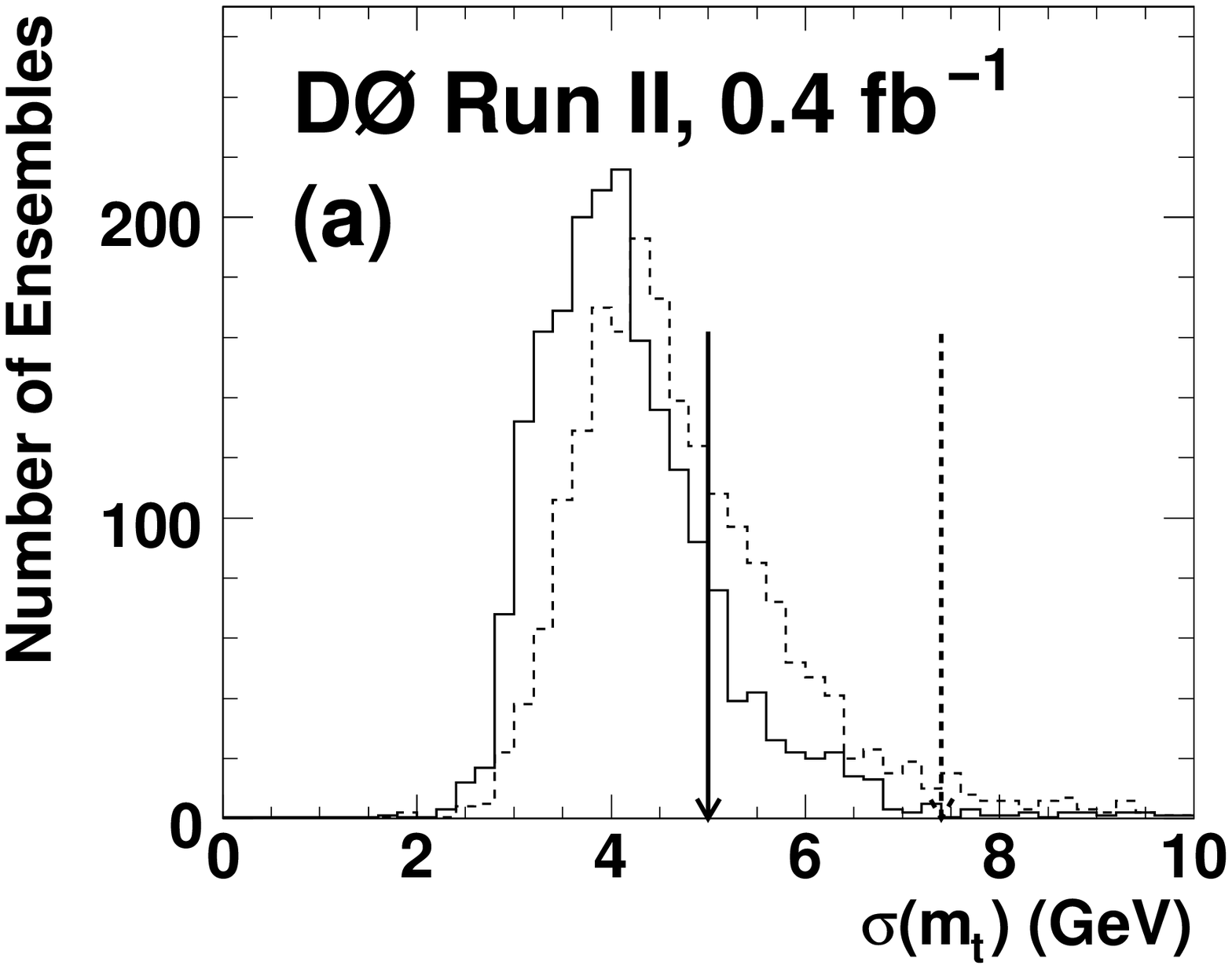}
\includegraphics[width=\figbtagwidth\textwidth]{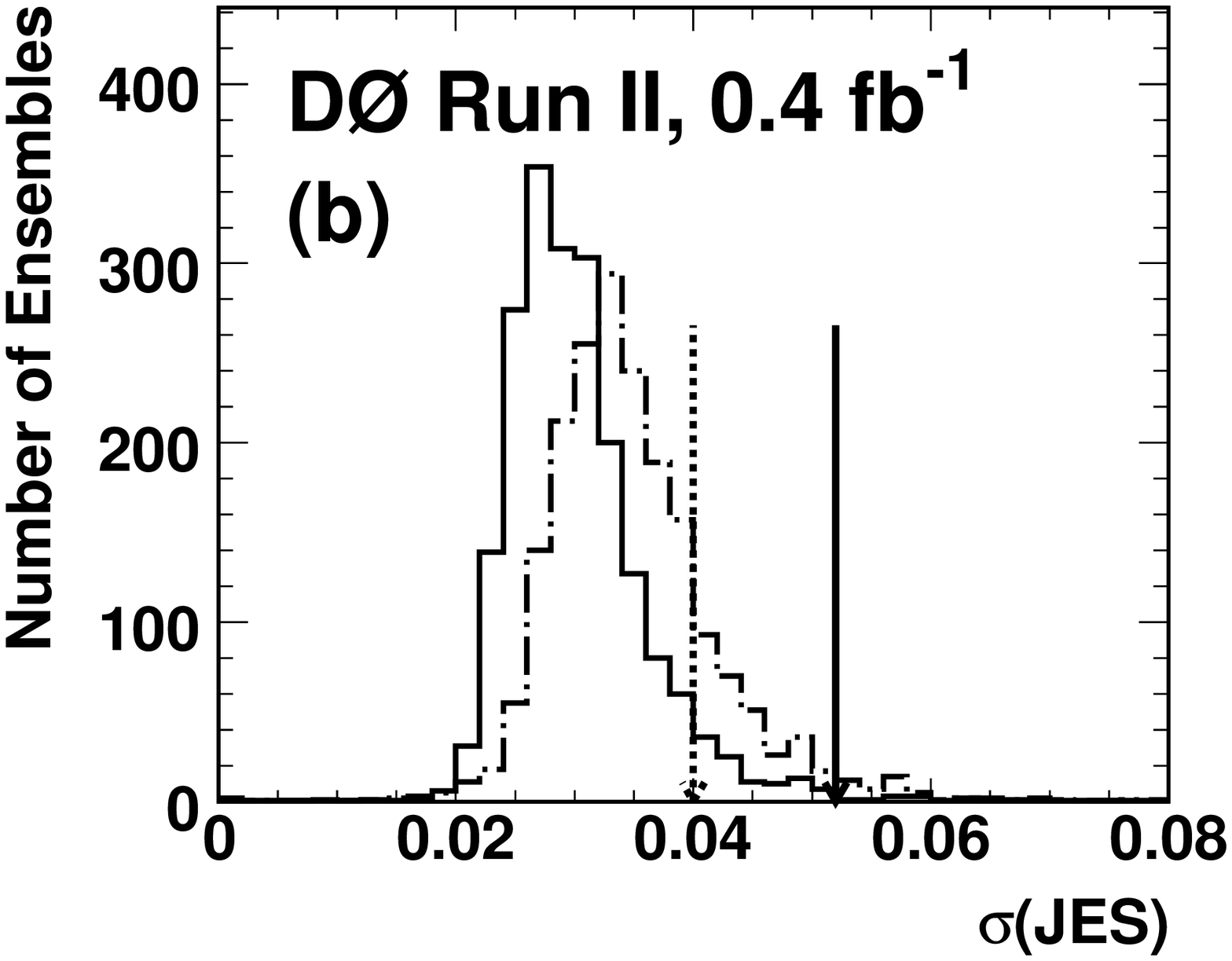}
\caption{\captionfont\label{fig:MEresult-topo-error-withftopdn}\dzero Matrix Element analysis in the \ljets channel:
  Effect of the \ttbar fraction \ftop on the expected fit
  uncertainties.  The uncertainties on (a) \mtop 
  and (b) $JES$ obtained by \dzero in the topological Matrix Element analysis
  when a sample composition according to the central \ftop
  value is assumed is shown by the solid histogram~\cite{bib-me}.
  Pseudo-experiments with \ftop varied down by one standard deviation
  yield the distributions of uncertainties shown by the 
  dash-dotted histogram.
  The upper and lower uncertainties observed in the data are indicated by the solid 
  and dashed arrows, respectively.}
\end{center}                                                                    
\end{figure}

The \ftop calibration curve for the CDF Matrix Element analysis in the
\ljets channel is shown in Figure~\ref{fig:CDFljetsMEftopcalib}.
The raw value $\ftop^{raw}$ is smaller than the true value.
This is due to the fact that a leading-order matrix element is used to describe the 
\ttbar process, while higher-order effects are included in the full simulation:
In the simulation, about
$20-30\,\%$ of \ttbar events have jets and partons that
cannot be unambiguously matched, i.e.\ at least one of the four
reconstructed jets cannot be assigned to a
parton from the \ttbar decay.
These events yield poor top quark mass information and degrade the uncertainty
estimate of the likelihood fit.
Figure~\ref{fig:MEpsgnandpbkg} shows a \dzero study which illustrates that
jet-parton matched \ttbar events tend to have a higher signal than background
likelihood, which is how the mass fit identifies them as
signal-like.
There is no such separation for signal events in which one or more
jets cannot be matched to a parton, so that these events contribute much 
less mass information to the final likelihood.

\begin{figure}
\begin{center}
\includegraphics[width=0.45\textwidth]{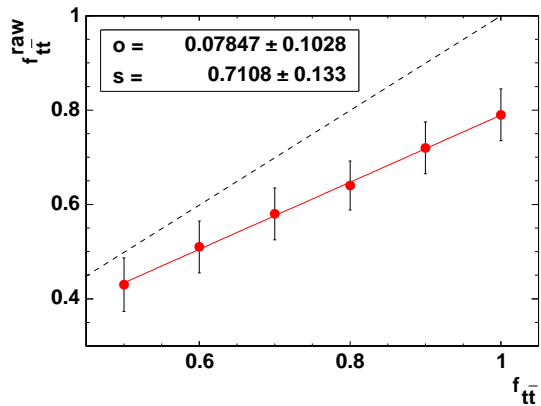}
\caption{\captionfont\label{fig:CDFljetsMEftopcalib}Calibration of the \ftop determination in the 
  CDF Matrix Element analysis in the \ljets channel~\cite{bib-CDFljetsme}.
  The points with error bars show the raw fitted \ftop value for various true \ttbar
  fractions in the pseudo-experiments.  The linear parametrization of these
  points is shown, and the values of the slope and offset (at $\ftop=0$) are 
  indicated in the inset.  To guide the eye, the dashed line shows the 
  line $\ftop^{raw} = \ftop$.}
\end{center}
\end{figure}

\begin{figure}
\begin{center}
\includegraphics[width=0.45\textwidth]{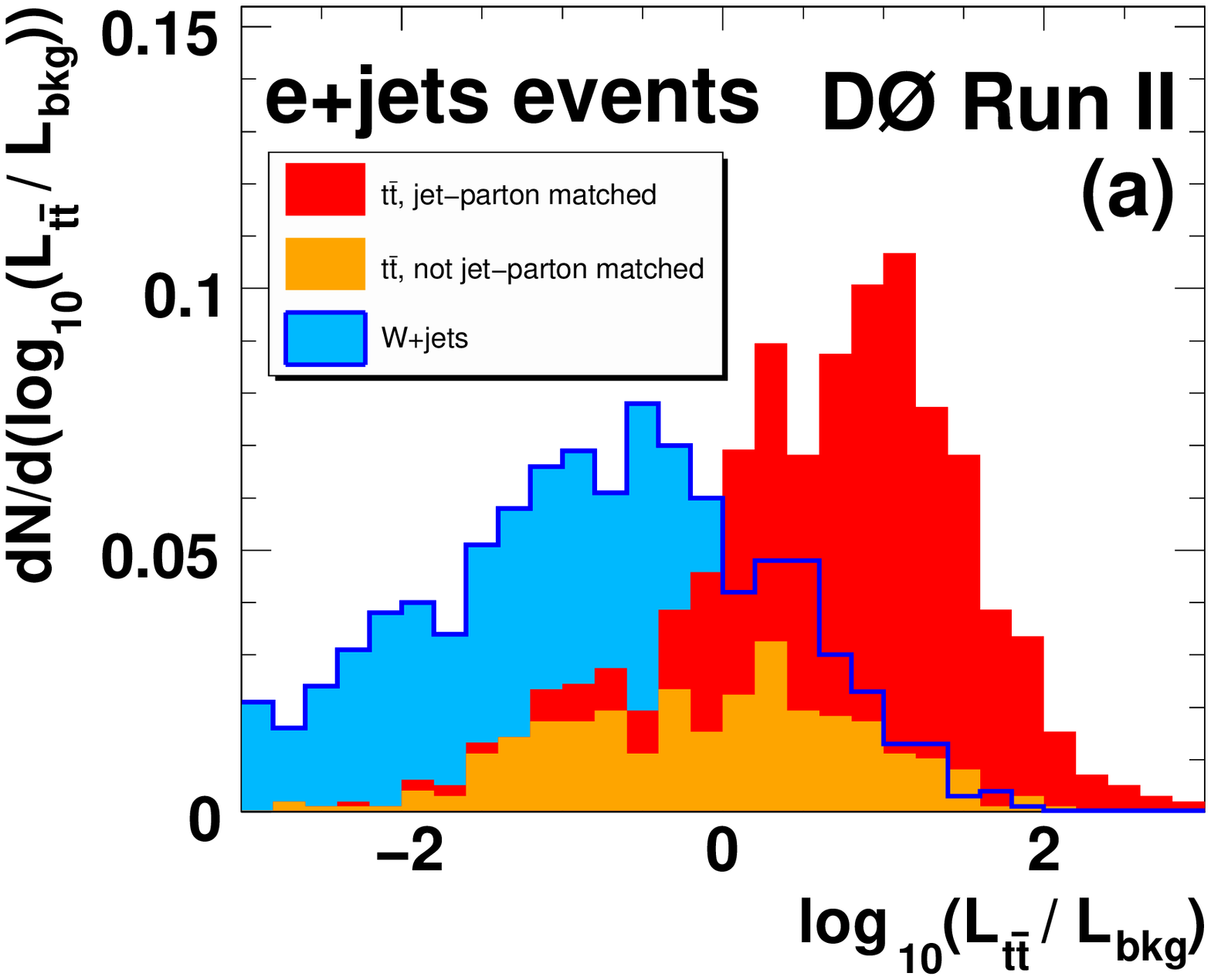}
\includegraphics[width=0.45\textwidth]{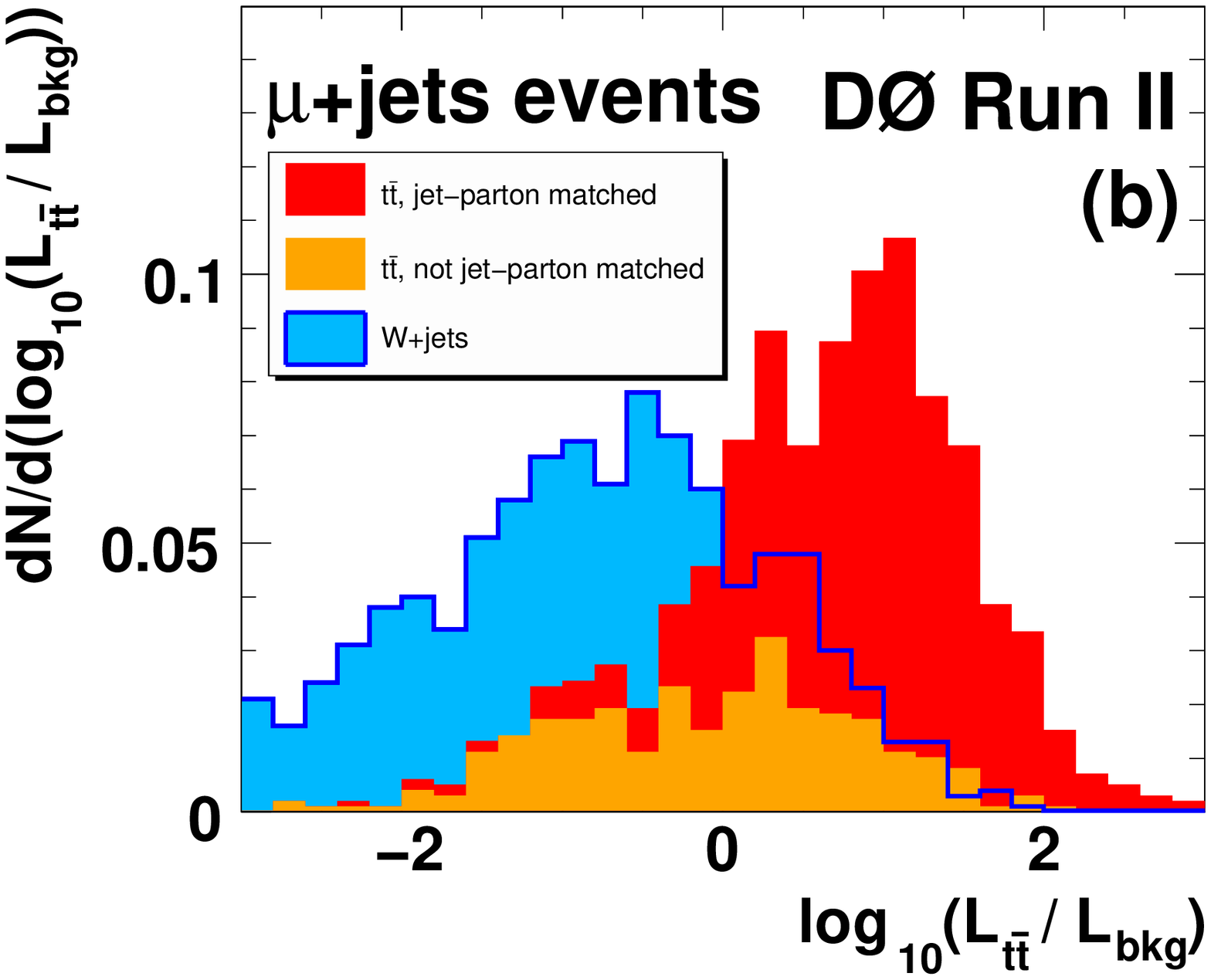}
\caption{\captionfont\label{fig:MEpsgnandpbkg}\dzero Matrix Element analysis in the \ljets
  channel~\cite{bib-me}:  Distributions of $\log_{10}(\psgn/\pbkg)$
  for \ttbar events with 
  $\mtop=175\,\GeV$ (red and orange areas) and \wjets events (dark blue lines) for
  (a) \ejets events and (b) \mujets events.
  The \psgn values are calculated for the assumption $\mtop=175\,\GeV$.
  The distributions for signal and background events are
  normalized individually.  Those \ttbar events
  where all jets can be matched to partons are shown in red, while
  \ttbar events that fail
  this requirement give rise to the orange distributions.}
\end{center}
\end{figure}

\subsection{Fit Results}
\label{massfit.fitresults.sec}
In this section, the fit results of the CDF 
template~\cite{bib-CDFmtop_ljetstemplate_prel} 
and \dzero Matrix Element~\cite{bib-me} analyses in the \ljets channel are described.
These measurements have been chosen in order to give one example 
for each of the two fitting techniques.

The reconstructed $\mtop^{\rm reco}$ and $m_{jj}$
estimator distributions in data are shown in Figures~\ref{CDFljetstemplatedatafitmtop.fig} 
and~\ref{CDFljetstemplatedatafitmw.fig}, respectively, for the CDF template
measurement.
The parametrized template distributions corresponding to the 
fitted parameters are overlaid.

\begin{figure}[htbp]
\begin{center}
\includegraphics[width=0.8\textwidth]{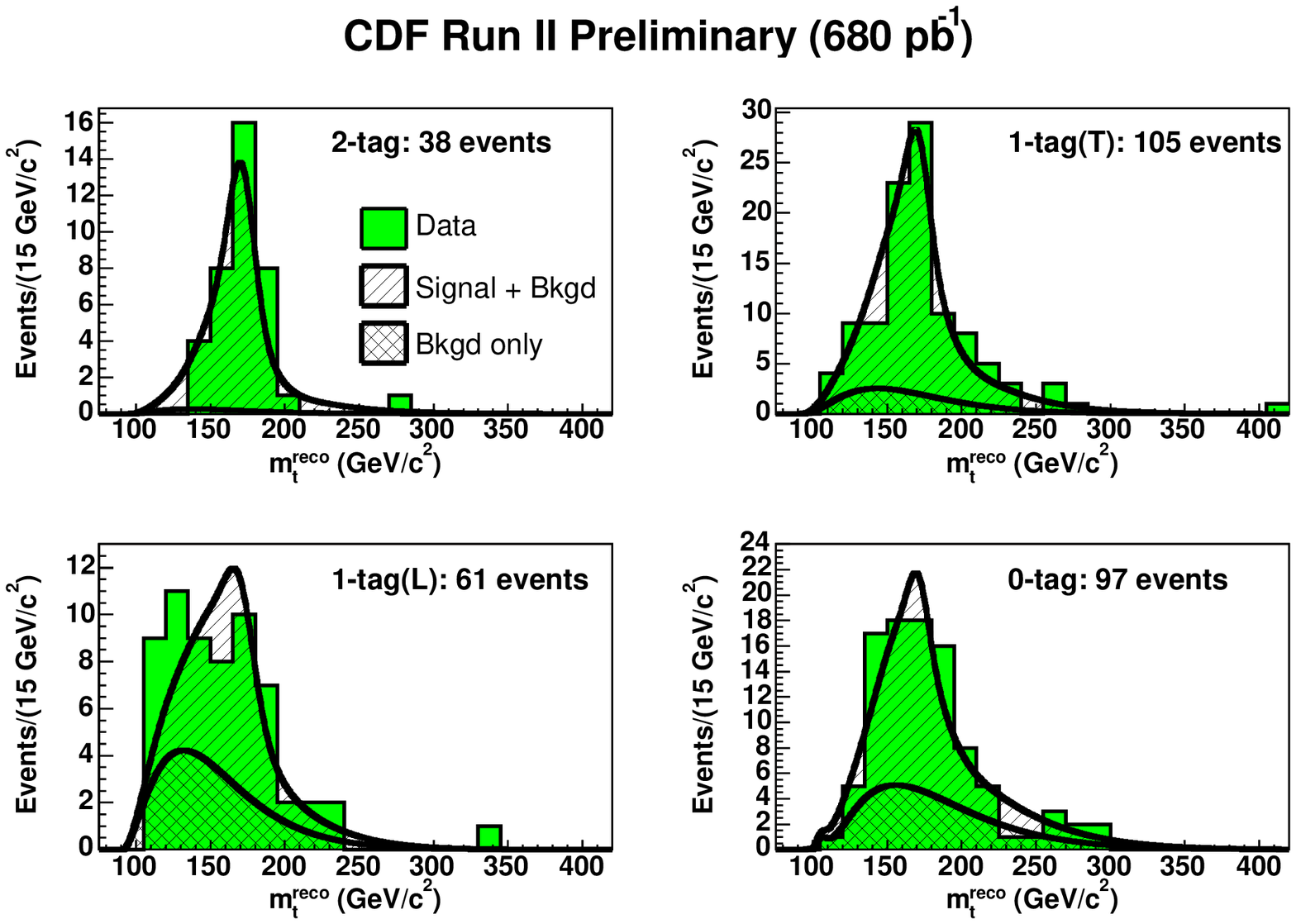}
\caption{\captionfont\label{CDFljetstemplatedatafitmtop.fig}CDF lepton+jets template measurement~\cite{bib-CDFmtop_ljetstemplate_prel}:
Data $\mtop^{\rm reco}$ distributions in the (a)~2-tag,
(b)~1-tag(T), (c)~1-tag(L), and (d)~0-tag event categories,
together with the parametrized template distributions 
for signal+background and background only that correspond to the 
fitted parameters.}
\end{center}
\end{figure}

\begin{figure}[htbp]
\begin{center}
\includegraphics[width=0.8\textwidth]{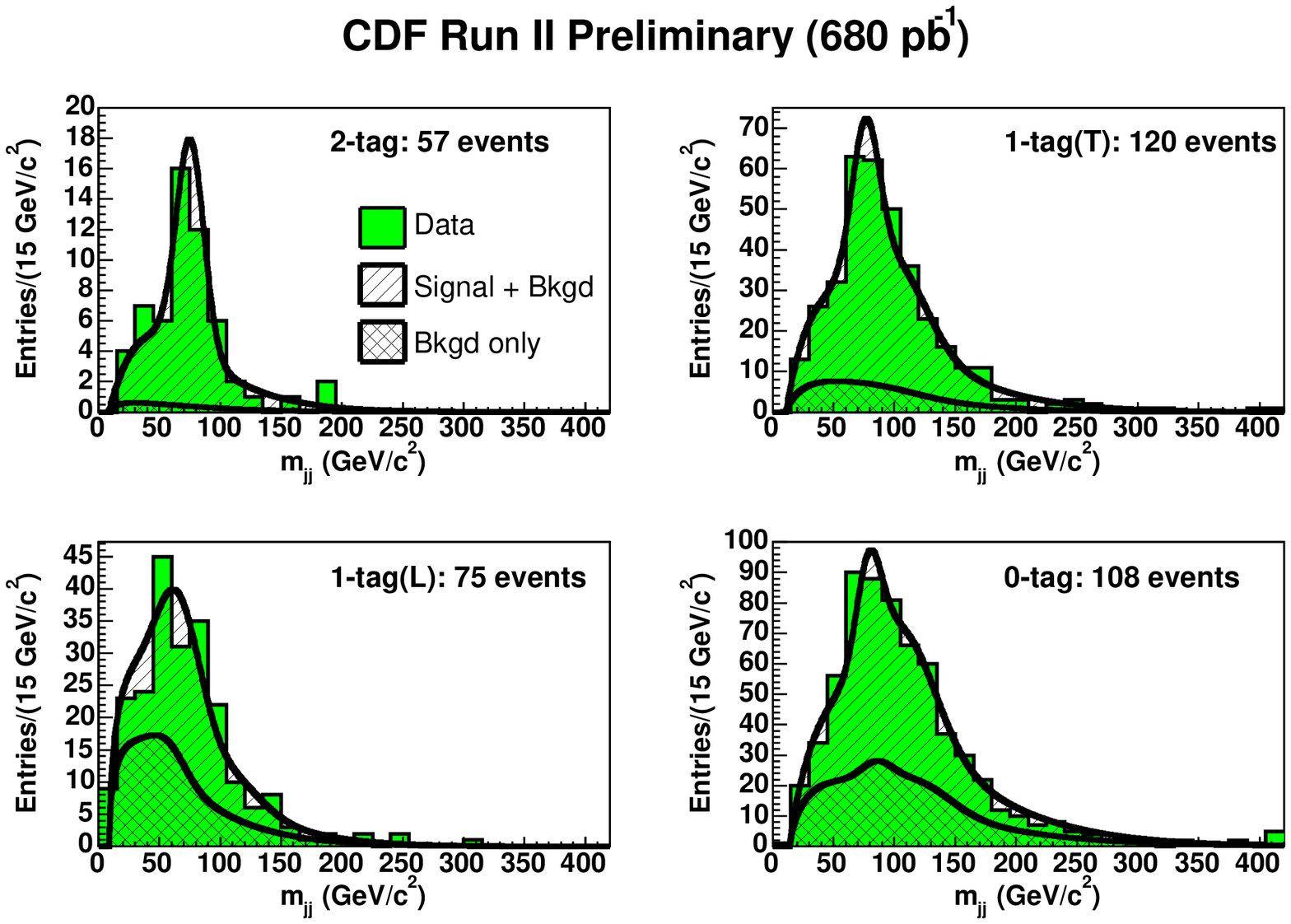}
\caption{\captionfont\label{CDFljetstemplatedatafitmw.fig}CDF lepton+jets template 
measurement~\cite{bib-CDFmtop_ljetstemplate_prel}:
Data $m_{jj}$ distributions in the (a)~2-tag,
(b)~1-tag(T), (c)~1-tag(L), and (d)~0-tag event categories,
together with the parametrized template distributions
for signal+background and background only that correspond to the 
fitted parameters.}
\end{center}
\end{figure}

In the template measurement, the combined fit to the estimator distributions
yields the likelihood as a function of assumed \mtop and \jes values.
In the Matrix Element technique, this information is determined from the 
individual event likelihoods.
These results, including the statistical uncertainties, are visualized
in Figure~\ref{result_lhood2d.fig} for the two measurements.
Contours are shown corresponding to $\Delta\ln L = 0.5$, $2.0$, $4.5$,
and $8.0$ relative to the minimum $-\ln L$ value, where $L$ denotes the 
likelihood for the event sample.
The calibrations for \mtop and \jes derived as discussed
in the previous section are taken into account.
\begin{figure}[htbp]
\begin{center}
\includegraphics[width=0.45\textwidth]{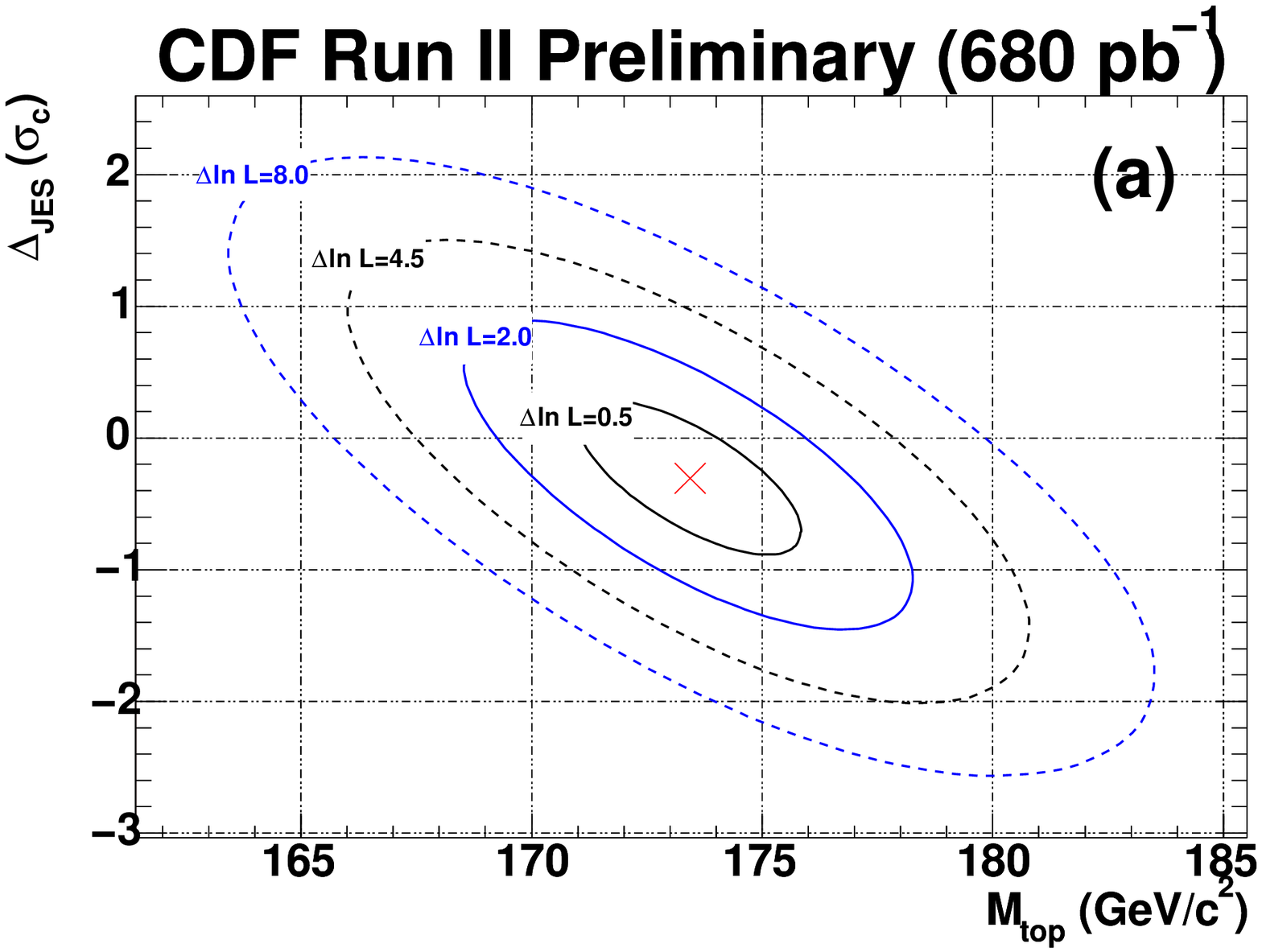}
\includegraphics[width=0.45\textwidth]{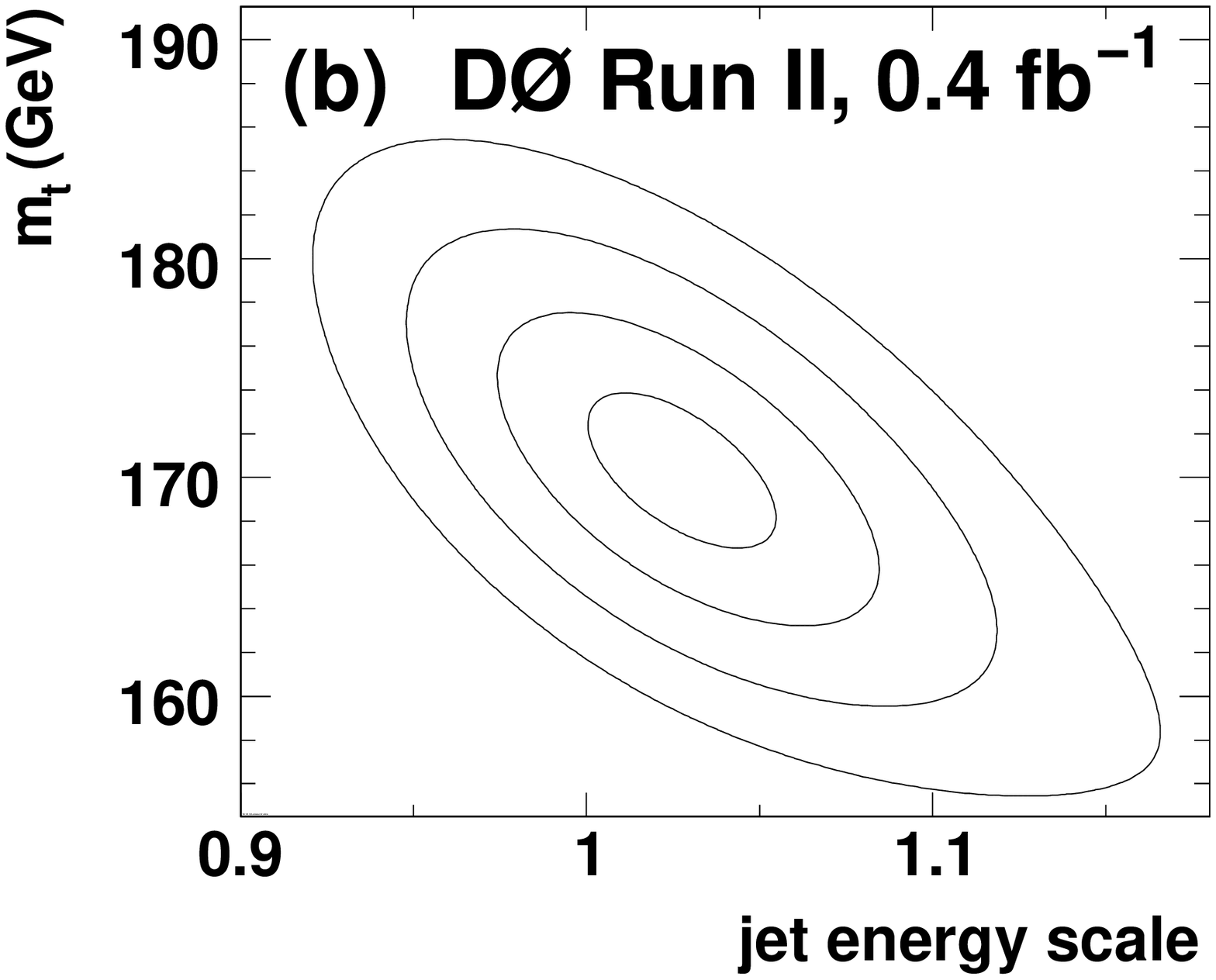}
\caption{\captionfont\label{result_lhood2d.fig}Results of the fits to determine the 
  top quark mass and jet energy scale.  (a) CDF template measurement in the \ljets 
  channel~\cite{bib-CDFmtop_ljetstemplate_prel}.  (b) \dzero Matrix Element measurement in the \ljets
  channel~\cite{bib-me}.  In both cases, the 
  contours corresponding to $\Delta\ln L = 0.5$, $2.0$, $4.5$,
  and $8.0$ relative to the minimum are shown.  Note that in (a) the jet energy
  scale (vertical axis) is measured in units of the uncertainty $\sigma_c$ of the 
  external calibration, while in (b) (horizontal axis) the multiplicative scale 
  factor for jet energies is given.
}
\end{center}
\end{figure}

The results quoted by the \dzero collaboration are obtained from the 
projection of the likelihood onto the \mtop and \jes axes as described
in Section~\ref{massfit.procedure.dzerome.sec}.
These projections are shown in Figure~\ref{result_lhood1d.fig}
together with the fitted curves.
The central values and 68\% confidence level intervals are also
indicated.
\begin{figure}[htbp]
\begin{center}
\includegraphics[width=0.45\textwidth]{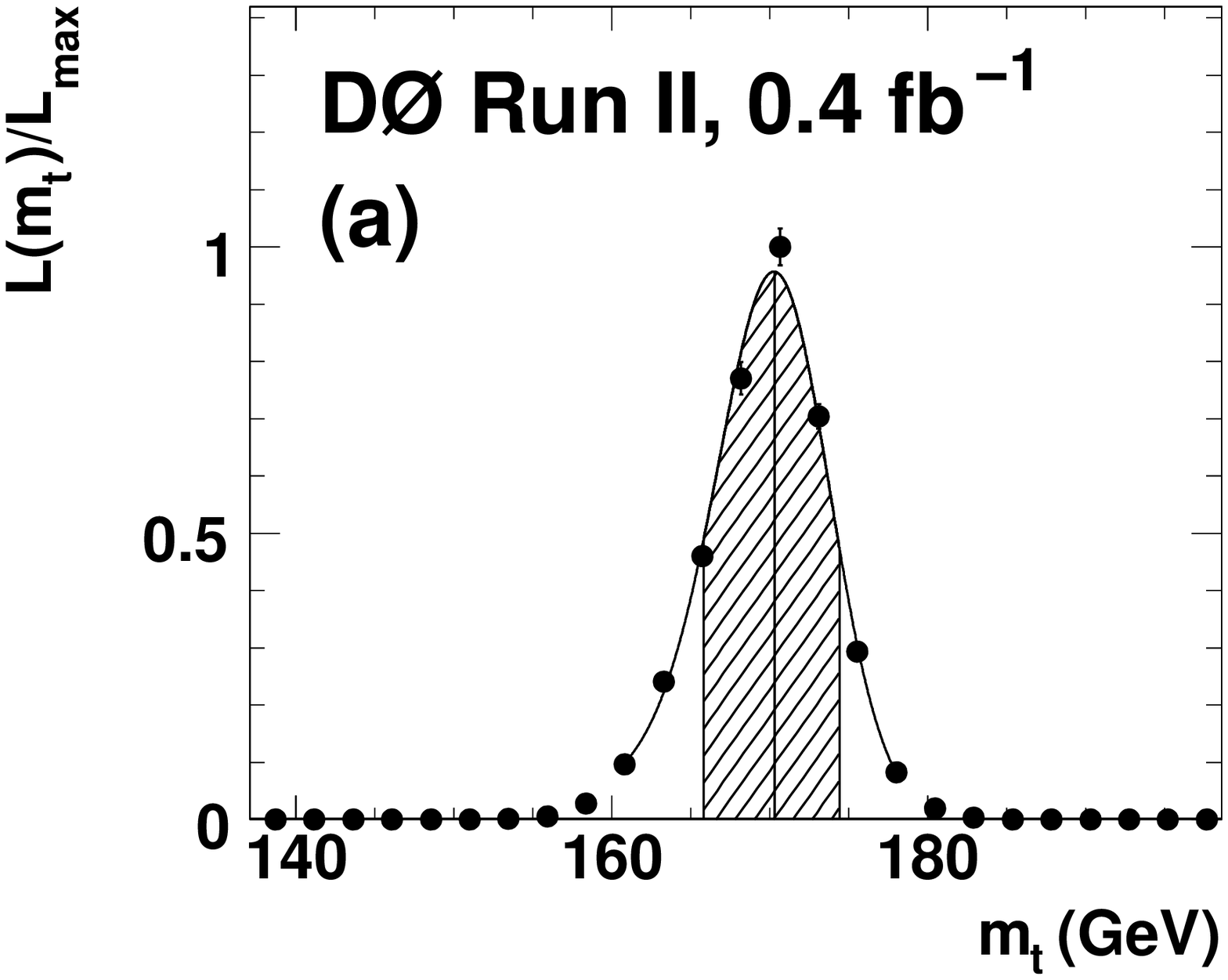}
\includegraphics[width=0.45\textwidth]{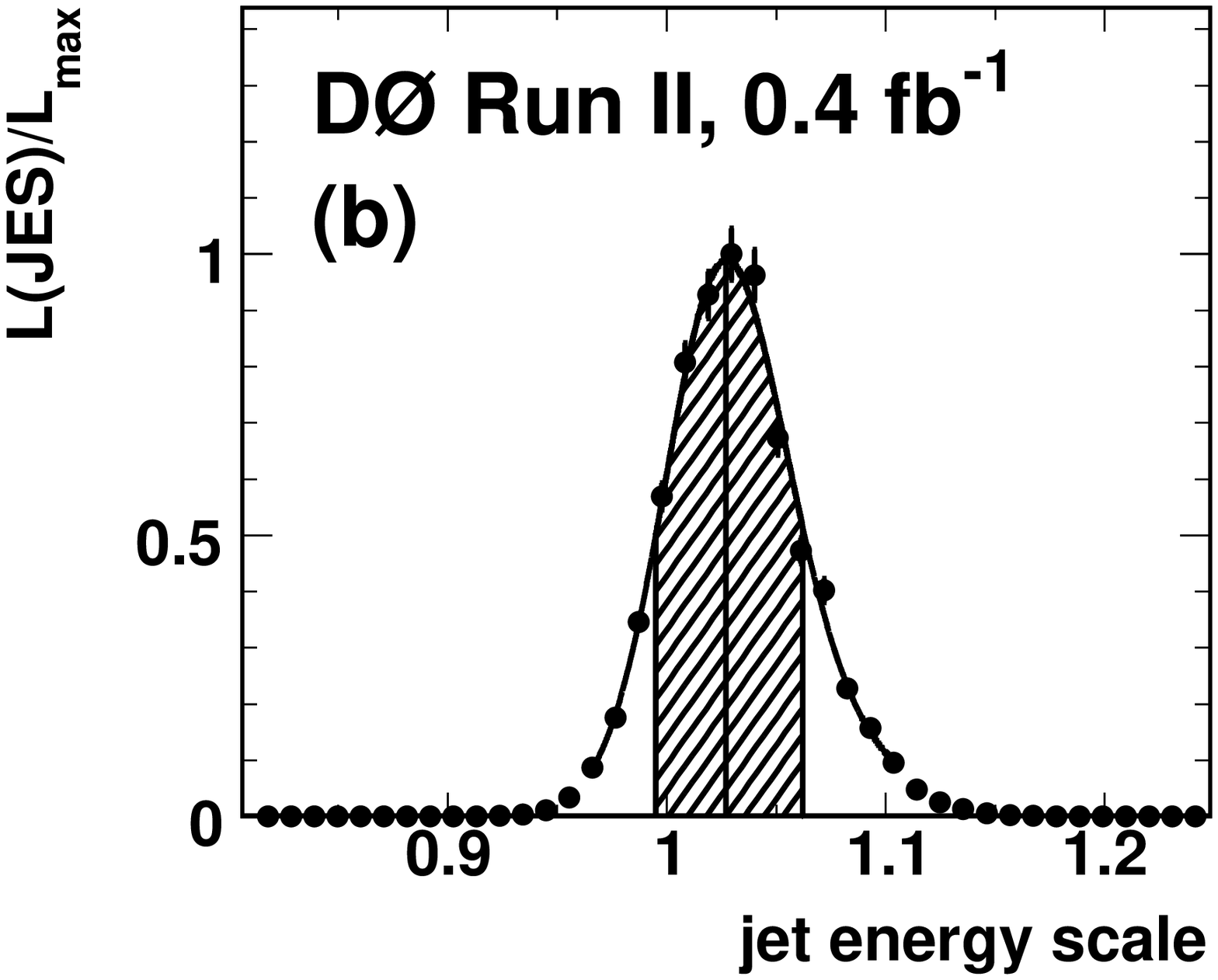}
\caption{\captionfont\label{result_lhood1d.fig}One-dimensional projections of the likelihood 
obtained in the \dzero Matrix Element measurement in the \ljets channel shown in 
Figure~\ref{result_lhood2d.fig}(b)~\cite{bib-me}.
Plot (a) shows the likelihood as a function of 
assumed top quark mass.  The correlation with the jet energy scale
is taken into account.
The fitted curve is shown, as well as the 
most likely value and the 68\% confidence level region.
The corresponding plot for the $JES$ parameter is shown in (b).}
\end{center}
\end{figure}

The comparison of the fitted uncertainties with the expectation from pseudo-experiments
is discussed in the previous section.
The statistical uncertainty includes the 
uncertainty from the absolute jet energy scale.
The contribution of the absolute jet energy scale to the total
statistical uncertainty can be estimated by repeating the fit
with the \jes parameter fixed.
It should however be noted that 
fitting for one overall factor
does not cover the entire systematic uncertainty due 
to the jet energy 
scale, cf.\ Section~\ref{systuncs.detectormodeling.jes.sec}.

\clearpage
\section{Systematic Uncertainties}
\label{systuncs.sec}
\begin{center}
\begin{tabular}{p{15cm}}
{\it The previous section described how the central measurement
  value of the top quark mass and the associated statistical uncertainty
  are determined.  To date, the world-average value for the top
  quark mass is already systematically limited.  This section 
  discusses the individual sources of systematic errors, describes the
  correlations among various measurements and how they are handled
  in the combination, and indicates where systematic uncertainties
  may be reduced in the future.}
\end{tabular}
\end{center}

With the increasing size of the datasets collected at \runii of the Tevatron,
the precision of the world-average value of the top quark mass has
already become limited by systematic uncertainties.
This is in spite of the fact that 
the measurement techniques have been improved
during the past years.
In particular, the determination of the jet energy scale
from the same data that is used to measure the top quark mass
has reduced the systematic uncertainty due to the
detector calibration.
Thus, the initial expectations for 
\runii of the Tevatron have already been surpassed.
At the LHC, systematic effects will become even more dominant.

In this section, the different sources of systematic uncertainties are 
discussed together with the way they are typically evaluated.
Systematic correlations between measurements or experiments,
which will tend to reduce the beneficial effect of combining 
several measurements, are mentioned.
Also indicated are ideas for future improvements, as well as 
limitations.

Systematic uncertainties can be broadly classified into three categories: 
modeling of the physics processes for \ttbar production and background, 
modeling of the detector performance, 
and uncertainties in the measurement methods.
The following discussion is ordered along the lines of this classification.

In Table~\ref{systuncs.table} an overview of systematic uncertainties
is given, quoting both the uncertainties on the world-average top
quark mass~\cite{bib-TEVEWWG} (which is only available in broad
categories) and on one individual measurement~\cite{bib-me}.
\begin{table}[htbp]
\begin{center}
\begin{tabular}{l@{\ \ \ \ \ }
                c@{\ \ \ \ \ }c}
\hline
\hline
  Source of Uncertainty
& 
  \begin{tabular}{@{}c@{}}World\\ Average\end{tabular}
&
  \begin{tabular}{@{}c@{}}\dzero, Lepton+Jets\\ Channel\end{tabular} 
\\
\hline
\phantom{.}\vspace{-1ex}\\
Statistical uncertainty            & \WAerrstat       
                                   & \MEberrstatnojeslong  \\
\phantom{.}\vspace{-1.5ex}\\
{\em Physics modeling:}            & \WAerrphysics    
                                   &                  \\
\quad PDF uncertainty              &                  
                                   & \MEberrpdf       \\
\quad ISR/FSR modeling             &                  
                                   & \MEberrsgnmod    \\
\quad $b$ fragmentation            &                  
                                   & \MEberrbjes      \\
\quad $b$/$c$ semileptonic decays  &                  
                                   & \MEberrbclepbr   \\
\quad \wjets background modeling   &                  
                                   & \MEberrbkgmod    \\
\quad QCD contamination            &                  
                                   & \MEberrqcd       \\
\phantom{.}\vspace{-1.5ex}\\
{\em Detector modeling:}           & \WAerrdetector   
                                   &                  \\
\quad Absolute jet energy scale    &                  
                                   & \MEberrjeslong   \\
\quad $JES$ $\pt$ dependence       &                  
                                   & \MEberrjespt     \\ 
\quad $b$ response (h/e)           &                  
                                   & \MEberrbresp     \\
\quad Trigger                      &                  
                                   & \MEberrtrg       \\
\quad $b$ tagging                  &                  
                                   & \MEberrbtagging  \\
\quad Noise, multiple interactions &                  
                                   & ---              \\
\phantom{.}\vspace{-1.5ex}\\
{\em Method:}                      & \WAerrfit        
                                   &                  \\
\quad Signal fraction              &                  
                                   & \MEberrftop      \\
\quad MC calibration               &                  
                                   & \MEberrmccalib   \\
\phantom{.}\vspace{-1.5ex}\\
Total uncertainty                  & \WAerrtotal      
                                   & \MEberrtotallong \\
\vspace{-2ex}\\
\hline
\hline
\end{tabular}
\caption{\captionfont\label{systuncs.table}Summary of uncertainties on the top
  quark mass.  
  All values are quoted in GeV.  
  Uncertainties on the world-average value~\cite{bib-TEVEWWG}
  are quoted in the second column.
  Only values corresponding to a broad classification of error
  sources are available for the world average.  
  Some values from~\cite{bib-TEVEWWG} have been combined to reflect
  the categories used here.  
  The detector modeling uncertainty is dominated by that on the
  absolute value \JES of the jet energy scale.  
  The right column shows uncertainties
  for the \dzero Matrix Element measurement in the \ljets
  channel~\cite{bib-me}.  
  For asymmetric uncertainties the upper and lower
  errors are quoted separately.
  The uncertainty from the absolute \JES value has been listed together
  with the systematic uncertainties in the right column even
  though it is determined with in situ calibration and scales
  with statistics.}
\end{center}
\end{table}

\subsection{Physics Modeling}
\label{systuncs.physicsmodeling.sec}
Many different processes can lead to the \ttbar event candidates
selected for a top quark mass measurement, and not all of them
can be taken into account in the simulation.
In addition, the description of the processes that are accounted for
may still be subject to uncertainties.
This type of uncertainties is discussed in this section, while
effects not arising from a single hard interaction (multiple
interactions) are treated in 
Section~\ref{systuncs.detectormodeling.sec}.

The top quark decay properties are well-known in the Standard Model, 
including the subsequent decay of the \W
boson into partons, since these decays are governed by the weak
interaction and the top quark does not hadronize.
The top quark width as a function of its
mass is known~\cite{bib-pdg},
and the branching fraction of the decay
$\tquark\to\W\bquark$ is $100\%$ for practical purposes.
Furthermore, the mass, width, and branching fractions of the $\W$ 
are known precisely~\cite{bib-pdg} and the associated uncertainties
can be neglected.

In contrast, significant uncertainties do arise from 
the production of the $\ttbar$ pair (modeling of the parton 
distribution functions and of initial-state radiation) 
and the formation of final-state jets (final-state 
radiation, fragmentation, and hadronization modeling).
Usually, the Monte Carlo simulation of \ttbar events is based
on the leading-order matrix element for the processes
$\qqbar\to\ttbar$ and $\glueglue\to\ttbar$.
However, next-to-leading-order Monte Carlo simulation is already
available~\cite{bib-MCatNLO}.
The PDF parametrization and the modeling of initial- and final-state
radiation have to be
matched with the description of the hard-scattering process
accordingly.
The general description below remains however valid in both cases.

\subsubsection{PDF Uncertainty}
\label{systuncs.physicsmodeling.PDF.sec}
Parton distribution functions (PDFs) parametrize the probability 
to find a parton of a given flavor and momentum fraction inside the 
proton or antiproton, and thus the kinematic distributions of signal
and background events depend on the PDFs.
The Tevatron experiments have agreed on a common procedure to evaluate
the top quark mass uncertainty related to PDF 
modeling, which is described for example in~\cite{bib-me}.
Typically, the simulated events used to calibrate the measurements
(cf.\ Section~\ref{massfit.validationandcalibration.sec}) are based
on a leading-order PDF set like CTEQ5L~\cite{bib-CTEQ5L}.
Systematic variations are however only provided for the PDF set
CTEQ6M~\cite{bib-CTEQ6Mvar}.
Therefore, the top quark mass is recomputed with a calibration
based on the central
CTEQ6M PDF set, and the differences between that value and the ones
obtained with the systematic variations of the CTEQ6M PDF
are added in quadrature and assigned as a systematic uncertainty.
Note that the difference between top quark masses evaluated
with the calibrations based on the CTEQ5L and central CTEQ6M PDF sets is 
not included in the uncertainty.
The difference between the results obtained with the CTEQ5L and 
MRST leading-order PDF sets is taken as another uncertainty.
Finally, the effect from using MRST PDF sets based on different
assumed $\alpha_s$ values is determined.
These three individual systematic uncertainties are summed in 
quadrature, the variation of CTEQ6M parameters yielding the 
dominant contribution.

For the determination of the world-average top quark mass, 
the resulting error is taken as $100\%$ correlated between 
individual measurements.
The size of the uncertainty is given in Table~\ref{systuncs.table}
for the \dzero measurement in the lepton+jets channel.
The individual contributions are
\begin{center}
  \begin{tabular}{l@{\ }c}
      CTEQ6M variations:        & $+0.12$ $-0.38$ GeV,
    \\
      difference MRST$-$CTEQ5L: & $\pm0.09$ GeV,
    \\
      variation of $\alpha_s$:  & $+0.06$ $-0.03$ GeV.
  \end{tabular}
\end{center}

Since this systematic error is correlated between all measurements,
a common procedure for its evaluation like the one described above
is important.
Improvements of the above scheme are however still desirable and 
possible:
\begin{list}{$\bullet$}{\setlength{\itemsep}{0.5ex}
                        \setlength{\parsep}{0ex}
                        \setlength{\topsep}{0ex}}
\item
The use of a leading-order matrix element together with 
a PDF set intended for processes in next-to-leading order is not
consistent.
The calibration of future measurements of the top quark mass should be
based on next-to-leading-order Monte Carlo simulation using CTEQ6M
(or updated PDF sets for which systematic variations are available),
which would naturally resolve this inconsistency.
\item
Different top quark mass measurements may be more or less sensitive
to variations of individual parameters describing the PDF set.
For example, depending on kinematic event selection cuts, the 
relative importance of the gluon PDF may vary even considering only 
Tevatron analyses; this will become a more important issue when 
measurements at the LHC are included as well, where the
$\glueglue\to\ttbar$ process dominates.
Consequently, the quadratic sum resulting from the variations of all PDF 
parameters should not be taken as
100\% correlated between measurements, but 
top quark mass shifts should be quoted for each individual PDF
parameter variation.
This will then allow for a more refined computation of the uncertainty
on the world average, potentially slightly reducing the overall systematic 
error due to PDF uncertainties.
(Note that no extra systematic uncertainties will have to be
evaluated, only a more refined report of individual variations is
needed.)
\item
The comparison of top quark masses obtained with leading-order CTEQ and
MRST PDF sets aims to quantify potential uncertainties 
arising from different PDF fitting procedures.
However, these PDF sets are not based on
exactly the same inputs, leading to additional differences that should
already be covered by the variation of CTEQ parameters.
Since the systematic error arising from the CTEQ/MRST comparison
is small, this is currently not an important issue.
\end{list}

Currently, the systematic top quark mass error related to PDF
uncertainties does not dominate the world average, 
cf.\ Table~\ref{systuncs.table}.
Because it is correlated between individual measurements, 
it may become important in the future, 
but only if no further improvements on PDF uncertainties
are assumed.

\subsubsection{Initial- and Final-State Radiation}
\label{systuncs.physicsmodeling.isrfsr.sec}
Radiation off the incoming and outgoing partons may affect the top
quark mass measurement.
Such radiation changes the kinematics of the \ttbar decay products
in the final state; for example, the transverse momentum of the \ttbar
system is not zero when initial-state radiation (isr) takes
place.  
Final state radiation (fsr) changes the momenta of the \ttbar
decay products and thus affects the shapes of templates or the signal
probability assigned to an event.  
Also, isr or fsr may lead to jets
which can be misidentified as \ttbar decay products.

Initial- and final-state radiation 
are governed by the same equations and are modeled in the 
shower evolution in the Monte Carlo simulation.
(Interference between isr and fsr cannot be taken into account in this
simulation, only when next-to-leading matrix elements are used.)
The CDF experiment has shown how the details of the radiation process 
can be studied with Drell-Yan
events~\cite{Abulencia:2005aj,bib-CDFljetsDLMmass}.
In Drell-Yan events only isr is present (photon radiation off charged
leptons is assumed to be well-modeled, so only QCD radiation is considered here); it
can lead to a non-zero transverse momentum \pt of the dilepton system.
The mean dilepton \pt is shown to have a linear dependence on the
logarithm of the dilepton invariant mass, which is reproduced by
{\pythia} simulation with standard parameter settings, 
cf.\ Figure~\ref{dileptonpT.fig}. 
The study of Drell-Yan events also motivates
two alternative \pythia parameter sets 
leading to more or less isr and fsr activity, which 
are used to evaluate the systematic uncertainty on the top quark mass.
The parameters changed are
\lambdaqcd and the scale factor $k$ to the 
transverse momentum scale for isr showering; settings of
$\lambdaqcd=292\,\MeV$ and $k=0.5$ are used for 
the sample with increased isr activity, while
$\lambdaqcd=73\,\MeV$ and
$k=2.0$ are taken for the sample with less isr.
The resulting mean dilepton \pt values are also indicated in
Figure~\ref{dileptonpT.fig}.

\begin{figure}
\begin{center}
\includegraphics[width=0.48\textwidth]{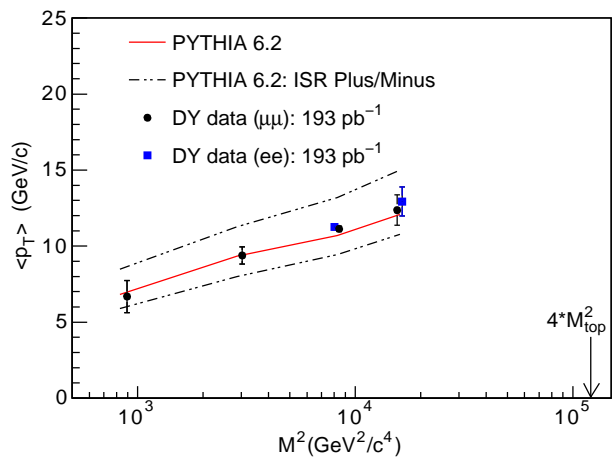}
\caption{\captionfont\label{dileptonpT.fig}The average \pt of the dilepton 
  system in Drell-Yan events, which is a measure of the level
  of isr activity, as a function of the dilepton invariant mass squared
  (note the logarithmic horizontal
  scale)~\cite{Abulencia:2005aj,bib-CDFljetsDLMmass}. 
  The points with error bars indicate CDF measurements, while the
  solid and dashed lines show predictions by the {\pythia}
  generator (standard parameter settings and variations for systematic
  error evaluation, respectively).}
\end{center}
\end{figure}

The approach followed by \dzero in~\cite{bib-me} is instead to vary
directly the fraction of events with significant radiation.
The calibration of the measurement is repeated based on events where
a \ttbar pair is produced together with an additional parton.
Since the cross-section for \ttbar production is 30\% larger in
next-to-leading order than in leading order, 30\% of the observed
difference between the top quark masses measured with the default and
this alternative calibration is assigned as systematic uncertainty.

In the \ljets channel, the CDF experiment quotes a systematic uncertainty
of 0.5~GeV for their template
measurement~\cite{bib-CDFmtop_ljetstemplate_prel}, while the Matrix
Element measurement is
more sensitive to the modeling and finds a 1.0~GeV 
uncertainty~\cite{bib-CDFljetsme} (adding
isr and fsr uncertainties in quadrature).
For the \dzero Matrix Element measurement in the \ljets channel, a
systematic error of 0.5~GeV has been evaluated~\cite{bib-me} 
(see Table~\ref{systuncs.table}) using 
a different technique as described above.
Similar uncertainties have been obtained in the 
dilepton~\cite{bib-CDFdileptonme,bib-CDFdileptonmewithbtagging,bib-CDFdileptontemplateprel} 
(0.4~GeV, 0.6~GeV, and 0.7~GeV, respectively) and
\alljets channels~\cite{bib-CDFallhadtemplate} (0.7~GeV), where
the values quoted are from the measurements using $1\,\ifb$ of data.

In the future, the uncertainty in isr and fsr modeling may become a
dominant systematic error since it is correlated between all
measurements.
For consistency, it would therefore be highly desirable to arrive at an agreement
between experiments on how to evaluate it, as is the case for the
PDF error, see Section~\ref{systuncs.physicsmodeling.PDF.sec}.
In addition, more precise studies of isr and fsr should be carried
out.
The analysis of Drell-Yan events that the CDF experiment has published 
in~\cite{Abulencia:2005aj,bib-CDFljetsDLMmass} 
can be repeated with much more data and 
thus extended to larger invariant dilepton masses, so that the
extrapolation to \ttbar events becomes smaller.
This may allow to decrease the width of the error band shown in
Figure~\ref{dileptonpT.fig}.
In addition, an examination of LEP/SLC results on hadronic $\Z$ decays may
yield independent experimental information on fsr.
Finally, it may soon become worthwhile to optimize the measurement
techniques not only in view of the statistical error, but to also
keep systematic effects in mind.
An idea developed for the LHC is to consider events in
which the top and antitop quarks have large \pt, which means that
their decay products are found in two separate
event hemispheres; the mass of the top quark with the hadronic
\W decay could then be reconstructed from the individual hadrons,
making jet reconstruction superfluous and rendering the measurement
mostly insensitive to final-state
radiation~\cite{Borjanovic:2004ce}.
But already the simultaneous \JES fit in the \ljets Matrix Element
analyses at the Tevatron has proven to reduce the sensitivity of the 
result to radiation modeling~\cite{bib-meprivatecommunication}, and 
an additional integration over the \ttbar transverse momentum, as used
in~\cite{bib-CDFdileptonMEpaper}, may further reduce
this uncertainty.

\subsubsection{Fragmentation}
\label{systuncs.physicsmodeling.fragmentation.sec}
Related to final-state radiation are the formation of jets in the 
final state and the spectra of hadrons within the jets.
The fragmentation and hadronization of $\bquark$-quark jets is
particularly important: in dilepton events, only $\bquark$-quark jets
are expected (except for jets from isr or fsr), and in \ljets and
\alljets events, in situ calibration of the jet energy scale can
largely absorb the dependence on the modeling of light ($\uquark$,
$\dquark$, $\squark$, $\cquark$) quark jets.
Simulations based on different fragmentation and hadronization models
may predict different average energy fractions
contained within the reconstructed jet; this leads to an uncertainty on the 
relation between jet and parton energies and thus on the measured
top quark mass.
In addition, $\bquark$-quark
fragmentation also affects the efficiency of $\bquark$-jet
identification: for a given \bquark-quark energy,
an increase of the average energy fraction $\langle x \rangle$
carried by the bottom hadron 
will lead to a higher probability to detect a well-separated secondary
decay vertex even for low-energy \bquark quarks and will thus
affect the kinematic
distribution of the selected events.

Data from LEP and SLC on $\Z\to\bbbar$ decays constrain $\bquark$
fragmentation models and yield for example a precise determination
of the mean energy fraction $\langle x_b\rangle$ of the
weakly-decaying bottom hadron in \Z decays~\cite{bib-Zbible}.
To extrapolate to \ttbar decays, 
different fragmentation models that are consistent with \Z data are 
used to simulate \ttbar events and the corresponding distribution 
in top quark decays (References~\cite{Cacciari:2002re-Corcella:2001hz} 
define $\langle x_B\rangle$
as the bottom hadron energy divided by the maximum possible
$\bquark$-quark energy).
To evaluate the uncertainty on the top quark mass,
the calibration of the measurement is determined using these different
models, and the observed differences in the top quark mass are assigned as 
systematic error.

Uncertainties in the decay of bottom (and charm)
hadrons can also play a role.
In particular, jets containing a semileptonic decay of a heavy hadron will on average
be reconstructed with a smaller energy due to the escaping neutrino.
Thus the top quark mass depends on the rate and modeling of
semileptonic heavy hadron decays.
To assess uncertainties in the decay model, 
the semileptonic branching fractions of heavy hadrons in \bquark-quark
jets are varied within the bounds from measurements in 
\Z decays~\cite{bib-Zbible}.

Like the other systematic errors related to physics modeling, 
the resulting uncertainties are correlated between measurements.
The semileptonic branching fractions are known so precisely that the 
associated systematic error is negligible; however, the \bquark-quark
fragmentation uncertainty  may become a dominating uncertainty in the
future, see Table~\ref{systuncs.table}.
As a first step, a common scheme for evaluating this uncertainty 
should be agreed on.
This could be the definition of a set of fragmentation models
and parameters (like the ones studied in~\cite{bib-me})
on which the evaluation of uncertainties is based for each measurement.
Such a common definition would not only lead to a consistent evaluation of 
uncertainties, but also allow for a correct determination of 
systematic correlations between individual measurements.
As a next step, measurement techniques with reduced sensitivity to the
details of \bquark-quark fragmentation could be developed.
The technique based on high-\pt top quarks mentioned
in Section~\ref{systuncs.physicsmodeling.isrfsr.sec}, which does not
rely on conventional jet finding, may serve as an
example, but will need to be refined to optimize the overall uncertainty.

\subsubsection{Top Quark Mass Definition}
\label{systuncs.physicsmodeling.definition.sec}
A top quark mass measurement based solely on invariant mass 
reconstruction from the momenta of the decay products corresponds to a
measurement of the pole mass.
All results available today are based mainly on properties of the
\ttbar decay products which in turn depend on the top quark (pole)
mass; the current measurements can therefore be regarded as
pole mass measurements to a good approximation.
However, calculations of effects involving the top quark mass
like the ones described in Sections~\ref{resinterp.interp.sm.sec}
and~\ref{resinterp.interp.mssm.sec} are typically not performed using
the pole mass.
The transformation into the \msbar scheme is known to three
loops and is e.g.\ given 
in~\cite{bib-pdg}; such a transformation introduces an
uncertainty when interpreting the top quark mass.

Measurements in the dilepton channel necessarily include 
other information as well since the kinematics of the \ttbar system is
underconstrained; also measurements in the \ljets and \alljets channels make
use of additional information from the \ttbar production process to
a varying degree in order to
reduce the statistical measurement uncertainty.
It still remains to be studied to what extent this fact leads to an
uncertainty in the interpretation of the measurement results.

In addition to the above, the pole mass itself is not
defined to arbitrary accuracy for a colored particle like the top quark,
and there is necessarily some additional
color flow involved in the creation of the colorless final state
measured in the detector.
It has been shown in~\cite{bib-scottwillenbrock} that this introduces an
intrinsic uncertainty of the order of \lambdaqcd on the pole mass.
The exact size of the uncertainty depends on the details of the
measurement, and detailed studies of this effect are only starting.

\subsubsection{Color-Reconnection Effects}
\label{systuncs.physicsmodeling.CR.sec}
Apart from the intrinsic uncertainty on the pole mass
of a colored particle, color-reconnection effects
between the final-state products may lead to additional effects.
Corresponding studies for the measurement of the \W
boson mass at LEP2 are described in~\cite{bib-LEPEWWG}.
The effect in \ww production at LEP2 is small compared to 
uncertainties on the top quark mass (a $35\,\MeV$ systematic error is
quoted in the \alljets \ww final state) in the present and near future.
The \alljets \ww final state may be considered similar to 
\ljets \ttbar events; however, the kinematics are different; the
colored beam remnants may well play an additional role in \ttbar events at
hadron colliders, and it is not clear how in situ calibration of the
jet energy scale is affected.
It is expected that the results of first studies of the size of 
color-reconnection effects will be published soon~\cite{bib-skandswicke}.

\subsubsection{Bose-Einstein Correlations}
\label{systuncs.physicsmodeling.be.sec}
The LEP experiments have determined the effects from
Bose-Einstein correlations between particles in \ww
events~\cite{bib-LEPEWWG}.
The resulting uncertainty on the top quark mass
has not yet been studied, but it can be expected to be
of the same order as that assigned to the \W mass measured in the 
\alljets \ww final state ($7\,\MeV$~\cite{bib-LEPEWWG}).
Such an uncertainty would be negligible for the top quark mass.

\subsubsection{Underlying Event}
\label{systuncs.physicsmodeling.ue.sec}
In principle, particles produced from the remnants of the colliding
hadrons may contribute energy to the jets reconstructed in the
detector.
It is therefore necessary to measure the average contribution and
subtract it from the jet energies.
This is done as part of the jet energy calibration.
The resulting uncertainty is small, as shown in
Figure~\ref{CDFjesunc.fig}, and included in the jet energy scale
uncertainty (even though it is in principle correlated between experiments).

\subsubsection{Background Modeling}
\label{systuncs.physicsmodeling.background.sec}
All physics uncertainties (except for the top quark mass definition)
discussed in the previous sections affect
the modeling of both signal and background events.
In this section, additional uncertainties that are specific to the
background model are discussed, separately for \ljets, dilepton, and
\alljets events.

\paragraph{Lepton+Jets Channel:}
\label{systuncs.physicsmodeling.background.ljets.sec}
The two main backgrounds in the \ljets channel are leptonically
decaying \W bosons produced in association with jets (\wjets events)
and multijet events containing a wrongly identified isolated 
lepton (QCD events).
Both CDF (see e.g.\ \cite{Abulencia:2005aj,bib-CDFljetsme}) and \dzero
(\cite{bib-me}) find that the main uncertainty related to the
modeling of \wjets background comes from a
variation of the factorization scale \factorizationscalesquared
used in the generation of these events.
An additional contribution comes from the variation of the flavor
composition of the jets in \wjets
events~\cite{Abulencia:2005aj,bib-CDFljetsme}.

Both CDF and \dzero base the estimate of QCD background on data.
It is not straightforward to define a sample for this estimation
that is kinematically unbiased and does not contain a
sizeable \ttbar component.
Therefore, the QCD background estimate is replaced with \wjets events,
and the resulting difference is conservatively quoted as systematic 
error.

The CDF values quoted for the uncertainty from modeling of \wjets and
QCD events are 
0.2~GeV~\cite{bib-CDFljetsme} and 0.5~GeV~\cite{Abulencia:2005aj}, but
cannot be compared directly to the \dzero value of
$0.4\oplus0.3\;\GeV = 0.5\;\GeV$~\cite{bib-me} (see Table~\ref{systuncs.table}) 
since CDF and \dzero consider different
factorization scales, and \dzero does not vary the heavy flavor content in
\wjets events.

\paragraph{Dilepton Channel:}
\label{systuncs.physicsmodeling.background.dilepton.sec}
The main backgrounds in the dilepton channel come from diboson 
(\ww, \wz) or Drell-Yan production
($\Z/\gamma^*\to\epem,\mupmum,\tauptaum$)
in association with jets, and from events with a
mis-identified electron (e.g., $\W(\to\mu\nu)+3\,{\rm jets}$ with a
jet faking an electron).
To estimate the systematic uncertainty, the number of expected events
from each source is varied independently within its error, and the
resulting top quark mass shifts are added in 
quadrature~\cite{bib-Dzerodileptonneutrinoweighting,bib-Dzerodileptonmatrixweighting,bib-CDFdileptonme,bib-CDFdileptonmewithbtagging}.
The systematic uncertainties assigned in individual top quark mass measurements
vary between $\pm 0.3\,\GeV$ and 
$\pm 1.0\,\GeV$ (and even $\pandm{0.3}{1.9}\,\GeV$).
In addition, systematic variations of the background shapes yield
another uncertainty of up to $\pm 1.0\,\GeV$.
Given the fact that the background contribution to the dilepton 
event samples is small while the statistical uncertainty is still large, 
it seems that some of these preliminary estimates are very
conservative and that a much smaller uncertainty will be quoted in the
future.

\paragraph{All-Jets Channel:}
\label{systuncs.physicsmodeling.background.allhadronic.sec}
In this channel, the dominant background is from QCD multijet
production.
In the CDF analysis~\cite{bib-CDFallhadtemplate} the background
is estimated from the data using a parametrization of the $b$-tagging
efficiency.
The overall normalization of this background estimate and
the residual signal contribution are varied and each contribute a
systematic error of $0.5\,\GeV$ on the top quark mass.
The accuracy of the background estimator is checked with signal 
depleted event samples, and no additional shape uncertainty is 
assigned.

\bigskip

The error on the world-average top quark mass that is due to 
background-specific uncertainties only amounts to 
$0.3\,\GeV$~\cite{bib-TEVEWWG}.
It will be possible to select \ljets and dilepton \ttbar
samples for top quark mass measurements with much smaller backgrounds
at the LHC~\cite{Borjanovic:2004ce,bib-CMSTDRVolII} 
because of the larger \ttbar cross section and better detector
resolution, 
so that it can be
expected that the uncertainty from background modeling
will further diminish in the future.

\subsection{Modeling of the Detector Response}
\label{systuncs.detectormodeling.sec}
For most individual measurements of the top quark mass, the dominant
error is due to uncertainties in the detector response, most notably
the jet energy measurement (see Table~\ref{systuncs.table}).
Even though these errors are only correlated between measurements of
one experiment, and despite the possibility of in situ \JES
calibration, the absolute jet energy scale uncertainty still dominates
the world average.

Because simulated events are used to calibrate the mass measurements, 
it is not the uncertainty on the absolute detector response that 
matters, but the uncertainty on the relative difference between the 
data and the simulation.
Contributions can in principle arise from any aspect of the data
related with the event selection and/or top quark mass reconstruction,
ranging from uncertainties in the modeling of an energy dependence
of event quality cuts, reconstruction or selection efficiencies,
to the calibration of the reconstruction of the final-state leptons
and jets.

\subsubsection{Jet and Charged Lepton Energy Scales}
\label{systuncs.detectormodeling.jes.sec}
In practice, by far the largest uncertainty arises from the uncertainty
on the ratio of absolute jet energy scales in the data and simulation.
Therefore, in situ calibration techniques are applied in the \ljets
channel as described in Sections~\ref{templatemeasurements.sec},
\ref{memeasurements.sec}, and~\ref{idmeasurements.sec}.
This means that the uncertainty on the absolute jet energy scales
with the statistical error.
The uncertainty obtained by \dzero with
$0.4\,\ifb$ is $\MEberrjes\,\GeV$~\cite{bib-me}; the CDF
experiment quotes $2.5\,\GeV$ using
$0.68\,\ifb$~\cite{bib-CDFmtop_ljetstemplate_prel}.

Without this technique, external measurements of the jet energy 
scale as described in Section~\ref{detcalib.jets.sec} have to be 
used, and the uncertainty on the ratio between data and simulation 
propagated to the final result.
The resulting systematic error is currently between $3$ and
$5\,\GeV$ in the \ljets~\cite{bib-CDFljetsDLMmass}, 
dilepton~\cite{bib-Dzerodileptonneutrinoweighting,bib-Dzerodileptonmatrixweighting,bib-CDFdileptonme,bib-CDFdileptonmewithbtagging,bib-CDFdileptontemplateprel},
and \alljets channels~\cite{bib-CDFallhadID,bib-CDFallhadtemplate}
and is correlated between all measurements
at the same experiment.

Even with in situ calibration, only one overall jet energy 
scale factor is determined.
Any discrepancy between data and simulation other than such a global 
scale difference may lead to an additional uncertainty on the top 
quark mass, which is however much smaller than that arising from the overall
absolute calibration.
Uncertainties on residual $|\eta|$ and $\pt$
dependencies of the jet energy scale are taken from the external 
calibration and are typically estimated to be below $0.5\,\GeV$, see for example 
References~\cite{bib-me,bib-CDFljetsme} and Table~\ref{systuncs.table}.

The second-largest detector modeling uncertainty in the \ljets
channel is the uncertainty on the double ratio between the 
jet energy scales for \bquark-quark and light jets in the data and 
simulation.
This error is due to differences between the calorimeter response to 
electromagnetic and hadronic showers and the uncertainty on the 
electromagnetic/hadronic energy ratio in $\bquark$-quark jets.
The CDF collaboration has evaluated it to be
$\pm0.6\,\GeV$~\cite{Abulencia:2005aj,bib-CDFljetsme}, and
the \dzero experiment has obtained
$\pandm{0.6}{1.4}\,\GeV$~\cite{bib-me}.

In comparison with the energy scale for jets, the absolute energies of
charged leptons are calibrated precisely using leptonic \Z decays.
The uncertainty has been found to be negligible at
\dzero~\cite{bib-meprivatecommunication}; the CDF experiment quotes
an uncertainty of $0.1\,\GeV$ in the dilepton 
measurements~\cite{bib-CDFdileptonme,bib-CDFdileptonmewithbtagging}.

In the future, information from the overall jet energy scale
calibration described in Section~\ref{detcalib.jets.jes.sec},
which is not 
used in measurements with in situ calibration, can be introduced as an
additional constraint to improve the world average.
The uncertainty related to the \bquark- to
light-quark jet energy scale ratio may become a limiting systematic error
in the mid-term future.
Even though it is related to detector response, it is correlated
between all measurements.
The measurement of $\Z\to\bbbar$ events has proven very
difficult at the Tevatron, and event samples with 
a \bquark jet balanced by a photon or \Z decay are limited in statistics.
Ideas for the an in situ calibration of this energy scale ratio would
therefore be very helpful; otherwise radical techniques like a
top quark mass measurement based on secondary vertex decay length
information~\cite{bib-CDFLxy} or leptonic $\jpsi$ decays in top quark
events~\cite{Kharchilava:1999yj} can be employed using the
large-statistics samples at the LHC.

\subsubsection{Event Selection}
\label{systuncs.detectormodeling.evtsel.sec}
Uncertainties in the event selection efficiency, notably
energy-dependent effects, can lead to systematic effects on the top
quark mass.
For example, the trigger efficiency is measured in the data using 
reference triggers, and the uncertainty on the dependence on charged
lepton and jet energies is propagated to the top quark mass result.
Similarly, the $b$-tagging efficiencies are determined from the data
and varied within their uncertainties.
Recent measurements in the \ljets channel quote systematic
uncertainties of not more than a few hundred
\MeV~\cite{Abulencia:2005aj,bib-me}. 
Since the event selection efficiencies are calibrated using the data,
it can be expected that the associated uncertainty will further
diminish in the future.

\subsubsection{Multiple Interactions}
\label{systuncs.detectormodeling.pileup.sec}
Bunch crossings with more than one hard interaction may lead to events
where the \ttbar decay products cannot be easily identified, or with 
additional energy contributions to the jets from the \ttbar final 
state.
As long as such events are modeled accurately, these effects can be
taken into account in the calibration.
However, uncertainties on the instantaneous luminosity and the properties of 
the additional hard interaction lead to a systematic uncertainty on
the top quark mass.
Recent CDF measurements quote a $0.05\,\GeV$~\cite{bib-CDFljetsme} 
to $0.2\,\GeV$~\cite{bib-CDFdileptonme,bib-CDFdileptonmewithbtagging,bib-CDFdileptontemplateprel} 
uncertainty.

Overlay of calorimeter energy from subsequent bunch crossings
was an issue at \dzero \runi but is no longer significant due to a
change in readout electronics~\cite{bib-me}.

\subsection{Uncertainties Related to the Measurement Method}
\label{systuncs.method.sec}
Since the calibration of a measurement method is based on 
simulated events, limited Monte Carlo statistics gives rise to a
systematic uncertainty on the top quark mass.
There may be other systematic errors inherent to a specific method.
An example is the \dzero Matrix Element measurement in the \ljets channel
where the calibration depends slightly on the \ttbar fraction in the
selected event sample; the uncertainty on this fraction then leads
to a systematic error on the top quark mass.
Uncertainties of this type are normally uncorrelated between
individual measurements, and are not dominant.

\subsection{Summary}
\label{systuncs.summary.sec}
The precision of the world-average top quark mass is already limited
by systematic errors~\cite{bib-TEVEWWG}.
Currently, the single largest uncertainty is due to the absolute jet energy
scale.
With in situ calibration using the hadronic \W mass, this error will
be reduced with larger data sets.
Until the startup of the LHC, physics modeling uncertainties (which
are correlated between all measurements) will become dominant.
In particular, work on the consistent evaluation (and reduction) of
the uncertainties due to isr/fsr modeling, \bquark-quark
fragmentation, and the \bquark/light jet energy scale ratio is very
desirable in the near future.

\clearpage
\section{Results, their Interpretation, and Future Prospects}
\label{resinterp.sec}
\begin{center}
\begin{tabular}{p{15cm}}
{\it This section gives an overview of the most recent measurements of
  the top quark mass and how they contribute to the world average.
  The current knowledge of the top quark mass is then set into
  perspective by discussing its implications for the Standard Model of
  particle physics, notably for consistency tests and indirect
  constraints on the mass of the Higgs boson.  Finally, the prospects
  for future improvements of top quark mass measurements are outlined.}
\end{tabular}
\end{center}

The Tevatron experiments have employed various methods to measure the
top quark mass, as described in 
Sections~\ref{methods.sec}-\ref{massfit.sec}.
The most relevant individual measurements are combined by the Tevatron
Electroweak Working Group, taking correlations into account as already
outlined in Section~\ref{systuncs.sec}.
In Section~\ref{resinterp.measurements.sec}
the individual measurement results are summarized, the
combination procedure is described, and its current results are
presented.
Section~\ref{resinterp.interp.sec} gives an interpretation of these
results in the framework of the Standard Model and also discusses
implications for the Minimal Supersymmetric Standard Model (MSSM).
Finally, an overview of improvements to be expected with the startup of the 
LHC and a future linear \epem collider (ILC) is given in
Section~\ref{resinterp.future.sec}.

\subsection{Measurement Results and Their Combination}
\label{resinterp.measurements.sec}
A large number of measurements of the top quark mass has been performed
to date at the Tevatron, using data in 
the \ljets, dilepton, and \alljets decay channels and applying a
wide variety of measurement techniques~\cite{bib-CDFupdates,bib-Dzeroupdates}.
Table~\ref{resultsoverview.table} summarizes the results.

\begin{table}[htbp]
\begin{center}
\begin{tabular}{c @{\ \ } c c c c@{\,}c@{\,}c @{\ \ } c @{\ \ } c}
\hline
\hline
  \begin{tabular}{@{}c@{}}Decay\\ Channel\end{tabular}
&
  \begin{tabular}{@{}c@{}}Measurement\\ Technique\\ (Section where described)\end{tabular} 
&
  \begin{tabular}{@{}c@{}}Exp./\\ Run\end{tabular}
&
  \begin{tabular}{@{}c@{}}Int.\\ Lumi.\\ $[\ifb]$ \end{tabular} 
&
  \multicolumn{3}{@{}c@{}}{Result [GeV]}
&
  Ref.
&
  Weight
\\
\hline
\phantom{.}\vspace{-2ex}\\
    \ljets
  & T, mass reco. (\ref{templatemeasurements.kinrec.sec})
  & \begin{tabular}{@{}c@{}} CDF I\phantom{I}\\
                             CDF II
    \end{tabular}
  & \begin{tabular}{@{}c@{}} $0.106$\\
                             $0.68\enspace$
    \end{tabular}
  & \begin{tabular}{@{}c@{}} $176.1$\\
                             $173.4$
    \end{tabular}
  & \begin{tabular}{@{}c@{}} $\enspace\pm 5.1$\\
                             $\enspace\pm 2.5$
    \end{tabular}
  & \begin{tabular}{@{}c@{}} $\enspace\pm 5.3$\\
                             $\enspace\pm 1.3$
    \end{tabular}
  & \begin{tabular}{@{}c@{}} \cite{Affolder:2000vy}\\
                             \cite{bib-CDFmtop_ljetstemplate_prel}
    \end{tabular}
  & \begin{tabular}{@{}c@{}} $\enspace-3.1\%\phantom{^\dag}$\\
                             \phantom{.}
    \end{tabular}
\\
    \ljets
  & T, multivariate
  & CDF II
  & $0.162$
  & $179.6$ & $\enspace\pandmenspaceul{6.4}{6.3}$ & $\enspace\pm 6.8$
  & \cite{CDFmultivariatetemplate}
\\ 
    \ljets
  & T, decay length (\ref{templatemeasurements.nocalo.sec})
  & CDF II
  & $0.695$
  & $180.7$ & $\enspace\pandm{15.5}{13.4}$ & $\enspace\pm 8.6$
  & \cite{bib-CDFLxy}
  & $\enspace+0.9\%^\dag$
\\ 
    \ljets
  & ME, topological (\ref{memeasurements.sec})
  & \begin{tabular}{@{}c@{}} \dzero I\phantom{I}\\
                             \dzero II
    \end{tabular}
  & \begin{tabular}{@{}c@{}} $0.125$\\
                             $0.4\enspace\enspace$
    \end{tabular}
  & \begin{tabular}{@{}c@{}} $180.1$\\
                             $169.2$
    \end{tabular}
  & \begin{tabular}{@{}c@{}} $\enspace\pm 3.6$\\
                             $\enspace\pandmenspaceul{5.0}{7.4}$
    \end{tabular}
  & \begin{tabular}{@{}c@{}} $\enspace\pm 3.9$\\
                             $\enspace\pandmenspaceul{1.5}{1.4}$
    \end{tabular}
  & \begin{tabular}{@{}c@{}} \cite{bib-nature}\\
                             \cite{bib-me}
    \end{tabular}
  & \begin{tabular}{@{}c@{}} $\enspace+8.0\%\phantom{^\dag}$\\
                             \phantom{.}
    \end{tabular}
\\ 
    \ljets
  & ME, \bquark tagging (\ref{memeasurements.sec})
  & \begin{tabular}{@{}c@{}} CDF II\\ 
                             \dzero II
    \end{tabular}
  & \begin{tabular}{@{}c@{}} $0.94\enspace$\\
                             $0.4\enspace\enspace$
    \end{tabular}
  & \begin{tabular}{@{}c@{}} $170.9$\\
                             $170.3$
    \end{tabular}
  & \begin{tabular}{@{}c@{}} $\enspace\pm 2.2$\\ 
                             $\enspace\pandmenspaceul{4.1}{4.5}$
    \end{tabular}
  & \begin{tabular}{@{}c@{}} $\enspace\pm 1.4$\\
                             $\enspace\pandmenspaceul{1.2}{1.8}$
    \end{tabular}
  & \begin{tabular}{@{}c@{}} \cite{bib-CDFljetsme}\\
                             \cite{bib-me}
    \end{tabular}
  & \begin{tabular}{@{}c@{}} $+61.7\%\phantom{^\dag}$\\
                             $+18.9\%\phantom{^\dag}$
    \end{tabular}
\\ 
    \ljets
  & DL (\ref{memeasurements.sec})
  & CDF II
  & $0.318$
  & $173.2$ & $\enspace\pandmenspaceul{2.6}{2.4}$ & $\enspace\pm 3.2$
  & \cite{bib-CDFljetsDLMmass}
\\ 
    \ljets
  & ID (\ref{idmeasurements.sec})
  & \dzero II
  & $0.4\enspace\enspace$
  & $173.7$ & $\enspace\pm 4.4$ & $\enspace\pandmenspaceul{2.1}{2.0}$
  & \cite{bib-ID}
\\
\vspace{-2.2ex}\\
\hline
\phantom{.}\vspace{-2ex}\\
    dilepton
  & T, $\pt(\nu)$ (\ref{templatemeasurements.dilepton.neutrinoweighting.sec})
  & \begin{tabular}{@{}c@{}} CDF I\phantom{I}\\
                             CDF II\\
                             \dzero II
    \end{tabular}
  & \begin{tabular}{@{}c@{}} $0.109$\\
                             $0.359$\\
                             $0.835$\\
    \end{tabular}
  & \begin{tabular}{@{}c@{}} $167.4$\\
                             $170.7$\\
                             $171.6$\\
    \end{tabular}
  & \begin{tabular}{@{}c@{}} $\pm 10.3$\\
                             $\enspace\pandmenspaceul{6.9}{6.5}$\\
                             $\enspace\pm 7.9$\\
    \end{tabular}
  & \begin{tabular}{@{}c@{}} $\enspace\pm 4.8$\\
                             $\enspace\pm 4.6$\\
                             $\enspace\pandmenspaceul{5.1}{4.0}$\\
    \end{tabular}
  & \begin{tabular}{@{}c@{}} \cite{Abe:1998bf}\\
                             \cite{bib-CDFcombineddileptontemplatepaper}\\
                             \cite{bib-Dzerodileptonneutrinoweighting}\\
    \end{tabular}
  & \begin{tabular}{@{}c@{}} $\enspace-0.6\%\phantom{^\dag}$\\
                             \phantom{.}\\
                             \phantom{.}\\
    \end{tabular}
\\
    dilepton
  & T, $\phi(\nu)$ (\ref{templatemeasurements.dilepton.neutrinophiweighting.sec})
  & CDF II
  & $0.34\enspace$
  & $169.7$ & $\enspace\pandmenspaceul{8.9}{9.0}$ & $\enspace\pm 4.0$
  & \cite{bib-CDFcombineddileptontemplatepaper}
\\
    dilepton
  & T, $\pz(\ttbar)$ (\ref{templatemeasurements.dilepton.fullkinematicanalysis.sec})
  & CDF II
  & $1.02\enspace$
  & $168.1$ & $\enspace\pandmenspaceul{5.6}{5.5}$ & $\enspace\pm 4.0$
  & \cite{bib-CDFdileptontemplateprel}
\\
    dilepton
  & T, matrix weighting (\ref{templatemeasurements.dilepton.matrixweighting.sec})
  & \dzero II
  & $0.835$
  & $177.7$ & $\enspace\pm 8.8$ & $\enspace\pandmenspaceul{3.7}{4.5}$
  & \cite{bib-Dzerodileptonmatrixweighting}
  & \enspace$-1.1\%^\dag$
\\
    dilepton
  & \begin{tabular}{@{}c@{}}
      T, $\pt(\nu)$ (\ref{templatemeasurements.dilepton.neutrinoweighting.sec}) and\\
      T, matrix weighting (\ref{templatemeasurements.dilepton.matrixweighting.sec})
    \end{tabular}
  & \dzero I\phantom{I}
  & $0.125$
  & $168.4$ & $\pm 12.3$ & $\enspace\pm 3.6$
  & \cite{Abbott:1998dn}
  & \enspace$+0.6\%\phantom{^\dag}$
\\
    dilepton
  & ME, topological (\ref{memeasurements.sec})
  & CDF II
  & $1.03\enspace$
  & $164.5$ & $\enspace\pm 3.9$ & $\enspace\pm 3.9$
  & \cite{bib-CDFdileptonme}
  & $\enspace+4.8\%\phantom{^\dag}$
\\
    dilepton
  & ME, \bquark tagging (\ref{memeasurements.sec})
  & CDF II
  & $0.955$
  & $167.3$ & $\enspace\pm 4.6$ & $\enspace\pm 3.8$
  & \cite{bib-CDFdileptonmewithbtagging}
\\
    dilepton
  & DL (\ref{memeasurements.sec})
  & CDF II
  & $0.34\enspace$
  & $166.6$ & $\enspace\pandmenspaceul{7.3}{6.7}$ & $\enspace\pm 3.2$
  & \cite{bib-CDFdileptonDLMmass}
\\
\vspace{-2.2ex}\\
\hline
\phantom{.}\vspace{-2ex}\\
    \alljets
  & T (\ref{templatemeasurements.allhad.sec})
  & \begin{tabular}{@{}c@{}} CDF I\phantom{I}\\
                             CDF II
    \end{tabular}
  & \begin{tabular}{@{}c@{}} $0.109$\\
                             $1.02\enspace$
    \end{tabular}
  & \begin{tabular}{@{}c@{}} $186\phantom{.0}$\\
                             $174.0$
    \end{tabular}
  & \begin{tabular}{@{}c@{}} $\pm 10\phantom{.0}$\\
                             $\enspace\pm 2.2$
    \end{tabular}
  & \begin{tabular}{@{}c@{}} $\pm 12\phantom{.0}$\\
                             $\enspace\pm 4.8$
    \end{tabular}
  & \begin{tabular}{@{}c@{}} \cite{Abe:1997rh}\\
                             \cite{bib-CDFallhadtemplate}
    \end{tabular}
  & \begin{tabular}{@{}c@{}} $\enspace-0.3\%\phantom{^\dag}$\\
                             $+10.3\%\phantom{^\dag}$
    \end{tabular}
\\
    \alljets
  & ID (\ref{idmeasurements.sec})
  & CDF II
  & $0.31\enspace$
  & $177.1$ & $\enspace\pm 4.9$ & $\enspace\pm 4.7$
  & \cite{bib-CDFallhadID}
\\
\vspace{-2.2ex}\\
\hline
\phantom{.}\vspace{-2ex}\\
    $\etmiss+{\rm jets}$
  & T 
  & CDF II
  & $0.31\enspace$
  & $172.3$ & $\enspace\pandmenspacel{10.8}{9.6}$ & $\pm 10.8$
  & \cite{bib-CDFmetjetstemplate}
\\
\vspace{-2ex}\\
\hline
\hline
\end{tabular}
\caption{\captionfont\label{resultsoverview.table}Overview of top quark mass
  measurements.
  Analyses are grouped according to the \ttbar decay channel
  listed in the leftmost column.  
  The symbol ``$\etmiss+{\rm jets}$''
  denotes a selection based on \etmiss and jets
  only, yielding a sample enriched in events
  with a $\W\to\tau\nu$ decay.
  The list is further ordered according to the analysis technique 
  (T: template based; ME: Matrix Element;
  DL: Dynamical Likelihood; ID: Ideogram), given 
  in the second column together with the section describing it.
  All recent CDF and \dzero analyses of \runii data and
  those \runi measurements that are included in the world
  average~\cite{bib-TEVEWWG} are listed.
  The experiment and integrated luminosity are given, and the
  top quark mass results are quoted with their statistical
  and systematic uncertainties.
  For measurements using in situ calibration, the uncertainty from the overall
  jet energy scale is included in the first quoted error
  as it will scale with statistics in future
  updates.
  The rightmost column lists the weight given to measurements in the
  world average value.
  For the measurements marked with a $^\dag$ sign, an earlier result
  is used in the combination, while the most recent value is given in 
  the table.}
\end{center}
\end{table}

No significant
deviations are apparent between the top quark masses measured in 
individual decay channels, with different measurement techniques, 
by the two experiments, or at the two Tevatron center-of-mass energies of
$1.8\,\TeV$ (\runi) or $1.96\,\TeV$ (\runii).
However, many of the individual results are 
systematically and also statistically correlated.
To quantify these statements, a consistent combination of results
is performed by the Tevatron Electroweak Working
Group~\cite{bib-TEVEWWG} based on 
the best linear unbiased estimator
(BLUE)~\cite{bib-BLUE1,bib-BLUE2}.
The procedure takes systematic correlations into account by 
treating individual systematic uncertainties as uncorrelated or 100\%
correlated between measurements as discussed in
Section~\ref{systuncs.sec}.
More detailed studies are in general needed to evaluate the
statistical correlation between measurements using the same dataset
and decay channel.
As an example, the statistical correlation between the top quark mass
values determined at \dzero in the topological Matrix Element
analysis and the Ideogram measurement (which uses \bquark tagging) has
been found to be only
$+40\%$~\cite{bib-ID,bib-meprivatecommunication}.
Since Tevatron \runii analyses are still evolving, such a study is not yet
available in many cases.
Therefore, a combination of \runi values and only the most precise 
\runii measurements in each channel is performed.
The correlations between these measurements are
close to zero unless a correlation arises via common jet energy scale
uncertainties; in that case correlation coefficients are typically 
of the order
of $30\%$, the largest being $56\%$ between the CDF \runi \ljets
and \runii all-jets measurements.

The combination yields average top quark masses in the
individual channels of
\begin{eqnarray}
    \nonumber
    \mtop({\text\ljets})
  & =
  & 171.3 \pm 2.2 \,\GeV
    \, ,
  \\
    \mtop({\rm dilepton})
  & =
  & 167.0 \pm 4.3 \,\GeV
    \, ,\ {\rm and}
  \\
    \nonumber
    \mtop({\text\alljets})
  & =
  & 173.4 \pm 4.3 \,\GeV
    \, ,
\end{eqnarray}
where the uncertainties include both statistical and systematic
errors.
The correlations ${\cal C}$ and resulting $\chi^2$ consistency values
(for one degree of freedom) have been determined as
\begin{equation}
  \begin{array}{rr@{\ }l}
    {\cal C}({\text\ljets},\ {\rm dilepton}) = +37\%
    \, ,
  &
    \chi^2({\text\ljets},\ {\rm dilepton}) = 1.2
    \, ,\enspace
  \\
    {\cal C}({\text\ljets},\ {\text\alljets}) = +29\%
    \, ,
  &
    \chi^2({\text\ljets},\ {\text\alljets}) = 0.24
    \, ,
  & {\rm and}
  \\
    {\cal C}({\rm dilepton},\ {\text\alljets}) = +46\%
    \, ,
  &
    \chi^2({\rm dilepton},\ {\text\alljets}) = 2.1
    \, .\enspace
  \end{array}
\end{equation}

Since the values for all three channels are consistent with each
other, one overall combined top quark mass value is computed.
It is found to be
\begin{equation}
  \label{worldaveragevalue.eqn}
  \mtop = 171.4 \pm 2.1 \,\GeV
  \, .
\end{equation}
The weights with which the individual top quark mass measurements 
contribute to this average are indicated in the
last column of Table~\ref{resultsoverview.table}.
Individual measurements may be assigned a negative weight in case
of large correlations; as long as the weight is non-zero, the
measurement still improves the average.
This effect is explained very clearly and intuitively
in~\cite{bib-BLUE1}.
The $\chi^2$ for the average
is $10.6$ for $10$ degrees of freedom, and the largest
single pull of any measurement that enters the combination 
is $1.8$, indicating good consistency of all $11$ measurements.

The \ljets, dilepton, and \alljets channels contribute with
weights of $86.4\%$, $3.7\%$, and $10.0\%$ to the world average, 
respectively.
These weights are indicative of the experimental situation at the
Tevatron, with limited statistics in the dilepton and large
backgrounds in the \alljets channel.
With increasing data sets at Tevatron \runii, the relative importance
of the dilepton channel may increase.
Precise measurements of the top quark mass in the \alljets channel
have only become possible after detailed studies of the background and
its evaluation from the data.
When in situ calibration techniques are applied in this channel, too,
its weight may further increase.

\subsection{Interpretation of the Top Quark Mass Measurement}
\label{resinterp.interp.sec}
As mentioned in Section~\ref{resinterp.measurements.sec}, the
top quark masses obtained in the \ljets, dilepton, and \alljets 
channels are consistent with each other.
Moreover, the cross section for production of \ttbar events at the 
Tevatron is consistent with the (Standard Model) expectation computed for
the combined top quark mass value given in
Equation~(\ref{worldaveragevalue.eqn}).
No average value of all Tevatron \runii measurements of the \ttbar
cross section exists yet; however, the CDF experiment has performed a
combination of CDF measurements~\cite{bib-cdfxscombination}.
The \dzero measurements can be found
in~\cite{bib-D0ljetsbtagxspaper,bib-dzeroxscombination,bib-Dzeroupdates}. 
Figure~\ref{xsmassconsistency.fig} shows the combined CDF result 
and the recent \dzero measurement from
Reference~\cite{bib-D0ljetsbtagxspaper} 
together with the dependencies of the \ttbar cross 
section measurements on the value of
the top quark mass.
Also shown are calculations of the \ttbar cross section 
in next-to-leading order~\cite{bib-xsprediction} as a function of the top
quark mass.
The measured cross sections agree well with the Standard Model
prediction when assuming the world-average top quark
mass value.

\begin{figure}
\begin{center}
\includegraphics[width=0.633\textwidth]{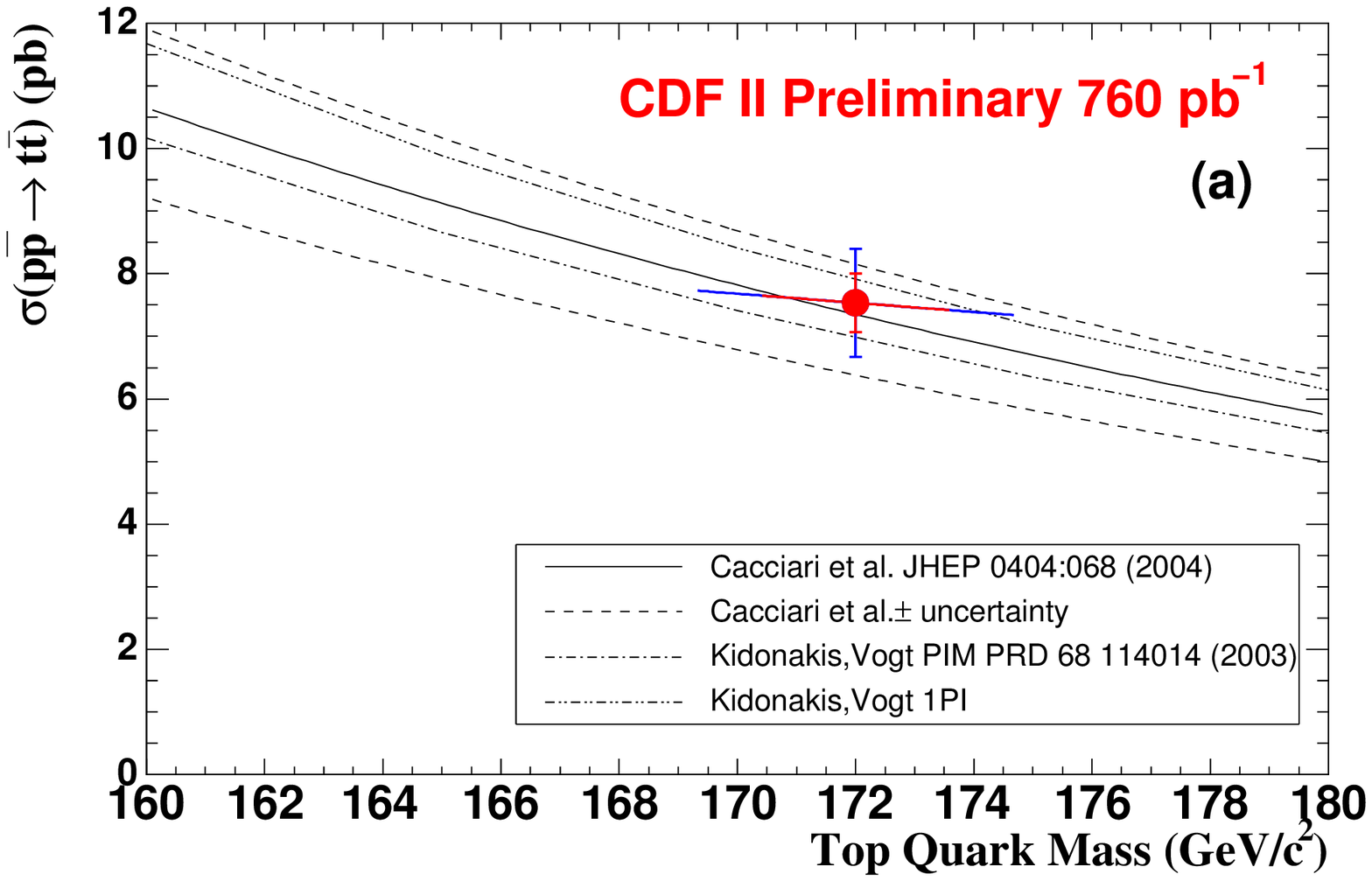}
\includegraphics[width=0.675\textwidth]{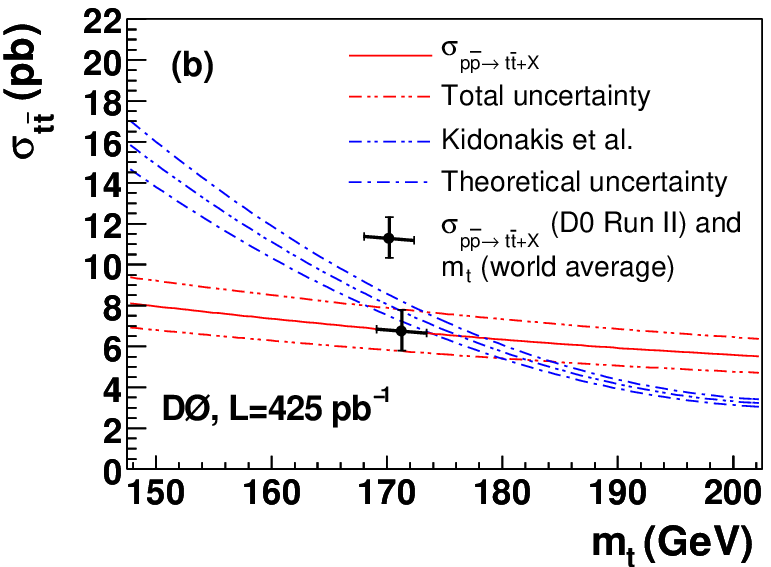}
\caption{\captionfont\label{xsmassconsistency.fig}The average value of the \ttbar
  production cross section in \ppbar collisions at $\sqrt{s}=1.96\,\TeV$
  as measured by the CDF experiment~\cite{bib-cdfxscombination} is
  shown in (a).
  The vertical error bar indicates the \ttbar cross section and its
  uncertainty evaluated at the CDF average value of the top quark
  mass.
  The dependence of the cross section measurement on the assumed top
  quark mass value is shown by the slope of the other error bar (its 
  projection onto the horizontal axis corresponds to the uncertainty
  on the top quark mass using CDF measurements only).
  Also shown are NLO calculations of the Standard Model \ttbar cross
  section, including threshold
  corrections from soft gluons~\cite{Cacciari:2003fi,Kidonakis:2003qe}. 
  The uncertainty on these predictions is shown, too; it is dominated
  by the uncertainty on the gluon PDF.
  Consequently, no PDF uncertainty is included in the experimental
  result.
  Similarly, the result of a recent \dzero
  measurement~\cite{bib-D0ljetsbtagxspaper} is shown in (b) by the red
  lines, with the theoretical prediction from~\cite{Cacciari:2003fi}
  overlaid.
  The combined information from the \dzero \ttbar cross section
  measurement and the world-average value of the top quark mass from is 
  indicated by the crossed error bars.
  Note the different scales on the horizontal axes of the two plots.
}
\end{center}
\end{figure}

Also other top quark measurements like the relative cross sections for
the various decay channels or differential cross sections agree well with
Standard Model predictions~\cite{bib-CDFupdates,bib-Dzeroupdates}.
Since there is no sign of effects beyond the Standard Model, it is
appropriate to use the world-average top quark mass value in a
consistency check of the Standard Model and, if consistency can be
established, to extract information on Standard Model parameters.
The interpretation within the Standard Model (SM) is discussed in 
Section~\ref{resinterp.interp.sm.sec}.
Analogously, the measurements can of course also be used to constrain the parameters
of any other model that describes them.
Particular attention has been devoted to supersymmetric models.
The interpretation within the Minimal Supersymmetric Standard Model 
(MSSM) and the differences between MSSM and SM predictions are 
described in Section~\ref{resinterp.interp.mssm.sec}.
While it is currently not yet possible to distinguish between the
SM and MSSM based on indirect precision measurements, further
improvements of these measurements may make this
possible and thus provide information e.g.\ to help interpret
potential future signals of new physics.

\subsubsection{Interpretation within the Standard Model}
\label{resinterp.interp.sm.sec}
An overall fit of Standard Model parameters is performed by the 
LEP Electroweak Working Group~\cite{bib-LEPEWWG}.
The general conclusion is that the Standard Model describes the 
measurements well and that there is no significant evidence for
phenomena beyond the Standard Model.

\begin{figure}
\begin{center}
\begin{tabular}{ll}
\hspace{45pt}{\bf (a)} & \hspace{45pt}{\bf (b)}\vspace{-27pt}\\
\includegraphics[width=0.48\textwidth]{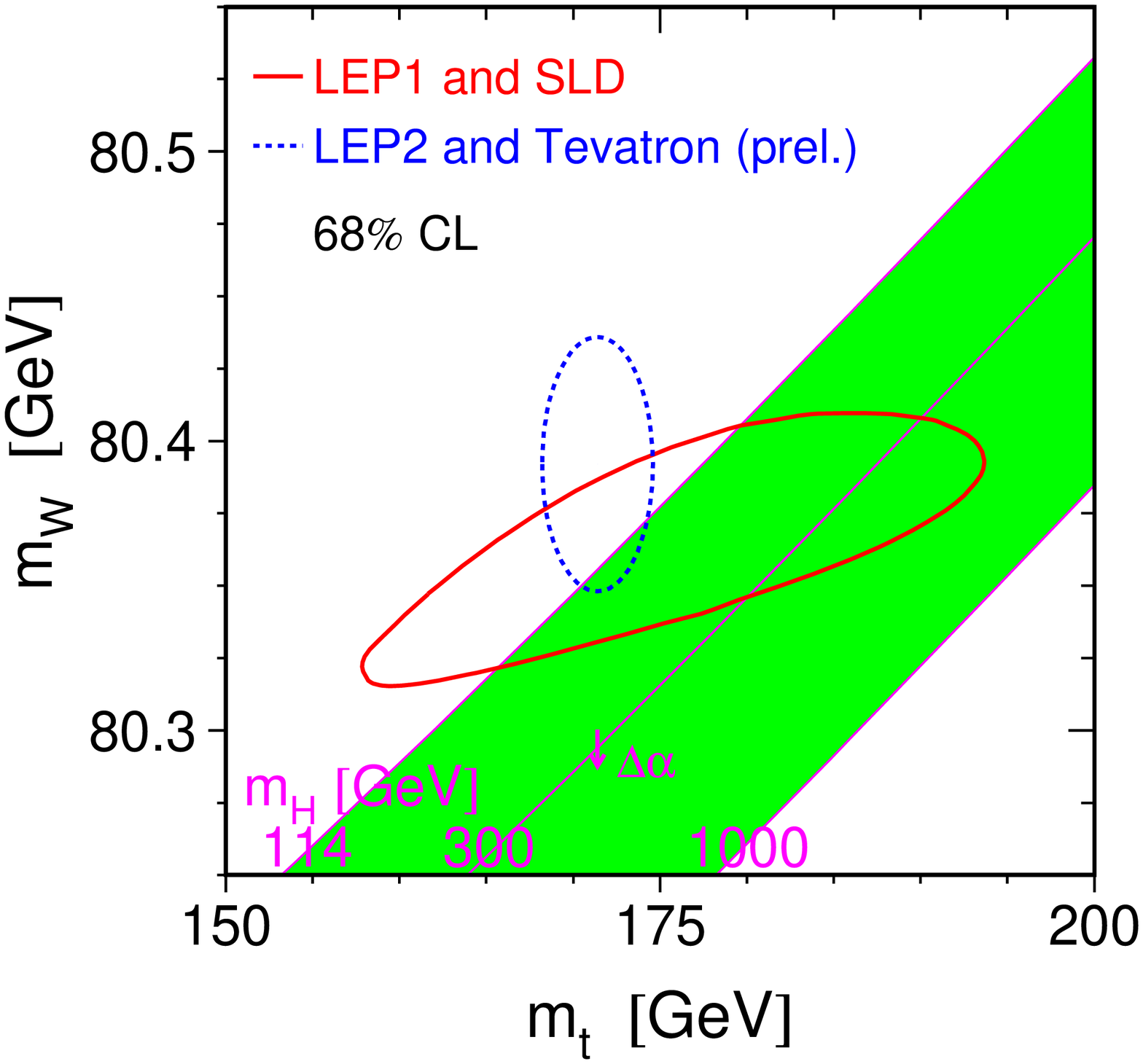}
&
\includegraphics[width=0.48\textwidth]{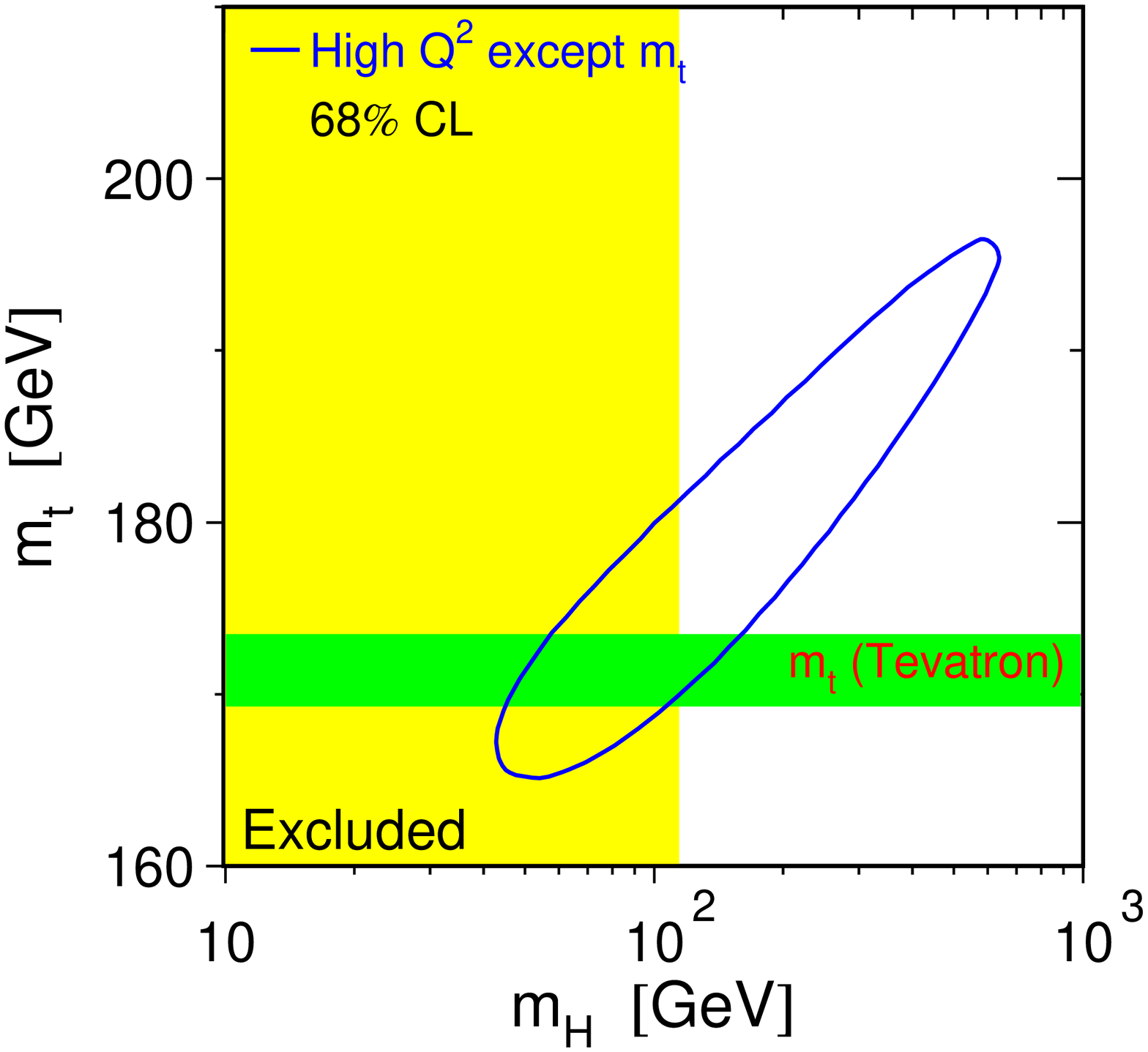}
\vspace{-20pt}\\
\hspace{45pt}{\bf (c)} & \hspace{45pt}{\bf (d)}\vspace{-27pt}\\
\includegraphics[width=0.48\textwidth]{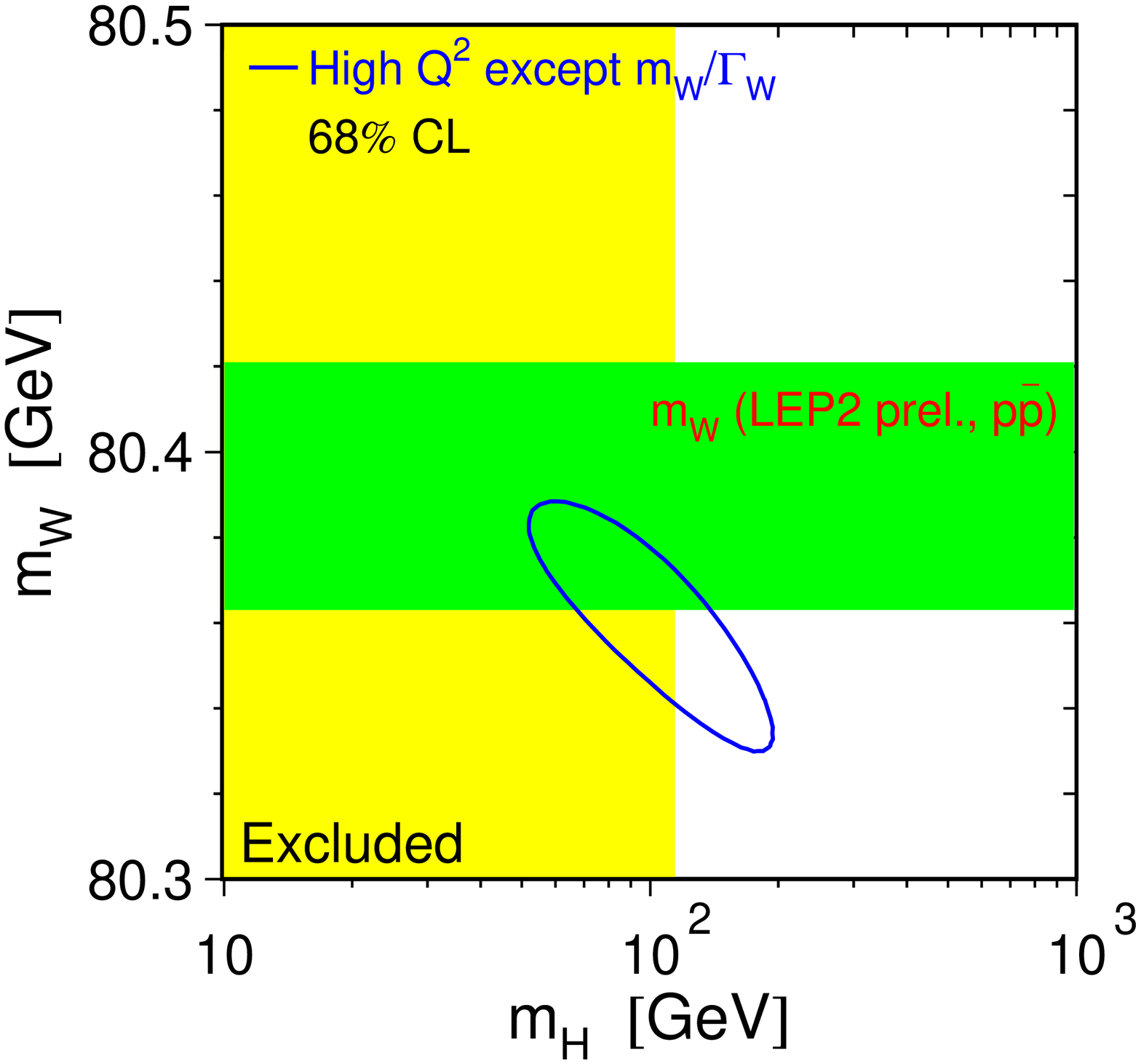}
&
\includegraphics[width=0.48\textwidth]{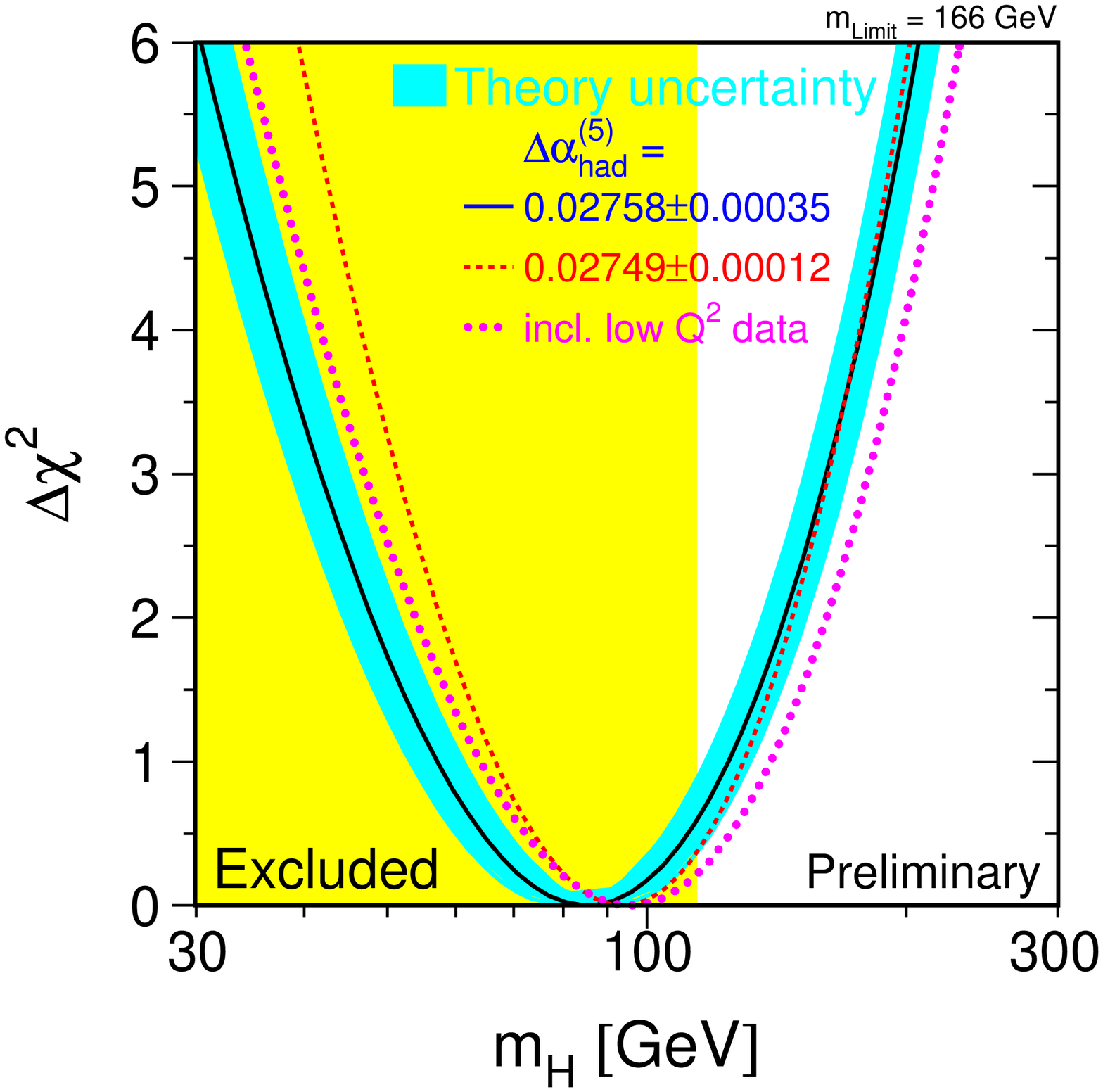}
\end{tabular}
\vspace{-20pt}
\caption{\captionfont\label{mWmtmH.fig}A comparison of direct measurements
  and indirect constraints within the Standard Model on the top
  and \W masses, as well as indirect constraints on the Standard
  Model Higgs mass~\cite{bib-LEPEWWG}.
  (a) Direct measurements (blue dashed contour) of and indirect
  constraints (red solid contour) on \mtop and \mW in the \mtop-\mW
  plane, together with the Standard Model prediction of the relation
  between \mtop and \mW for various assumed Higgs masses.
  (b) Direct measurement of \mtop (green band) and indirect
  constraints on \mtop and \mH, excluding the direct \mtop measurement
  (blue contour).
  (c) Direct measurement of \mW (green band) and indirect
  constraints on \mW and \mH, excluding the direct \mW measurement
  (blue contour).
  (d) Indirect constraint on \mH: $\Delta\chi^2$ with respect to the 
  best fit as a function of assumed Standard Model Higgs boson mass.
  The light blue band indicates the uncertainty from higher-order
  corrections not included in the calculation.
  Also shown are fits including the NuTeV \mW result (pink dotted curve)
  or based on a value of $\alpha(\mZ)$ obtained with additional
  theoretical input (red dashed curve).
  In (b), (c), and (d) the yellow area shows the region
  of Standard Model Higgs masses excluded by direct searches.}
\end{center}
\end{figure}

Using the Standard Model relations, it is possible to infer
information even on those parameters that have not (yet) been directly
measured.
Of particular interest is the constraint on the mass of the
Higgs boson.
As outlined in Section~\ref{theory.sec}, within the Standard Model the 
mass of the \W boson depends quadratically on the top quark mass and 
logarithmically on the mass of the Higgs boson.
This dependence is visualized in Figure~\ref{mWmtmH.fig}.
Figure~\ref{mWmtmH.fig}(a) shows the agreement between direct measurements of the 
\W and top quark masses from LEP2 and the Tevatron, shown in blue, 
and indirect constraints that are valid within the Standard Model
(red contour).
Also shown is the Standard Model relation between \mW and \mtop for
various assumed values of the Higgs mass; the green band covers the 
range $114\,\GeV<\mH<1000\,\GeV$.
The lower value of $\mH=114\,\GeV$ corresponds to the direct exclusion
limit from LEP searches.

Figure~\ref{mWmtmH.fig}(b) shows how the top quark mass
measurement contributes to the indirect constraint on the Higgs mass. 
In the \mH-\mtop plane, the blue contour depicts the information on 
these two parameters obtained from the Standard Model fit, where the
direct \mtop measurement is not used as input.
The projection of the blue contour onto the vertical axis thus
corresponds to the indirect constraint on the top quark mass within
the Standard Model of $\mtop=178\,\pandmenspacel{12}{9}\,\GeV$ (note 
that a projection of the 
68\%~C.L.\ contour from two dimensions to one does not correspond to
one-dimensional 68\% confidence limits).
The green band corresponds to the direct top quark mass measurement,
which is in good agreement with the indirect prediction.
The band visualizes how this information, given the \W mass and other
measurements, excludes large values of the Higgs mass within the
Standard Model.
Further improvements of the precision of the top quark mass
measurement will improve the indirect constraint on the Higgs
mass, but are unlikely to push this constraint into the region of 
mass values that has already been excluded at LEP, shown in yellow.

Similarly, the information on the Higgs mass obtained from the \W mass
measurement is shown in Figure~\ref{mWmtmH.fig}(c).
The direct measurements of the \W mass, shown as the green band,
are in agreement with the blue contour showing the indirect constraints.
The contour from all
measurements but the \W mass does not extend to high \mH values in
this plot 
since the top quark mass information is already included.
With a significant improvement of the \W mass uncertainty 
the region of Standard Model self-consistency might be
significantly reduced even before direct
Higgs searches become sensitive beyond the current limit.

All indirect information on the Higgs boson mass is summarized in 
Figure~\ref{mWmtmH.fig}(d), where the black curve shows the
$\Delta\chi^2$ within the Standard Model as a function of assumed 
Higgs mass relative to the minimum value.
The light blue band around it shows an estimate of the uncertainty
from higher-order corrections that were not included in the calculation.
Taking the information from this curve and including these theoretical
uncertainties, the one-sided 95\%~C.L.\ upper limit on the Standard Model
Higgs mass is $166\,\GeV$.
When the lower limit from direct searches is included, the upper limit
shifts to $199\,\GeV$.

In summary, the Standard Model yields a good description of
experimental data; for example the top quark mass measurement is in
good agreement with indirect constraints valid within the Standard
Model. 
The top quark mass measurement is an important ingredient to 
fits in which indirect information on the mass of the Standard Model
Higgs boson can be obtained.
With the precision of the top quark mass value
achieved with the techniques described in this report 
it is possible to place stringent upper bounds on the mass of the
Higgs boson within the Standard Model.

\subsubsection{Interpretation within the Minimal Supersymmetric Standard Model}
\label{resinterp.interp.mssm.sec}
Even though there is no compelling experimental evidence of physics
effects beyond the Standard Model from collider experiments, 
it is instructive to interpret 
precision electroweak measurements also in extended models.
As the top quark contributes via loop diagrams to the predictions for 
electroweak parameters, it is mandatory to know these contributions 
(and therefore the top quark mass) precisely to pin down any potential effects from 
additional, yet unknown, particles.
In particular, a detailed study has been performed that compares
the predictions of the Minimum Supersymmetric Standard
Model (MSSM)~\cite{bib-mssm} with those of the Standard Model (SM) in
view of the 
precision measurements of the top quark and \W boson 
masses~\cite{bib-SMvsMSSM}.
This study is summarized here.

After calculating contributions from loop diagrams involving
supersymmetric particles, it is possible to compare the  
predictions of the SM and MSSM with each other and with the
experimental data, as shown in Figure~\ref{heinemeyerweiglein.fig}.
The two model predictions lie within bands in the \mtop-\mW plane, with
only a narrow overlap region.
Apart from the fact that the
red and blue regions correspond to a variation of the Standard Model Higgs
mass between $114\,\GeV$ and only $400\,\GeV$, the information is 
equivalent to the predictions shown in Figure~\ref{mWmtmH.fig}(a) where the
upper value of the Higgs mass is set to $1000\,\GeV$ (and the axes are
scaled differently).
The green and blue areas indicate the allowed region for the
MSSM (in the region above the green area, at least one of the mass
ratios $m_{{\tilde t}_2}/m_{{\tilde t}_1}$ and 
$m_{{\tilde b}_2}/m_{{\tilde b}_1}$ is larger than $2.5$, where
in both cases the lighter mass state is denoted by the index 1).
The MSSM allowed region was obtained by varying supersymmetry
parameters independently from each other.
In the Standard Model, the blue area which is allowed in both models
corresponds to the case of a light Higgs boson within the range
allowed in the MSSM, while in the MSSM, it corresponds to the case
where all superparticles are so heavy that the theory becomes
effectively equivalent to the Standard Model. 
The current direct measurements of the top quark and \W
boson masses are shown by the blue ellipse.
The black and red contours indicate rough estimates of the precision
that can be achieved at the LHC and a future linear \epem collider (ILC),
respectively (for the LHC, uncertainties on \mtop and \mW of
$1\,\GeV$ and $15\,\MeV$ have been taken, respectively, while values
of $0.1\,\GeV$ and $7\,\MeV$ have been assumed for the ILC).
The current central measurement values have been used to place these
contours.

\begin{figure}
\begin{center}
\includegraphics[width=0.48\textwidth]{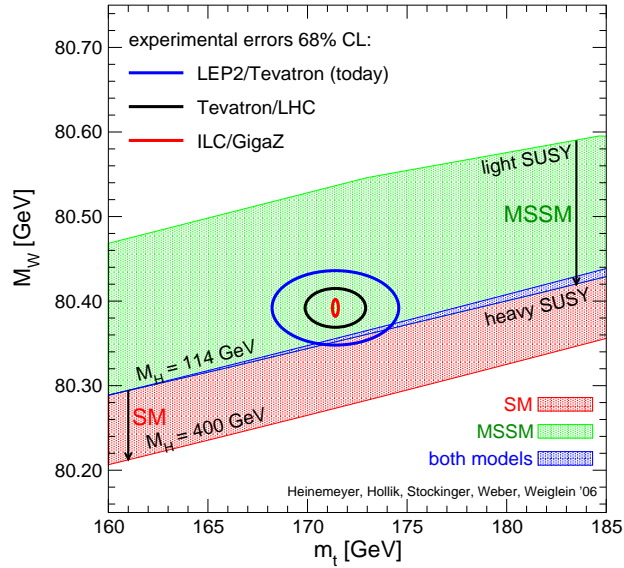}
\caption{\captionfont\label{heinemeyerweiglein.fig}Comparison of 
  predictions of the SM and MSSM with current and potential future
  direct measurements of \mtop and \mW.
  The figure is from~\cite{bib-SMvsMSSM} and has been prepared
  including calculations described in~\cite{Djouadi:1996pa}.
  The area allowed in the SM corresponds to Higgs masses within the
  range $114\,\GeV<\mH<400\,\GeV$, while the MSSM region has been
  obtained in a parameter scan.
  The blue ellipse shows the current direct measurements, while the 
  sizes of the black and red contours indicate potential future
  improvements of the uncertainties with
  data from the LHC and a future linear \epem collider, respectively
  (the central values for these contours are arbitrary).
}
\end{center}
\end{figure}

Even though the central measurement values of \mtop and \mW are not
within the SM allowed region, based on the current data 
it is not possible to
distinguish between the SM and MSSM.
Nevertheless, it is evident that with increasing precision on \mtop
and \mW, a comparison with model predictions may provide important
constraints on the model parameters --- or provide a cross-check of
models to help decide which one is correct should physics effects 
beyond the Standard Model be discovered in the future.

Similar to the consistency check performed within the SM described in
Section~\ref{resinterp.interp.sm.sec}, it is possible to evaluate
for models beyond the Standard Model which sets of parameter values 
are most likely.
Such an analysis has for example been carried out in~\cite{Ellis:2006ix}
for various constrained versions of the MSSM.
The results tend to favor a relatively low MSSM scale, which would
make the discovery of light supersymmetric particles possible
at the LHC or even the Tevatron; an actual determination of
parameter values of the MSSM can however not be performed with the
current data.

\subsection{Potential for Improved Top Quark Mass Measurements}
\label{resinterp.future.sec}
Given the interpretation of the top quark mass measurement outlined in
Section~\ref{resinterp.interp.sec} above, it is clear that a
further improvement of the experimental precision is desirable.
The uncertainty on the current world average discussed in
Section~\ref{resinterp.measurements.sec} is already
dominated by systematic uncertainties.
In the future, the focus will therefore have to shift from an
optimization of the statistical uncertainty to a detailed study of the
systematics listed in Section~\ref{systuncs.sec}.
In this quest, larger event samples will still help in two ways:
First, some of the systematic errors are expected to improve with
increasing sample sizes, and second, large event samples will allow
to select small subsamples which are less prone to systematics than
the rest.

In the following, the evolving situation at the Tevatron experiments
is discussed first.
Second, the prospects for measuring the top quark mass at the LHC are 
described.
Finally, an outline of the potential for top quark mass measurements at a
future linear \epem collider (ILC) is given.

\subsubsection{Future Top Quark Mass Measurements at the Tevatron}
\label{resinterp.future.tev.sec}
With the \runii{}a dataset not even fully analyzed, the Tevatron
experiments have
already surpassed the expectation that a combined top quark mass uncertainty 
of $2$-$3\,\GeV$ would be possible with the full \runii dataset.
This shows how important the newly developed techniques (Matrix
Element method, in situ calibration) are, and how delicate it is to 
extrapolate from the current situation into the future.
An extrapolation from the current combined result is particularly difficult
as the world average combines different types of measurements based on data sets 
corresponding to different integrated luminosities.
A general picture can however still be obtained from an analysis of
how various measurements contribute to the current world average and
of how the individual uncertainties of the most sensitive measurements
will evolve.

From Table~\ref{resultsoverview.table} it is obvious that the \runii
measurements in the \ljets channel carry by far the largest weight
(the two measurements that exploit full event reconstruction have a 
combined weight of $80.6\%$).
The statistical sensitivities of the CDF~\cite{bib-CDFljetsme}
and \dzero~\cite{bib-me} Matrix Element measurements in the \ljets
channel are quite similar; the
difference in uncertainties 
comes from the difference in the size of the data sets analyzed so far
and also
from the fact that the observed \dzero error is slightly larger than 
expected from simulations.
Also, the systematic uncertainties are similar.
Figure~\ref{d0expectation.fig} shows how the uncertainties of the
\dzero measurement will evolve with statistics if the analysis is
unchanged.
All errors are assumed to remain constant except the statistical 
and \jes uncertainties which will be reduced with larger statistics.
At the end of \runii, signal modeling and the \bquark-jet energy scale 
will give rise to the limiting uncertainties.
To estimate the reach of a combination of CDF and \dzero
analyses it is a good approximation to read off the diagram at the sum of integrated
luminosities analyzed:
The signal modeling uncertainties are fully correlated, and even the 
error due to the \bquark-jet energy scale will be partly correlated as
it is a combination of calorimeter response ($e/h$) and hadronization
uncertainties.

\begin{figure}
\begin{center}
\includegraphics[width=0.98\textwidth]{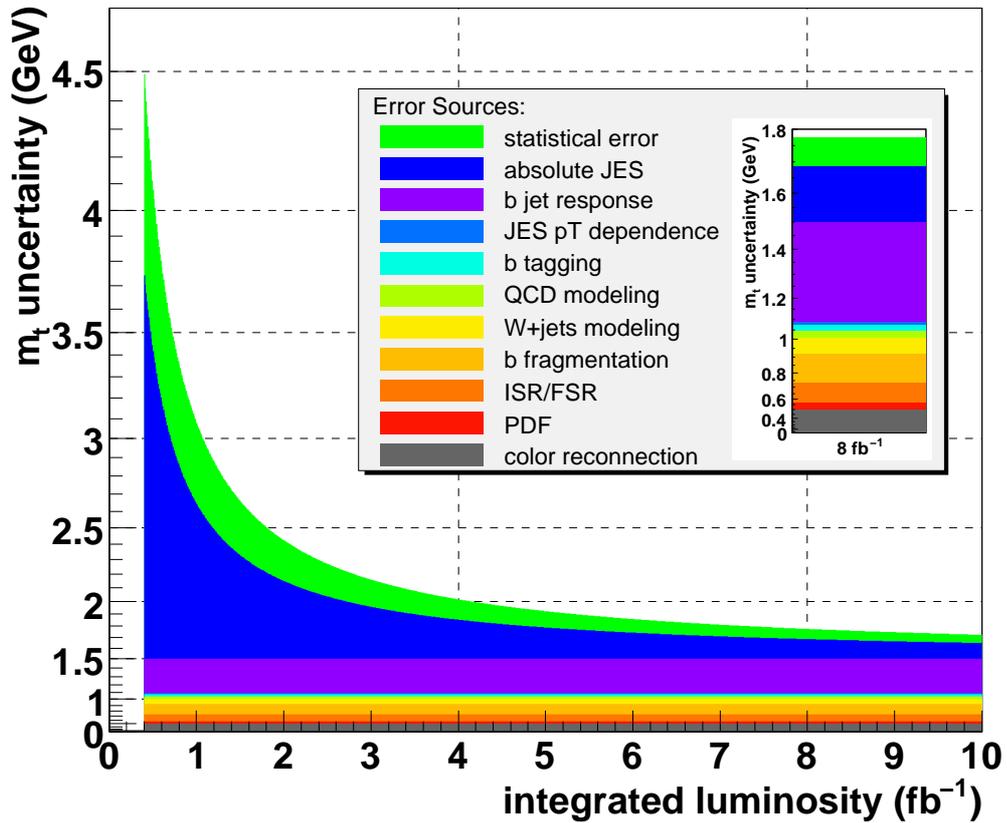}
\caption{\captionfont\label{d0expectation.fig}The composition of the uncertainty
  in the \dzero Matrix Element measurement in the \ljets
  channel~\cite{bib-me}, which is based on an integrated luminosity of
  $0.4\,\ifb$, and the
  expected evolution of the uncertainty with integrated luminosity
  when the measurement technique is kept unchanged.
  An uncertainty of $0.5\,\GeV$ to cover color reconnection effects, 
  which have been
  shown in~\cite{bib-scottwillenbrock} to give rise to an effect 
  of the order of \lambdaqcd, has conservatively
  been added (grey band), while the errors from semileptonic \bquark- or
  \cquark-hadron decays and from trigger efficiencies are
  negligible and have been omitted.
  The color code for the individual error contributions is explained in
  the figure.
  The widths of the colored bands indicate the individual squared
  uncertainties; the vertical axis is therefore non-linear and
  accounts for their quadratic addition.
  The two vertical dashed lines indicate the range of expectations for the
  integrated luminosity delivered to each Tevatron experiment by the
  end of \runii.
  The inset shows the expected uncertainties if the method is 
  applied unchanged to an $8\,\ifb$ dataset.
  The figure is only intended to visualize the relative importance of
  various sources of uncertainty; it cannot provide accurate
  predictions for future measurements of the top quark mass.}
\end{center}
\end{figure}

While it is not justified to make more precise extrapolations into the
future, the above indicates the areas where improvements are most needed.
On the one hand, it will become important to develop 
measurement strategies that are less sensitive to the
\bquark-jet energy scale.
An extreme example is the measurement based on the secondary vertex
decay length~\cite{bib-CDFLxy}.
One could also envisage for example a measurement based on the 
Matrix Element method which minimizes the
combined statistical and systematic error by artificially worsening
the \bquark-jet energy resolution used in the probability computation, 
or which is extended to determine
a \bquark-jet energy response factor.
On the other hand, it will be possible to repeat the studies of the \Z
\pt spectrum described in
Section~\ref{systuncs.physicsmodeling.isrfsr.sec} with 
much larger samples, extend the invariant mass
range, and obtain a more precise extrapolation to \ttbar events.
Dilepton measurements will be less affected by uncertainties on final-state
radiation than those in the other channels.
Currently, the dilepton channel contributes less than $5\%$ to the
world average, which is mainly because no in situ calibration of the
jet energy scale is used here.
It would be very worthwhile to check if this is possible, as it would 
yield information on the $\bquark$-jet energy scale.
Even without in situ calibration, the CDF measurement in the
\alljets channel~\cite{bib-CDFallhadtemplate} contributes to the world
average with a weight of $10\%$.
While events in this channel are well constrained kinematically, it 
remains to be seen if uncertainties due to hadronization and color
reconnection can be kept under control.

In summary, it appears feasible that the total uncertainty on the top
quark mass will be reduced from the current value of $2.1\,\GeV$ to
about $1.5\,\GeV$ by the end of Tevatron \runii.
The actual precision reachable will depend more on further innovative ideas
on the treatment of systematics than on the exact integrated
luminosity delivered.

\subsubsection{Future Top Quark Mass Measurements at the LHC}
\label{resinterp.future.lhc.sec}
The Tevatron measurements of the top quark mass
have only been possible with a very good 
understanding of the detectors.
After the startup of the LHC, it will still take some time 
until the LHC experiments will be able to improve the combined
Tevatron result significantly.
On the other hand, the physics of 
\ttbar production will be well-understood from the Tevatron,
and large samples will be selected, which can be used for the
commissioning and calibration of the detectors -- most notably, to 
determine the absolute jet energy scale and the \bquark-jet
identification efficiency.

Consequently, studies for the LHC experiments focus on two aspects:
Detector commissioning with \ttbar
events~\cite{bib-atlascommissioningwithtop,bib-CMSTDRVolI}
and innovative measurements with reduced top quark mass systematics that are not
feasible with Tevatron
statistics~\cite{Borjanovic:2004ce,bib-CMSTDRVolII,Kharchilava:1999yj}.
Surely, top quark mass measurements will become a field of precision
studies of systematic effects, but
it is difficult to say today exactly what precision will finally be
reached at the LHC for the top quark pole mass, 
as that depends on techniques that are only being
developed now and will be developed further when the data is being taken.

An interesting proposal
has been made in~\cite{bib-diffractivetopphysics} to 
identify double-diffractive \ttbar events at the LHC.
In these events, the \ttbar center-of-mass energy could be
measured from the reconstructed protons, and a measurement of the 
\ttbar cross section as a function of this center-of-mass energy would 
lead to a determination of the top quark mass.
This technique would be
complementary to measurements of the top quark pole mass from the 
properties of the decay products, and would rather be similar
to a \ttbar threshold scan at an \epem collider, 
which is outlined in the following section. 
If the cross section and integrated luminosity are large enough and 
the experimental challenges are solved, this measurement technique might be
a way to overcome the principal theoretical limitations of top
quark pole mass measurements at a hadron collider.

\subsubsection{Future Top Quark Mass Measurements at the ILC}
\label{resinterp.future.ilc.sec}

When a top-antitop cross section measurement is compared to predictions, this 
comparison yields a measurement of the parameter ``top quark mass''
that was used in the calculation of the prediction.
Because of its large width, the top quark does not hadronize, but 
the $\epem\to\ttbar$ production cross section still rises steeply at
the energy corresponding to a 1S resonance, and the top quark mass can
be determined from the energy where this rise is observed.
(At larger energies, the cross section varies much less rapidly with
center-of-mass energy, leading to a larger uncertainty when
interpreted in terms of the top quark mass.)
Suitable definitions of the top quark mass for such calculations
(so-called ``threshold mass'' definitions) are discussed 
in~\cite{bib-topmassdefinitionsforilc}.

At an \epem collider, \ttbar events will have a striking experimental
signature and can be selected with very low 
backgrounds.
This means that the event selection can be kept simple enough so that 
it does not (or only very marginally, thus not introducing large
uncertainties) depend on the exact properties of the top quark decay
products --- which would otherwise result in a measurement of the top quark 
pole mass, as discussed before in Section~\ref{theory.mtopdef.sec}.
In addition to the clean signature by which \ttbar events can be selected,
another prerequisite for this type of measurement is that the initial state
is well-known.
This is the case for an \epem collider (where only initial-state
photon radiation has to be taken into account), but not for a hadron collider
where the partons that initiate the hard interaction are only a part of the 
colliding hadrons.

References~\cite{Martinez:2002st,Brandenburg:2003aa} quote
experimental uncertainties of $20$-$30\,\MeV$ on the 1S top quark
mass.
An uncertainty of $\Delta \alpha_s(\mZ) = 0.001$ corresponds to an
uncertainty of $70\,\MeV$ in the conversion of the threshold mass to
the \msbar scheme~\cite{Hoang:2000yr}, leading to an overall
uncertainty on the \msbar top quark mass of less than $100\,\MeV$.

In addition to the threshold scan, measurements of the top quark pole
mass will of course also be possible at the ILC above the \ttbar
threshold; while the statistical and experimental systematic
uncertainties may be small, the interpretation of such measurements
will be limited by an additional uncertainty of order \lambdaqcd as
discussed before.
Thus complex analysis techniques like the Matrix Element method will no
longer be needed for the measurement of the top quark mass for which
they were originally developed (but this does not invalidate them as a
means of minimizing the statistical uncertainty in any other measurement based
on few events whose kinematic properties are well-understood).

The possibility to determine the top quark mass via a threshold scan
at the ILC corresponds to an order of magnitude improvement of the 
current uncertainty that will to current knowledge not be possible 
via explicit mass reconstruction from the decay products.
The resulting constraints on the Standard Model or models beyond it
will be very precise, as shown in Figure~\ref{heinemeyerweiglein.fig},
and since the parametric uncertainties in the model predictions
resulting from the top quark mass will be much smaller than today, stringent
consistency tests of the models will be possible, as outlined for
example in~\cite{Heinemeyer:2003ud}.

\clearpage
\section{Summary and Conclusions}
\label{conclusion.sec}

A measurement of the top quark mass is interesting per se
because the top quark is by far the heaviest known
elementary fermion.
It is interesting also because the top quark mass is needed as an input
parameter to calculations of electroweak precision variables --
measurements of which can then be used to perform consistency tests of
models or to obtain indirect information on as yet unmeasured
parameters like the Higgs boson mass.

To date, top quarks can only be produced at the Fermilab Tevatron
collider.
The physics of top-antitop pair production and the resulting event
topologies have been outlined.
The reconstruction of the events has been described, it has been
discussed how an accurate calibration of the detectors is
indispensable for the measurement of the top quark mass, and the
calibration procedures applied at the Tevatron experiments have been 
introduced.
The determination of the absolute calorimeter energy scale is
particularly challenging but also of particular importance for the
measurement of the top quark mass.

Since the discovery of the top quark at Tevatron \runi, our 
understanding of \ttbar production at hadron colliders has matured, 
much more integrated luminosity has been accumulated, and
sophisticated techniques to measure the top quark mass have been
developed that have led to an unanticipatedly large reduction of the
uncertainty.
These experimental techniques have been described in detail.
On the one hand, this allows the reader to understand the details of
the measurements of the top quark mass.
On the other hand, this report is also intended as a reference for the 
methods, which can be used
in the future for other measurements of similar experimental nature.

The current world-average value of the top quark mass is already
dominated by systematic uncertainties.
Their various sources have been discussed to identify the current
limitations and to point out possible future improvements.

The interpretation of our current knowledge of the top quark mass
within the Standard Model of particle physics has been presented.
Current results of precision electroweak measurements are in striking
agreement with Standard Model predictions, and thus the top quark mass can
serve as an input to calculations with which constraints on the
mass of the Standard Model Higgs boson can be placed.
A similar interpretation can also be performed within extended models,
and an analysis in the Minimal
Supersymmetric Standard Model has been shown.
Within the Standard Model, a light Higgs boson is clearly favored,
and the data is also consistent with predictions within the 
Minimal Supersymmetric Standard Model.

Finally, the prospects for future measurements at the LHC and a linear 
\epem collider have been outlined.
Many studies have been performed of how to measure the top quark mass 
based on the reconstruction of the final state in \ttbar events at the
LHC.
At the LHC much larger event samples will be available than
at the Tevatron.
The LHC experiments will thus be able to improve the Tevatron results:
Given the large event samples, systematic
uncertainties related to detector calibration can be reduced, and 
other uncertainties may be improved by a careful selection of 
special subsamples of \ttbar events for the mass measurement.
Measurements of the top quark pole mass will however always be limited
by an intrinsic uncertainty of \lambdaqcd.
This uncertainty can be overcome in a threshold scan by comparing the
measured \ttbar production cross section as a function of the \ttbar
center-of-mass energy with predictions calculated as a function of the 
top quark mass (where calculations are not done in terms of the pole
mass).
This measurement technique, applied at a future \epem collider,
will allow for an order of
magnitude improvement of the uncertainty on the top quark mass
measurement.

Measurements of the top 
quark mass have already become so precise that interesting constraints can 
be placed on the Standard Model.
It is foreseeable how the precision will further improve in
the future.
The measurement of the top quark mass will remain an important input
for stringent consistency checks of the Standard Model or of models
describing potential new discoveries.

\clearpage
\section*{Acknowledgements}
\label{acknowledgements.sec}
This work would not have been possible without the
support, advice, help, and (last but not least) friendship of many
people, to whom I feel greatly indebted.
Moreover, it is these people that made work in general very enjoyable.
Being aware that these acknowledgements will necessarily be
incomplete, I would like to explicitly thank Dorothee
Schaile, Otmar Biebel, Arnold Staude, Philipp Schieferdecker, 
Alexander Grohsjean, Petra Haefner, Chris Tully, Arnulf Quadt, Ivor
Fleck, Martin Faessler, and Gaston Gutierrez.
Many thanks to
Albert Engl,
Alexander Brandt,
Alexander Grohsjean,
Alexander Mlynek,
Arnold Staude,
Attila Varga,
Bal\'azs Ujv\'ari,
Benjamin Ruckert,
Britta Tiller,
C\'edric Serfon,
Christian Kummer,
Daniela G\"orisch,
Doris Merkl,
Dorothee Schaile,
Felix Rauscher,
Fritz Vollmer,
Gaby Reiter,
Gernot Krobath,
G\"unter Duckeck,
Hartmut Steffens,
Herta Franz,
Jana Traupel,
Johannes Elmsheuser,
John Kennedy,
J\"org Dubbert,
Madjid Boutemeur,
Marc Rykaczewski,
Marion Lambacher,
Markus Stoye,
Martin Lamprecht,
Matthias Obermaier,
Matthias Schott,
Meta Binder,
Michael Bu{\ss}mann,
Oliver Kortner,
Otmar Biebel,
Patricia M\'endez Lorenzo,
Petra Haefner,
Philippe Calfayan,
Philipp Schieferdecker,
Raimund Str\"ohmer,
Ralf Hertenberger,
Raphael Mameghani,
R\'obert V\'ertesi,
Sofia Chouridou,
Tariq Mahmoud,
Tatjana Unverhau,
Thomas M\"uller,
Thomas Nunnemann,
Tim Christiansen,
and
Wolfram Stiller
of the experimental particle physics group at Munich University,
to the many others who contributed, and to you.

\medskip

\noindent
But finally and above all, what would I do if it were not for you,
Grit, Lukas, and Julia?


\clearpage
\begin{thebibliography}{99}
\bibitem{bib-Zbible}
  The ALEPH, DELPHI, L3, OPAL, and SLD Collaborations, the 
  LEP Electroweak Working Group, SLD Electroweak Group, 
  and SLD Heavy Flavour Group,
  Phys.\ Rept.\  {\bf 427}, 257 (2006).

\bibitem{bib-LEPEWWG}
  The LEP Collaborations ALEPH, DELPHI, L3, OPAL, and the LEP
  Electroweak Working Group,
  {\it A Combination of Preliminary Electroweak Measurements and
  Constraints on the Standard Model},
  [arXiv:hep-ex/0612034], {\tt http://www.cern.ch/LEPEWWG}~.

\bibitem{bib-pdg}  
  W.~M.~Yao {\it et al.}  [Particle Data Group],
  J.\ Phys.\ G {\bf 33} (2006) 1.

\bibitem{bib-topdiscovery}
  F.~Abe {\it et al.}  [CDF Collaboration],
  Phys.\ Rev.\ Lett.\  {\bf 74} (1995) 2626
  [arXiv:hep-ex/9503002];\\
  S.~Abachi {\it et al.}  [D0 Collaboration],
  Phys.\ Rev.\ Lett.\  {\bf 74} (1995) 2632
  [arXiv:hep-ex/9503003].

\bibitem{bib-topmassruni}
  F.~Abe {\it et al.}  [CDF Collaboration],
  Phys.\ Rev.\ Lett.\  {\bf 80} (1998) 2767
  [arXiv:hep-ex/9801014];\\
  F.~Abe {\it et al.}  [CDF Collaboration],
  Phys.\ Rev.\ Lett.\  {\bf 80} (1998) 2779
  [arXiv:hep-ex/9802017];\\
  F.~Abe {\it et al.}  [CDF Collaboration],
  Phys.\ Rev.\ Lett.\  {\bf 82} (1999) 271
  [Erratum-ibid.\  {\bf 82} (1999) 2808]
  [arXiv:hep-ex/9810029];\\
  A.~A.~Affolder {\it et al.}  [CDF Collaboration],
  Phys.\ Rev.\  D {\bf 63} (2001) 032003
  [arXiv:hep-ex/0006028];\\
  S.~Abachi {\it et al.}  [D0 Collaboration],
  Phys.\ Rev.\ Lett.\  {\bf 79} (1997) 1197
  [arXiv:hep-ex/9703008];\\
  B.~Abbott {\it et al.}  [D0 Collaboration],
  Phys.\ Rev.\ Lett.\  {\bf 80} (1998) 2063
  [arXiv:hep-ex/9706014];\\
  B.~Abbott {\it et al.}  [D0 Collaboration],
  Phys.\ Rev.\  D {\bf 58} (1998) 052001
  [arXiv:hep-ex/9801025];\\
  B.~Abbott {\it et al.}  [D0 Collaboration],
  Phys.\ Rev.\  D {\bf 60} (1999) 052001
  [arXiv:hep-ex/9808029];\\
  V.~M.~Abazov {\it et al.}  [D0 Collaboration],
  Nature {\bf 429} (2004) 638
  [arXiv:hep-ex/0406031];\\
  V.~M.~Abazov {\it et al.}  [D0 Collaboration],
  Phys.\ Lett.\  B {\bf 606} (2005) 25
  [arXiv:hep-ex/0410086].

\bibitem{bib-topmassrunii}
  A.~Abulencia {\it et al.}  [CDF Collaboration],
  Phys.\ Rev.\  D {\bf 73} (2006) 032003
  [arXiv:hep-ex/0510048];\\
  A.~Abulencia {\it et al.}  [CDF Collaboration],
  Phys.\ Rev.\ Lett.\  {\bf 96} (2006) 022004
  [arXiv:hep-ex/0510049];\\
  A.~Abulencia {\it et al.}  [CDF Collaboration],
  Phys.\ Rev.\ Lett.\  {\bf 96} (2006) 152002
  [arXiv:hep-ex/0512070];\\
  A.~Abulencia {\it et al.}  [CDF Collaboration],
  Phys.\ Rev.\  D {\bf 73} (2006) 092002
  [arXiv:hep-ex/0512009];\\
  A.~Abulencia {\it et al.}  [CDF Collaboration],
  Phys.\ Rev.\  D {\bf 73} (2006) 112006
  [arXiv:hep-ex/0602008];\\
  A.~Abulencia {\it et al.}  [CDF Collaboration],
  Phys.\ Rev.\  D {\bf 74} (2006) 032009
  [arXiv:hep-ex/0605118];\\
  A.~Abulencia {\it et al.}  [CDF Collaboration],
  arXiv:hep-ex/0612060 (2006), submitted to Phys.\ Rev.\ Lett.;\\
  A.~Abulencia {\it et al.}  [CDF Collaboration],
  arXiv:hep-ex/0612061 (2006), submitted to Phys.\ Rev.\ D;\\
  V.~M.~Abazov {\it et al.}  [D0 Collaboration],
  Phys.\ Rev.\ D {\bf 74} (2006) 092005
  [arXiv:hep-ex/0609053];\\
  V.~M.~Abazov {\it et al.}  [D0 Collaboration],
  arXiv:hep-ex/0609056 (2006), submitted to Phys.\ Rev.\ Lett.;\\
  V.~M.~Abazov {\it et al.}  [D0 Collaboration],
  arXiv:hep-ex/0702018 (2006), submitted to Phys.\ Rev.\ D.

\bibitem{bib-CDFupdates}
Updates of CDF measurements can be found at\\
{\tt http://www-cdf.fnal.gov/physics/new/top/top.html}~.

\bibitem{bib-Dzeroupdates}
Updates of \dzero measurements can be found at\\
{\tt http://www-d0.fnal.gov/Run2Physics/top/top\_public\_web\_pages/top\_public.html}~.

\bibitem{bib-TEVEWWG}  
  E.~Brubaker {\it et al.}  [Tevatron Electroweak Working Group],
  arXiv:hep-ex/0608032,\\
  {\tt http://tevewwg.fnal.gov}~.

\bibitem{bib-CDFruniiTDR}  
  R.~Blair {\it et al.}  [CDF Collaboration],
  FERMILAB-PUB-96-390-E.

\bibitem{bib-ttbarxsruni}
  A.~A.~Affolder {\it et al.}  [CDF Collaboration],
  Phys.\ Rev.\ D {\bf 64} (2001) 032002
  [Erratum-ibid.\ D {\bf 67} (2003) 119901]
  [arXiv:hep-ex/0101036];\\
  B.~Abbott {\it et al.}  [D0 Collaboration],
  Phys.\ Rev.\ Lett.\  {\bf 83} (1999) 1908
  [arXiv:hep-ex/9901023];\\
  B.~Abbott {\it et al.}  [D0 Collaboration],
  Phys.\ Rev.\ D {\bf 60} (1999) 012001
  [arXiv:hep-ex/9808034];\\
  V.~M.~Abazov {\it et al.}  [D0 Collaboration],
  Phys.\ Rev.\ D {\bf 67} (2003) 012004
  [arXiv:hep-ex/0205019].

\bibitem{bib-ttbarxsrunii}
  A recent combination of CDF \runii results can be found in:\\
  The CDF Collaboration,
  {\it Combination of CDF top quark pair production cross section
  measurements with up to $760\,\ipb$},
  CDF note 8148 (2006);\\
  a combination of \dzero \runii results is given in:\\
  The \dzero Collaboration,
  {\it Combined \ttbar Production Cross Section at
  $\sqrt{s}=1.96\,\TeV$ in the Lepton+Jets and Dilepton Final States
  using Event Topology},
  \dzero note 4906 (2005);\\
  for individual measurements and updates 
  see~\cite{bib-CDFupdates,bib-Dzeroupdates}.

\bibitem{bib-xsprediction}  
  N.~Kidonakis and R.~Vogt,
  Phys.\ Rev.\ D {\bf 68} (2003) 114014
  [arXiv:hep-ph/0308222];\\
  M.~Cacciari, S.~Frixione, M.~L.~Mangano, P.~Nason and G.~Ridolfi,
  JHEP {\bf 0404} (2004) 068
  [arXiv:hep-ph/0303085].

\bibitem{bib-searchesfornonsmeffects-runi}
  A.~A.~Affolder {\it et al.}  [CDF Collaboration],
  Phys.\ Rev.\ Lett.\  {\bf 85} (2000) 2062
  [arXiv:hep-ex/0003005];\\
  V.~M.~Abazov {\it et al.}  [D0 Collaboration],
  Phys.\ Rev.\ Lett.\  {\bf 92} (2004) 221801
  [arXiv:hep-ex/0307079];\\
  A.~A.~Affolder {\it et al.}  [CDF Collaboration],
  Phys.\ Rev.\ Lett.\  {\bf 87} (2001) 102001;\\
  B.~Abbott {\it et al.}  [D0 Collaboration],
  Phys.\ Rev.\ D {\bf 58} (1998) 052001
  [arXiv:hep-ex/9801025];\\
  S.~Abachi {\it et al.}  [D0 Collaboration],
  Phys.\ Rev.\ Lett.\  {\bf 79} (1997) 1197
  [arXiv:hep-ex/9703008].

\bibitem{bib-Whelicity}
  Measurements of the helicity of \W bosons in top quark decay have
  been published in\\
  A.~Abulencia {\it et al.}  [CDF Collaboration],
  arXiv:hep-ex/0608062 (2006), submitted to Phys.\ Rev.\ Lett.;\\
  A.~Abulencia {\it et al.}  [CDF Collaboration],
  Phys.\ Rev.\ D {\bf 73} (2006) 111103
  [arXiv:hep-ex/0511023];\\
  D.~Acosta {\it et al.}  [CDF Collaboration],
  Phys.\ Rev.\ D {\bf 71} (2005) 031101
  [Erratum-ibid.\ D {\bf 71} (2005) 059901]
  [arXiv:hep-ex/0411070];\\
  V.~M.~Abazov {\it et al.}  [D0 Collaboration],
  Phys.\ Rev.\  D {\bf 75} (2007) 031102
  [arXiv:hep-ex/0609045];\\
  V.~M.~Abazov {\it et al.}  [D0 Collaboration],
  Phys.\ Rev.\ D {\bf 72} (2005) 011104
  [arXiv:hep-ex/0505031];\\
  for updates 
  see~\cite{bib-CDFupdates,bib-Dzeroupdates}.

\bibitem{bib-othersearchesfornonsmeffects}
  D.~Acosta {\it et al.}  [CDF Collaboration],
  Phys.\ Rev.\ Lett.\  {\bf 95} (2005) 022001
  [arXiv:hep-ex/0412042];\\
  A.~Abulencia {\it et al.}  [CDF Collaboration],
  Phys.\ Lett.\ B {\bf 639} (2006) 172
  [arXiv:hep-ex/0510063];\\
  D.~Acosta {\it et al.}  [CDF Collaboration],
  Phys.\ Rev.\ Lett.\  {\bf 95} (2005) 102002
  [arXiv:hep-ex/0505091];\\
  A.~Abulencia {\it et al.}  [CDF Collaboration],
  Phys.\ Rev.\ Lett.\  {\bf 96} (2006) 042003
  [arXiv:hep-ex/0510065];\\
  V.~M.~Abazov {\it et al.}  [D0 Collaboration],
  arXiv:hep-ex/0608044 (2006), submitted to Phys.\ Rev.\ Lett.;\\
  V.~M.~Abazov {\it et al.}  [D0 Collaboration],
  Phys.\ Lett.\ B {\bf 639} (2006) 616
  [arXiv:hep-ex/0603002];\\
  for updates and preliminary results of other searches
  see~\cite{bib-CDFupdates,bib-Dzeroupdates}.

\bibitem{bib-arnulf}
  A.~Quadt,
  Eur.\ Phys.\ J.\ C {\bf 48} (2006) 835.

\bibitem{bib-scottwillenbrock}  
  M.~C.~Smith and S.~S.~Willenbrock,
  Phys.\ Rev.\ Lett.\  {\bf 79} (1997) 3825
  [arXiv:hep-ph/9612329].

\bibitem{bib-pythia}  
  T.~Sjostrand, S.~Mrenna and P.~Skands,
  JHEP {\bf 0605} (2006) 026
  [arXiv:hep-ph/0603175].

\bibitem{bib-alpgen}  
  M.~L.~Mangano, M.~Moretti, F.~Piccinini, R.~Pittau and A.~D.~Polosa,
  JHEP {\bf 0307} (2003) 001
  [arXiv:hep-ph/0206293].

\bibitem{bib-topmassdefinitionsforilc}
  A.~H.~Hoang {\it et al.},
  Eur.\ Phys.\ J.\ directC {\bf 2} (2000) 1
  [arXiv:hep-ph/0001286] and references therein; see also\\
  A.~H.~Hoang, A.~V.~Manohar, I.~W.~Stewart and T.~Teubner,
  Phys.\ Rev.\ D {\bf 65} (2002) 014014
  [arXiv:hep-ph/0107144] and\\
  O.~I.~Yakovlev and S.~Groote,
  Phys.\ Rev.\ D {\bf 63} (2001) 074012
  [arXiv:hep-ph/0008156].

\bibitem{bib-CTEQ5L}  
  H.~L.~Lai {\it et al.}  [CTEQ Collaboration],
  Eur.\ Phys.\ J.\ C {\bf 12} (2000) 375
  [arXiv:hep-ph/9903282].

\bibitem{bib-singletop}  
  V.~M.~Abazov {\it et al.}  [D0 Collaboration],
  arXiv:hep-ex/0612052 (2006), submitted to Phys.\ Rev.\ Lett.

\bibitem{bib-ttbarxsatLHC}  
  R.~Bonciani, S.~Catani, M.~L.~Mangano and P.~Nason,
  Nucl.\ Phys.\ B {\bf 529} (1998) 424
  [arXiv:hep-ph/9801375].

\bibitem{bib-singletoptevatronkidonakis}  
  N.~Kidonakis,
  Phys.\ Rev.\ D {\bf 74} (2006) 114012
  [arXiv:hep-ph/0609287].

\bibitem{bib-singletoplhckidonakis}  
  N.~Kidonakis,
  arXiv:hep-ph/0701080.

\bibitem{bib-CTEQ6Mvar}  
  J.~Pumplin, D.~R.~Stump, J.~Huston, H.~L.~Lai, P.~Nadolsky and W.~K.~Tung,
  JHEP {\bf 0207} (2002) 012
  [arXiv:hep-ph/0201195].  

\bibitem{bib-MRST}  
  A.~D.~Martin, R.~G.~Roberts, W.~J.~Stirling and R.~S.~Thorne,
  Eur.\ Phys.\ J.\ C {\bf 28} (2003) 455
  [arXiv:hep-ph/0211080];\\
  A.~D.~Martin, R.~G.~Roberts, W.~J.~Stirling and R.~S.~Thorne,
  Eur.\ Phys.\ J.\ C {\bf 35} (2004) 325
  [arXiv:hep-ph/0308087].

\bibitem{bib-singletopatnlo}  
  Z.~Sullivan,
  Phys.\ Rev.\ D {\bf 70} (2004) 114012
  [arXiv:hep-ph/0408049];\\
  J.~Campbell, R.~K.~Ellis and F.~Tramontano,
  Phys.\ Rev.\ D {\bf 70} (2004) 094012
  [arXiv:hep-ph/0408158];\\
  J.~Campbell and F.~Tramontano,
  Nucl.\ Phys.\ B {\bf 726} (2005) 109
  [arXiv:hep-ph/0506289].

\bibitem{bib-LHCtopreport}  
  M.~Beneke {\it et al.},
  arXiv:hep-ph/0003033.

\bibitem{bib-annheinson}
  The figure has been provided by Ann Heinson for the \dzero collaboration.

\bibitem{bib-CDFdet}
  D.~Acosta {\it et al.}  [CDF Collaboration],
  Phys.\ Rev.\ D {\bf 71} (2005) 032001
  [arXiv:hep-ex/0412071].

\bibitem{bib-Dzerodet}
  V.~M.~Abazov {\it et al.}  [D0 Collaboration],
  Nucl.\ Instrum.\ Meth.\ A {\bf 565} (2006) 463
  [arXiv:physics/0507191];\\
  V.~M.~Abazov {\it et al.},
  Nucl.\ Instrum.\ Meth.\ A {\bf 552} (2005) 372
  [arXiv:physics/0503151];\\
  S.~Abachi {\it et al.}  [D0 Collaboration],
  Nucl.\ Instrum.\ Meth.\ A {\bf 338} (1994) 185.

\bibitem{bib-CDFdetectordrawing}
  {\tt http://www-cdf.fnal.gov/upgrades/tdr/doc/cdfelev.ps}, to be
  found in\\
  {\tt http://www-cdf.fnal.gov/upgrades/upgrades.html}~.

\bibitem{bib-Dzerodetectordrawing}
  {\tt http://www-d0.fnal.gov/Run2Physics/displays/presentations/gallery/\\
  patwa\_dzero\_2d\_view.eps},
  to be found in\\
  {\tt http://www-d0.fnal.gov/Run2Physics/displays/presentations/}~.

\bibitem{Abulencia:2005uq}
A.~Abulencia {\it et al.}  [CDF Collaboration],
Phys.\ Rev.\ Lett.\  {\bf 96}, 152002 (2006)
[arXiv:hep-ex/0512070].

\bibitem{Abazov:2005yt}
V.~M.~Abazov {\it et al.}  [D0 Collaboration],
Phys.\ Lett.\ B {\bf 626} (2005) 55
[arXiv:hep-ex/0505082].

\bibitem{Abulencia:2005aj}
A.~Abulencia {\it et al.}  [CDF Collaboration],
Phys.\ Rev.\ D {\bf 73}, 032003 (2006)
[arXiv:hep-ex/0510048].

\bibitem{bib-me}  
  V.~M.~Abazov {\it et al.}  [D0 Collaboration],
  Phys.\ Rev.\ D {\bf 74} (2006) 092005
  [arXiv:hep-ex/0609053].

\bibitem{bib-CDFallhadxs}
  The CDF Collaboration,
  {\it Measurement of the \ttbar production cross section in the
  all-hadronic channel ($1.02\,\ifb$)},
  CDF note 8402 (2006).

\bibitem{bib-D0allhadxs}
  V.~M.~Abazov  [D0 Collaboration],
  arXiv:hep-ex/0612040 (2006), submitted to Phys.\ Rev.\ D.

\bibitem{Acosta:2005zd}
D.~Acosta {\it et al.}  [CDF Collaboration],
Phys.\ Rev.\ D {\bf 72} (2005) 032002
[arXiv:hep-ex/0506001].

\bibitem{bib-CDFalljetsxspaper}  
  A.~Abulencia {\it et al.}  [CDF Collaboration],
  Phys.\ Rev.\ D {\bf 74} (2006) 072005
  [arXiv:hep-ex/0607095].

\bibitem{bib-D0ljetsbtagxspaper}  
  V.~M.~Abazov {\it et al.}  [D0 Collaboration],
  Phys.\ Rev.\  D {\bf 74} (2006) 112004
  [arXiv:hep-ex/0611002].

\bibitem{bib-CDFljetstopoxspaper}  
  D.~Acosta {\it et al.}  [CDF Collaboration],
  Phys.\ Rev.\ D {\bf 72} (2005) 052003
  [arXiv:hep-ex/0504053].

\bibitem{bib-CDFljetsbtagxspaper}  
  D.~Acosta {\it et al.}  [CDF Collaboration],
  Phys.\ Rev.\ D {\bf 71} (2005) 052003
  [arXiv:hep-ex/0410041].

\bibitem{Abe:1991ui}
  F.~Abe {\it et al.}  [CDF Collaboration],
  Phys.\ Rev.\ D {\bf 45} (1992) 1448.

\bibitem{bib-D0conejetalgorithm}
  \dzero uses the iterative, seed-based cone algorithm including
  midpoints, as described on page 47 in G.~C.~Blazey {\it et al.},
  Proceedings of the Workshop {\it QCD and Weak Boson Physics in
  Run\,II}, edited by U.~Baur, R.~K.~Ellis, and D.~Zeppenfeld,
  FERMILAB-PUB-00-297 (2000).

\bibitem{bib-herwig}
  G.~Marchesini and B.~R.~Webber,
  Nucl.\ Phys.\ B {\bf 310} (1988) 461;\\
  G.~Corcella {\it et al.},
  JHEP {\bf 0101} (2001) 010
  [arXiv:hep-ph/0011363].

\bibitem{bib-evtgen}  
  D.~J.~Lange,
  Nucl.\ Instrum.\ Meth.\ A {\bf 462} (2001) 152.

\bibitem{bib-QQ}
  P.~Avery, K.~Read, and G.~Trahern (1985), CLEO Report CSN-212 (unpublished).

\bibitem{bib-tauola}  
  S.~Jadach, Z.~Wa\hspace{-0.4ex},\hspace{-0.1ex}s, R.~Decker and J.~H.~Kuhn,
  Comput.\ Phys.\ Commun.\  {\bf 76} (1993) 361.

\bibitem{bib-geant}
  R.~Brun and F.~Carminati, CERN Programming Library Long Writeup 
  {\bf W5013} (1993).

\bibitem{bib-generalMC}  
  T.~Sj\"ostrand,
  {\it Monte Carlo generators},
  arXiv:hep-ph/0611247.

\bibitem{bib-topquarkMC}  
  S.~R.~Slabospitsky,
  PoS {\bf TOP2006} (2006) 019
  [arXiv:hep-ph/0603124].

\bibitem{bib-matrixmethod}  
  B.~Abbott {\it et al.}  [D0 Collaboration],
  Phys.\ Rev.\ D {\bf 61} (2000) 072001
  [arXiv:hep-ex/9906025].

\bibitem{bib-CDFljetsDLMmass}  
  A.~Abulencia {\it et al.}  [CDF Collaboration],
  Phys.\ Rev.\ D {\bf 73} (2006) 092002
  [arXiv:hep-ex/0512009].

\bibitem{bib-CDFallhadID}
  The CDF Collaboration,
  {\it Measurement of the top quark mass in the all hadronic channel
    using the Ideogram method},
  CDF note 8233 (2006).

\bibitem{bib-cdfwmass}
  The CDF Collaboration,
  {\it First Measurement of the \W Boson Mass with CDF in \runii},
  CDF note 8665 (2007).

\bibitem{bib-d0zmumu}
  The \dzero Collaboration,
  {\it Measurement of the Cross Section for Inclusive \Z Production in
    Di-Muon Final States at $\sqrt{s}=1.96~\TeV$}, 
  \dzero note 4573 (2004).

\bibitem{Abulencia:2005ix}
  A.~Abulencia {\it et al.}  [CDF Collaboration],
  arXiv:hep-ex/0508029 (2005), submitted to Phys.\ Rev.\ D.

\bibitem{bib-Dzerotagandprobe}  
  V.~M.~Abazov {\it et al.}  [D0 Collaboration],
  Phys.\ Lett.\ B {\bf 626} (2005) 45
  [arXiv:hep-ex/0504043].

\bibitem{bib-CDFruniijes}  
  A.~Bhatti {\it et al.},
  Nucl.\ Instrum.\ Meth.\ A {\bf 566} (2006) 375
  [arXiv:hep-ex/0510047].  

\bibitem{bib-D0runijes}  
  B.~Abbott {\it et al.}  [D0 Collaboration],
  Nucl.\ Instrum.\ Meth.\ A {\bf 424} (1999) 352
  [arXiv:hep-ex/9805009].

\bibitem{bib-nature}  
  V.~M.~Abazov {\it et al.}  [D0 Collaboration],
  Nature {\bf 429} (2004) 638
  [arXiv:hep-ex/0406031].

\bibitem{bib-CDFljetsme}
  The CDF Collaboration,
  {\it Measurement of the Top Quark Mass using the Matrix Element
    Analysis Technique in the Lepton+Jets Channel with In-Situ
    $\W\to jj$ Calibration},
  CDF note 8375 (2006).


\bibitem{bib-CDFmtop_ljetstemplate_prel}
  The CDF Collaboration,
  {\it Measurement of the Top Quark Mass using the Template Method in the
    Lepton plus Jets Channel With In Situ $\W\to jj$ Calibration at CDF-II},
  CDF note 8125 (2006).

\bibitem{bib-CDFLxy}  
  A.~Abulencia {\it et al.}  [CDF Collaboration],
  arXiv:hep-ex/0612061 (2006), submitted to Phys.\ Rev.\ D.

\bibitem{bib-CDFallhadtemplate}
  The CDF Collaboration,
  {\it Measurement of the top mass in the all-hadronic channel using the
    Template Method with $1.02\,\ifb$},
  CDF note 8420 (2006).

\bibitem{bib-runineutrinoweighting}
  B.~Abbott {\it et al.}  [D0 Collaboration],
  Phys.\ Rev.\ D {\bf 60} (1999) 052001
  [arXiv:hep-ex/9808029];\\
  F.~Abe {\it et al.}  [CDF Collaboration],
  Phys.\ Rev.\ Lett.\  {\bf 82} (1999) 271
  [Erratum-ibid.\  {\bf 82} (1999) 2808]
  [arXiv:hep-ex/9810029].

\bibitem{bib-CDFcombineddileptontemplatepaper}  
  A.~Abulencia {\it et al.}  [CDF Collaboration],
  Phys.\ Rev.\ D {\bf 73} (2006) 112006
  [arXiv:hep-ex/0602008].

\bibitem{bib-Dzerodileptonneutrinoweighting}
  V.~M.~Abazov {\it et~al.}  [D0 Collaboration],
  arXiv:hep-ex/0609056 (2006), submitted to Phys.\ Rev.\ Lett.;\\
  The \dzero Collaboration,
  {\it Measurement of \mtop in \emu Events with Neutrino Weighting in
    \runii at \dzero},
  \dzero note 5171 (2006).

\bibitem{bib-Dzerodileptonmatrixweighting}
  V.~M.~Abazov {\it et~al.}  [D0 Collaboration],
  arXiv:hep-ex/0609056 (2006), submitted to Phys.\ Rev.\ Lett.;\\
  The \dzero Collaboration,
  {\it Measurement of the Top Quark Mass in the $\emu$ Channel Using the
    Matrix Weighting Method at \dzero},
  \dzero note 5200 (2006).

\bibitem{bib-CDFdileptonMEpaper}  
  A.~Abulencia {\it et al.}  [CDF Collaboration],
  Phys.\ Rev.\ D {\bf 74} (2006) 032009
  [arXiv:hep-ex/0605118];\\
  A.~Abulencia {\it et al.}  [CDF Collaboration],
  Phys.\ Rev.\ Lett.\  {\bf 96} (2006) 152002
  [arXiv:hep-ex/0512070].

\bibitem{bib-CDFdileptonme}  
  A.~Abulencia {\it et al.}  [CDF Collaboration],
  arXiv:hep-ex/0612060 (2006), submitted to Phys.\ Rev.\ Lett.

\bibitem{bib-CDFdileptonmewithbtagging}
  The CDF Collaboration,
  {\it Measurement of the Top Quark Mass using a Matrix Element Method
    in a $\bquark$-Tagged Dilepton Sample},
  CDF note 8401 (2006).

\bibitem{bib-CDFdileptonDLMmass}
  Ryo Tsuchiya,
  {\it Measurement of the Top Quark Mass by Dynamical Likelihood
  Method using the Dilepton events with the Collider Detector at
  Fermilab},
  PhD thesis, Waseda University, Tokyo (2006).

\bibitem{bib-mahlonparke}  
  G.~Mahlon and S.~J.~Parke,
  Phys.\ Lett.\ B {\bf 411} (1997) 173
  [arXiv:hep-ph/9706304].

\bibitem{Lepage:1977sw}
  G.~P.~Lepage,
  J.\ Comput.\ Phys.\  {\bf 27} (1978) 192.
 
\bibitem{Lepage:1980dq}
  G.~P.~Lepage,
  Cornell preprint CLNS:80-447 (1980).

\bibitem{bib-vecbos}  
  F.~A.~Berends, H.~Kuijf, B.~Tausk and W.~T.~Giele,
  Nucl.\ Phys.\ B {\bf 357} (1991) 32.

\bibitem{Abreu:1997ic}
  P.~Abreu {\it et al.}  [DELPHI Collaboration],
  Eur.\ Phys.\ J.\ C {\bf 2} (1998) 581.

\bibitem{bib-ID}  
  V.~M.~Abazov {\it et al.}  [D0 Collaboration],
  arXiv:hep-ex/0702018 (2006), submitted to Phys.\ Rev.\ D.
\bibitem{bib-minuit}
  F.~James,
  {\it MINUIT: Function Minimization and Error Analysis, Reference Manual},
  CERN Program Library Long Writeup D506.

\bibitem{bib-madgraph}  
  F.~Maltoni and T.~Stelzer,
  JHEP {\bf 0302} (2003) 027
  [arXiv:hep-ph/0208156].

\bibitem{bib-philipp}
  P.~Schieferdecker,
  {\it Measurement of the Top Quark Mass at \dzero \runii with the
  Matrix Element Method in the Lepton+Jets Final State},
  FERMILAB-THESIS 2005-46.

\bibitem{bib-MCatNLO}  
  S.~Frixione and B.~R.~Webber,
  ``The MC@NLO 3.2 event generator,''
  arXiv:hep-ph/0601192.

\bibitem{bib-CDFdileptontemplateprel}
  The CDF Collaboration,
  {\it Measurement of the top mass using full kinematic template method
    in dilepton channel at CDF with $1\,\ifb$},
  CDF note 8554 (2006).

\bibitem{Borjanovic:2004ce}
  I.~Borjanovic {\it et al.},
  Eur.\ Phys.\ J.\ C {\bf 39S2} (2005) 63
  [arXiv:hep-ex/0403021].

\bibitem{bib-meprivatecommunication}
  V.~M.~Abazov {\it et al.}  [D0 Collaboration],
  Phys.\ Rev.\ D {\bf 74} (2006) 092005
  [arXiv:hep-ex/0609053]; private communication with the authors of
  the analysis.

\bibitem{Cacciari:2002re-Corcella:2001hz}
  M.~Cacciari, G.~Corcella and A.~D.~Mitov,
  JHEP {\bf 0212} (2002) 015
  [arXiv:hep-ph/0209204];\\
  G.~Corcella and A.~D.~Mitov,
  Nucl.\ Phys.\ B {\bf 623} (2002) 247
  [arXiv:hep-ph/0110319].

\bibitem{bib-skandswicke}
  P.~Skands and D.~Wicke,
  work to be published.

\bibitem{bib-CMSTDRVolII}
  The CMS Collaboration,
  {\it CMS Physics Technical Design Report, Volume II},
  CERN/LHCC 2006-021 (2006).

\bibitem{Kharchilava:1999yj}
  A.~Kharchilava,
  Phys.\ Lett.\ B {\bf 476} (2000) 73
  [arXiv:hep-ph/9912320].

\bibitem{Affolder:2000vy}
  A.~A.~Affolder {\it et al.}  [CDF Collaboration],
  Phys.\ Rev.\ D {\bf 63} (2001) 032003
  [arXiv:hep-ex/0006028].

\bibitem{CDFmultivariatetemplate}
  CDF Collaboration,
  {\it Top Mass Measurement in the Lepton+Jets Channel using a
    Multivariate Template Method},
  CDF note 7102 (2004).

\bibitem{Abe:1998bf}
  F.~Abe {\it et al.}  [CDF Collaboration],
  Phys.\ Rev.\ Lett.\  {\bf 82} (1999) 271
  [Erratum-ibid.\  {\bf 82} (1999) 2808]
  [arXiv:hep-ex/9810029].

\bibitem{Abbott:1998dn}
  B.~Abbott {\it et al.}  [D0 Collaboration],
  Phys.\ Rev.\ D {\bf 60} (1999) 052001
  [arXiv:hep-ex/9808029].

\bibitem{Abe:1997rh}
  F.~Abe {\it et al.}  [CDF Collaboration],
  Phys.\ Rev.\ Lett.\  {\bf 79} (1997) 1992.

\bibitem{bib-CDFmetjetstemplate}
  The CDF Collaboration,
  {\it Measurement of the top quark mass in the missing \et + jets
    channel},
  CDF note 8573 (2006).

\bibitem{bib-BLUE1}  
  L.~Lyons, D.~Gibaut and P.~Clifford,
  Nucl.\ Instrum.\ Meth.\ A {\bf 270} (1988) 110.

\bibitem{bib-BLUE2}  
  A.~Valassi,
  Nucl.\ Instrum.\ Meth.\ A {\bf 500} (2003) 391.

\bibitem{bib-cdfxscombination}
  The CDF Collaboration,
  {\it Combination of CDF top quark pair production cross section
  measurements with up to $760\,\ipb$},
  CDF note 8148 (2006).

\bibitem{Cacciari:2003fi}
  M.~Cacciari, S.~Frixione, M.~L.~Mangano, P.~Nason and G.~Ridolfi,
  JHEP {\bf 0404} (2004) 068
  [arXiv:hep-ph/0303085].

\bibitem{Kidonakis:2003qe}
  N.~Kidonakis and R.~Vogt,
  Phys.\ Rev.\ D {\bf 68} (2003) 114014
  [arXiv:hep-ph/0308222].

\bibitem{bib-dzeroxscombination}
  The \dzero Collaboration,
  {\it Combined \ttbar Production Cross Section at
  $\sqrt{s}=1.96\,\TeV$ in the Lepton+Jets and Dilepton Final States
  using Event Topology},
  \dzero note 4906 (2005).

\bibitem{bib-mssm}
  H.~P.~Nilles,
  Phys.\ Rept.\  {\bf 110} (1984) 1;\\
  H.~E.~Haber and G.~L.~Kane,
  Phys.\ Rept.\  {\bf 117} (1985) 75.

\bibitem{bib-SMvsMSSM}
  S.~Heinemeyer, W.~Hollik, D.~Stockinger, A.~M.~Weber and G.~Weiglein,
  JHEP {\bf 0608} (2006) 052
  [arXiv:hep-ph/0604147];\\
  S.~Heinemeyer, W.~Hollik and G.~Weiglein,
  Phys.\ Rept.\  {\bf 425} (2006) 265
  [arXiv:hep-ph/0412214];\\
  the plot is taken from 
  {\tt http://quark.phy.bnl.gov/\~{}heinemey/uni/plots/} .

\bibitem{Djouadi:1996pa}
  A.~Djouadi, P.~Gambino, S.~Heinemeyer, W.~Hollik, C.~J\"unger and G.~Weiglein,
  Phys.\ Rev.\ Lett.\  {\bf 78} (1997) 3626
  [arXiv:hep-ph/9612363].

\bibitem{Ellis:2006ix}
  J.~R.~Ellis, S.~Heinemeyer, K.~A.~Olive and G.~Weiglein,
  JHEP {\bf 0605} (2006) 005
  [arXiv:hep-ph/0602220] 
  and references therein.

\bibitem{bib-atlascommissioningwithtop}
  S.~Bentvelsen and M.~Cobal,
  {\it Top studies for the Atlas detector commissioning},
  ATLAS note ATL-PHYS-PUB-2005-024 (2005).

\bibitem{bib-CMSTDRVolI}
  The CMS Collaboration,
  {\it CMS Physics Technical Design Report, Volume I},
  CERN/LHCC 2006-001 (2006); a study on jet energy calibration using
  reconstructed hadronic \W decays in \ttbar events is presented
  in Section 11.6.5 of this document.

\bibitem{bib-diffractivetopphysics}  
  M.~Boonekamp, J.~Cammin, R.~Peschanski and C.~Royon,
  {\it Threshold scans in central diffraction at the LHC},
  arXiv:hep-ph/0504199 (2005), submitted to Nucl.\ Phys.\ B.

\bibitem{Martinez:2002st}
  M.~Martinez and R.~Miquel,
  Eur.\ Phys.\ J.\ C {\bf 27} (2003) 49
  [arXiv:hep-ph/0207315].

\bibitem{Brandenburg:2003aa}
  A.~Brandenburg,
  arXiv:hep-ph/0308094.

\bibitem{Hoang:2000yr}
  A.~H.~Hoang {\it et al.},
  Eur.\ Phys.\ J.\ directC {\bf 2} (2000) 1
  [arXiv:hep-ph/0001286].

\bibitem{Heinemeyer:2003ud}
  S.~Heinemeyer, S.~Kraml, W.~Porod and G.~Weiglein,
  JHEP {\bf 0309} (2003) 075
  [arXiv:hep-ph/0306181].

\end{thebibliography}
\end{document}